\documentclass[manuscript, screen]{acmart}

\AtBeginDocument{%
  }

\usepackage{graphicx}%
\usepackage{multirow}%

\usepackage{amsmath,amssymb,amsfonts}%
\usepackage{amsthm}%
\usepackage{mathrsfs}%
\usepackage[title]{appendix}%
\usepackage{xcolor}%
\usepackage{textcomp}%
\usepackage{manyfoot}%
\usepackage{booktabs}%
\usepackage{algorithm}%
\usepackage{algorithmicx}%
\usepackage{algpseudocode}%
\usepackage{listings}%
\newcommand\rev[1]{\textcolor{black}{#1}}

\usepackage{enumitem}
\usepackage{tabularx}
\usepackage{forest}
\usepackage{longtable}
\usepackage{pdflscape}
\usepackage{wrapfig}
\usepackage{url}

\begin{document}

\title[Neural Gaze]{Guidelines for Gaze-based Neural Preliminary Diagnosis}

\author{Mayar Elfares}
\email{mayar.elfares@vis.uni-stuttgart.de}
\orcid{1234-5678-9012}
\affiliation{%
  \institution{Depatment of Computer Science, University of Stuttgart }
  \country{Germany}
}

\author{Salma Younis}
\authornote{Work done while interning at the University of Stuttgart.}
\affiliation{%
  \institution{Department of Behavioral and Cognitive Neuroscience, University of Maryland}
  \country{USA}
}
\email{syounis@terpmail.umd.edu}

\author{Pascal Reisert}
\affiliation{%
  \institution{Depatment of Computer Science, University of Stuttgart }
  \country{Germany}
}
\email{pascal.reisert@sec.uni-stuttgart.de}

\author{Ralf Küsters}
\affiliation{%
  \institution{Depatment of Computer Science, University of Stuttgart }
  \country{Germany}
}
\email{ralf.kuesters@sec.uni-stuttgart.de}

\author{Tobias Renner}
\affiliation{%
  \institution{Department of Child and Adolescent Psychiatry, University Hospital Tübingen}
  \country{Germany}}
\email{tobias.renner@med.uni-tuebingen.de}

\author{Andreas Bulling}
\affiliation{%
  \institution{Depatment of Computer Science, University of Stuttgart }
  \country{Germany}
}
\email{andreas.bulling@vis.uni-stuttgart.de}

\renewcommand{\shortauthors}{Elfares et al.}

\begin{abstract}
  \rev{Neural} disorders refer to \rev{any condition affecting the nervous system and} that influence how individuals perceive and interact with the world. Traditional \rev{neural} diagnoses rely on cumbersome, time-consuming, or subjective methods, such as clinical interviews, behavioural observations, or medical imaging. Eye tracking is an attractive alternative because analysing eye movements, such as fixations and saccades, can provide more objective insights into brain function and cognitive processing \rev{by capturing non-verbal and unconscious responses}. Despite its potential, existing gaze-based studies presented seemingly contradictory findings. They are dispersed across diverse fields, requiring further research to standardise protocols and expand their application, particularly \rev{as a preliminary indicator of neural processes for } differential diagnosis. Therefore, this paper outlines the main agreed-upon findings and provides a systematisation of knowledge and key guidelines towards advancing gaze-based \rev{neural} preliminary diagnosis.
\end{abstract}



\keywords{Gaze, Eye Tracking, Neural Disorders, Diagnosis}



\begin{teaserfigure}
    \includegraphics[width=\textwidth]{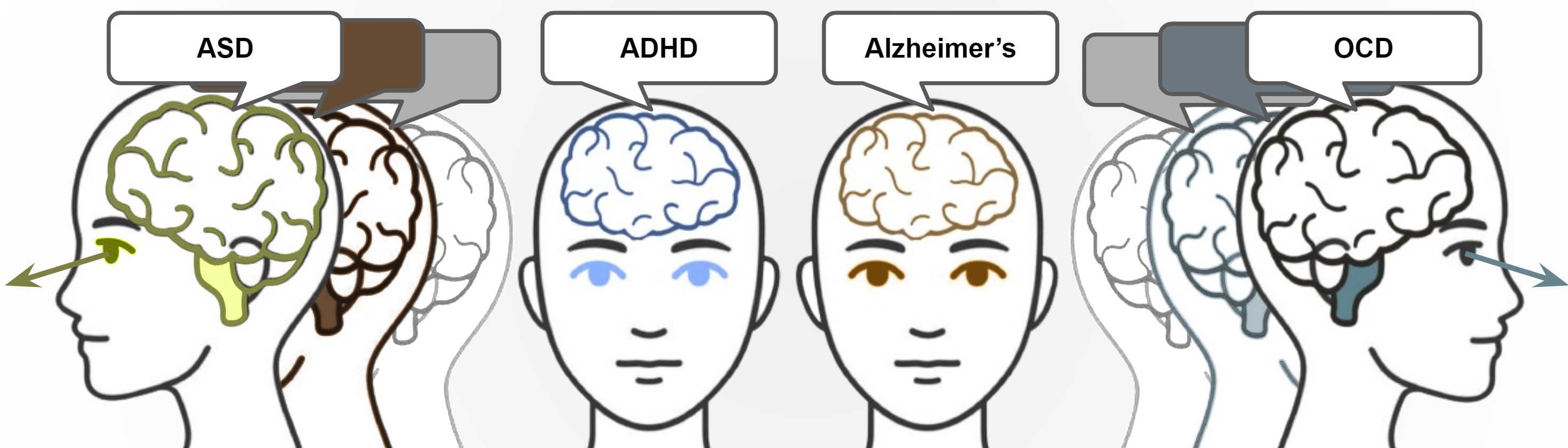}
    \caption{"Eyes are the mirror of the mind" when it comes to diagnosing \rev{neural} disorders. The subtle patterns in eye movements, e.g. fixations, saccades, and blinks, reflect underlying \rev{neural} processes and cognitive functions. Studying these patterns allows us to uncover early signs of \rev{neural} differences without invasive procedures.}
  \end{teaserfigure}
  
\maketitle

\section{Introduction}
\label{sec:intro}

\begin{figure}
    \centering\includegraphics[width=0.8\textwidth]{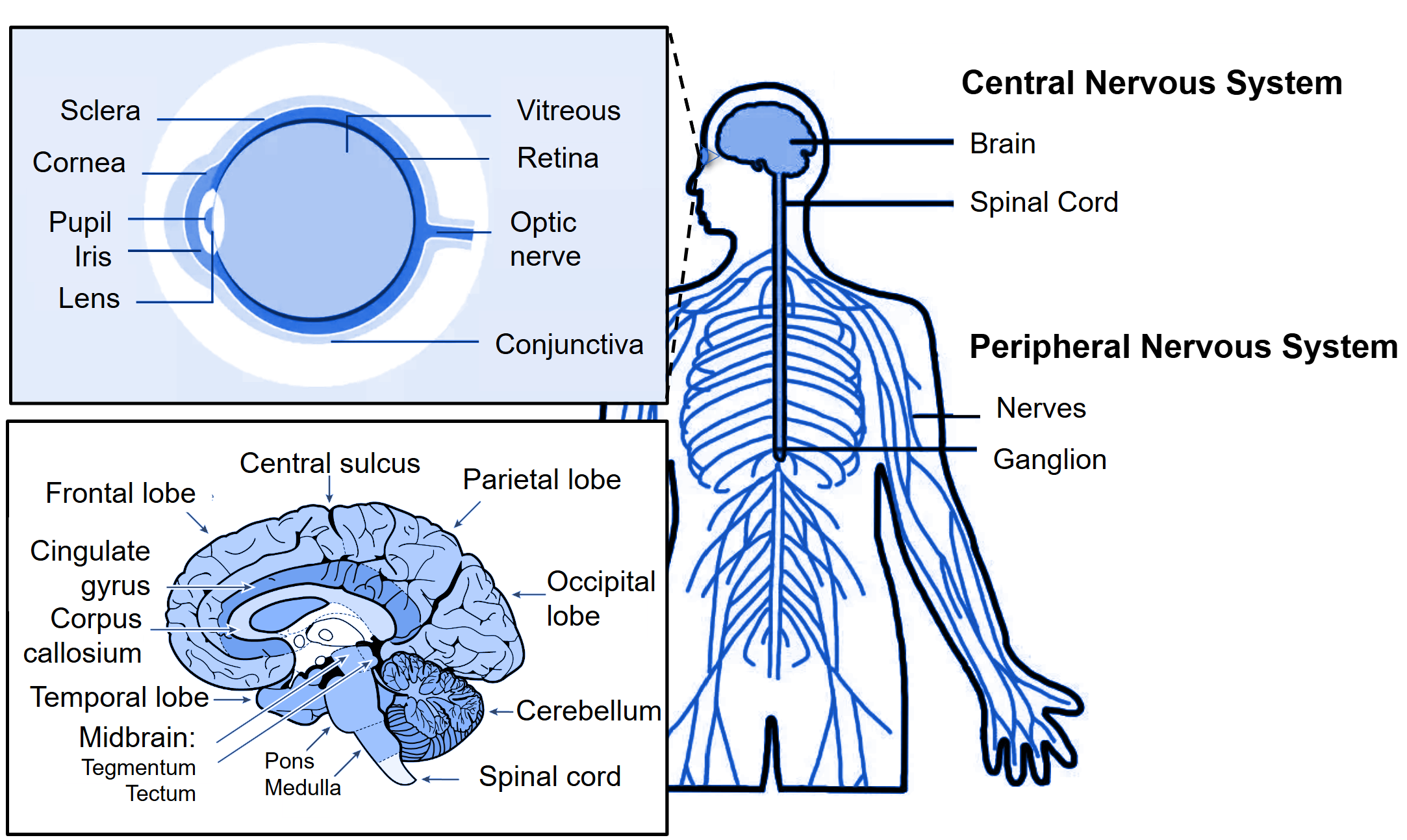}
    \caption{\rev{The human nervous system (right) includes the central nervous system (i.e. brain and spinal cord) and the peripheral nervous system (nerves and ganglia). The eye anatomy is shown in top-left and the brain anatomy in bottom-left.}}
    \label{fig: nervous}
\end{figure}

\rev{Neural} disorders, such as Autism Spectrum Disorder (ASD), Attention Deficit Hyperactivity Disorder (ADHD), dyslexia, Alzheimer's, and Parkinson's, help in understanding differences in the way individuals perceive and interact with the world \cite{dwyer2022neurodiversity, rosqvist2020neurodiversity}.
These differences result from variations in the human brain and can profoundly impact \rev{the individual's} communication, learning, and behaviour.
More specifically, \rev{neural} disorders affect the \rev{central nervous system (i.e. brain and spinal cord)} or the peripheral nervous system \rev{(i.e. nerves and ganglia)}, as shown in Figure \ref{fig: nervous}, leading to deficits in motor, sensory, or cognitive functions \cite{world2006neurological}.
These differences can bring both challenges and strengths. For example, people with ASD may have heightened attention to detail \cite{travers2011attention}, while those with ADHD might excel in environments requiring quick thinking or creativity \cite{salari2023global}.

Unfortunately, conventional \rev{neural} diagnosis involves a combination of tedious, time-consuming, or invasive screening processes, such as (i) subjective assessments based on clinical interviews, questionnaires, and behavioural observations influenced by factors like the patient’s memory and communication ability \cite{taylor2014brief, broder2018improving}, 
(ii) invasive testing like MRI and CT scans that are expensive and can be physically uncomfortable for patients \cite{tafasca2023ai4autism},
(iii) lengthy processing of progressive diseases that can take months or even years \cite{torres2013autism, castellani2010alzheimer},
(iv) involving multi-disciplinary specialists such as neurologists, psychiatrists, psychologists, and speech therapists \cite{dereu2010screening, janvier2016screening}, and 
(v) having difficulties in early diagnosis since many \rev{neural} conditions manifest with subtle or vague early symptoms \cite{janvier2016screening, larsen2018identification}. Without objective, real-time metrics, early diagnosis remains highly dependent on subjective reporting and clinical expertise.

Fortunately, \rev{processes in brain activation} also manifest through changes in eye movement patterns, fixations (i.e. steady gaze), 
saccades (i.e. rapid movements between points of focus), smooth pursuit movements (tracking moving objects)
and other gaze metrics \cite{tafasca2023ai4autism, vidal12_comcom, hokken2024eyes}.
Hence, eye-tracking research becomes particularly valuable in diagnosing and studying \rev{neural} disorders because the brain and central nervous system tightly control eye movements. 
Around half of the brain's neural pathways are dedicated to vision and control of eye movements \cite{bell2023vision}.
In addition, gaze tracking provides a non-invasive, objective 
\rev{indicator of neural processes}
to study how individuals perceive and respond to their environment due to its ability to capture non-verbal, unconscious responses that are less likely to be influenced by subjective bias or compensatory strategies \cite{kumar2016smarteye, kullmann2021portable, chang2020accurate, li2020eye, wolf2021contribution}.

\begin{figure}
    \includegraphics[width=\textwidth]{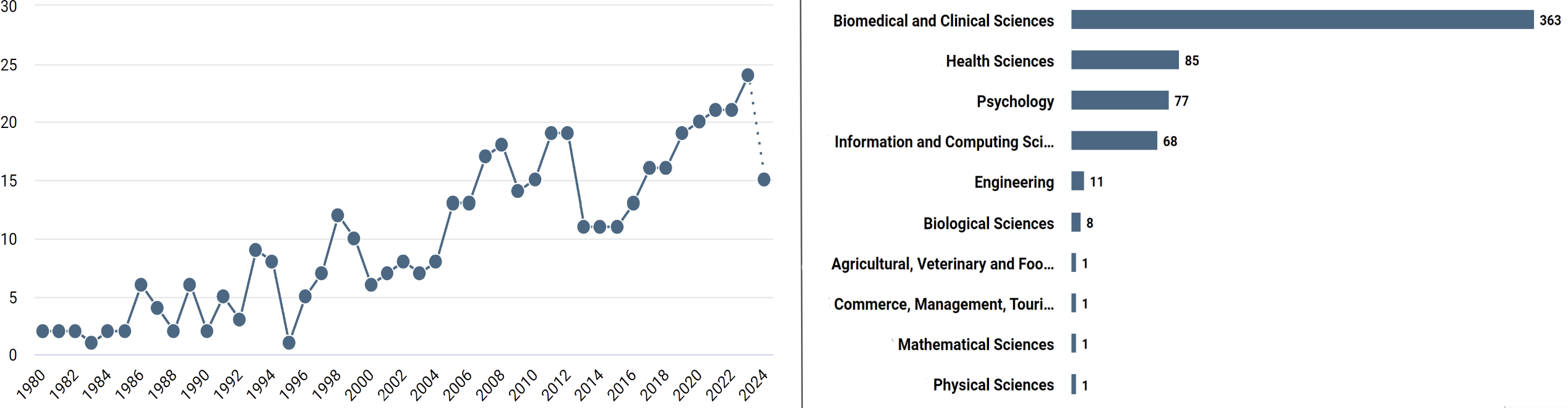}
    \caption{The number of publications per year (left) and per the main research categories (right) for eye tracking in \rev{neural} diagnosis sourced from dimensions.ai \cite{dimensions}
    }
    \label{fig: stats}
\end{figure}

\paragraph{Significance} Despite the potential of eye-tracking research in diagnosing \rev{neural} disorders and its growing attention (c.f. figure \ref{fig: stats}), the field remains in its infancy and further research is needed to refine gaze-tracking protocols.
Hence, we present this systemization of knowledge (SoK) paper to offer an overview of \rev{the neural diagnostics supported by gaze analysis} based on the key agreed-upon guidelines that are supported by a broad consensus across multiple studies and convergence of findings across different fields as presented in several peer-reviewed publications (c.f. figure \ref{fig: stats}).
We believe this paper \rev{contributes to the standardisation of gaze-based neural diagnosis methods and their application} across diverse clinical populations, mostly to get a preliminary \rev{neural} diagnosis.
Therefore, readers can use this SoK to 
(i) get equipped with a foundational knowledge of the field, 
(ii) understand the potential best practices, cautionary guidelines, and expected findings for future studies, 
(iii) design ethical studies that consider the well-being of participants, 
(iv) save time and effort through our compiled structure of multidisciplinary works,
(v) facilitate the collection of high-quality gaze data, addressing the current gap in research; 
(vi) analyse and validate the behaviour and performance of computational methods, such as AI models, by comparing their outputs against known verifiable facts as a reference; 
(vii) leverage guidelines as input for knowledge-based learning frameworks such as meta-learning and rule-based learning models, and ultimately,
(viii) create a generic diagnostic tool that distinguishes between the different \rev{neural} disorders and enables the decoupling of comorbid factors.

\paragraph{Contributions} Therefore, in this paper, we aim to advance research on eye tracking in \rev{neural} diagnosis based on the well-established knowledge and the agreed-upon findings across the different studies we surveyed. We compiled the accepted procedures and conclusions and made the following contributions:
\begin{itemize}[leftmargin = *]
    \item We survey and present the different \rev{neural} theories and the main causes and consequences of contradicting findings.
    \item We provide generic guidelines that can be applied to different disorders to help standardise gaze-based \rev{neural} protocols and ensure ethical, valid, and scientifically sound results.
    \item In addition to the generic guidelines, we provide further disorder-specific guidelines by compiling different findings from the fields of neuroscience, biomedical sciences, psychology, sociology, eye-tracking, and computer science.
    \item Finally, we present the current status, advantages, and limitations of gaze-based \rev{neural} diagnosis.
\end{itemize}

\paragraph{Methodology}
We compiled a list of 1,692 papers on gaze and \rev{neural} disorders through dimensions.ai \cite{dimensions}.
After only including studies that are (i) peer-reviewed, (ii) written in English, (iii) utilizing gaze-based measures (e.g., fixations, saccades, pupil dilation) to evaluate \rev{neural} functions or cognitive processes, and (iv) investigating gaze behaviour in relation to diagnostic procedures, excluding other papers on prognosis (e.g. \cite{silva2020prognostic, bacon2020identifying}), treatment (e.g. \cite{clendaniel2010effects, wyder2015gaze, carelli2022gaze, goodwin2021interstaars}), the use of gaze as an assistive input tool (e.g.  \cite{biswas2011new, ngo2024eeg, fernandes2023digital}), and gaze behaviour of \rev{neural} experts (e.g. \cite{crowe2018new, venjakob2012radiologists, orlosky2017emulation}),
we carefully investigated 464 research papers. We classified them into the different disorders and then into six main \rev{neural} classes according to \cite{svenaeus2014diagnosing}
(some comorbidity studies have also been investigated, c.f. sections \ref{sec:generic} and \ref{sec:specific}).

\paragraph{Structure}

In Section \ref{sec:gaze}, we start by presenting the relation between gaze and \rev{neural} disorders, why gaze is \rev{an informative tool for diagnosis}, and the main causes and consequences of contradicting findings in gaze-based \rev{neural} disorders.
In Section \ref{sec:generic}, we present the generic guidelines that can be used with studies involving \textit{any} disorder throughout the eye-tracking pipeline, such as the participant selection and screening, experiment design, data collection and equipment, data analysis and processing, reporting and reproducibility, and ethics and privacy considerations.
Then, in Section \ref{sec:specific}, we present the disorder-specific guidelines categorised by the \rev{neural} class (e.g. neurodevelopmental or neurodegenerative disorders).
For each disorder, we present an overview of the literature, the main eye-tracking focus, guidelines for task design, considerations, and key metrics with the expected gaze behaviour.
Finally, in Section \ref{sec:discussion}, we summarise the current status and limitations of gaze-based \rev{neural} diagnosis.

\section{Gaze \rev{as an indicator of neural processes}}
\label{sec:gaze}

The concept of \rev{neural} disorders offers distinct frameworks for understanding how individual variations affect perception, behaviour, and cognition \cite{walker2014neurodiversity}.
\rev{The spectrum of neural disorders differ in terms of their severity, symptoms, underlying causes, and functional impact on individuals. The concept of a spectrum recognizes that these disorders are not binary (having or not having the disorder) but exist along a continuum with varying degrees of impairment or presentation \cite{svenaeus2014diagnosing}, reflecting the complexity of understanding neural conditions.}
Further research continues to refine \rev{the diagnostic procedures}, especially in light of new technologies like eye-tracking, which offer additional insights into brain function and behaviour \cite{kumar2016smarteye, 
 kullmann2021portable}.

\subsection{Eyes and the brain}
The connection between \rev{neural} disorders and eye tracking is grounded in the neurological control of eye movements and how visual processing reflects brain function \cite{hallowell2012exploiting, reichle2013neurophysiological, hubel1995eye}.

\rev{Neural} disorders affect various brain regions responsible for sensory processing, motor control, cognition, and emotions. For instance, Alzheimer's disease affects the hippocampus, leading to memory deficits; Parkinson's disease involves dopamine depletion, affecting motor control; and ADHD involves dysregulation of attention and impulsivity in the prefrontal cortex \cite{world2006neurological, svenaeus2014diagnosing}.

\rev{The aetiology (causal origins) of neural disorders \cite{stewart2018pathology, stolze2004falls} involves complex interactions between genetic, neurobiological, environmental, and cognitive factors. Understanding these causes can help refine gaze-based indicators for specific diagnoses (c.f. Section \ref{sec:specific} for more details). (i) In neural disorders with genetic and neurodevelopmental basis (e.g. ASD), genes influencing synaptic plasticity, neurotransmitter function, and brain connectivity contribute to atypical gaze behaviours.
(ii) For disorders caused by structural and functional brain abnormalities (e.g. Alzheimer's and Parkinson’s), damage to specific brain regions controlling gaze and attention results in abnormal eye movement patterns. (iii) For disorders caused by neurotransmitter dysregulation (e.g. ADHD), imbalances in neurotransmitters like dopamine, serotonin, and norepinephrine affect attention, impulsivity, and decision-making, which correlate with abnormal gaze dynamics. (iv) 
For disorders caused by cognitive and attentional control deficits (e.g. OCD), deficits in executive function, attention, and cognitive control also affect gaze behaviour.
(v) Environmental and early-life factors (e.g. stress, trauma, or prenatal exposure to toxins) can lead to deficits in gaze processing (e.g. anxiety disorders). (vi) Other disorders (e.g. dyslexia) are caused by sensorimotor and visual processing abnormalities, affecting eye coordination and leading to gaze instability and reading difficulties.}

\paragraph{Anatomy of the eye}
\rev{The eye is a sensory organ that captures visual information and transmits it to the brain. As shown in Figure \ref{fig: nervous}, the eye key structures include: (i)
the cornea and lens, which focus light onto the retina, (ii) the retina, which contains photoreceptor cells (rods and cones) that convert light into electrical signals, (iii) the optic nerve, 
which transmits visual signals to the brain, and (iv) the fovea, which is a specialized region in the retina responsible for sharp central vision and high visual acuity.}

\paragraph{Visual pathways}
The brain's visual pathways and eye movement systems are deeply intertwined \cite{hallowell2012exploiting, reichle2013neurophysiological, hubel1995eye}. Eye movements such as fixations, saccades, pupil dilation, and smooth pursuits (c.f. Section \ref{sec:generic}) are controlled by different brain regions, including the visual cortex, frontal eye fields, cerebellum, and brainstem. 
\rev{The visual information from the retina follows the following pathway: The electrical signals from retinal ganglion cells travel through the optic nerve 
, which exits the back of the eye.
The optic nerves from both eyes meet at the optic chiasm, located at the base of the brain.
At the optic chiasm, nerve fibers partially cross over; the nasal (inner) retinal fibers cross to the opposite hemisphere, and the temporal (outer) retinal fibers remain on the same side. This crossing ensures that the left visual field is processed in the right hemisphere, and the right visual field is processed in the left hemisphere. After the optic chiasm, the reorganized visual information continues along the optic tracts to the lateral geniculate nucleus (LGN) in the thalamus. The LGN is a relay centre that processes and refines visual signals before sending them to the cortex. The LGN has six layers and organizes information based on the eye of origin (left vs. right eye) and the type of visual information (color, movement, fine detail).
The LGN transmits the processed signals to the primary visual cortex (i.e. V1 or the striate cortex) in the occipital lobe at the back of the brain. V1 processes:
edges, contrast, and orientation of objects, as well as the spatial organization of the visual field (retinotopic mapping). From V1, visual information is sent to higher-order visual areas for more advanced interpretation:
(i) The dorsal stream (i.e. the 'Where Pathway') travels from V1 to the parietal lobe to process spatial awareness, motion detection,
object location, and depth perception, which are crucial for guiding actions (e.g., reaching for an object). (ii) The ventral stream (i.e. the 'What Pathway') travels from V1 to the temporal lobe to process object recognition (faces, shapes, colors), and semantic meaning of visual stimuli, which are crucial for recognizing familiar faces, reading text, and identifying objects.}

For example, saccades are coordinated by the frontal eye fields \rev{(in the frontal cortex} and superior colliculus \rev{(midbrain)}, and fixations and smooth pursuit movements involve the cerebellum and parietal cortex \cite{noda1991cerebellar, mcdougal2015autonomic}.
\rev{In addition, gaze patterns reflect some brain functions such as cognitive and perceptual processing regulated by the parietal and frontal lobes, memory and recognition through the medial temporal lobe (including the hippocampus), and decision-making through the prefrontal cortex. Furthermore, social interactions are reflected in gaze patterns, e.g. the superior temporal sulcus (STS) processes eye contact and social cues, and the amygdala interprets emotional expressions in gaze patterns.}
Therefore, since the neural pathways closely regulate eye movements, disruptions in eye movement patterns can serve as an indicator for neural conditions \cite{terao2023patients, tao2020eye} (c.f. Section \ref{sec:specific} for more details).

\subsection{Gaze as a diagnostic tool}
Eye-tracking  can be a suitable tool for \rev{neural} diagnosis since it is 
(i) a non-invasive tool, making it particularly suitable for vulnerable populations such as children or individuals with severe \rev{neural} impairments with no harmful side effects or risks.
(ii) Gaze is a \rev{suitable} tool for early detection since gaze patterns can reveal subtle differences in \rev{neural} functioning before more noticeable symptoms arise \cite{kullmann2021portable, wolf2021contribution}. 
(iii) Eye-tracking provides quantitative and objective measurements (e.g., fixation duration, saccadic movements) that can be standardized and compared across individuals. This eliminates the subjectivity of many traditional behavioural assessments and allows for consistent measurements over time or between different populations \cite{tao2020eye}.
(iv) Gaze tracking can assess various \rev{neural} conditions by identifying impairments in areas like attention, memory, motor control, and emotional processing \cite{kumar2016smarteye, kullmann2021portable}. 
(v) Eye-tracking data can be captured in real-time, making it a powerful tool for diagnosing and monitoring progress in \rev{neural} disorders. This can provide immediate feedback to clinicians and inform adjustments to therapies or medications \cite{vajs2023accessible, grossman2019facetime}.
(vi) With advances in mobile eye-tracking technology, there’s potential for home-based diagnosis and monitoring, which could benefit patients who cannot frequently visit clinics or live in remote areas \cite{chang2020accurate, allen2007use, de2023novel, de2023oculomotor, adhanom2023eye, stokes2022measuring, koch2024eye, jiang2024auxiliary, barahim2024self, de2023oculomotor}.

\subsection{\rev{The spectrum of challenges}}
\rev{Research findings in gaze behaviour in neural disorders differ due to} complex factors, including differences in research methods, the multifaceted nature of \rev{neural} disorders, and the interdisciplinary nature of the different fields. Below are the main causes and consequences of these conflicting viewpoints that we compiled from the literature:

\paragraph{\textbf{Causes}}
\rev{Neural} disorders are heterogeneous, meaning they can present with widely varying symptoms across individuals \cite{haider2013neurological, espay2018current}. For example, individuals with ASD can exhibit different degrees of social \rev{communication} or sensory sensitivity, leading to differing interpretations of gaze behaviours \cite{stuart2023eye}. The variability in symptoms \rev{within and across disorders} causes researchers to focus on different aspects, such as social deficits, sensory sensitivities, or cognitive processes, sometimes leading to contradictory conclusions. 
\newline
In addition, differences in experimental design can also contribute to such discrepancies. Studies may vary in the age group \cite{holmqvist2023retracted, froment2022assess}, type of stimuli \cite{grossman2019perceptions,giarelli2014sensory,trevisan2017adults, bowman2004gaze}, environmental conditions \cite{holmqvist2011eye}, or tasks used to measure gaze \cite{macinska2024visual, putra2021identifying}. For example, gaze studies in ASD may involve free-viewing tasks, structured social interactions, or specific object-based tasks, each of which may lead to different conclusions about the role of gaze in the disorder \cite{stuart2023eye}. Different technological tools and data collection methods, such as head-mounted eye trackers vs. screen-based trackers, can yield varying results, further complicating consensus \cite{fiedler2020guideline, holmqvist2023retracted}.
\newline
Furthermore, gaze research intersects with multiple disciplines, including psychology, \rev{psychiatry,} neurology, cognitive science, and computer science, each of which may approach the data from a different theoretical framework. For instance, psychologists may interpret gaze avoidance in social situations as a behavioural issue, while neurologists might focus on sensory processing differences \cite{walker2014neurodiversity, world2006neurological}.
\newline
Moreover, eye movements are influenced by multiple neural systems, including those responsible for visual attention, motor control, and social cognition \cite{hallowell2012exploiting, reichle2013neurophysiological, hubel1995eye}. As a result, abnormal gaze patterns may be interpreted differently depending on which neural system researchers believe is most affected by a disorder. In ADHD, for example, some researchers emphasize deficits in motor control (leading to unstable gaze) \cite{selaskowski2023gaze, mauriello2022dysfunctional, sweere2022clinical}, while others highlight attentional regulation as the primary issue \cite{tye2013neurophysiological, mauriello2022dysfunctional, motomura2023effects, uono2023reduced}. This divergence stems from the fact that gaze is modulated by multiple underlying systems, making it difficult to attribute irregularities to a single cause.
\newline
Finally, gaze behaviour can be highly dependent on contexts (e.g. particular situation's social or cognitive demands) and \rev{individual parameters (e.g. sleep deprivation or mood)} \cite{eriksson1997eye, hopstaken2016shifts, wang2019assessment, yousefi2022stress, hirt2020stress, jyotsna2018eye}. This means that findings from studies using different tasks or social settings can yield contrasting results, even when studying the same disorder. For instance, gaze behaviour in individuals with anxiety may differ depending on whether they are in a low-stress or high-stress environment, leading to seemingly contradictory conclusions about the role of gaze in anxiety \cite{holmes2006anxiety, rosa2014effects}.

\paragraph{\textbf{Consequences}}

\rev{The above-mentioned challenges makes it difficult for} researchers trying to reach a consensus on the role of eye movements in diagnosing \rev{neural} disorders. This can lead to fragmentation within the research community, with different groups working under distinct assumptions, hindering the advancement of unified diagnostic and therapeutic frameworks. As a result, the progress toward developing reliable tools for \rev{neural} disorders is slowed, and the broader field struggles to establish clear clinical applications for gaze-based research \cite{band2024advancements, li2020eye}.
\newline
In addition, while some researchers advocate for the use of eye-tracking as an objective \rev{indicator} for diagnosing conditions  \cite{stuart2023eye,klin2020affording, walton2020brief}, others argue that gaze behaviours are too context-dependent to serve as reliable diagnostic criteria, especially when used without other modalities \cite{ko2023protocol}. This variability complicates the development of standardized clinical guidelines for using gaze data to diagnose \rev{neural} disorders. This further makes clinicians hesitate to rely on eye-tracking as a predictive measure, especially with the lack of standardization across studies, further complicating efforts to implement gaze-based diagnostics in clinical settings \cite{dawson2023could, blondon2015use}.

In brief, while the diversity of findings can foster innovation and deeper inquiry, it also highlights the need for more integrated research considering the multifaceted nature of gaze behaviours across \rev{neural} conditions. Hence, this paper provides guidelines according to the consensus-based findings and commonly accepted procedures in the literature to standardise and advance gaze-based \rev{neural} diagnosis.

\section{Generic guidelines for gaze-based \rev{neural} research}
\label{sec:generic}

Conducting research on \rev{neural} disorders using eye-tracking technology requires adherence to several guidelines to ensure ethical, valid, and scientifically sound results.
Here, we compiled the commonly accepted key guidelines from the literature, focusing on participant well-being, accurate data collection, and methodological rigour. 
For all-purpose eye-tracking guidelines, we refer the reader to \cite{holmqvist2011eye}. This paper specifically focuses on guidelines and considerations relevant to \rev{neural} diagnosis.
Below, we present some generic key guidelines. Then, in section \ref{sec:specific}, we present specific guidelines for each condition.

\subsection{Participant Selection and Screening}
\paragraph{Appropriate Sampling:} When studying \rev{neural} disorders, including a representative sample of the disorder in question is important \cite{stuart2023eye, guillon2014visual}. For example, in research on autism, include participants across the spectrum of severity \cite{keles2022atypical, michalek2019predicting}.
Sampling by age group is essential for ensuring that neurological and cognitive developmental changes are accurately captured \cite{holmqvist2023retracted, froment2022assess, maruta2017visual}. The brain and cognitive processes vary significantly by age, and \rev{neural} disorders can present differently in children, adolescents, adults, and the elderly \cite{zaino2023different, kullmann2021portable, larsen2018identification, kullmann2021portable}. If possible, use age stratification, i.e., include narrow age ranges or age cohorts to analyse developmental trajectories or how symptoms manifest differently across life stages \cite{pitigoi2024attentional, moghadami2021investigation, kynast2020mindreading}.
In addition, gender differences play a critical role in \rev{neural} disorders since certain disorders may manifest differently across genders, and this should be reflected in both sampling and data analysis \cite{pitigoi2024attentional, moghadami2021investigation, kullmann2021portable}. 
For example, boys are diagnosed with ASD at a higher rate than girls, but girls may present with less subtle symptoms \cite{putnam2023effects, bao2024gender}. 
Finally, collect participants’ current treatment status since it can influence their performance in studies, especially those involving cognitive or behavioural tasks \cite{cstefcsnescu2024eye, sweere2024efficacy}. Consider comparing medicated vs unmedicated groups when analysing data \cite{holmqvist2023retracted, fujita2023istradefylline, sweere2022clinical}.
Similarly, family history and genetics can influence the likelihood of developing a disorder and its severity and the inheritability of certain conditions \cite{froment2022assess, dubey2023psychosocial}.

\paragraph{Control Group Inclusion:} A matched control group (e.g. age-matched or gender-matched) is a necessary baseline to compare gaze behaviour and draw meaningful conclusions about the studied disorder \cite{li2020eye, lima2022comprehensive, hodgson2019eye}. Comparing patient data with a normative dataset allows for more accurate detection of abnormalities and diagnosis \cite{hasse2015eye, kullmann2021portable, kullmann2021normative}.

\paragraph{Co-morbidities and Confounding Factors:} Take into account co-occurring conditions (e.g., ADHD with ASD or OCD with anxiety) as they can influence eye-tracking data and adjust the study design to account for these variables \cite{tye2013neurophysiological, yerys2013separate, aydin2023face, ioannou2020comorbidity}. For certain studies, you may want to stratify participants based on single diagnoses vs. comorbid diagnoses \cite{tye2013neurophysiological, groom2017atypical, mostofsky2001oculomotor}.

\subsection{Experiment Design}
\paragraph{Task Selection:} 
Ensure tasks are appropriate for the cognitive and sensory capabilities of the participants \cite{li2020eye}. For example, individuals with ASD may struggle with tasks involving high social or emotional content \cite{boyd2022manipulating, giarelli2014sensory}, and those with dyslexia may face difficulty in reading tasks \cite{fischer2000stability, wang2019gaze}. Design tasks that accommodate these limitations while still collecting meaningful data (c.f. section \ref{sec:specific}).

\paragraph{Stimuli:} Choose visual stimuli that are relevant to the disorder, such as fearful or anxiety-inducing scenarios, facial expressions or social scenes \cite{grossman2019facetime}, static and dynamic stimuli \cite{grossman2019perceptions}, or colours and transitions \cite{giarelli2014sensory}.
Avoid overly complex visual stimuli that lead to sensory overload \cite{trevisan2017adults, bowman2004gaze}.
In addition, if your study involves tasks with verbal components, consider language barriers (e.g. native vs non-native speakers) and the need for translation or language accommodations \cite{poletti2017eye, sui2023geco, prystauka2024online}.

\paragraph{Duration and Breaks:} Participants with \rev{neural} disorders might have limited attention spans \cite{selaskowski2023gaze, mauriello2022dysfunctional} or suffer from muscle fatigue/weakness \cite{bekteshi2023towards, gandolfi2024window}. Design shorter tasks and provide regular breaks to maintain focus and minimize fatigue \cite{leary1996moving, torres2013autism}.

\subsection{Data collection and equipment}

\paragraph{Choice of Eye-Tracking Device:} 
For high-quality data, use high-precision eye-tracking systems that offer a good sampling rate (e.g., 60–120 Hz for general use, 250+ Hz for saccadic studies); this responsiveness ensures that no movements are missed, especially for disorders where small deviations in gaze can provide significant insights (e.g., subtle differences in social attention in ASD or reading patterns in dyslexia) \cite{fiedler2020guideline, holmqvist2023retracted, prabha2020predictive, wang2019gaze}. 
The higher resolution of these devices captures more detailed data about eye movements, including microsaccades (tiny, rapid movements) and smooth pursuits (when the eye follows a moving object), allowing a thorough fine-grained analysis \cite{holmqvist2011eye}. Note that lower-precision eye trackers can suffer from noise (random inaccuracies in measurements) and drift (gradual inaccuracy over time). High-precision trackers minimize these issues, leading to more reliable long-term tracking and making it easier to interpret results over extended periods \cite{holmqvist2012eye, feit2017toward}.

\paragraph{Data Quality Control:} Implement robust data cleaning techniques, as individuals with \rev{neural} disorders might produce noisy data due to difficulty maintaining focus or motor control issues \cite{wojcik2016eye, helmchen2003eye}. For instance, algorithms to handle blinking \cite{morris2002blink, bhaskar2003blink, krolak2012eye}, head movements \cite{al2013eye, larsson2016head}, or gaze drift \cite{barbara2023real, drewes2014smaller} may be necessary. Also note that external factors, such as fatigue \cite{eriksson1997eye, hopstaken2016shifts, wang2019assessment} or stress \cite{yousefi2022stress, hirt2020stress, jyotsna2018eye} can influence eye-tracking results, making it important to control these variables in experimental designs.

\paragraph{Setup} Use a suitable setup for the target group. For example, children or participants with mobility issues should use non-invasive, comfortable devices (e.g., screen-based eye trackers) \cite{fiedler2020guideline, torres2013autism}. For some participants, especially those with anxiety or sensory processing difficulties, ensure that the environment is non-threatening or familiar to avoid triggering abnormal behaviour unrelated to the task \cite{trevisan2017adults, duncan1991gaze, cludius2019attentional, basel2023attention}, cf. Section \ref{sec:specific} for disorder-specific considerations.

\subsection{Data Analysis and Processing}

\paragraph{Custom Metrics for Disorders:} Use specialized eye-tracking measures tailored to assess specific cognitive, motor, or perceptual deficits that capture the disorder-specific gaze behaviours \cite{froment2022assess, tao2020eye, martinez2013impact}. For instance, in autism research, reduced fixation on eye regions of faces is a key metric, while in dyslexia, measures of saccades and fixation duration during reading tasks are critical (cf. Table \ref{tab:metrics} and Section \ref{sec:specific} for more details).

\begin{longtable}[H]{ p{.15\textwidth}  p{.25\textwidth}   p{.20\textwidth}  p{.30\textwidth} }
     \toprule
      \textbf{Metric} & \textbf{Definition} & \textbf{Behaviour} & \textbf{Example}\\
     \midrule
     Fixations & A period when the gaze remains relatively still, allowing the eyes to focus on a specific point or object, commonly with a duration of 200-300ms & Sustained attention, engagement, cognitive processing (cognitive load or difficulty in processing information), interest or preference & Prolonged fixations can indicate difficulties in processing information or heightened attention to detail in Alzheimer's disease and ASD. In contrast, shorter fixations suggest difficulties maintaining attention as in ADHD.
\\
     \midrule
     Saccades & Rapid eye movements between fixations, used to quickly shift attention from one point to another, commonly with a duration of 30-80 ms, an amplitude of 4-20°, and a velocity of 30-500°/s & Visual search and scanning, cognitive load, impulsivity and attention disorders & Impaired (delayed or slow) saccadic eye movements are often linked to motor control deficits in Parkinson’s disease and Huntington’s disease while erratic or rapid saccades may suggest hyperactivity or excessive scanning of the visual field as in ADHD and anxiety disorders.
     \\
     \midrule
     Pupil dilation & Changes in the size of the pupil, often in response to light but also as a reaction to cognitive and emotional stimuli & Cognitive load (task complexity and mental effort), emotional arousal (stress, excitement, or fear), selective attention &
     Reduced and slower pupil dilation/constriction can be a biomarker for Parkinson’s while exaggerated dilation during cognitive tasks can be linked to Alzheimer’s. Larger baseline pupil size in response to social stimuli is linked to ASD.
     
     \\
     \midrule
     Blink rate & The frequency of spontaneous blinking & Cognitive load, sustained attention, stress, anxiety, fatigue, dopaminergic activity, &  Increased blinking can reflect heightened stress or obsessive-compulsive tendencies, as in OCD and anxiety disorders. In contrast, a lower blink rate is associated with impaired motor control or cognitive decline, as in Parkinson’s disease and Alzheimer’s disease.
     \\
     \midrule
     Microsaccades & Small, involuntary saccades that occur during fixation to correct gaze drift, commonly for a duration of 10-30 ms, an amplitude of 10-40' (minutes where 1°=60'), and a velocity of 15-50°/s & Attention stability, cognitive fatigue or stress & More frequent microsaccades can indicate difficulty focusing or excessive scanning of the environment as in ADHD and anxiety disorders. In contrast, decreased microsaccades may signal cognitive decline or attentional deficits in Alzheimer's disease.
     \\
     \midrule
     Smooth pursuit & Eye movements that enable the tracking of moving objects smoothly, commonly with a velocity of 10-30°/s & Motor control, attentional engagement, oculomotor coordination &  Difficulties in tracking moving objects are common in Multiple Sclerosis (MS) due to deficits in coordination. Jerky smooth pursuit can signal motor system dysfunction in Dystonia and essential tremor. \\
     \bottomrule
    
    \caption{Most common metrics for gaze-base neurological behaviour as defined by Holmqvist et al.  \cite{holmqvist2011eye}. Metrics offer a structured way to interpret the findings. Each metric should be assessed alongside others, and often, multiple metrics together provide a more accurate diagnosis (cf. Section \ref{sec:specific} for more details).}
    
    \label{tab:metrics}
\end{longtable}

\paragraph{Longitudinal Studies:} \rev{Neural} disorders often evolve over time, so consider conducting longitudinal studies to track changes in gaze behaviour as the disorder progresses or in response to treatments \cite{proudfoot2016eye, de2023oculomotor}. Note that eye gaze is likely critical for developing long-term social skills and higher-order social-cognitive abilities, such as theory of mind (ToM) and perspective taking  \cite{clarke2012assessing, snowden2003social}.

\paragraph{Multimodal Integration:} 
Eye-tracking can reflect cognitive load, emotional responses, and attention shifts, but it doesn’t provide direct information on underlying brain activity \cite{hallowell2012exploiting, reichle2013neurophysiological, hubel1995eye}. In addition, since gaze as a diagnostic tool is \rev{a relatively new research field} and neural disorders often involve complex and overlapping symptoms, relying on a single diagnostic tool, with the current research state, can lead to incomplete or misleading interpretations \cite{quatieri2017multimodal, barone2023potential}. 
Combining multiple modalities allows for cross-validation, reducing the chances of false positives or negatives. If abnormalities are observed and \rev{cross-referenced} in both eye-tracking data and another measure (e.g., EEG, MRI, \rev{subjects' and doctors' reports}), the likelihood of those results being meaningful and linked to a specific \rev{neural} condition is higher, especially in complex or comorbid cases where symptoms of different disorders overlap \cite{moghadami2021investigation, black2017mechanisms}. 
Therefore, consider combining eye-tracking data with other physiological or cognitive measures (e.g., EEG, MRI, behavioural observations) to gain a holistic understanding of how gaze behaviour relates to the \rev{neural} condition \cite{ko2023protocol, barone2023potential, hu2023brain, krishna2014long}.

\paragraph{Statistical Analysis and Hypothesis Testing} 

Once gaze data has been collected, statistical analysis is crucial to extract meaningful patterns and test hypotheses about the \rev{neural} disorder. The goal is to determine whether there are significant differences in gaze behaviour between groups (e.g., patients vs healthy controls or medicated vs unmedicated patients) or over time (e.g., disease progression) \cite{li2020eye, hodgson2019eye, hasse2015eye, kullmann2021normative}.
For this, develop clear, testable hypotheses based on the research question \cite{raschke2014visual, ghose2020pytrack, holmqvist2011eye, rucker2021dysfunctional}, such as "Patients with Parkinson’s disease will have longer saccade latencies compared to healthy controls.". Then, select the appropriate statistical test based on the data type and research design; here, we refer the reader to and highlight key guidelines following \cite{mandel2012statistical, anderson2012new}:
\begin{itemize}[leftmargin = *]
    \item Parametric Tests: For normally distributed data, including:
    \begin{itemize}[leftmargin = *]
        \item T-Tests: e.g. compare the average fixation duration between patients with Alzheimer's disease and healthy controls during a memory recall task to check for significant differences between the two groups (i.e. two-sample t-test) \cite{tokushige2023early}.
        \item ANOVA (Analysis of Variance): e.g. compare gaze patterns across multiple groups (e.g., Parkinson’s patients, Alzheimer’s patients, and healthy controls), a one-way ANOVA could be used to determine if there are significant differences in saccade velocity between the groups, with Parkinson's patients showing the slowest velocities due to motor deficits \cite{archibald2013visual}.
        \item Repeated-Measures ANOVA: e.g. assess how fixation duration changes over time during a cognitive task in the same group of participants, a repeated-measures ANOVA could account for within-subject variations. For instance, fixation duration could increase over time as participants become more fatigued during a long, attention-heavy task \cite{hummel2017influence}.
    \end{itemize}
   
    \item Non-Parametric Tests: For data that does not meet normality assumptions, including:
    \begin{itemize}
        \item Mann-Whitney U-Test: E.g. if fixation durations between two groups (e.g., healthy controls and stroke patients) are highly skewed or don’t follow a normal distribution, the Mann-Whitney U-test can be used to compare the distributions between groups \cite{chan2023examining}.
        \item Kruskal-Wallis Test: e.g. for more than two groups (e.g., different subtypes of autism), when fixation duration is not normally distributed, a Kruskal-Wallis test can determine whether the groups differ significantly.
        For example, different subtypes of autism could display varying gaze patterns, with some subtypes showing a preference for non-social stimuli (e.g., objects instead of faces) \cite{amestoy2015developmental}.
    \end{itemize}
    
    \item Mixed-Effects Models: For repeated measures data (e.g., where gaze data is collected at multiple time points or across multiple conditions).
    For example, in a longitudinal study where gaze data is collected from the same patients with multiple sclerosis (MS) at different stages of the disease, a linear mixed-effects model can be used to account for both fixed effects (e.g., disease stage) and random effects (e.g., individual variability) \cite{mallery2018visual}. Another example is if gaze data is nested within tasks (e.g., multiple visual tasks completed by the same subject), mixed-effects models can also handle these nested structures. For instance, gaze data from various attention tasks for each participant with ADHD can be modelled to see if task difficulty affects gaze measures differently across subjects \cite{stokes2022measuring}.

    \item Correlation Analysis: To investigate relationships between gaze metrics and other variables, such as cognitive test scores or brain activity, including:
    \begin{itemize}
        \item Pearson's Correlation: E.g. when examining the relationship between average fixation duration and cognitive test scores (e.g., memory recall performance) in Alzheimer's patients, Pearson's correlation can be used for normally distributed data to test if there will be a negative correlation between fixation duration and memory test performance, with longer fixations indicating poorer cognitive function \cite{tokushige2023early}.
        \item Spearman's Rank Correlation: If the data is non-normally distributed (e.g., fixation duration data is skewed), Spearman's rank correlation can be used to assess the relationship between fixation duration and disease severity scores \cite{hassan2022approach}. E.g. in Parkinson’s patients, Spearman's Rank is used to check if there will be a positive correlation between fixation duration and disease severity, with more severe motor symptoms being associated with longer fixation durations \cite{stuart2016accuracy}.
    \end{itemize}
    
    \item Correct for Multiple Comparisons: Eye-tracking experiments often involve a large number of gaze metrics (e.g., fixation duration, saccade velocity, etc.). Applying a correction method (e.g., Bonferroni correction) is necessary to reduce the risk of false positives. For example, after running several independent t-tests for each gaze metric, the p-value threshold can be adjusted using the Bonferroni correction, which divides the significance level by the number of comparisons (e.g., three tests) \cite{laurens2019spatial}.
\end{itemize}

In addition to p-values, reporting effect sizes (Cohen’s delta, eta squared) and confidence intervals provide information on the practical significance of the findings, which is especially important in clinical settings \cite{wetzel2018eye, mazidi2021time, suslow2020attentional}.

It is also essential to analyze gaze data in segments depending on the type of stimulus (e.g., visual scene, text, social stimuli) to assess which cognitive or visual processes are impaired \cite{chan2023examining, pitigoi2024attentional, moghadami2021investigation}. 

With complex (high-dimensional) non-linear patterns that may be difficult to identify through traditional statistical methods, machine learning models can be applied \cite{sun2022novel, ahmed2022eye, przybyszewski2023machine}.

\paragraph{Data visualization}
Data visualization is a supplementary step in understanding and interpreting neurological gaze data. It transforms raw values into visual formats that highlight patterns, anomalies, or trends in gaze behaviour, making it easier to communicate results to both clinicians and researchers \cite{raschke2014visual, yoo2021gaze}.
Common visualizations in eye-tracking analysis include:
\begin{itemize} [leftmargin = *]
    \item Heatmaps: These visualize where participants are looking most frequently. Areas with more fixations are represented with "hotter" colours (e.g., red), and areas with fewer fixations are shown with "cooler" colours (e.g., blue) \cite{vspakov2007visualization}. For example, in a study on autism spectrum disorder (ASD), a heat map could show whether participants with ASD focus less on the eyes in images of human faces, indicating social attention deficits \cite{keles2022atypical, santiago2021visual}.

    \item Scanpaths: These trace the sequence and direction of eye movements, showing how a person visually explores a scene or object \cite{posner1980orienting}. For example, scanpaths could be used in Alzheimer's research to show how patients' gaze patterns become more erratic when navigating visual scenes, reflecting a cognitive decline in spatial awareness \cite{boz2023examination}.

    \item Time-Series Plots: These visualize how gaze metrics (e.g., pupil dilation, fixation duration) change over time during a task \cite{kasprowski2016gaze}. For example, in a cognitive load experiment, a time-series plot might show that Alzheimer’s patients exhibit greater fluctuations in pupil dilation over time, indicating increased difficulty in maintaining attention \cite{sturm2011mutual}.

    \item Gaze-Overlaid Videos: By overlaying gaze data on videos of stimuli (e.g., a movie scene or interactive interface), the participant’s attention shifts can be tracked over time \cite{tien2012measuring}. For example, for assessing visual attention deficits in ADHD, a gaze-overlaid video can illustrate the participant's difficulty in maintaining focus on relevant stimuli \cite{bast2023sensory}.
\end{itemize}

\subsection{Reporting}


\paragraph{Transparency:} Provide clear documentation of study protocols, participant characteristics, data preprocessing steps, and analytical techniques to ensure applicability \cite{dunn2024minimal, fiedler2020guideline}.

\paragraph{Pre-registration:} Pre-registration refers to the practice of registering a research study's methodology, hypotheses, and analysis plans in a public registry before data collection begins to promote transparency, reduce the risks of bias, and improve the credibility of scientific findings \cite{nosek2018preregistration, hardwicke2023reducing}. Pre-registration is becoming increasingly recognized as a "golden standard" in research reporting, especially in fields prone to exploratory analyses \cite{wagenmakers2012agenda} or p-hacking \cite{wasserstein2016asa}, e.g. gaze-based \rev{neural} studies \cite{tipples2019closer, fosterpre, prein2024variation}. Studies are commonly pre-registered on platforms like ClinicalTrials.gov \cite{clinicaltrials2024} or the Open Science Framework (OSF) \cite{osf_project2024}.

\paragraph{Reproducability} Wherever possible, share the code or software used for data analysis (e.g., Python, MATLAB, R scripts) on public repositories like GitHub \cite{github2024} or OSF \cite{osf_project2024}. This allows others to reproduce the exact analysis pipeline \cite{dunn2024minimal, molina2024eye}. In addition, whenever possible, make the gaze data available on public repositories, following ethical guidelines (e.g., via Dryad \cite{dryad2024}, Zenodo \cite{zenodo2024}), allowing other researchers to replicate the findings \cite{fiedler2020guideline, holmqvist2011eye}.


For further reporting guidelines for general eye-tracking studies, we refer the reader to \cite{fiedler2020guideline}.

\subsection{Ethics and Privacy Considerations}
In addition to the general ethics and privacy considerations for general eye tracking studies \cite{etra2024}, further considerations arise for \rev{neural} applications.

\paragraph{Informed Consent:} 
Participants, especially those with \rev{neural} disorders, must be fully informed about the research goals, procedures, and potential risks before agreeing to participate \cite{holmqvist2011eye, fiedler2020guideline}. Ensure that proper consent is obtained, possibly through guardians or caregivers, when working with vulnerable populations, such as children or individuals with cognitive impairments (e.g., ASD or Alzheimer’s) \cite{zurek2024can, johansson2021methodological, poletti2017eye}. Participants should also have the right to withdraw their consent and request the deletion of their personal data at any time \cite{gdpr2016}, e.g. commonly achieved through an end-user license agreement (EULA) \cite{ahuja2016commercial}. 
\rev{Consents could be generated for eye tracking experiments using} \cite{Ehinger}. \rev{For specific use-cases,  Open Brain Consent} \cite{openbrainconsent} \rev{provides several consent forms for different jurisdictions, languages, and/or guidelines that could be adapted to the specific experiment \footnote{\rev{For stronger privacy guarantees, privacy-preserving mechanisms could be adopted, e.g. differential privacy \cite{steil2019privacy, liu2019differential, li2021kalvarepsilonido, david2022your}, federated learning \cite{elfares2023federated, fenoglio2023federated}, and secure computation protocols \cite{ozdel2024privacy, elfares2024privateyes}.}}.}

\fbox{
    \parbox{14cm}{
    \textbf{\rev{Consent example of a free-viewing task} \cite{Ehinger}.}\\
    \textbf{Replace the [\textit{text between brackets}] with your experiment details:}\\
    Dear participant,\\  
    You have volunteered to participate in this study. Here you will now receive some information about your rights and the procedure of the following experiment. Please read the following sections carefully. \\
    \textbf{1) Purpose of the study}\\
    In this study, we will investigate [\textit{how humans perceive images by recording eye movements using an eye tracker}]. \\
    \textbf{2) Study procedure}\\
    The study will proceed as follows:
    We perform [\textit{an eye test (visual acuity and dominant eye)}]. \\
    The eye tracker needs to be calibrated, for this, a moving dot will appear on the monitor which you will follow with your gaze. \\
    We will present [\textit{different pictures on the screen. Your task is to freely explore these pictures}]. \\
    Including the questionnaires and preparations, the experiment takes about [\textit{one}] hour.\\
    \textbf{3) Risks and side effects}\\
    According to current knowledge, this study is harmless and painless for the participants. By participating in this study, you are not exposing yourself to any particular risks and there are no known side effects. However, because this study is new in its entirety, the occurrence of as-yet-unknown side effects cannot be completely ruled out. \\
    Important: Please inform the experimenter immediately if you [\textit{have a neural disease, if strong (light) stimulation can trigger migraine, or if you have had an epileptic seizure}]. If you have any questions about this, please contact the experimental investigator. \\
   \textbf{4) Termination of the experiment}\\
   Participation in the study is voluntary. You may withdraw your consent to participate in this study at any time without giving any reason and without any disadvantages. Even if you terminate the study prematurely you will be remunerated accordingly for the time spent up to that point. \\
   If you experience [\textit{headaches or any other kind of discomfort}] during the experiment, please inform the experimental investigator immediately.\\
   \textbf{5) Confidentiality} \\
   Your data will only be stored in pseudonymized form (e.g. "sub-003"). [\textit{The mapping file will be deleted after the completion of the data collection. The data will be made available for scientific publications, but also as open scientific data for third parties.}]\\
    \textbf{6) Remuneration} \\
    The study will be remunerated with [\textit{15 euros}] per hour. Partial half hours will be rounded up. \\
    \textbf{7) Declaration of consent} \\
    I hereby confirm that I have understood the participant information described above and that I agree to the stated conditions of participation, in particular: \\
    \textit{[an unlimited storage of my pseudonymized data,
    the usage of my pseudonymized data for the current research project and for other exclusively scientific purposes,
    a publication of my pseudonymized data as open data.]}
    \newline
    .......................................\\
    Place, date, signature
    }
}

\paragraph{Ethical Approval and Institutional Review Boards (IRBs)}
Before conducting studies, research involving human subjects needs to be approved by an IRB or ethics committee \cite{hhs_irbs_assurances, hhs_45cfr46}, which ensures the study complies with ethical and legal standards, e.g. the General Data Protection Regulation (GDPR) \cite{gdpr2016}, or the Medical Device Regulation (MDR) \cite{mdr2017} in the EU and/or the Health Insurance Portability and Accountability Act (HIPAA) \cite{hipaa1996}, the California Consumer Privacy Act (CCPA) \cite{ccpa2018}, or the Food and Drug Administration (FDA) \cite{fda2024} in the US. Ethical boards evaluate whether data collection processes respect patient privacy, minimize harm, and maintain the confidentiality of sensitive gaze and \rev{neural} data.


\section{Specific guidelines for gaze-based \rev{neural} research}
\label{sec:specific}

\rev{Neural} disorders can be categorised into a diverse group of conditions that affect the nervous system, which includes the brain, spinal cord, and nerves. 
They can result from structural, biochemical, or electrical variations and can impact movement, communication, cognition, and behaviour.

\rev{We emphasize that it is crucial to determine whether gaze patterns are merely an epiphenomenon or a direct reflection of neural function \cite{rossano2012gaze, wang2002neural, emery2000eyes, mancini2021lived}.
An epiphenomenon is a phenomenon that occurs alongside another condition but does not necessarily cause or result from it. For example, in dyslexia, eye movement abnormalities (e.g., longer fixations, more regressions) are often observed, however, these do not directly cause reading difficulties; rather, they emerge as a byproduct of phonological and language processing deficits.
In this case, gaze abnormalities are not necessarily indicative of a core neural deficit, but rather reflect higher-level cognitive processing difficulties.
In contrast, eye movements can directly reflect the functional state of neural networks, particularly in disorders affecting attention control, executive function, or motor planning. For example, in attention disorders (e.g., ASD and ADHD), gaze behaviour is tightly linked to neural mechanisms governing attention allocation, inhibition, and sensory integration. These processes involve specific brain networks, such as the fronto-parietal network (for voluntary attention and saccade control), the dorsal attention network (for top-down attentional guidance), and ventral attention network (for reorienting attention to unexpected stimuli). Hence, in these cases,
gaze abnormalities are not just symptoms but are rooted in underlying neural dysfunctions.
}

\rev{We also recommend adopting a transdiagnostic framework, recognizing that many neural disorders involve overlapping but distinct disruptions in neural circuits (c.f. Section \ref{sec:generic}). A transdiagnostic freamework focuses on functional (or dysfunctional) neural networks that cut across traditional diagnostic boundaries. In other words, since many neural disorders share overlapping dysfunctions in brain circuits that control attention, perception, and cognitive control, rather than studying each disorder in isolation (e.g., ASD vs. ADHD), a network-based approach considers which neural systems are affected and how they vary across individuals. Therefore, instead of treating disorders categorically, gaze analysis should be contextualized within broader neural dysfunctions to improve diagnostic accuracy and treatment outcomes.
}

\begin{figure} 
  \centering

\begin{forest}
  for tree={
    font=\footnotesize,
    child anchor=west,
    parent anchor=east, 
    grow'=east,
    text width=6.5cm,
    draw,
    anchor=south,
    edge path={
      \noexpand\path[\forestoption{edge}]
        (.child anchor) -| +(-5pt,0) -- +(-5pt,0) |-
        (!u.parent anchor)\forestoption{edge label};
    },
  }
    [\textbf{Neural Disorders}, rotate=90, parent anchor=south
        [\textbf{Neurodevelopmental Disorders:}
        A group of mental conditions that appear in early childhood and affect the development of the nervous system.
            [Autism Spectrum Disorder (ASD)]
            [Attention Deficit Hyperactivity Disorder (ADHD)]
            [Dyslexia]
            [Cerebral Palsy (CP)]  
        ]
        [\textbf{Neurodegenerative Disorders:}
        A group of conditions caused by the progressive loss of neurons in the brain.
            [Alzheimer's Disease (AD)]
            [Parkinson's Disease (PD)]
            [Amyotrophic Lateral Sclerosis (ALS)]
            [Huntington’s Disease (HD)]
            [Multiple Sclerosis (MS)]
        ]
        [\textbf{Movement Disorders:}
        A group of clinical syndromes characterised by either a set of voluntary and involuntary movements or an excess of movement.
            [Tourette Syndrome (TS)]
            [Essential Tremor (ET)]
            [Dystonia]
        ]
        [\textbf{Seizure Disorders:}
        The manifestation of an excessive\, abnormal\, and synchronized electrical discharge in the neurons.
            [Epilepsy] 
        ]
        [\textbf{Cerebrovascular Disorders:}
        A set of medical conditions that affect the blood vessels of the brain and the cerebral circulation\, characterized by damaged arteries that supply oxygen and nutrients to the brain.
            [Stroke]
        ]
        [\textbf{Mental Health Disorders:}
        A group of patterns causing distress or impairment of personal functioning\, such as cognition\, emotional regulation\, or behaviour\, especially in social contexts.
            [Anxiety Disorders]
            [Obsessive-Compulsive Disorder (OCD)]
        ]
        [\textbf{Other Disorders:}
        Neurological disorders are not limited to the above-mentioned examples. However\, these neurological disorders have largely been neglected in the eye-tracking community but could potentially be diagnosed similarly.
        ]
    ]
\end{forest}
\caption{Categories and definitions of neural disorders adapted from the American Psychiatric Association Diagnostic and Statistical Manual of Mental Disorders, Fifth Edition, (DSM-5) \cite{svenaeus2014diagnosing}. Note that the same disorder can have different categories in different fields; in this paper, we classify disorders according to their primary class.}\label{fig:tree}
\end{figure}
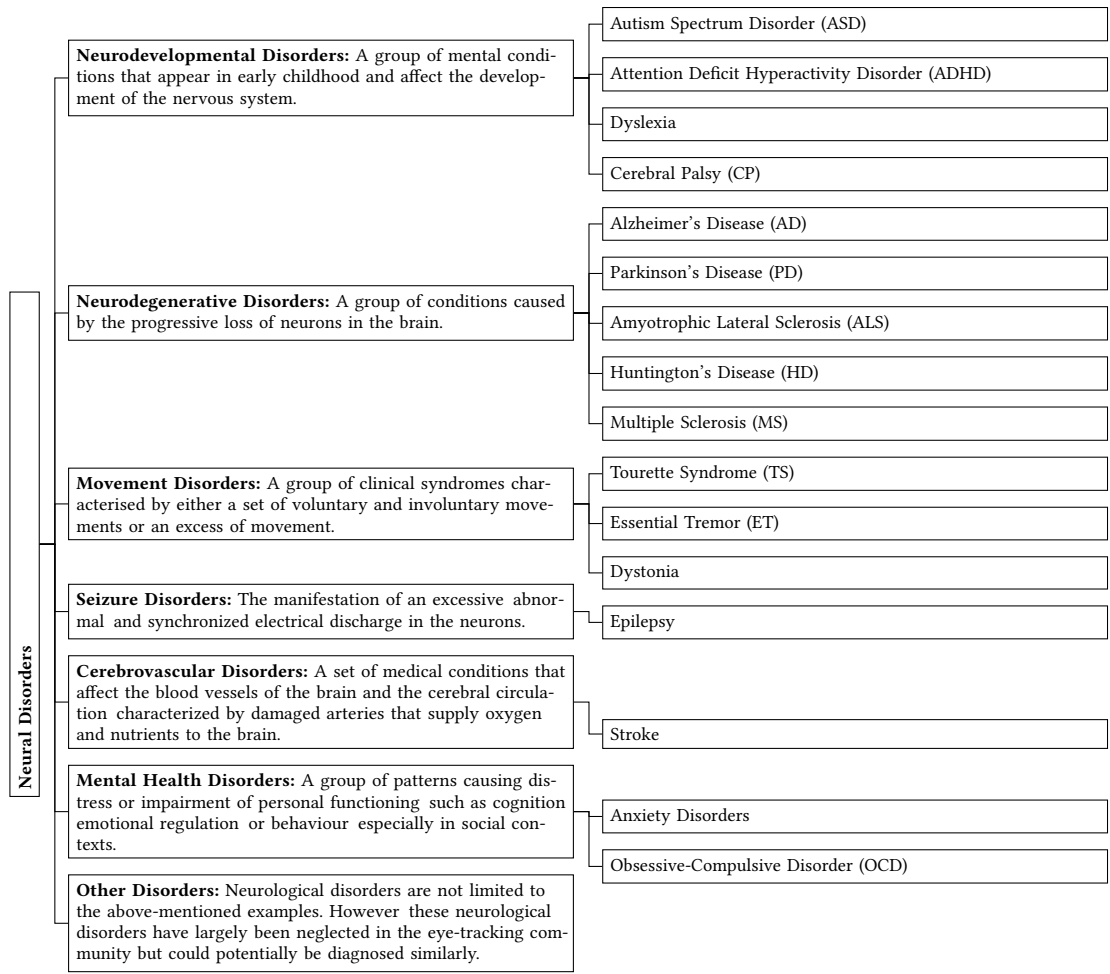

In the following, as shown in Figure \ref{fig:tree}, we introduce a taxonomy for the main categories along with specific guidelines for conducting gaze-based diagnosis research.
For each disorder, we compiled the agreed-upon key eye-tracking focus, task design, considerations, and metrics that were widely employed or recommended in the literature.

\subsection{Neurodevelopmental Disorders} 
    \paragraph{\textbf{Autism Spectrum Disorder (ASD):}} \label{para:asd}
    ASD affects social communication and behaviour, with a wide range of symptoms and severity levels. 
    Most commonly, ASD is qualitatively characterized by deficits in social interaction, restrictive and repetitive patterns of behaviour or interests, attentional disengagement, and impairments in communication \cite{constantino2021clinical, wass2015shorter, johnson2021cognitive, chien2022game}. Impairments across sensory modalities (touch, smell, vision, hearing, taste) \rev{could be} correlated with an inability to respond to social cues, poor social problem-solving, and increased isolation in ASD \cite{klin2020affording, walton2020brief, rudling2024infant, jonsdottir2023autistic, palmer2018autistic}. However, the most reliably replicated sign of ASD is differences in social attention, found in attenuated eye gaze \cite{stuart2023eye,klin2020affording, walton2020brief, bast2023sensory}. Such atypical gaze patterns are consistent across age groups, suggesting that gaze-tracking could be used as an early \rev{indicator for ASD diagnosis}, potentially identifying ASD-related behaviours in infancy before other symptoms are noticeable \cite{rudling2024infant, de2023eye}. Furthermore, the assumption that individuals with ASD exhibit 'impaired' attentional orientation and engagement behaviours has been both corroborated and disputed in various contexts due to experimental differences mentioned above (c.f. section \ref{sec:gaze}). Some studies  \cite{jones2008absence, speer2007face, guillon2014visual} have shown an association between reduced visual fixation on the eye region and greater social difficulties. Other studies \cite{schauder2019initial, best2010gender, pantelis2017deconstructing, bal2010emotion, wieckowski2017eye, jording2024impaired} have indicated that ASD individuals were indeed slower to initially fixate and maintain fixation on faces. They spend significantly less time looking at faces relative to non-social stimuli in comparison to typically developing (TD) individuals; nonetheless, they showed optimal facial identification, gender discrimination, and emotional recognition abilities, which corroborate those of other studies \cite{baker2006neurodiversity, gillespie2013atypical, thorup2016altered, viktorsson202318}, which propose that deficits related to facial processing in ASD do not begin with the initial eye movement, but may begin more downstream and that individuals with ASD might use facial information differently (e.g. increased gaze towards the mouth more than both eyes).
    
    \begin{itemize}[leftmargin =*]
        \item Key Eye-Tracking Focus: Social attention, gaze fixation on faces, and responses to social stimuli.
        \item Task Design: (i) Use dynamic social scenes (e.g., videos of people interacting) to measure how individuals with ASD focus on social cues like facial expressions, eye regions, and gestures \cite{grossman2019perceptions, grossman2019facetime, macinska2024visual}. (ii) In addition to social stimuli, non-social stimuli are used as a control to contrast gaze patterns. For instance, show participants neutral objects to see if their attention differs from social stimuli \cite{stefanelli2024pupillary, mo2019shifting}. (iii) Avoid overly complex visual stimuli, as sensory overload can easily overstimulate ASD participants \cite{trevisan2017adults, bowman2004gaze}.
        \item Considerations: (i) ASD participants may have difficulty sitting still, so shorter tasks or frequent breaks are important to maintain engagement \cite{leary1996moving, torres2013autism}. (ii) Some individuals with ASD have heightened sensitivity to bright lights or sudden movements, so use gentle transitions in stimuli \cite{boyd2022manipulating, giarelli2014sensory}. 
        \item Metrics: (i) Reduced fixation on the eye region of faces is often a key indicator in ASD studies \cite{dalton2005gaze, stuart2023eye}. (ii) Saccadic patterns when looking at social stimuli can help measure social engagement. i.e. quick eye shifts can indicate disengagement \cite{zalla2018saccadic, stuart2023eye}. (iii) Pupil dilation can indicate cognitive load or emotional response, but it is not a conclusive biomarker for ASD \cite{camero2021gaze, galazka2024pupil}. (iv) Additionally, heatmaps and scan path visualization can indicate how attention is distributed during specific tasks, often showing less focus on socially relevant areas (i.e. ROI) in ASD participants \cite{klin2020affording, walton2020brief}. (v) Delayed or reduced blinking might correlate with difficulties in social engagement,  interpreting social cues, or sensory processing, while heightened blink frequency could be linked to increased stress or discomfort \cite{laycock2020blink, amirault2009alteration}.
    \end{itemize}

    \paragraph{\textbf{Attention Deficit Hyperactivity Disorder (ADHD):}}  \label{para:adhd}
    ADHD is characterized by inattention, hyperactivity, and impulsivity \cite{svenaeus2014diagnosing}.
    One of the most notable gaze-related differences in individuals with ADHD is reduced sustained attention to task-relevant stimuli \cite{tye2013neurophysiological, mauriello2022dysfunctional, motomura2023effects, uono2023reduced}. ADHD participants may exhibit shorter fixation durations and more frequent saccadic movements, indicating difficulty in maintaining attention. Eye-tracking studies \cite{tye2013neurophysiological, shen2023evaluating, mauriello2022dysfunctional, airdrie2018facial, berger2014effect, stokes2022measuring} have demonstrated that individuals with ADHD tend to shift their gaze more frequently and are less able to remain focused on a single point of interest, especially during tasks that require prolonged attention such as reading or visual search tasks, where sustained gaze on relevant areas is critical for performance.
    Additionally, eye-tracking research in ADHD often focuses on impulsivity \rev{(less goal-oriented eye movements as opposed to attention, indicated by shorter fixation durations, higher saccade frequency and amplitudes, increased blink rate and microsaccades, and fluctuating pupil size)}\cite{selaskowski2023gaze, mauriello2022dysfunctional, sweere2022clinical}, as measured by premature gaze shifts or difficulties in controlling eye movements. For instance, in tasks like the antisaccade task, where participants are instructed to suppress automatic eye movements toward a distractor and instead look in the opposite direction, individuals with ADHD tend to show higher error rates, reflecting impaired inhibitory control. This aligns with the impulsivity and difficulty with self-regulation seen in ADHD. Other gaze-related metrics used in ADHD research include pupil dilation where some studies \cite{rojas2019pupil, kleberg2021increased, varela2019clinical} suggest that children and adults with ADHD may show abnormal patterns of pupil dilation when faced with cognitive demands, reflecting differences in the way their brains allocate attention and resources. Gaze variability is another important metric, where individuals with ADHD display greater variability in where they direct their attention, indicating challenges in maintaining a stable focus on task-relevant areas \cite{michalek2019predicting, cazzato2019non}.
    
    \begin{itemize}[leftmargin =*]
        \item Key Eye-Tracking Focus: Sustained attention, impulsive saccades, and visual exploration.
        \item Task Design: (i) Use tasks that require visual attention over time, such as continuous performance tasks (CPT), to observe lapses in attention \cite{lev2022eye}. (ii) Implement tasks that challenge impulse control, such as "go/no-go" tasks where participants must inhibit responses to certain stimuli \cite{putra2021identifying, yerys2013separate}.
        \item Considerations: (i) ADHD participants may struggle with longer tasks, so break up the sessions and avoid tasks that require sustained effort without rewards \cite{selaskowski2023gaze, mauriello2022dysfunctional}. (ii) Incentives like gamified tasks can help keep participants engaged throughout the study \cite{lavalle4713241gamifying, craven2015computer, ikuse2022effects}.
        \item Metrics: (i) Track fixation durations and saccade frequencies. Short fixations and frequent saccades may indicate attentional difficulties \cite{tye2013neurophysiological, shen2023evaluating, mauriello2022dysfunctional, airdrie2018facial, lee2015saccadic}. (ii) Blink rate can be an additional measure, as higher rates may correlate with inattention \cite{fried2014adhd, mauriello2022dysfunctional, armstrong2003attentional}. (iii) Individuals with ADHD, with a tendency to have smaller baseline pupil sizes, may exhibit a delayed or exaggerated pupil dilation response to cognitive tasks or emotional stimuli \cite{rojas2019pupil, kleberg2021increased, varela2019clinical}. (iv) Individuals with ADHD often exhibit poorer smooth pursuit than \rev{TD} individuals and might demonstrate increased saccadic intrusions (i.e., jerky eye movements) during pursuit \cite{ross2000smooth, tajik2011smooth, lee2010pursuit}. (v) Microssacades are increased and can have altered frequency, potentially linked to difficulties in visual fixation and attentional control (i.e. reduced microsaccadic inhibition) \cite{fried2014adhd, panagiotidi2017increased}.
    \end{itemize}

    \paragraph{\textbf{Dyslexia:}}  \label{para:dyslexia}
    Dyslexia is a learning disorder that affects reading and language processing \cite{svenaeus2014diagnosing}.
    Dyslexia is primarily characterized by difficulties with phonological processing and word decoding, but studies \cite{fischer2000stability, wang2019gaze, saluja2019analyzing, bellocchi2013can, whitford2023eye, neruvsil2021eye} indicate that eye movement patterns during reading can offer valuable insights into the cognitive challenges faced by individuals with dyslexia. 
    Compared to TD, individuals with dyslexia tend to have longer fixation durations, reflecting the increased cognitive load required to decode words \cite{prabha2020predictive, vagge2015evaluation, zhou2023pathogenesis}. Additionally, dyslexic readers exhibit more frequent regressive saccades, indicating difficulties in word recognition and comprehension, as they often need to re-scan previously read words to make sense of the text \cite{bucci2008poor, fischer2000stability}. Studies also highlight that dyslexic readers may show more variable saccadic amplitudes, meaning their eye movements between words are less consistent, further slowing down reading speed and reducing comprehension \cite{bucci2008poor, prabha2020predictive}. Dyslexic readers typically show increased pupil dilation during reading tasks, suggesting that their brains are working harder to process the same visual and phonological information as non-dyslexic readers (i.e. cognitive effort) \cite{parikh2018predicting, wang2019gaze}.
    
    \begin{itemize}[leftmargin =*]
        \item Key Eye-Tracking Focus: Reading difficulties, word recognition, and saccadic movements during reading tasks.
        \item Task Design: (i) Use text-based tasks where participants read sentences or paragraphs, allowing researchers to track eye movements during reading \cite{fischer2000stability, wang2019gaze, saluja2019analyzing, bellocchi2013can, neruvsil2021eye}. (ii) Include different reading difficulty levels, such as phonemically complex words or sentences, to evaluate how gaze changes with increasing reading challenge \cite{mcchesney2018gaze, bellocchi2013can}. (iii) Include non-text control tasks (e.g., images or symbols) to assess if dyslexia-related gaze behaviours are specific to reading or extend to other visual tasks \cite{el2022predicting, kim2014investigating, temelturk2022binocular}.
        \item Considerations: (i) Tasks should be adapted to the participant's reading level (e.g. age), as dyslexia may result in significant variation in reading ability \cite{fischer2000stability}. 
        \item Metrics: (i) Fixation duration and regressions (returning to previously viewed text) are important indicators of reading difficulty \cite{prabha2020predictive, saluja2019analyzing}. (ii) Measure saccade length — shorter saccades and more frequent regressions typically indicate reading difficulty \cite{prabha2020predictive, wang2019gaze}.
    \end{itemize}

    \paragraph{\textbf{Cerebral Palsy (CP):}}  \label{para:cp}
    A motor disability caused by damage to the developing brain, affecting movement and posture and impacting the oculomotor system, leading to eye movement difficulties, reduced visual tracking ability, and gaze stability issues \cite{svenaeus2014diagnosing}. Gaze-based research in CP focuses on assessing visual, cognitive, and motor control impairments that are often associated with this condition. Research \cite{katayama1987saccadic, maioli2019visuospatial} shows that individuals with CP may have slower saccades, reduced accuracy in their eye movements, and difficulties with smooth pursuit, particularly in tasks that require precise gaze coordination with physical movements. These challenges are linked to motor control issues affecting the muscles responsible for eye movement, which can also hinder overall visual perception \cite{saavedra2009eye}. In addition, individuals with CP may struggle to maintain consistent attention on visual stimuli, often exhibiting longer fixation durations and delayed response times to visual cues \cite{maioli2019visuospatial, sargent2013use, lampe2014eye}. These abnormalities are thought to stem from a combination of motor impairments and cognitive difficulties, such as challenges with executive function and attention regulation \cite{sargent2013use, stadskleiv2017executive}. 

    \begin{itemize}[leftmargin =*]
        \item Key eye-tracking focus: Oculomotor impairments include gaze stability, smooth pursuit, controlling saccades, visual attention, and cognitive response. 
        \item Task design: (i) Use smooth pursuit to measure motor coordination between eyes and head \cite{gauthier1983visual, almutairi2018vestibular, ballester2018cognitive}. (ii) Design tasks that require maintaining gaze on a fixed target for a sustained period to assess fixation stability and motor control over eye movements \cite{katayama1987saccadic, maioli2019visuospatial}.  (iii) Display different objects or faces and ask participants to focus on specific ones to assess visual recognition and attention, a valuable way to evaluate memory and comprehension in individuals with severe CP \cite{clarke2012assessing, ballester2018cognitive}.
        \item Considerations: (i) Individuals with CP often have involuntary movements that can interfere with head-mounted or screen-based eye trackers \cite{saavedra2009eye}. Recalibrate the system frequently and use systems that allow for greater movement tolerance and use head stabilization techniques. (ii) Some individuals with CP may also have cognitive delays or sensory impairments (e.g., cortical visual impairment) \cite{sargent2013use, stadskleiv2017executive, saavedra2009eye}. Adjust stimuli to ensure they are visually simple and easy to comprehend, minimising cognitive overloads, such as large, contrasting shapes or images. (iii) Calibration can be difficult for individuals with motor impairments who struggle to focus on specific points for extended periods \cite{saavedra2009eye, ballester2018cognitive}. Use simplified calibration processes or allow more time for calibration to ensure accurate tracking. (iv) Where possible, incorporate adaptive technology like gaze-controlled communication devices for participants with severe motor impairments, enabling them to express cognitive responses without physical input \cite{galante2012gaze}.
        \item Metrics: (i) Measure the latency (time delay) in initiating saccades and the accuracy in reaching visual targets. Individuals with CP may have delays or inaccuracies due to motor control issues \cite{saavedra2009eye}. (ii) Saccade velocity (speed of eye movement) and saccade amplitude (distance of eye movement) can also be indicative of motor impairments \cite{katayama1987saccadic}. (iii) Assess how long participants can maintain a steady fixation on a target without drifting \cite{saavedra2009eye, ballester2018cognitive}. Fixation instability may reflect motor control challenges that are common in CP. (iv) Measure total fixation time on relevant stimuli, especially in tasks where attention to a specific target is required \cite{maioli2019visuospatial, sargent2013use}. (v) Smooth pursuit gain measures how closely the eyes track a moving object. In individuals with CP, reduced smooth pursuit gain (i.e., inability to follow a moving target smoothly) can reflect both motor and coordination difficulties \cite{gauthier1983visual, almutairi2018vestibular, ballester2018cognitive}. (vi) Use scanpath analysis to evaluate how participants explore visual stimuli. Irregular scanpaths may indicate difficulties in planning or controlling eye movements \cite{lampe2014eye, tenenbaum2021distance}. 
    \end{itemize}

\subsection{Neurodegenerative Disorders}

    \paragraph{\textbf{Alzheimer’s Disease (AD):}}  \label{para:ad}
    A progressive disorder causing memory loss, confusion, and cognitive decline due to brain cell death. It further impacts visual processing and oculomotor control \cite{svenaeus2014diagnosing}. 
    In individuals with AD, gaze-based studies typically reveal longer fixation durations and increased difficulty in shifting gaze from one visual target to another \cite{uomori1993analysis, insch2017gaze, antoniades2015ocular}. These oculomotor impairments are linked to the degeneration of brain regions responsible for executive function, spatial orientation, and visual attention \cite{castellani2010alzheimer, svenaeus2014diagnosing}. For instance, tasks that require visual search or attention to multiple objects are often challenging for Alzheimer’s patients, who tend to have slower and less accurate saccades, reflecting difficulties in efficiently scanning their environment \cite{yamada2024distinct, pereira2020visual, li2024automating, lopis2019investigating, mapstone2001dynamic}. Additionally, people with AD often show reduced attention to important visual cues, such as faces or emotionally relevant stimuli, which correlates with their broader cognitive and memory deficits \cite{sturm2011mutual, ramzaoui2018alzheimer, insch2017gaze, polet2022eye, comon2022impaired}. 
    A particular area of interest is eye movement behaviour in reading tasks \cite{groznik2021gaze, fernandez2015diagnosis, guantay2024accounting, heidarzadeh2023eye}. Alzheimer’s patients exhibit increased fixation times on words and more frequent regressive saccades (backward eye movements), indicating difficulties in decoding text and retaining information across sentences. These reading difficulties are tied to the memory and comprehension deficits (i.e. language processing and cognitive decline) that define the disease \cite{fernandez2014lack, lopis2019eye}.
   
    \begin{itemize}[leftmargin =*]
        \item Key Eye-Tracking Focus: Visual attention, memory recall, and fixation patterns during memory tasks.
        \item Task Design: (i) Use tasks involving object or face recognition to assess memory and attention, such as presenting familiar versus unfamiliar faces and tracking gaze \cite{sturm2011mutual, polet2022eye}. (ii) Implement visual search tasks to measure how participants scan and process visual information, such as finding a target among distractors \cite{pereira2020visual, mapstone2001dynamic}.
        \item Considerations: (i) Dementia can affect attention span and comprehension. Tasks should be simple and visually clear to prevent confusion \cite{readman2021potential}. (ii) Avoid complex stimuli or rapid changes, as they may overwhelm or confuse participants with cognitive impairment \cite{yamada2024distinct}.
        \item Metrics: (i) Measure fixation duration and time to first fixation on familiar objects or faces versus novel ones \cite{insch2017gaze, sturm2011mutual}. (ii) Track scanpaths to analyze how participants explore visual stimuli, as irregular patterns can indicate cognitive impairment \cite{ramzaoui2018alzheimer, readman2021potential, fletcher1986saccadic}. (iii) Pupil dilation is another measure that is studied, as the reduced response to stimuli can indicate changes in cognitive effort or arousal during tasks. Individuals with Alzheimer’s tend to have abnormal pupil responses, reflecting the increased cognitive load as they attempt to process information \cite{ramzaoui2018alzheimer, fotiou2007pupil, granholm2017pupillary}. (iv) Alzheimer's patients typically exhibit a reduced blink rate, which may reflect cognitive slowing and decreased sensory engagement \cite{d2021blink, mohammadian2015blink, srivastava2020blink}. (v) 
        Alzheimer's patients often show altered oblique microsaccadic behaviour. This may reflect impaired visual attention and gaze stability, potentially indicating that the brain is attempting to compensate for visual or cognitive decline by making more frequent small eye movements \cite{kapoula2014distinctive, alexander2018microsaccade}. (vi) Alzheimer's patients may also demonstrate reduced accuracy in smooth pursuit. Patients may find it difficult to maintain fixation on moving objects, leading to more frequent corrective saccades to reposition the gaze \cite{fletcher1988smooth, zaccara1992smooth, frei2021abnormalities}.
    \end{itemize}

    \paragraph{\textbf{Parkinson’s Disease (PD):}}  \label{para:pd}
    A disorder affecting movement, characterized by tremors, stiffness, and slow movement. It further impacts areas of the brain responsible for eye movement and visual processing \cite{fernandez2015diagnosis, kuechenmeister1977eye}. A key feature of PD is bradykinesia (slowness of movement), which is often reflected in slower saccadic movements and longer fixation durations \cite{dejong1971akinesia, hodgson2002abnormal, tsitsi2021fixation, li2023abnormal, jung2019abnormal, shibasaki1979oculomotor, cstefcsnescu2024eye}. Individuals with Parkinson’s may struggle to initiate saccades or exhibit hypometric saccades, meaning their eyes undershoot visual targets, requiring multiple corrective movements \cite{shaikh2019saccades, pretegiani2017eye, kimmig2002pathological, cstefcsnescu2024eye}. These abnormalities can be particularly noticeable in tasks that require shifting attention, such as visual search or reading, suggesting a disruption in the coordination between cognitive planning and motor execution in PD \cite{hodgson2024gaze, gotardi2022parkinson, terao2023patients, falaki2023multi}.
    In addition, Parkinson’s patients frequently show deficits in smooth pursuit, particularly in tracking objects with predictable motion, often leading to jerky or interrupted eye movements \cite{wu2018eye, helmchen2012role, pinkhardt2009comparison, marino2010effect, white1983ocular}. Furthermore, patients with PD may struggle with visual attention tasks, particularly those that involve divided attention or responding to multiple stimuli \cite{stuart2017direct, hodgson2002abnormal, briand2001automatic, cstefcsnescu2024eye}.
    
    \begin{itemize}[leftmargin =*]
        \item Key Eye-Tracking Focus: Eye movement control, including saccadic movements, smooth pursuit, and fixation stability.
        \item Task Design: (i) Use tasks that require smooth pursuit to assess motor control related to eye movements \cite{wu2018eye, helmchen2012role, white1983ocular}. (ii) Include gaze-following tasks to assess whether eye-tracking can identify difficulties in attention and motor integration \cite{hodgson2002abnormal, tsitsi2021fixation}. (iii) Use antisaccadic tasks to understand the inhibitory mechanisms \cite{hapakova2024antisaccadic, brien2023classification, li2023abnormal, waldthaler2021antisaccades}.
        \item Considerations: (i) Parkinson’s participants may struggle with motor control, so ensure that the eye-tracking system can account for potential head movements or tremors \cite{simieli2017gaze}.
        \item Metrics: (i) Measure saccade latency, saccade amplitude, and fixation duration, as motor delays can manifest in these aspects \cite{simieli2017gaze, shaikh2019saccades, pretegiani2017eye, kimmig2002pathological, likitgorn2021freezing}. (ii) Assess smooth pursuit gain (the ratio of eye velocity to target velocity), as reduced gain is common in Parkinson’s \cite{wu2018eye, helmchen2012role}. (iii) Measure pupil dilation to measure cognitive effort during tasks since Parkinson’s patients often show abnormal pupil responses \cite{tsitsi2021fixation, ranchet2017pupillary}. (iv) People with PD often exhibit less stable fixations, with increased fixation duration and more difficulty in maintaining a steady gaze (increased drifts) \cite{tsitsi2021fixation, pinnock2010exploration, revankar2020ocular}. (v) PD patients often show a reduction in the frequency of microsaccades due to the corrective behaviour \cite{otero2013saccades, mcinnis2014microsaccades}.
    \end{itemize}

    \paragraph{\textbf{Amyotrophic Lateral Sclerosis (ALS):}}  \label{para:als}
    Also known as Lou Gehrig’s disease, ALS leads to the gradual degeneration of motor neurons, affecting voluntary muscle control \cite{svenaeus2014diagnosing}. While ALS typically spares the muscles responsible for eye movement until the later stages, gaze-based studies provide valuable insights into the cognitive deficits and communication challenges faced by ALS patients \cite{aust2024impairment}. Eye-tracking studies have revealed slower saccades and difficulty in maintaining steady fixation on visual targets, which may reflect the spread of motor neuron degeneration to the muscles controlling eye movement \cite{proudfoot2016eye, becker2019saccadic, zaino2023different, burrell2013early, antoniades2015ocular}. These oculomotor impairments, although usually mild, can be used to monitor disease progression and help distinguish ALS from other neurodegenerative disorders, such as Parkinson’s or Alzheimer’s disease, where eye movement abnormalities are more pronounced earlier in the disease course \cite{sharma2011oculomotor}.
    In addition, while ALS is primarily a motor disorder, up to 50\% of patients experience some form of cognitive impairment, often in the form of executive dysfunction or frontotemporal dementia (FTD) \cite{saxon2020cognition, nanning2023altered}. Eye-tracking tasks that measure visual attention, working memory, and decision-making can help detect these cognitive deficits \cite{kane2017let, severens2014comparing, nonoka2024study}. For example, ALS patients with cognitive impairment may show longer fixation times or difficulty in shifting gaze during tasks that require decision-making or response inhibition, indicating problems with cognitive control and flexibility \cite{proudfoot2016eye, burrell2013early, leveille1982eye, keller2015eye}. 
    
    \begin{itemize}[leftmargin =*]
        \item Key Eye-Tracking Focus: Oculomotor Control
        \item Task Design: (i) Use antisaccade tasks to assess higher-order executive functions and control over involuntary motor movements \cite{proudfoot2016eye, becker2019saccadic, sharma2011oculomotor}. (ii) Include smooth pursuit tasks to measure the progression of ALS over time \cite{leveille1982eye, guo2022eye, jacobs1981eye}. (iii) Fixation stability tasks can further detect motor impairments in ALS \cite{proudfoot2016eye, becker2019saccadic}. (iv) ALS patients may experience frontotemporal dementia, leading to cognitive decline \cite{saxon2020cognition}. Designing tasks that measure both motor and cognitive performance separately.
        \item Considerations: (i) Ensure that patients with severe motor impairment can participate in the tasks (c.f. \ref{sec:generic}).
        \item Metrics: (i) Measure fixation duration; irregular fixation durations may indicate difficulties in motor control, which can reflect ALS progression \cite{proudfoot2016eye, zaino2023different, burrell2013early}. (ii) Measure saccadic latency and amplitude; increased saccadic latency (slower reaction times) or irregular saccade amplitudes (deviations in the distance of eye movements) are key metrics in identifying motor control issues in ALS patients \cite{becker2019saccadic, leveille1982eye, nakamagoe2023saccadic}. (iii) Assess smooth pursuit gain; lower pursuit gain could suggest impaired motor coordination, which becomes prominent as ALS progresses \cite{leveille1982eye, guo2022eye}. (iv) Measure the error rate in antisaccade tasks; high error rates (looking towards a stimulus instead of away) can reveal deficits in cognitive control and executive function, often associated with ALS-related cognitive impairments \cite{proudfoot2016eye, becker2019saccadic, sharma2011oculomotor}. (v) ALS patients can show abnormalities in pupil responses, leading to altered responses to light or cognitive tasks. Some ALS patients exhibit reduced pupil reactivity, potentially because of dysautonomia, which may impact the speed and extent of dilation and constriction \cite{jlassi2022analysis, oprisan2023dysautonomia}. (vi) ALS may lead to reduced microsaccade frequency and amplitude, especially as the disease progresses. These changes likely arise from the degeneration of motor neurons that impact the precision of small eye movements \cite{becker2019saccadic, guo2022eye, nakamagoe2023saccadic}.
    \end{itemize}

    \paragraph{\textbf{Huntington’s Disease (HD):}}  \label{para:hd}
    A genetic disorder causing the progressive breakdown of nerve cells in the brain, affecting movement, cognition, and \rev{causing psychiatry conditions} \cite{svenaeus2014diagnosing}. A prominent oculomotor symptom in HD is the impairment of voluntary saccadic eye movements \cite{becker2009eye, porciuncula2020quantifying, blekher2006saccades, sierra2024deciphering, grabska2024reflexive}. Individuals with Huntington’s often show slowed saccades and increased latency in initiating them \cite{reyes2024saccades, kassavetis2022eye, leigh1983abnormal, peltsch2008saccadic}. Studies using eye-tracking have found that HD patients take longer to initiate saccades and frequently require multiple attempts to reach visual targets accurately \cite{tian1991saccades, olivetti2021abnormal, lasker1987saccades}. In addition to saccadic abnormalities, Huntington’s disease also affects smooth pursuit; studies have revealed that HD patients have impaired smooth pursuit, often exhibiting jerky and fragmented eye movements \cite{starr1967disorder, patel2022evolution, kennard2011disorders, antoniades2015ocular}. These deficits are early indicators of the motor control dysfunctions seen in HD.
    Cognitive dysfunction is another key feature of HD. For instance, individuals with HD often show prolonged fixation durations and difficulty in shifting attention between different visual stimuli, reflecting impairments in cognitive flexibility and decision-making \cite{becker2009eye, blekher2006saccades, olivetti2021abnormal, snowden2003social, kordsachia2017abnormal}. These cognitive symptoms often appear before significant motor decline. They can be tracked using eye movements, which may provide early diagnostic clues and help monitor the cognitive aspects of HD over time \cite{rubin1993quantitative, rupp2011abnormal, snowden2003social}.
    
    \begin{itemize}[leftmargin =*]
        \item Key Eye-Tracking Focus: Oculomotor dysfunction and cognitive decline.
        \item Task Design: (i) The most commonly used tasks for HD patients involve testing the accuracy and speed of saccadic eye movements. Both reflexive (in response to a suddenly-appearing stimulus) and voluntary saccades (self-initiated look at a stimulus on command) can be impaired in HD \cite{becker2009eye, porciuncula2020quantifying, blekher2006saccades, turner2011behavioral, patel2012reflexive}. (ii) Employ antisaccade tasks; HD patients often have trouble suppressing the reflex to look at the stimulus, revealing both motor and cognitive control issues \cite{becker2009eye, blekher2006saccades, peltsch2008saccadic, hapakova2024antisaccadic}. (iii) Smooth pursuit tasks are also recommended since HD patients tend to show lower smooth pursuit gain \cite{starr1967disorder, patel2022evolution, kennard2011disorders, attoni2016abnormal}. (iv) Use fixation tasks as patients with HD often require corrective eye movements to hold their gaze \cite{becker2009eye, porciuncula2020quantifying, attoni2016abnormal}.
        \item Considerations: (i) Check early vs. advanced stages through longitudinal studies: In the early stages, HD patients may have relatively subtle oculomotor impairments before HD becomes clinically evident \cite{rubin1993quantitative}. (ii) Since HD affects both motor and cognitive functions, task designs should account for both types of deficits. For example, tasks that require simultaneous movement and decision-making can be valuable for identifying early impairments \cite{kordsachia2018visual, porciuncula2020quantifying, olivetti2021abnormal}.
        \item Metrics: (i) Increased latency and reduced accuracy of saccadic eye movements are hallmark features of HD \cite{bollen1986horizontal, reyes2024saccades, kassavetis2022eye, leigh1983abnormal}. (ii) Measure smooth pursuit gain as it typically decreases in HD patients \cite{starr1967disorder, patel2022evolution, kennard2011disorders}. (iii) High error rates in antisaccade tasks are also common in HD \cite{becker2009eye, blekher2006saccades, peltsch2008saccadic, hapakova2024antisaccadic}. (iv) Measure the fixation stability: HD patients may have difficulty maintaining fixation on a single point, with increased instability and corrective saccades \cite{becker2009eye, beenen1986diagnostic}. (v) Patients with HD often show an increased frequency of microsaccades during fixation, suggesting difficulty in maintaining a stable gaze \cite{willard2014ocular, reyes2024saccades}.
    \end{itemize}

    \paragraph{\textbf{Multiple Sclerosis (MS):}}  \label{para:ms}
    MS is a chronic condition where the immune system attacks the myelin sheath of nerve fibres, leading to a range of symptoms like numbness, fatigue, and difficulty walking \cite{svenaeus2014diagnosing}. Eye-tracking technology has proven valuable in studying these effects, as it can assess visual dysfunction, which is one of the earliest and most common symptoms of MS, such as delayed saccades, prolonged fixation durations, and increased saccadic latency \cite{manago2016gaze, loyd2022rehabilitation, garg2018gaze, de2023oculomotor}. Eye-tracking can detect internuclear ophthalmoplegia (INO - an oculomotor disorder common in MS) saccadic abnormalities by measuring the misalignment of gaze or impaired coordination between the eyes \cite{johkura2015gaze, muri1985clinical, guantay2024usefulness}. Furthermore, research shows that MS patients may exhibit increased microsaccades during attempts to maintain a steady gaze \cite{manago2016gaze, loyd2022rehabilitation, garg2018gaze}. Eye-tracking technology is also able to precisely quantify the amplitude and frequency of nystagmus (a disruption in visual stability that makes it difficult for individuals to focus on stationary objects) \cite{nerrant2017ocular, prasad2010eye, guantay2024usefulness, feys2008unsteady}. Moreover, eye-tracking has been used to assess smooth pursuit deficits, which often manifest as jerky or fragmented tracking \cite{jozefowicz2011evaluation, de2011pursuit, sheehy2018methods}. By analyzing smooth pursuit metrics, eye-tracking can offer insights into how MS affects sensorimotor integration and visual processing, helping to differentiate MS from other neurological conditions. Finally, MS frequently leads to cognitive dysfunction, particularly in areas such as working memory, executive function, and visual attention \cite{polet2023facial, gerardo2024abnormal}. Eye-tracking studies have shown that individuals with MS may have difficulty maintaining attention during visual tasks, leading to longer fixation times or difficulties shifting their gaze between targets \cite{manago2016gaze, williams2016divided}.

    \begin{itemize}[leftmargin =*]
        \item Key Eye-Tracking Focus: Oculomotor dysfunction (as deficits in smooth pursuit, saccadic movements, and fixation) and cognitive and motor decline.
        \item Task Design: (i) Use saccadic tasks; MS patients often show abnormal (reflexive and voluntary) saccadic behaviour, such as increased latency (reaction time) and hypermetric saccades (reduced saccade amplitude). Common tasks involve looking at a series of stimuli that appear in different locations to assess the speed and accuracy of saccadic movements \cite{sheehy2018methods, manago2016gaze, garg2018gaze}. (ii) Add smooth pursuit tasks; MS patients may show reduced smooth pursuit gain, which means they rely more on corrective saccades to follow the moving target \cite{jozefowicz2011evaluation, de2011pursuit, sheehy2018methods}. (iii) Fixation tasks: MS patients may exhibit difficulty maintaining stable fixation, and frequent involuntary corrective saccades (nystagmus) are common \cite{manago2016gaze, nerrant2017ocular, prasad2010eye, feys2008unsteady}. (iv) Since MS can affect cognitive functions like attention and processing speed, employ tasks that combine motor (eye movement) and cognitive challenges. For example, antisaccade tasks that require patients to inhibit reflexive saccades and look in the opposite direction of a stimulus help assess executive function \cite{manago2016gaze, williams2016divided}.
        \item Considerations: (i) MS symptoms vary widely between patients; patients may show early signs of oculomotor issues even if other symptoms are mild \cite{prasad2010eye, manago2016gaze}. (ii) MS progresses unpredictably, with periods of remission and relapse. Eye-tracking tests should be conducted over time to monitor fluctuations in symptom severity \cite{loyd2022rehabilitation, prasad2010eye, garcia2022toward}. (iii) Due to reduced blink rate or incomplete blinks, many MS patients experience dry eye symptoms, which may lead to compensatory blinking in response to eye irritation \cite{sorkhabi2024frequency}.
        \item Metrics: (i) Increased latency and hypometric saccades (shortened eye movements) are common in MS patients \cite{manago2016gaze, loyd2022rehabilitation, garg2018gaze}. (ii) Smooth pursuit gain is commonly reduced in MS patients \cite{jozefowicz2011evaluation, de2011pursuit, sheehy2018methods}. (iii) Check for nystagmus (involuntary eye movements), particularly during fixation tasks for eye stability \cite{nerrant2017ocular, prasad2010eye}. (iv) MS can lead to delays in pupil dilation and constriction. This can result in slower or reduced pupil responses to light and other stimuli \cite{de2017fatigue, moro2007recovery, surakka2008pupillary, jakobsen1990pupillary, de2001pupillary}. (v)  Microsaccades in MS are often less frequent and more variable in both amplitude and direction, which may reflect broader impairments in motor coordination. This variability can interfere with stable gaze maintenance, particularly during tasks requiring precise focus \cite{sheehy2020fixational, sheehy2018methods, sheehy2018validation}.
    \end{itemize}

\subsection{Movement Disorders}

    \paragraph{\textbf{Tourette Syndrome (TS):}}  \label{para:ts}
    TS is mainly characterized by repetitive, involuntary movements and vocalizations (tics) but also affects attention, cognitive control, and impulsivity \cite{svenaeus2014diagnosing}. Eye-tracking technology has been used to study how individuals with TS process visual stimuli and how their tics and associated cognitive challenges influence eye movement patterns and visual attention \cite{gaze2006co, georgiou1998directed, mostofsky2001oculomotor, jung2019abnormal, frankel1984neuro}.
    Individuals with TS may exhibit difficulties in sustaining attention or filtering out distractions, which can affect their ability to maintain stable fixation or perform tasks that require continuous eye movements \cite{georgiou1998directed, martino2012prevalence}. Eye-tracking can objectively measure these deficits, for instance, by tracking how long individuals with TS can fixate on a visual target or how efficiently they perform in visual search tasks \cite{shaikh2017fixational, ackermans2007vertical}. Furthermore, while most tics in TS are motor or vocal, some individuals may experience ocular tics or tics that interfere with their normal eye movements \cite{baizabal2022clinical, tatlipinar2001ophthalmic, kovacich2008tourette, tulen1999quantitative}. Eye tracking can help researchers better understand how these tics interact with normal visual processing. For example, studies have shown that tics can disrupt gaze stability and interfere with smooth pursuit movements, particularly when patients are trying to suppress their tics, leading to a compensatory increase in saccadic eye movements or other gaze abnormalities \cite{tajik2011smooth, dursun2000antisaccade, hapakova2024antisaccadic}. Additionally, individuals with TS may exhibit a higher frequency of premature saccades or errors in task performance when asked to withhold or redirect their gaze, reflecting difficulties in cognitive control \cite{tajik2013enhanced, levasseur2001control}.

    \begin{itemize}[leftmargin =*]
        \item Key Eye-Tracking Focus: Gaze control, including the relationship between tics and eye movements.
        \item Task Design: (i) Use tasks that challenge gaze stability, such as tasks where participants need to maintain focus on a fixed point \cite{georgiou1998directed, martino2012prevalence}. (ii) Include response inhibition tasks (e.g., antisaccadic tasks where participants must look away from a stimulus) to evaluate impulse control \cite{tajik2013enhanced, hapakova2024antisaccadic, dursun2000antisaccade}.
        \item Considerations: Tourette syndrome may involve tics that interrupt the study, so the tasks should be designed to accommodate potential interruptions \cite{gaze2006co, georgiou1998directed, mostofsky2001oculomotor, conelea2024automated}. (ii) Provide rest periods to manage fatigue and tic exacerbations \cite{tajik2013enhanced}.
        \item Metrics: (i) Measure saccadic control (ability to suppress unwanted eye movements) and gaze stability \cite{tajik2013enhanced, dursun2000antisaccade, hapakova2024antisaccadic, levasseur2001control}. (ii) Track whether tics coincide with interruptions in eye movements, such as loss of fixation during a task \cite{georgiou1998directed, martino2012prevalence}.
    \end{itemize}

    \paragraph{\textbf{Essential Tremor (ET):}}  \label{para:et}
    ET is the most common tremor disorder and primarily manifests as involuntary, rhythmic shaking of the hands, head, or voice \cite{svenaeus2014diagnosing}. Research has shown that individuals with ET may also experience cognitive symptoms and oculomotor abnormalities, making eye-tracking a useful tool for assessing the broader neurological impact of the disease \cite{wojcik2016eye, helmchen2003eye, jung2019abnormal}. Studies have found that individuals with ET may show mild impairments in saccadic control, such as slowed saccadic velocity and increased latency in initiating saccades \cite{gitchel2013slowed, blondiaux2023impairments}.
    In addition, eye-tracking studies suggest that individuals with ET may experience increased tremor-related instability during fixation, resulting in small, involuntary eye movements that disrupt gaze stability \cite{wojcik2016eye, helmchen2003eye}.  
    Fixation instability in ET patients may also contribute to difficulties in tasks requiring sustained attention, such as reading or tracking moving objects. Furthermore, research has shown that ET patients may exhibit impaired smooth pursuit, characterized by jerky or fragmented tracking of moving stimuli  \cite{wojcik2016eye, helmchen2003eye, gitchel2013slowed}. 
    Although ET is primarily known as a motor disorder, some individuals experience mild cognitive impairment, particularly in areas of executive function and visual processing. Eye-tracking tasks that measure response inhibition or attention-shifting can provide insights into these cognitive symptoms. For example, individuals with ET may show longer fixation durations or have difficulty shifting their gaze when required, indicating deficits in cognitive flexibility \cite{huys2021misdirected, springer2010essential, rekik2023eye}.

    \begin{itemize}[leftmargin =*]
        \item Key Eye-Tracking Focus: Oculomotor control, subtle cognitive impairments.
        \item Task Design: (i) Use fixation tasks since tremor-related involuntary eye movements (micro-saccades or nystagmus) may become evident as patients struggle to maintain a steady gaze \cite{wojcik2016eye, helmchen2003eye}. (ii) Use smooth pursuit tasks since ET patients might experience difficulty in smoothly following a moving target due to tremor involvement in the oculomotor system, leading to an increase in corrective saccades \cite{wojcik2016eye, helmchen2003eye, gitchel2013slowed}. (iii) Conduct reflexive and voluntary saccadic tasks since ET patients may exhibit normal saccadic function but might also display increased latency due to cognitive processing difficulties, which are sometimes associated with long-term ET \cite{gitchel2013slowed, blondiaux2023impairments}.
        (iv) Implement dynamic visual acuity tasks to assess the patient’s ability to focus on moving objects and help reveal oculomotor dysfunctions such as tremor-related issues with eye muscle control \cite{gitchel2013slowed, ko2016eye, gitchel2013slowed}.
        \item Considerations: (i) ET is often confused with Parkinson’s disease because of overlapping tremor symptoms. Eye-tracking may help differentiate the two by identifying distinct oculomotor patterns such as smooth pursuit impairments (more common in Parkinson's patients than in ET patients) \cite{wojcik2016eye, macaskill2020pervasive, zhou2022oculomotor,  kassavetis2022eye}. (ii) The severity of tremors can fluctuate, and eye-tracking metrics should be interpreted in light of the tremor's current state. Repeated measurements over time may be necessary to assess tremor variability \cite{macaskill2020pervasive, rekik2023eye}.
        \item Metrics: (i) Analysing the fixation stability regarding frequency and amplitude of corrective saccades can quantify tremor severity in the oculomotor system \cite{wojcik2016eye, helmchen2003eye}. (ii) Measure the smooth pursuit gain (reduced in ET patients, indicated by more corrective saccades) \cite{wojcik2016eye, helmchen2003eye, gitchel2013slowed}. (iii) Measure the saccadic latency and accuracy to detect any subtle cognitive involvement or differentiate ET from similar disorders like Parkinson’s disease \cite{gitchel2013slowed, blondiaux2023impairments}. (iv) Check for the presence of small, involuntary microsaccades or nystagmus during fixation tasks to indicate tremor-related dysfunction in the eye muscles \cite{kassavetis2022eye, celebisoy2016evaluation, muller19nystagmus}.
    \end{itemize}

    \paragraph{\textbf{Dystonia:}}  \label{para:dystonia}
    Dystonia involves sustained or repetitive muscle contractions that cause twisting and abnormal postures \cite{svenaeus2014diagnosing}. 
    Dystonia primarily affects motor functions, but recent studies have explored how eye movements and gaze behaviours are impacted, particularly in focal dystonias, such as blepharospasm (affecting the eyes) \cite{coles1977signs, jankovic1982blinking, ferrazzano2015botulinum, demer1990ocular} and cervical dystonia (affecting the neck) \cite{shaikh2016cervical, hutchinson2014cervical}.
    For example, individuals with blepharospasm often exhibit increased blink rates and longer blink durations, leading to interruptions in gaze fixation and disrupted visual input \cite{coles1977signs, jankovic1982blinking, ferrazzano2015botulinum, demer1990ocular}. 
    Furthermore, individuals with cervical dystonia may experience difficulty in maintaining a steady gaze or performing smooth pursuit movements, as their abnormal head postures affect the alignment of the eyes and visual field \cite{shaikh2016cervical, hutchinson2014cervical}. Eye-tracking can measure these deficits, including delayed saccades or increased saccadic latency, as patients struggle to maintain or shift their gaze in the intended direction due to muscle contractions in the neck \cite{mahajan2021impaired, hutchinson2014cervical}.
    Gaze-based research is also valuable for studying sensorimotor integration deficits in dystonia. For example, studies \cite{desrochers2019sensorimotor, tramonti2023pointing, hutchinson2014cervical} have found that dystonia patients may exhibit impaired anticipatory gaze movements, suggesting that the brain has difficulty integrating visual information with motor planning. This impairment in sensorimotor integration is particularly evident in tasks that require precise coordination of eye and hand movements, such as tracking a moving target or reaching for an object while maintaining visual fixation. Similarly, for task-specific dystonias, such as writer’s cramp or musician’s dystonia, where involuntary muscle contractions occur only during specific tasks, eye-tracking has been used to measure how visual attention and gaze stability are affected during the performance of these tasks \cite{jhunjhunwala2017abnormalities, maguire2020normal}. For instance, individuals with musician’s dystonia may experience difficulties maintaining steady fixation on musical notes or sheet music, as their motor symptoms interfere with gaze control and visual processing during performance \cite{lim2001focal}. Finally, note that since dystonia primarily affects voluntary muscles, eye movements are not always as prominently impacted as in other movement disorders like Parkinson’s disease or Huntington’s disease \cite{deuschl1998spontaneous}.

    \begin{itemize}[leftmargin =*]
        \item Key Eye-Tracking Focus:
         Oculomotor dysfunction (e.g. eye movement coordination, abnormal saccadic movements, and smooth pursuit), fixation stability, and blink rate.
        \item Task Design: (i) Design simple visual fixation tasks to detect issues with maintaining a steady gaze due to muscle spasms or involuntary contractions \cite{shaikh2016cervical, hutchinson2014cervical}. (ii) Add smooth pursuit tasks to identify problems with tracking and oculomotor control \cite{hirsig2020oculomotor, kassavetis2022eye}. (iii) Measure reflexive and voluntary saccades to assess how well dystonia patients can initiate and control rapid eye movements \cite{mahajan2021impaired, hutchinson2014cervical}.
        \item Considerations: (i) In cases of cervical dystonia, head movements might interfere with gaze tracking. Special equipment like chin rests or head restraints can help isolate eye movements for more accurate measurement \cite{shaikh2016cervical, beck2018reduced}. (ii) Dystonia often leads to muscle fatigue, so task design should account for this by providing breaks to prevent discomfort and poor data quality \cite{pesenti2001transient}.
        \item Metrics: (i) Measure saccadic latency, which may be delayed in dystonia patients \cite{mahajan2021impaired, hutchinson2014cervical}. (ii) Measure the smooth pursuit gain, often diminished in dystonia \cite{hirsig2020oculomotor, shaikh2016cervical, hutchinson2014cervical}. (iii) Measure fixation duration since unstable or prolonged fixation durations are common in dystonia \cite{shaikh2016cervical, hutchinson2014cervical}. (iv) Measure the blink frequency as either reduced or excessive blinking can signal dystonic eye dysfunction \cite{deuschl1998spontaneous, pauletti1993blink, deuschl1998spontaneous}. (v) Finally, tiny and involuntary eye movements (microsaccades) and tremors may be particularly relevant to detecting subtle forms of dystonia that affect the ocular muscles \cite{mahajan2021impaired, hirsig2020oculomotor}.
    \end{itemize}

\subsection{Seizure Disorders} 

    \paragraph{\textbf{Epilepsy:}}  \label{para:epilepsy}
    Epilepsy is a disorder characterized by recurrent seizures, which are abnormal electrical activity in the brain \cite{svenaeus2014diagnosing}. The relationship between eye tracking and epilepsy can provide insights into the visual processing deficits, cognitive impairments, and oculomotor abnormalities that often accompany this disorder, especially in individuals with focal seizures originating in brain regions related to vision and attention, such as show delays in saccadic eye movements, difficulties with gaze fixation, and abnormal smooth pursuit \cite{huang2024reduced, asato2011deficits}. For instance, studies have found that people with epilepsy exhibit increased saccadic latency due to disruptions in neural networks responsible for coordinating eye movements as a biomarker for attention\cite{guerrero2021antisaccades, thurtell2009evidence, lunn2016saccadic}. 
    In addition to saccades, eye-tracking can also measure fixation stability and gaze shifts during epileptic events or interictal periods (the time between seizures) \cite{huang2024reduced, metternich2022eye, hunter2021impaired, gomez2014recognition}. 
    Some individuals with epilepsy experience oculomotor seizures, where eye movements are directly affected by abnormal brain activity. This can include sustained upward gaze or sideward eye deviation \cite{asato2011deficits, thurston1985epileptic}. 
    Additionally, during postictal states (after a seizure), individuals may show impaired gaze control or prolonged fixations, reflecting temporary deficits in cognitive processing and attention recovery \cite{wyllie1986lateralizing, fisher2000postictal, izadyar2018comparison}. 
    Furthermore, people with epilepsy often show deficits in tasks requiring sustained visual attention or complex decision-making, and eye-tracking metrics like fixation duration and visual search patterns can provide insights into these cognitive challenges \cite{lunn2016saccadic, hunter2021impaired, okruszek2017gaze}. For example, individuals with epilepsy may exhibit longer fixation times or difficulty disengaging attention from irrelevant stimuli, indicating impairments in cognitive flexibility and attentional control. Furthermore, by tracking eye movements during visual field tests, researchers can assess how epilepsy impacts visual perception and the ability to scan the visual environment, especially for patients who report visual disturbances or difficulty navigating their surroundings \cite{duncan1991gaze, fisher2022visually, laurent2014visual, huang2024reduced}.

    \begin{itemize}[leftmargin =*]
        \item Key Eye-Tracking Focus:
        Seizure-related eye movements, postictal gaze behaviour, photosensitivity, and gaze deviation.
        \item Task Design: (i) Design gaze-following tasks for detecting irregular saccades and smooth pursuit dysfunctions caused by seizures \cite{guerrero2021antisaccades, thurtell2009evidence}. (ii) Add stimulus-induced tasks to trigger epileptic episodes through visual stimuli (e.g., flickering lights) for patients with photosensitive epilepsy \cite{duncan1991gaze, fisher2022visually}. (ii) Use visual attention tasks to measure visual distraction, fixations, and response times to stimuli caused by seizure-related brain changes \cite{ fisher2022visually, laurent2014visual, huang2024reduced}. (iii) Track pre- and post-seizure movements to map how the brain's control over eye movement fluctuates with seizure activity \cite{fisher2000postictal, izadyar2018comparison}.
        \item Considerations: (i) Experimental design should account for how frequent and long seizures last. Shorter tasks may be needed to avoid triggering discomfort or seizures during testing \cite{wilkins1984neurological, thurston1985epileptic}. (ii) In photosensitive epilepsy patients, careful consideration must be given to the intensity and type of visual stimuli to avoid inducing seizures \cite{fischer2000stability, harding2008wind}.        
        \item Metrics: (i) Measure the saccadic accuracy since misfiring neurons during seizures can disrupt accurate eye movements \cite{lunn2016saccadic, mrabet2024study, guerrero2021antisaccades}. (ii) Expect unsteady or prolonged fixations, particularly after seizures, indicating reduced control over gaze stability \cite{huang2024reduced, metternich2022eye}. (iii) Seizures, particularly photosensitive ones, may cause abnormal pupil dilation responses \cite{ali2022response, fu2024recognition}. (iv) Smooth pursuit gain helps identify coordination problems during seizure events \cite{guerrero2021antisaccades, thurtell2009evidence}. 
    \end{itemize}

\subsection{Cerebrovascular Disorders} 

    \paragraph{\textbf{Stroke:}}  \label{para:stroke}
    A sudden interruption in the blood supply to the brain, caused by a clot (ischemic stroke) or rupture of a blood vessel (hemorrhagic stroke), leading to brain damage. Strokes can lead to a wide range of neurological impairments depending on the location and severity of the brain damage, and visual dysfunction is one of the common consequences, as strokes frequently disrupt the neural pathways responsible for eye movements and visual processing \cite{svenaeus2014diagnosing}.
    Individuals who experience strokes often suffer from abnormal saccadic behaviour, including prolonged saccadic latency, decreased saccadic accuracy, and hypometric saccades \cite{rowe2013profile, alves2016eye, rizzo2019eye, hassan2022approach}. 
    In addition, eye-tracking studies have shown that patients with stroke-related damage exhibit increased fixation instability and involuntary gaze shifts, which can impact their ability to process visual information effectively, especially for tasks that require sustained attention, such as reading or tracking moving objects \cite{alves2014eye, rowe2013profile, alves2016eye, istance2017supporting}. Another key focus of gaze-based stroke research is on visual neglect, a common post-stroke condition, especially after right-hemisphere strokes. Patients with hemispatial neglect may fail to perceive or respond to stimuli on one side of their visual field despite having normal vision \cite{ting2011visual, maxton2013don, takamura2016intentional}. In other words, stroke patients with neglect may exhibit asymmetrical scanning behaviour, where their gaze is heavily biased toward the non-neglected side. Furthermore, stroke survivors may exhibit difficulties with visual attention and executive function, such as a longer time to shift their gaze between visual targets, which is indicative of slowed cognitive processing and reduced flexibility in attentional control \cite{takamura2016intentional, alves2016eye}. Furthermore, patients may show fragmented pursuit or delays in initiation of smooth pursuit, particularly when the stroke affects areas of the brain involved in motion detection or motor control of the eyes \cite{rowe2013profile, alves2014eye, alves2016eye}.

    \begin{itemize}[leftmargin =*]
        \item Key Eye-Tracking Focus: Visual field deficits, attention recovery, and oculomotor control post-injury.
        \item Task Design: (i) Use visual search tasks to evaluate spatial neglect or deficits in visual attention (e.g., finding targets on a screen) \cite{takamura2016intentional, alves2016eye, ting2011visual}. (ii) Include object tracking tasks to measure how well participants follow moving stimuli \cite{rowe2013profile, alves2014eye, alves2016eye}.
        \item Considerations: (i) Depending on the severity of the stroke or brain injury, participants may experience visual field loss or difficulty controlling eye movements, so tasks should be adjusted accordingly \cite{rowe2013profile, lech2019human}. (ii) Design tasks assessing visual and cognitive recovery over time \cite{pimenta2017effects}.
        \item Metrics: (i) Track fixation stability, reaction times, and scan path efficiency (the ability to search and find relevant information in a visual field) \cite{chan2023examining, ionescu2023correlating}. (ii) Measure visual field exploration to detect any neglect or inattention to one side of the visual field \cite{ting2011visual, maxton2013don, takamura2016intentional}. (iii) Measure saccadic patterns: short saccades often indicate limited visual exploration or spatial neglect, while long saccades may suggest compensatory strategies or difficulty in motor control \cite{rowe2013profile, alves2016eye, rizzo2019eye, hassan2022approach}. (iv) Strokes can cause asymmetrical or abnormal pupil dilation responses. This can manifest as sluggish or incomplete dilation in response to light or other stimuli \cite{maza2020visual, kim2020neurological}. (v) Microsaccade frequency and amplitude may be diminished. The may be irregular in direction and amplitude, reflecting disrupted motor coordination and impaired control over small, corrective eye movements \cite{gao2017microsaccade, alexander2018microsaccade}.
    \end{itemize}

\subsection{Mental Health Disorders with Neurological Components}
    \paragraph{\textbf{Anxiety:}}  \label{para:anxiety}
    Anxiety disorders, which include generalized anxiety disorder (GAD), social anxiety disorder, panic disorder, and specific phobias, often result in abnormal patterns of eye movements as individuals with anxiety tend to exhibit hypervigilance toward perceived threats \cite{svenaeus2014diagnosing}. People with anxiety disorders frequently display a tendency to fixate on negative or threatening cues in their environment, such as angry or fearful faces, dangerous objects, or even ambiguous stimuli \cite{laretzaki2011threat, weeks2013gaze, michalska2017anxiety, holmes2006anxiety, chen2017gaze, toh2017attentional, burra2023association}. Eye-tracking studies have shown that anxious individuals exhibit faster initial fixations and longer gaze durations on threatening stimuli than neutral or positive stimuli (i.e. threat hypervigilance), leading to an overestimation of threats in their environment.  After an initial fixation on a threatening stimulus, many anxiety sufferers exhibit gaze avoidance, meaning they quickly divert their gaze away from the source of anxiety \cite{weeks2013gaze, weeks2019fear, garner2006orienting, chen2024does}. This pattern reflects both a heightened awareness of potential threats and a strategy to reduce the discomfort associated with prolonged exposure to these stimuli. Moreover, eye-tracking studies indicate that individuals with anxiety tend to scan their environments differently, displaying more disorganized or scattered visual search patterns \cite{gregory2019attention, li2024eye, murray2003anxiety, ellmers2020evidence}. This may be due to an inability to disengage from threatening stimuli or a heightened state of alertness that makes it difficult to focus on non-threatening aspects of their surroundings. Furthermore, high levels of anxiety are associated with impaired cognitive flexibility and difficulties in focusing attention. Eye-tracking technology has demonstrated that anxious individuals often have trouble shifting their attention away from negative stimuli, which can lead to rumination and repetitive negative thinking \cite{calvo2005time, gregory2019attention, mathews2003face, capriola2020gaze, holmes2006anxiety}.

    \begin{itemize}[leftmargin =*]
        \item Key Eye-Tracking Focus: Attentional bias toward threat, gaze avoidance, and hypervigilance.
        \item Task Design: (i) Use tasks with both threatening and neutral stimuli (e.g., images of faces showing fear versus neutral expressions) to assess gaze patterns in response to potential threats \cite{holmes2006anxiety, gregory2019attention}. (ii) Include emotional Stroop tasks where participants respond to emotional and neutral words or images to measure attentional biases \cite{holmes2006anxiety, toh2017attentional}.
        \item Considerations: (i) Participants with anxiety may find certain stimuli (e.g., images of fearful or angry faces) distressing. Ensure tasks are well-balanced to avoid overwhelming participants \cite{holmes2006anxiety, rosa2014effects}. (ii) Desensitization or anxiety management strategies might be needed before or after tasks involving emotional stimuli \cite{rosa2014effects}.
        \item Metrics: (i) Measure first fixation latency (how quickly a participant looks at a threat-related stimulus) and total gaze time on threatening versus neutral stimuli, fixations usually have longer durations \cite{weeks2013gaze, weeks2019fear, garner2006orienting, chen2024does, kitt2024using}. (ii) Gaze avoidance of threatening stimuli (such as quickly looking away from negative stimuli) is a key feature in social anxiety studies \cite{weeks2013gaze, weeks2019fear, garner2006orienting, chen2024does}.
        (iii) Other visual search patterns can be assessed through metrics such as saccadic latency, fixation density, and gaze entropy \cite{gregory2019attention, li2024eye, murray2003anxiety, ellmers2020evidence}. (iv)  Anxiety often leads to a larger baseline pupil size, reflecting increased arousal or alertness \cite{blini2024pupil, simpson1971effects, white1999differential}. (v) Anxiety is commonly associated with an elevated blink rate, potentially due to heightened arousal and stress, as well as dry eyes \cite{kojima2002blink, miranda2014anxiety}. (vi) Anxiety often leads to an increased frequency of microsaccades, as people with anxiety may have difficulty maintaining a steady gaze due to hypervigilance and scanning behaviours \cite{kashihara2020microsaccadic, mellor2021study}. (v) Finally, anxiety can interfere with smooth pursuit accuracy, leading to a less steady or slightly fragmented tracking of moving objects, especially if the object or task is anxiety-provoking \cite{smyrnis2007smooth, kattoulas2011predictive}.
    \end{itemize}

    \paragraph{\textbf{Obsessive-Compulsive Disorder (OCD):}}  \label{para:ocd}
    A mental health condition characterized by intrusive thoughts (obsessions) and repetitive behaviours (compulsions), often driven by an overwhelming need to reduce anxiety \cite{svenaeus2014diagnosing}. Individuals with OCD tend to exhibit hypervigilance and increased attention toward stimuli that are relevant to their obsessions (e.g., contamination fears, symmetry concerns) \cite{stein2019obsessive,jansen2020social, basel2023attention}. Eye-tracking studies have demonstrated that individuals with OCD often show longer fixation durations and difficulty disengaging from stimuli that trigger obsessive thoughts \cite{cludius2019attentional, basel2023attention, thierfelder23_pervasiveh}. For example, in tasks where participants are shown neutral versus contamination-related images, those with contamination fears may fixate longer on contaminated objects, indicating an attentional bias toward perceived threats \cite{tata1996attentional,armstrong2012attentional}. 
    In addition, after an initial fixation on obsession-related stimuli, some OCD patients may quickly avert their gaze as part of an avoidance strategy, attempting to reduce anxiety associated with the visual stimuli \cite{cludius2019attentional, fink2022you}. For example, individuals with OCD may exhibit gaze avoidance when viewing asymmetrical or disordered visual patterns. Eye-tracking can quantify these patterns by measuring the speed and frequency of gaze shifts away from anxiety-provoking stimuli, providing objective insights into the compulsive avoidance behaviours seen in OCD \cite{fox2001threatening, lazarov2016social}. Eye-tracking research has also been useful in examining the cognitive control deficits often seen in OCD, such as inhibitory control and cognitive flexibility \cite{clauss2022eye}. Individuals with OCD may have difficulty shifting attention between tasks or stimuli, leading to perseverative gaze behaviours, where they continue to focus on a particular stimulus even when it is no longer relevant \cite{basel2023attention, dalmaso2022direct, cludius2019attentional}. This can be particularly evident in visual search tasks, where individuals with OCD may exhibit slower response times and longer fixations on irrelevant or distracting stimuli, reflecting an inability to flexibly redirect attention \cite{liang2017sustained,armstrong2012eye}. 
    Furthermore, individuals with OCD often display rigid, repetitive search patterns, which may be linked to the compulsions to check or verify certain visual information repeatedly (e.g., ensuring objects are properly aligned or that an environment is clean) \cite{unoki1999attentional, hu2020investigating, basel2023attention}.

    \begin{itemize}[leftmargin =*]
        \item Key Eye-Tracking Focus: Fixation patterns on obsessive stimuli, avoidance behaviours, and gaze dwelling and scanning patterns.
        \item Task Design: (i) Present obsessive stimuli tasks with stimuli related to common OCD obsessions (e.g., cleanliness, contamination, threatening faces) to show how attention is biased toward certain stimuli. Similarly, OCD patients may be sensitive to symmetry and order, and tasks that involve visual arrays of objects can measure whether their gaze patterns reveal heightened attention to imbalance or disarray \cite{cludius2019attentional, basel2023attention, thierfelder23_pervasiveh}. (ii)  Add decision-making tasks since OCD is linked with indecision and the need for reassurance. Require participants to make choices or verify the information to assess how these cognitive patterns influence gaze behaviour, such as excessive revisiting of options \cite{unoki1999attentional, hu2020investigating, basel2023attention}.
        \item Considerations: (i) The frequency and nature of compulsive checking behaviours, both in real life and within experimental tasks, need to be carefully monitored, as these can lead to repetitive gaze patterns or scanning behaviour that needs to be distinguished from normal cognitive processing \cite{cludius2019attentional, basel2023attention, thierfelder23_pervasiveh}.
        \item Metrics: (i) Measure fixation duration as OCD patients may have longer fixations on objects of obsession, signalling excessive focus or rumination \cite{basel2023attention, bradley2016obsessive, cludius2019attentional}. (ii) Measure gaze dwell time: Total time spent focusing on distressing or obsession-related stimuli can provide a measure of the intensity of obsessive attention \cite{basel2023attention, armstrong2010orienting}. (iii) OCD can delay saccadic movements between fixations as patients deliberate or hesitate in their gaze shifts due to obsessive thoughts \cite{maruff1999abnormalities, spengler2006evidence, hardeman2006saccadic}. (iv) Repeated scanning of the same object or scene, as seen in compulsive checking behaviours, can be quantified through metrics like revisit count and scanning patterns \cite{bey2018impaired, chapman2023early, toh2011current}. (v) OCD patients may show changes in pupil dilation in response to obsession-related stimuli due to heightened arousal or stress, which can be measured with eye-tracking technology \cite{portnova2024temporal, pohlchen2021examining, pruneti2023systematic}.
    \end{itemize}

\paragraph{\textbf{Summary:}}
Based on the most common and agreed-upon methods and findings in the literature, these disorder-specific guidelines ensure that eye-tracking studies are tailored to the unique characteristics of each neurological condition, providing better data quality and more meaningful insights into the relationship between gaze behaviours and neurological dysfunction.

In Table \ref{tab:summary}, we summarized the key disorder-specific eye-tracking metrics discussed in Table \ref{tab:metrics}. Note that (i) certain metrics vary significantly between individuals, especially in disorders like anxiety and OCD, where fixation and blink rates may be affected by personal triggers or stress levels, (ii) some disorders may have overlapping biomarker characteristics, so multiple metrics are often used together to improve diagnostic precision, and (iii) in disorders like autism and ADHD, social context or task requirements may alter certain eye-tracking metrics (e.g., fixation duration in social vs. non-social contexts). Therefore, here, we present the primary eye-tracking biomarkers and highlight how each metric may differ by neurological condition for preliminary diagnosis. However, these metrics need to be further tailored depending on specific diagnostic needs or research goals.

\begin{landscape}
\begin{longtable}[H]
{ p{.08\textwidth}  p{.17\textwidth}   p{.17\textwidth}  p{.17\textwidth} p{.17\textwidth}  p{.17\textwidth} p{.17\textwidth}}
     \toprule
      \textbf{Disorder} & \textbf{Fixations} & \textbf{Saccades} & \textbf{Pupil Dilation} & \textbf{Blink Rate} & \textbf{Microsaccades}
      & \textbf{Smooth pursuit}
      
      \\
     \midrule
     \hyperref[para:asd]{ASD} & Abnormal durations, fewer on social cues & Erratic amplitude, atypical patterns &  Altered dilation to social cues & Abnormal in social situations & - & Difficulty tracking faces/social stimuli
\\
     \midrule
     \hyperref[para:adhd]{ADHD} & Shorter duration, impulsive shifting & Increased frequency, reduced inhibition & Hypo- or hyper-reactivity & High, associated with impulsivity & Reduced inhibition & Reduced tracking accuracy
     \\
     \midrule
     \hyperref[para:dyslexia]{Dyslexia} & Extended duration on text & Slower initiation & - & - & - & -
     \\
     \midrule
     \hyperref[para:cp]{CP} & Reduced stability, tremor in fixations & Limited range, slow initiation & - & - & - & Difficulty following smooth targets
     \\
     \midrule
     \hyperref[para:ad]{AD} & Increased duration, fewer fixations & Prolonged latency, reduced amplitude & Reduced response to stimuli & Decreased blink rate & Oblique behaviour & Reduced tracking ability
     \\
     \midrule
     \hyperref[para:pd]{PD} & Shorter duration, tremor presence & Increased latency, reduced peak velocity & Decreased reactivity & Reduced blink rate & Reduced frequency, increased latency & Saccadic intrusions, jerky tracking
     \\
     \midrule
     \hyperref[para:als]{ALS} & Irregular duration, difficulty maintaining & Slowed initiation, irregular amplitude & Normal but can vary & - & Reduced frequency and amplitude & Reduced tracking, effortful movements
     \\
     \midrule
     \hyperref[para:hd]{HD} & Increased fixation duration & Prolonged latency, jerky movements & - & - & Erratic, often exaggerated & Poor tracking, saccadic intrusions
     \\
     \midrule
     \hyperref[para:ms]{MS} & Unstable fixations & Delayed initiation, limited range & Slower reactivity & - & Altered frequency, instability & Reduced accuracy and steadiness
     \\
     \midrule
     \hyperref[para:ts]{TS} & Brief fixations, affected by tics & Normal but may be influenced by tics & - & Increased, associated with tics & Increased during tics & -
     \\
     \midrule
     \hyperref[para:et]{ET} & Stable but can show slight tremor & Normal latency, occasional tremors & - & - & Normal, some tremor effects & Mild issues with smoothness
     \\
     \midrule
     \hyperref[para:dystonia]{Dystonia} & Longer duration & Increased latency, restricted range & Reduced or excessive blinking & - & Irregular frequency and amplitude & Low smooth pursuit gain
     \\
     \midrule
     \hyperref[para:epilepsy]{Epilepsy} & Unsteady and prolonged & Inaccurate & Abnormal during seizures & - & Normal but can show suppression & Fluctuations during seizures
     \\
     \midrule
     \hyperref[para:stroke]{Stroke} & Longer duration, less stable & Irregular & Reduced response & - & Reduced frequency and amplitude & Reduced smooth pursuit accuracy
     \\
     \midrule
     \hyperref[para:anxiety]{Anxiety} & Longer duration, fixation on threats & Slower, smaller amplitude & Overreactivity to stimuli & Increased, especially in social anxiety & Increased during anxiety-provoking stimuli & Altered, may show instability
     \\
     \midrule
     \hyperref[para:ocd]{OCD} & Increased duration on specific areas & Delayed & - & Reduced blink rate & Repeated, erratic around specific areas & -
     \\

     \bottomrule
    
    \caption{Summary of Section \ref{sec:specific} of the primary eye-tracking metrics for specific disorders, A '-' indicates that the corresponding metric is not a biomarker for the disorder (i.e. unaffected).}
    \label{tab:summary}
\end{longtable}
\end{landscape}

\section{Conclusion}
\label{sec:discussion}

Taking the guidelines outlined in the paper into account, conducting research on neural disorders using eye tracking requires a careful balance of ethical considerations and appropriate experimental design, ensuring that the tasks, environment, and data collection methods are adapted to the capabilities and needs of individuals with neural conditions is key to obtaining reliable and useful insights.

\paragraph{Limitations:} However, several limitations exist: Gaze-based neural diagnosis is still in its infancy, and further research is needed to refine these methods and explore how they can be integrated into clinical practice.
While eye-tracking provides a wealth of data, interpreting gaze patterns in the context of specific neural disorders can be complex. For example, a specific gaze pattern (e.g., reduced fixation on faces) might be seen in several disorders like ASD, anxiety, or even depression, making it challenging to isolate a single diagnosis based on gaze alone. In addition, gaze behaviour varies significantly among individuals, even within the same diagnostic group. Some individuals with ASD, for example, might show typical eye movements, while others may show marked differences. This variability can make it difficult to develop standardized benchmarks that apply universally across all individuals within a disorder group. Furthermore, technical issues such as calibration errors, sensitivity to head movements, and lighting conditions can also affect data quality, especially when working with individuals with motor impairments (e.g., Parkinson’s or ALS). Moreover, eye-tracking results are often context-dependent, meaning that results can vary based on the task, environment, or stimuli presented.

\paragraph{Current State:} While gaze tracking can provide valuable insights into cognitive and attentional processes, it is currently insufficient for a standalone diagnosis. In its current status, it often needs to be used alongside other diagnostic tools (e.g., neuroimaging and neuropsychological assessments) to get a complete understanding of the disorder. Finally, eye-tracking primarily highlights symptoms rather than root causes, making it a valuable tool for preliminary assessment but not a definitive diagnostic method.
\newline

Therefore, in this paper, we provide the first guidelines for gaze-based neural diagnosis based on careful evaluation, comparison, and structuring of consensus-based methods and findings in the literature. We aim that our paper enables more structured eye tracking studies that can be used in clinical setups and can ultimately advance the preliminary diagnosis of neural disorders.

\begin{acks}
M. Elfares was funded by the Ministry of Science, Research and the Arts Baden-Württemberg in the Artificial Intelligence Software Academy (AISA). P. Reisert and R. Küsters were supported by the German Federal Ministry of Education and Research under Grant Agreement No. 16KIS1441 (CRYPTECS project) and the German Research Foundation under Grant No. 548713845. A. Bulling was funded by the European Research Council (ERC; grant agreement 801708).
\end{acks}

\bibliographystyle{ACM-Reference-Format}
\bibliography{sample-base}


\begin{thebibliography}{531}


\ifx \showCODEN    \undefined \def \showCODEN     #1{\unskip}     \fi
\ifx \showISBNx    \undefined \def \showISBNx     #1{\unskip}     \fi
\ifx \showISBNxiii \undefined \def \showISBNxiii  #1{\unskip}     \fi
\ifx \showISSN     \undefined \def \showISSN      #1{\unskip}     \fi
\ifx \showLCCN     \undefined \def \showLCCN      #1{\unskip}     \fi
\ifx \shownote     \undefined \def \shownote      #1{#1}          \fi
\ifx \showarticletitle \undefined \def \showarticletitle #1{#1}   \fi
\ifx \showURL      \undefined \def \showURL       {\relax}        \fi
\providecommand\bibfield[2]{#2}
\providecommand\bibinfo[2]{#2}
\providecommand\natexlab[1]{#1}
\providecommand\showeprint[2][]{arXiv:#2}

\bibitem[Ackermans et~al\mbox{.}(2007)]%
        {ackermans2007vertical}
\bibfield{author}{\bibinfo{person}{Linda Ackermans}, \bibinfo{person}{Yasin Temel}, \bibinfo{person}{Noel~JC Bauer}, \bibinfo{person}{Veerle Visser-Vandewalle}, \bibinfo{person}{Dutch-Flemish Tourette Surgery~Study Group}, {et~al\mbox{.}}} \bibinfo{year}{2007}\natexlab{}.
\newblock \showarticletitle{Vertical gaze palsy after thalamic stimulation for Tourette syndrome: case report}.
\newblock \bibinfo{journal}{\emph{Neurosurgery}} \bibinfo{volume}{61}, \bibinfo{number}{5} (\bibinfo{year}{2007}), \bibinfo{pages}{E1100}.
\newblock


\bibitem[Adhanom et~al\mbox{.}(2023)]%
        {adhanom2023eye}
\bibfield{author}{\bibinfo{person}{Isayas~Berhe Adhanom}, \bibinfo{person}{Paul MacNeilage}, {and} \bibinfo{person}{Eelke Folmer}.} \bibinfo{year}{2023}\natexlab{}.
\newblock \showarticletitle{Eye tracking in virtual reality: a broad review of applications and challenges}.
\newblock \bibinfo{journal}{\emph{Virtual Reality}} \bibinfo{volume}{27}, \bibinfo{number}{2} (\bibinfo{year}{2023}), \bibinfo{pages}{1481--1505}.
\newblock


\bibitem[Ahmed et~al\mbox{.}(2022)]%
        {ahmed2022eye}
\bibfield{author}{\bibinfo{person}{Ibrahim~Abdulrab Ahmed}, \bibinfo{person}{Ebrahim~Mohammed Senan}, \bibinfo{person}{Taha~H Rassem}, \bibinfo{person}{Mohammed~AH Ali}, \bibinfo{person}{Hamzeh Salameh~Ahmad Shatnawi}, \bibinfo{person}{Salwa~Mutahar Alwazer}, {and} \bibinfo{person}{Mohammed Alshahrani}.} \bibinfo{year}{2022}\natexlab{}.
\newblock \showarticletitle{Eye tracking-based diagnosis and early detection of autism spectrum disorder using machine learning and deep learning techniques}.
\newblock \bibinfo{journal}{\emph{Electronics}} \bibinfo{volume}{11}, \bibinfo{number}{4} (\bibinfo{year}{2022}), \bibinfo{pages}{530}.
\newblock


\bibitem[Ahuja(2016)]%
        {ahuja2016commercial}
\bibfield{author}{\bibinfo{person}{Neha Ahuja}.} \bibinfo{year}{2016}\natexlab{}.
\newblock \showarticletitle{Commercial creations: The role of end user license agreements in controlling the exploitation of user generated content}.
\newblock \bibinfo{journal}{\emph{J. Marshall Rev. Intell. Prop. L.}}  \bibinfo{volume}{16} (\bibinfo{year}{2016}), \bibinfo{pages}{ii}.
\newblock


\bibitem[Airdrie et~al\mbox{.}(2018)]%
        {airdrie2018facial}
\bibfield{author}{\bibinfo{person}{Jac~N Airdrie}, \bibinfo{person}{Kate Langley}, \bibinfo{person}{Anita Thapar}, {and} \bibinfo{person}{Stephanie~HM van Goozen}.} \bibinfo{year}{2018}\natexlab{}.
\newblock \showarticletitle{Facial emotion recognition and eye gaze in attention-deficit/hyperactivity disorder with and without comorbid conduct disorder}.
\newblock \bibinfo{journal}{\emph{Journal of the American Academy of Child \& Adolescent Psychiatry}} \bibinfo{volume}{57}, \bibinfo{number}{8} (\bibinfo{year}{2018}), \bibinfo{pages}{561--570}.
\newblock


\bibitem[Al-Rahayfeh and Faezipour(2013)]%
        {al2013eye}
\bibfield{author}{\bibinfo{person}{Amer Al-Rahayfeh} {and} \bibinfo{person}{Miad Faezipour}.} \bibinfo{year}{2013}\natexlab{}.
\newblock \showarticletitle{Eye tracking and head movement detection: A state-of-art survey}.
\newblock \bibinfo{journal}{\emph{IEEE journal of translational engineering in health and medicine}}  \bibinfo{volume}{1} (\bibinfo{year}{2013}), \bibinfo{pages}{2100212--2100212}.
\newblock


\bibitem[Alexander et~al\mbox{.}(2018)]%
        {alexander2018microsaccade}
\bibfield{author}{\bibinfo{person}{Robert~G Alexander}, \bibinfo{person}{Stephen~L Macknik}, {and} \bibinfo{person}{Susana Martinez-Conde}.} \bibinfo{year}{2018}\natexlab{}.
\newblock \showarticletitle{Microsaccade characteristics in neurological and ophthalmic disease}.
\newblock \bibinfo{journal}{\emph{Frontiers in neurology}}  \bibinfo{volume}{9} (\bibinfo{year}{2018}), \bibinfo{pages}{144}.
\newblock


\bibitem[Ali et~al\mbox{.}(2022)]%
        {ali2022response}
\bibfield{author}{\bibinfo{person}{Eman~N Ali}, \bibinfo{person}{Christian~J Lueck}, \bibinfo{person}{Corinne~F Carle}, \bibinfo{person}{Kate~L Martin}, \bibinfo{person}{Angela Borbelj}, {and} \bibinfo{person}{Ted Maddess}.} \bibinfo{year}{2022}\natexlab{}.
\newblock \showarticletitle{Response characteristics of objective perimetry in persons living with epilepsy}.
\newblock \bibinfo{journal}{\emph{Journal of the Neurological Sciences}}  \bibinfo{volume}{436} (\bibinfo{year}{2022}), \bibinfo{pages}{120237}.
\newblock


\bibitem[Allen et~al\mbox{.}(2007)]%
        {allen2007use}
\bibfield{author}{\bibinfo{person}{DP Allen}, \bibinfo{person}{JR Playfer}, \bibinfo{person}{NM Aly}, \bibinfo{person}{P Duffey}, \bibinfo{person}{A Heald}, \bibinfo{person}{SL Smith}, {and} \bibinfo{person}{DM Halliday}.} \bibinfo{year}{2007}\natexlab{}.
\newblock \showarticletitle{On the use of low-cost computer peripherals for the assessment of motor dysfunction in Parkinson's disease—quantification of bradykinesia using target tracking tasks}.
\newblock \bibinfo{journal}{\emph{IEEE Transactions on Neural Systems and Rehabilitation Engineering}} \bibinfo{volume}{15}, \bibinfo{number}{2} (\bibinfo{year}{2007}), \bibinfo{pages}{286--294}.
\newblock


\bibitem[Almutairi et~al\mbox{.}(2018)]%
        {almutairi2018vestibular}
\bibfield{author}{\bibinfo{person}{Anwar Almutairi}, \bibinfo{person}{Jennifer~Braswell Christy}, {and} \bibinfo{person}{Laura Vogtle}.} \bibinfo{year}{2018}\natexlab{}.
\newblock \showarticletitle{Vestibular and oculomotor function in children with cerebral palsy: a scoping review}. In \bibinfo{booktitle}{\emph{Seminars in hearing}}, Vol.~\bibinfo{volume}{39}. Thieme Medical Publishers, \bibinfo{pages}{288--304}.
\newblock


\bibitem[Alves et~al\mbox{.}(2016)]%
        {alves2016eye}
\bibfield{author}{\bibinfo{person}{J{\'u}lio Alves}, \bibinfo{person}{A Vourvopoulos}, \bibinfo{person}{A Bernardino}, {et~al\mbox{.}}} \bibinfo{year}{2016}\natexlab{}.
\newblock \showarticletitle{Eye Gaze correlates of motor impairment in VR observation of motor actions}.
\newblock \bibinfo{journal}{\emph{Methods of information in medicine}} \bibinfo{volume}{55}, \bibinfo{number}{01} (\bibinfo{year}{2016}), \bibinfo{pages}{79--83}.
\newblock


\bibitem[Alves et~al\mbox{.}(2014)]%
        {alves2014eye}
\bibfield{author}{\bibinfo{person}{J{\'u}lio Alves}, \bibinfo{person}{Athanasios Vourvopoulos}, \bibinfo{person}{Alexandre Bernardino}, {and} \bibinfo{person}{Sergi Berm{\'u}dez~i Badia}.} \bibinfo{year}{2014}\natexlab{}.
\newblock \showarticletitle{Eye gaze patterns after stroke: correlates of a VR action execution and observation task}. In \bibinfo{booktitle}{\emph{Proceedings of the 8th International Conference on Pervasive Computing Technologies for Healthcare}}. \bibinfo{pages}{339--342}.
\newblock


\bibitem[Amestoy et~al\mbox{.}(2015)]%
        {amestoy2015developmental}
\bibfield{author}{\bibinfo{person}{Anouck Amestoy}, \bibinfo{person}{Etienne Guillaud}, \bibinfo{person}{Manuel~P Bouvard}, {and} \bibinfo{person}{Jean-Ren{\'e} Cazalets}.} \bibinfo{year}{2015}\natexlab{}.
\newblock \showarticletitle{Developmental changes in face visual scanning in autism spectrum disorder as assessed by data-based analysis}.
\newblock \bibinfo{journal}{\emph{Frontiers in Psychology}}  \bibinfo{volume}{6} (\bibinfo{year}{2015}), \bibinfo{pages}{989}.
\newblock


\bibitem[Amirault et~al\mbox{.}(2009)]%
        {amirault2009alteration}
\bibfield{author}{\bibinfo{person}{Marion Amirault}, \bibinfo{person}{Kattalin Etchegoyhen}, \bibinfo{person}{Sandrine Delord}, \bibinfo{person}{Sandrine Mendizabal}, \bibinfo{person}{Caroline Kraushaar}, \bibinfo{person}{Isabelle Hesling}, \bibinfo{person}{Mich{\`e}le Allard}, \bibinfo{person}{Manuel Bouvard}, {and} \bibinfo{person}{Willy Mayo}.} \bibinfo{year}{2009}\natexlab{}.
\newblock \showarticletitle{Alteration of attentional blink in high functioning autism: A pilot study}.
\newblock \bibinfo{journal}{\emph{Journal of autism and developmental disorders}}  \bibinfo{volume}{39} (\bibinfo{year}{2009}), \bibinfo{pages}{1522--1528}.
\newblock


\bibitem[Anderson and Finn(2012)]%
        {anderson2012new}
\bibfield{author}{\bibinfo{person}{Theodore~Wilbur Anderson} {and} \bibinfo{person}{Jeremy~D Finn}.} \bibinfo{year}{2012}\natexlab{}.
\newblock \bibinfo{booktitle}{\emph{The new statistical analysis of data}}.
\newblock \bibinfo{publisher}{Springer Science \& Business Media}.
\newblock


\bibitem[Antoniades and Kennard(2015)]%
        {antoniades2015ocular}
\bibfield{author}{\bibinfo{person}{CA Antoniades} {and} \bibinfo{person}{C Kennard}.} \bibinfo{year}{2015}\natexlab{}.
\newblock \showarticletitle{Ocular motor abnormalities in neurodegenerative disorders}.
\newblock \bibinfo{journal}{\emph{Eye}} \bibinfo{volume}{29}, \bibinfo{number}{2} (\bibinfo{year}{2015}), \bibinfo{pages}{200--207}.
\newblock


\bibitem[Archibald et~al\mbox{.}(2013)]%
        {archibald2013visual}
\bibfield{author}{\bibinfo{person}{Neil~K Archibald}, \bibinfo{person}{Sam~B Hutton}, \bibinfo{person}{Michael~P Clarke}, \bibinfo{person}{Urs~P Mosimann}, {and} \bibinfo{person}{David~J Burn}.} \bibinfo{year}{2013}\natexlab{}.
\newblock \showarticletitle{Visual exploration in Parkinson’s disease and Parkinson’s disease dementia}.
\newblock \bibinfo{journal}{\emph{Brain}} \bibinfo{volume}{136}, \bibinfo{number}{3} (\bibinfo{year}{2013}), \bibinfo{pages}{739--750}.
\newblock


\bibitem[Armstrong and Munoz(2003)]%
        {armstrong2003attentional}
\bibfield{author}{\bibinfo{person}{IT Armstrong} {and} \bibinfo{person}{DP Munoz}.} \bibinfo{year}{2003}\natexlab{}.
\newblock \showarticletitle{Attentional blink in adults with attention-deficit hyperactivity disorder: Influence of eye movements}.
\newblock \bibinfo{journal}{\emph{Experimental Brain Research}}  \bibinfo{volume}{152} (\bibinfo{year}{2003}), \bibinfo{pages}{243--250}.
\newblock


\bibitem[Armstrong and Olatunji(2012)]%
        {armstrong2012eye}
\bibfield{author}{\bibinfo{person}{Thomas Armstrong} {and} \bibinfo{person}{Bunmi~O Olatunji}.} \bibinfo{year}{2012}\natexlab{}.
\newblock \showarticletitle{Eye tracking of attention in the affective disorders: A meta-analytic review and synthesis}.
\newblock \bibinfo{journal}{\emph{Clinical psychology review}} \bibinfo{volume}{32}, \bibinfo{number}{8} (\bibinfo{year}{2012}), \bibinfo{pages}{704--723}.
\newblock


\bibitem[Armstrong et~al\mbox{.}(2010)]%
        {armstrong2010orienting}
\bibfield{author}{\bibinfo{person}{Thomas Armstrong}, \bibinfo{person}{Bunmi~O Olatunji}, \bibinfo{person}{Shivali Sarawgi}, {and} \bibinfo{person}{Casey Simmons}.} \bibinfo{year}{2010}\natexlab{}.
\newblock \showarticletitle{Orienting and maintenance of gaze in contamination fear: Biases for disgust and fear cues}.
\newblock \bibinfo{journal}{\emph{Behaviour research and therapy}} \bibinfo{volume}{48}, \bibinfo{number}{5} (\bibinfo{year}{2010}), \bibinfo{pages}{402--408}.
\newblock


\bibitem[Armstrong et~al\mbox{.}(2012)]%
        {armstrong2012attentional}
\bibfield{author}{\bibinfo{person}{Thomas Armstrong}, \bibinfo{person}{Shivali Sarawgi}, {and} \bibinfo{person}{Bunmi~O Olatunji}.} \bibinfo{year}{2012}\natexlab{}.
\newblock \showarticletitle{Attentional bias toward threat in contamination fear: overt components and behavioral correlates.}
\newblock \bibinfo{journal}{\emph{Journal of abnormal psychology}} \bibinfo{volume}{121}, \bibinfo{number}{1} (\bibinfo{year}{2012}), \bibinfo{pages}{232}.
\newblock


\bibitem[Asato et~al\mbox{.}(2011)]%
        {asato2011deficits}
\bibfield{author}{\bibinfo{person}{Miya~R Asato}, \bibinfo{person}{Natalie Nawarawong}, \bibinfo{person}{Bruce Hermann}, \bibinfo{person}{Patricia Crumrine}, {and} \bibinfo{person}{Beatriz Luna}.} \bibinfo{year}{2011}\natexlab{}.
\newblock \showarticletitle{Deficits in oculomotor performance in pediatric epilepsy}.
\newblock \bibinfo{journal}{\emph{Epilepsia}} \bibinfo{volume}{52}, \bibinfo{number}{2} (\bibinfo{year}{2011}), \bibinfo{pages}{377--385}.
\newblock


\bibitem[Attoni et~al\mbox{.}(2016)]%
        {attoni2016abnormal}
\bibfield{author}{\bibinfo{person}{Tiago Attoni}, \bibinfo{person}{Rog{\'e}rio Beato}, \bibinfo{person}{Serge Pinto}, {and} \bibinfo{person}{Francisco Cardoso}.} \bibinfo{year}{2016}\natexlab{}.
\newblock \showarticletitle{Abnormal eye movements in three types of chorea}.
\newblock \bibinfo{journal}{\emph{Arquivos de Neuro-Psiquiatria}} \bibinfo{volume}{74}, \bibinfo{number}{9} (\bibinfo{year}{2016}), \bibinfo{pages}{761--766}.
\newblock


\bibitem[Aust et~al\mbox{.}(2024)]%
        {aust2024impairment}
\bibfield{author}{\bibinfo{person}{Elisa Aust}, \bibinfo{person}{Sven-Thomas Graupner}, \bibinfo{person}{Ren{\'e} G{\"u}nther}, \bibinfo{person}{Katharina Linse}, \bibinfo{person}{Markus Joos}, \bibinfo{person}{Julian Grosskreutz}, \bibinfo{person}{Johannes Prudlo}, \bibinfo{person}{Sebastian Pannasch}, {and} \bibinfo{person}{Andreas Hermann}.} \bibinfo{year}{2024}\natexlab{}.
\newblock \showarticletitle{Impairment of oculomotor functions in patients with early to advanced amyotrophic lateral sclerosis}.
\newblock \bibinfo{journal}{\emph{Journal of Neurology}} \bibinfo{volume}{271}, \bibinfo{number}{1} (\bibinfo{year}{2024}), \bibinfo{pages}{325--339}.
\newblock


\bibitem[Aydin et~al\mbox{.}(2023)]%
        {aydin2023face}
\bibfield{author}{\bibinfo{person}{{\"U}mit Aydin}, \bibinfo{person}{Roser Ca{\~n}igueral}, \bibinfo{person}{Charlotte Tye}, {and} \bibinfo{person}{Gr{\'a}inne McLoughlin}.} \bibinfo{year}{2023}\natexlab{}.
\newblock \showarticletitle{Face processing in young adults with autism and ADHD: An event related potentials study}.
\newblock \bibinfo{journal}{\emph{Frontiers in Psychiatry}}  \bibinfo{volume}{14} (\bibinfo{year}{2023}), \bibinfo{pages}{1080681}.
\newblock


\bibitem[Bacon et~al\mbox{.}(2020)]%
        {bacon2020identifying}
\bibfield{author}{\bibinfo{person}{Elizabeth~C Bacon}, \bibinfo{person}{Adrienne Moore}, \bibinfo{person}{Quimby Lee}, \bibinfo{person}{Cynthia Carter~Barnes}, \bibinfo{person}{Eric Courchesne}, {and} \bibinfo{person}{Karen Pierce}.} \bibinfo{year}{2020}\natexlab{}.
\newblock \showarticletitle{Identifying prognostic markers in autism spectrum disorder using eye tracking}.
\newblock \bibinfo{journal}{\emph{Autism}} \bibinfo{volume}{24}, \bibinfo{number}{3} (\bibinfo{year}{2020}), \bibinfo{pages}{658--669}.
\newblock


\bibitem[Baizabal-Carvallo and Jankovic(2022)]%
        {baizabal2022clinical}
\bibfield{author}{\bibinfo{person}{Jos{\'e}~Fidel Baizabal-Carvallo} {and} \bibinfo{person}{Joseph Jankovic}.} \bibinfo{year}{2022}\natexlab{}.
\newblock \showarticletitle{The clinical phenomenology and correlations of oculogyric tics}.
\newblock \bibinfo{journal}{\emph{Acta Neurologica Belgica}} \bibinfo{volume}{122}, \bibinfo{number}{4} (\bibinfo{year}{2022}), \bibinfo{pages}{925--930}.
\newblock


\bibitem[Baker(2006)]%
        {baker2006neurodiversity}
\bibfield{author}{\bibinfo{person}{Dana~Lee Baker}.} \bibinfo{year}{2006}\natexlab{}.
\newblock \showarticletitle{Neurodiversity, neurological disability and the public sector: notes on the autism spectrum}.
\newblock \bibinfo{journal}{\emph{Disability \& Society}} \bibinfo{volume}{21}, \bibinfo{number}{1} (\bibinfo{year}{2006}), \bibinfo{pages}{15--29}.
\newblock


\bibitem[Bal et~al\mbox{.}(2010)]%
        {bal2010emotion}
\bibfield{author}{\bibinfo{person}{Elgiz Bal}, \bibinfo{person}{Emily Harden}, \bibinfo{person}{Damon Lamb}, \bibinfo{person}{Amy~Vaughan Van~Hecke}, \bibinfo{person}{John~W Denver}, {and} \bibinfo{person}{Stephen~W Porges}.} \bibinfo{year}{2010}\natexlab{}.
\newblock \showarticletitle{Emotion recognition in children with autism spectrum disorders: Relations to eye gaze and autonomic state}.
\newblock \bibinfo{journal}{\emph{Journal of autism and developmental disorders}}  \bibinfo{volume}{40} (\bibinfo{year}{2010}), \bibinfo{pages}{358--370}.
\newblock


\bibitem[Ballester-Plan{\'e} et~al\mbox{.}(2018)]%
        {ballester2018cognitive}
\bibfield{author}{\bibinfo{person}{J{\'u}lia Ballester-Plan{\'e}}, \bibinfo{person}{Olga Laporta-Hoyos}, \bibinfo{person}{Alfons Macaya}, \bibinfo{person}{Pilar P{\'o}o}, \bibinfo{person}{Mar Melendez-Plumed}, \bibinfo{person}{Esther Toro-Tamargo}, \bibinfo{person}{Francisca Gimeno}, \bibinfo{person}{Ana Narberhaus}, \bibinfo{person}{Dolors Segarra}, {and} \bibinfo{person}{Roser Pueyo}.} \bibinfo{year}{2018}\natexlab{}.
\newblock \showarticletitle{Cognitive functioning in dyskinetic cerebral palsy: Its relation to motor function, communication and epilepsy}.
\newblock \bibinfo{journal}{\emph{European Journal of Paediatric Neurology}} \bibinfo{volume}{22}, \bibinfo{number}{1} (\bibinfo{year}{2018}), \bibinfo{pages}{102--112}.
\newblock


\bibitem[Band et~al\mbox{.}(2024)]%
        {band2024advancements}
\bibfield{author}{\bibinfo{person}{Tali~G Band}, \bibinfo{person}{Rotem~Z Bar-Or}, {and} \bibinfo{person}{Edmund Ben-Ami}.} \bibinfo{year}{2024}\natexlab{}.
\newblock \showarticletitle{Advancements in eye movement measurement technologies for assessing neurodegenerative diseases}.
\newblock \bibinfo{journal}{\emph{Frontiers in Digital Health}}  \bibinfo{volume}{6} (\bibinfo{year}{2024}).
\newblock


\bibitem[Bao et~al\mbox{.}(2024)]%
        {bao2024gender}
\bibfield{author}{\bibinfo{person}{Mengyi Bao}, \bibinfo{person}{Yingchao Ying}, \bibinfo{person}{Jia Ye}, \bibinfo{person}{Yiting Hu}, \bibinfo{person}{Lin Yao}, \bibinfo{person}{Wenyu Li}, {and} \bibinfo{person}{Kewen Jiang}.} \bibinfo{year}{2024}\natexlab{}.
\newblock \showarticletitle{Gender differences in attention bias in ADHD children’s cancellation testing: An eye movement tracking study}.
\newblock  (\bibinfo{year}{2024}).
\newblock


\bibitem[Barahim~Bastani et~al\mbox{.}(2024)]%
        {barahim2024self}
\bibfield{author}{\bibinfo{person}{Pouya Barahim~Bastani}, \bibinfo{person}{Ali~S Saber~Tehrani}, \bibinfo{person}{Shervin Badihian}, \bibinfo{person}{Hector Rieiro}, \bibinfo{person}{David Rastall}, \bibinfo{person}{Nathan Farrell}, \bibinfo{person}{Max Parker}, \bibinfo{person}{Jorge Otero-Millan}, \bibinfo{person}{Ahmed Hassoon}, \bibinfo{person}{David Newman-Toker}, {et~al\mbox{.}}} \bibinfo{year}{2024}\natexlab{}.
\newblock \showarticletitle{Self-Recording of Eye Movements in Amyotrophic Lateral Sclerosis Patients Using a Smartphone Eye-Tracking App}.
\newblock \bibinfo{journal}{\emph{Digital Biomarkers}} \bibinfo{volume}{8}, \bibinfo{number}{1} (\bibinfo{year}{2024}), \bibinfo{pages}{111--119}.
\newblock


\bibitem[Barbara et~al\mbox{.}(2023)]%
        {barbara2023real}
\bibfield{author}{\bibinfo{person}{Nathaniel Barbara}, \bibinfo{person}{Tracey~A Camilleri}, {and} \bibinfo{person}{Kenneth~P Camilleri}.} \bibinfo{year}{2023}\natexlab{}.
\newblock \showarticletitle{Real-time continuous EOG-based gaze angle estimation with baseline drift compensation under stationary head conditions}.
\newblock \bibinfo{journal}{\emph{Biomedical Signal Processing and Control}}  \bibinfo{volume}{86} (\bibinfo{year}{2023}), \bibinfo{pages}{105282}.
\newblock


\bibitem[Barone et~al\mbox{.}(2023)]%
        {barone2023potential}
\bibfield{author}{\bibinfo{person}{Valentina Barone}, \bibinfo{person}{Johannes~P van Dijk}, \bibinfo{person}{Mariette~HJA Debeij-van Hall}, {and} \bibinfo{person}{Michel~JAM van Putten}.} \bibinfo{year}{2023}\natexlab{}.
\newblock \showarticletitle{A potential multimodal test for clinical assessment of visual attention in neurological disorders}.
\newblock \bibinfo{journal}{\emph{Clinical EEG and Neuroscience}} \bibinfo{volume}{54}, \bibinfo{number}{5} (\bibinfo{year}{2023}), \bibinfo{pages}{512--521}.
\newblock


\bibitem[Basel et~al\mbox{.}(2023)]%
        {basel2023attention}
\bibfield{author}{\bibinfo{person}{Dana Basel}, \bibinfo{person}{Hadar Hallel}, \bibinfo{person}{Reuven Dar}, {and} \bibinfo{person}{Amit Lazarov}.} \bibinfo{year}{2023}\natexlab{}.
\newblock \showarticletitle{Attention allocation in OCD: A systematic review and meta-analysis of eye-tracking-based research}.
\newblock \bibinfo{journal}{\emph{Journal of Affective Disorders}}  \bibinfo{volume}{324} (\bibinfo{year}{2023}), \bibinfo{pages}{539--550}.
\newblock


\bibitem[Bast et~al\mbox{.}(2023)]%
        {bast2023sensory}
\bibfield{author}{\bibinfo{person}{Nico Bast}, \bibinfo{person}{Luke Mason}, \bibinfo{person}{Christine Ecker}, \bibinfo{person}{Sarah Baumeister}, \bibinfo{person}{Tobias Banaschewski}, \bibinfo{person}{Emily~JH Jones}, \bibinfo{person}{Declan~GM Murphy}, \bibinfo{person}{Jan~K Buitelaar}, \bibinfo{person}{Eva Loth}, \bibinfo{person}{Gahan Pandina}, {et~al\mbox{.}}} \bibinfo{year}{2023}\natexlab{}.
\newblock \showarticletitle{Sensory salience processing moderates attenuated gazes on faces in autism spectrum disorder: a case--control study}.
\newblock \bibinfo{journal}{\emph{Molecular Autism}} \bibinfo{volume}{14}, \bibinfo{number}{1} (\bibinfo{year}{2023}), \bibinfo{pages}{5}.
\newblock


\bibitem[Beck et~al\mbox{.}(2018)]%
        {beck2018reduced}
\bibfield{author}{\bibinfo{person}{Rebecca~B Beck}, \bibinfo{person}{Simone~L Kneafsey}, \bibinfo{person}{Shruti Narasimham}, \bibinfo{person}{Sean O’Riordan}, \bibinfo{person}{Tadashi Isa}, \bibinfo{person}{Michael Hutchinson}, {and} \bibinfo{person}{Richard~B Reilly}.} \bibinfo{year}{2018}\natexlab{}.
\newblock \showarticletitle{Reduced frequency of ipsilateral express saccades in cervical dystonia: probing the nigro-tectal pathway}.
\newblock \bibinfo{journal}{\emph{Tremor and Other Hyperkinetic Movements}}  \bibinfo{volume}{8} (\bibinfo{year}{2018}).
\newblock


\bibitem[Becker et~al\mbox{.}(2019)]%
        {becker2019saccadic}
\bibfield{author}{\bibinfo{person}{Wolfgang Becker}, \bibinfo{person}{Martin Gorges}, \bibinfo{person}{Doroth{\'e}e Lul{\'e}}, \bibinfo{person}{Elmar Pinkhardt}, \bibinfo{person}{Albert~C Ludolph}, {and} \bibinfo{person}{Jan Kassubek}.} \bibinfo{year}{2019}\natexlab{}.
\newblock \showarticletitle{Saccadic intrusions in amyotrophic lateral sclerosis (ALS)}.
\newblock \bibinfo{journal}{\emph{Journal of Eye Movement Research}} \bibinfo{volume}{12}, \bibinfo{number}{6} (\bibinfo{year}{2019}).
\newblock


\bibitem[Becker et~al\mbox{.}(2009)]%
        {becker2009eye}
\bibfield{author}{\bibinfo{person}{W Becker}, \bibinfo{person}{Reinhart J{\"u}rgens}, \bibinfo{person}{Jan Kassubek}, \bibinfo{person}{Daniel Ecker}, \bibinfo{person}{B Kramer}, {and} \bibinfo{person}{Bernhard Landwehrmeyer}.} \bibinfo{year}{2009}\natexlab{}.
\newblock \showarticletitle{Eye--head coordination in moderately affected Huntington’s Disease patients: do head movements facilitate gaze shifts?}
\newblock \bibinfo{journal}{\emph{Experimental brain research}}  \bibinfo{volume}{192} (\bibinfo{year}{2009}), \bibinfo{pages}{97--112}.
\newblock


\bibitem[Beenen et~al\mbox{.}(1986)]%
        {beenen1986diagnostic}
\bibfield{author}{\bibinfo{person}{N Beenen}, \bibinfo{person}{U B{\"u}ttner}, {and} \bibinfo{person}{HW Lange}.} \bibinfo{year}{1986}\natexlab{}.
\newblock \showarticletitle{The diagnostic value of eye movement recording in patients with Huntington's disease and their offspring}.
\newblock \bibinfo{journal}{\emph{Electroencephalography and clinical Neurophysiology}} \bibinfo{volume}{63}, \bibinfo{number}{2} (\bibinfo{year}{1986}), \bibinfo{pages}{119--127}.
\newblock


\bibitem[Bekteshi et~al\mbox{.}(2023)]%
        {bekteshi2023towards}
\bibfield{author}{\bibinfo{person}{Saranda Bekteshi}, \bibinfo{person}{Elegast Monbaliu}, \bibinfo{person}{Sarah McIntyre}, \bibinfo{person}{Gillian Saloojee}, \bibinfo{person}{Sander~R Hilberink}, \bibinfo{person}{Nana Tatishvili}, {and} \bibinfo{person}{Bernard Dan}.} \bibinfo{year}{2023}\natexlab{}.
\newblock \showarticletitle{Towards functional improvement of motor disorders associated with cerebral palsy}.
\newblock \bibinfo{journal}{\emph{The Lancet Neurology}} \bibinfo{volume}{22}, \bibinfo{number}{3} (\bibinfo{year}{2023}), \bibinfo{pages}{229--243}.
\newblock


\bibitem[Bell et~al\mbox{.}(2023)]%
        {bell2023vision}
\bibfield{author}{\bibinfo{person}{Carter~A Bell}, \bibinfo{person}{Scott~N Grossman}, \bibinfo{person}{Laura~J Balcer}, {and} \bibinfo{person}{Steven~L Galetta}.} \bibinfo{year}{2023}\natexlab{}.
\newblock \showarticletitle{Vision as a piece of the head trauma puzzle}.
\newblock \bibinfo{journal}{\emph{Eye}} \bibinfo{volume}{37}, \bibinfo{number}{12} (\bibinfo{year}{2023}), \bibinfo{pages}{2385--2390}.
\newblock


\bibitem[Bellocchi et~al\mbox{.}(2013)]%
        {bellocchi2013can}
\bibfield{author}{\bibinfo{person}{St{\'e}phanie Bellocchi}, \bibinfo{person}{Mathilde Muneaux}, \bibinfo{person}{Mireille Bastien-Toniazzo}, {and} \bibinfo{person}{St{\'e}phanie Ducrot}.} \bibinfo{year}{2013}\natexlab{}.
\newblock \showarticletitle{I can read it in your eyes: What eye movements tell us about visuo-attentional processes in developmental dyslexia}.
\newblock \bibinfo{journal}{\emph{Research in developmental disabilities}} \bibinfo{volume}{34}, \bibinfo{number}{1} (\bibinfo{year}{2013}), \bibinfo{pages}{452--460}.
\newblock


\bibitem[Berger and Cassuto(2014)]%
        {berger2014effect}
\bibfield{author}{\bibinfo{person}{Itai Berger} {and} \bibinfo{person}{Hanoch Cassuto}.} \bibinfo{year}{2014}\natexlab{}.
\newblock \showarticletitle{The effect of environmental distractors incorporation into a CPT on sustained attention and ADHD diagnosis among adolescents}.
\newblock \bibinfo{journal}{\emph{Journal of neuroscience methods}}  \bibinfo{volume}{222} (\bibinfo{year}{2014}), \bibinfo{pages}{62--68}.
\newblock


\bibitem[Best et~al\mbox{.}(2010)]%
        {best2010gender}
\bibfield{author}{\bibinfo{person}{Catherine~A Best}, \bibinfo{person}{Nancy~J Minshew}, {and} \bibinfo{person}{Mark~S Strauss}.} \bibinfo{year}{2010}\natexlab{}.
\newblock \showarticletitle{Gender discrimination of eyes and mouths by individuals with autism}.
\newblock \bibinfo{journal}{\emph{Autism Research}} \bibinfo{volume}{3}, \bibinfo{number}{2} (\bibinfo{year}{2010}), \bibinfo{pages}{88--93}.
\newblock


\bibitem[Bey et~al\mbox{.}(2018)]%
        {bey2018impaired}
\bibfield{author}{\bibinfo{person}{Katharina Bey}, \bibinfo{person}{Leonhard Lennertz}, \bibinfo{person}{Rosa Gr{\"u}tzmann}, \bibinfo{person}{Stephan Heinzel}, \bibinfo{person}{Christian Kaufmann}, \bibinfo{person}{Julia Klawohn}, \bibinfo{person}{Anja Riesel}, \bibinfo{person}{Inga Meyh{\"o}fer}, \bibinfo{person}{Ulrich Ettinger}, \bibinfo{person}{Norbert Kathmann}, {et~al\mbox{.}}} \bibinfo{year}{2018}\natexlab{}.
\newblock \showarticletitle{Impaired antisaccades in obsessive-compulsive disorder: Evidence from meta-analysis and a large empirical study}.
\newblock \bibinfo{journal}{\emph{Frontiers in psychiatry}}  \bibinfo{volume}{9} (\bibinfo{year}{2018}), \bibinfo{pages}{284}.
\newblock


\bibitem[Bhaskar et~al\mbox{.}(2003)]%
        {bhaskar2003blink}
\bibfield{author}{\bibinfo{person}{TN Bhaskar}, \bibinfo{person}{Foo~Tun Keat}, \bibinfo{person}{Surendra Ranganath}, {and} \bibinfo{person}{YV Venkatesh}.} \bibinfo{year}{2003}\natexlab{}.
\newblock \showarticletitle{Blink detection and eye tracking for eye localization}. In \bibinfo{booktitle}{\emph{TENCON 2003. Conference on Convergent Technologies for Asia-Pacific Region}}, Vol.~\bibinfo{volume}{2}. IEEE, \bibinfo{pages}{821--824}.
\newblock


\bibitem[Biswas and Langdon(2011)]%
        {biswas2011new}
\bibfield{author}{\bibinfo{person}{Pradipta Biswas} {and} \bibinfo{person}{Pat Langdon}.} \bibinfo{year}{2011}\natexlab{}.
\newblock \showarticletitle{A new input system for disabled users involving eye gaze tracker and scanning interface}.
\newblock \bibinfo{journal}{\emph{Journal of Assistive Technologies}} \bibinfo{volume}{5}, \bibinfo{number}{2} (\bibinfo{year}{2011}), \bibinfo{pages}{58--66}.
\newblock


\bibitem[Black et~al\mbox{.}(2017)]%
        {black2017mechanisms}
\bibfield{author}{\bibinfo{person}{Melissa~H Black}, \bibinfo{person}{Nigel~TM Chen}, \bibinfo{person}{Kartik~K Iyer}, \bibinfo{person}{Ottmar~V Lipp}, \bibinfo{person}{Sven B{\"o}lte}, \bibinfo{person}{Marita Falkmer}, \bibinfo{person}{Tele Tan}, {and} \bibinfo{person}{Sonya Girdler}.} \bibinfo{year}{2017}\natexlab{}.
\newblock \showarticletitle{Mechanisms of facial emotion recognition in autism spectrum disorders: Insights from eye tracking and electroencephalography}.
\newblock \bibinfo{journal}{\emph{Neuroscience \& Biobehavioral Reviews}}  \bibinfo{volume}{80} (\bibinfo{year}{2017}), \bibinfo{pages}{488--515}.
\newblock


\bibitem[Blekher et~al\mbox{.}(2006)]%
        {blekher2006saccades}
\bibfield{author}{\bibinfo{person}{T Blekher}, \bibinfo{person}{Shannon~A Johnson}, \bibinfo{person}{J Marshall}, \bibinfo{person}{K White}, \bibinfo{person}{S Hui}, \bibinfo{person}{Marjorie Weaver}, \bibinfo{person}{Jaqueline Gray}, \bibinfo{person}{R Yee}, \bibinfo{person}{Julie~C Stout}, \bibinfo{person}{Xabier Beristain}, {et~al\mbox{.}}} \bibinfo{year}{2006}\natexlab{}.
\newblock \showarticletitle{Saccades in presymptomatic and early stages of Huntington disease}.
\newblock \bibinfo{journal}{\emph{Neurology}} \bibinfo{volume}{67}, \bibinfo{number}{3} (\bibinfo{year}{2006}), \bibinfo{pages}{394--399}.
\newblock


\bibitem[Blini et~al\mbox{.}(2024)]%
        {blini2024pupil}
\bibfield{author}{\bibinfo{person}{Elvio Blini}, \bibinfo{person}{Giovanni Anobile}, {and} \bibinfo{person}{Roberto Arrighi}.} \bibinfo{year}{2024}\natexlab{}.
\newblock \showarticletitle{What pupil size can and cannot tell about math anxiety}.
\newblock \bibinfo{journal}{\emph{Psychological Research}} (\bibinfo{year}{2024}), \bibinfo{pages}{1--14}.
\newblock


\bibitem[Blondiaux et~al\mbox{.}(2023)]%
        {blondiaux2023impairments}
\bibfield{author}{\bibinfo{person}{Florence Blondiaux}, \bibinfo{person}{Louisien Lebrun}, \bibinfo{person}{Bernard~J Hanseeuw}, {and} \bibinfo{person}{Fr{\'e}d{\'e}ric Crevecoeur}.} \bibinfo{year}{2023}\natexlab{}.
\newblock \showarticletitle{Impairments of saccadic and reaching adaptation in essential tremor are linked to movement execution}.
\newblock \bibinfo{journal}{\emph{Journal of Neurophysiology}} \bibinfo{volume}{130}, \bibinfo{number}{5} (\bibinfo{year}{2023}), \bibinfo{pages}{1092--1102}.
\newblock


\bibitem[Blondon et~al\mbox{.}(2015)]%
        {blondon2015use}
\bibfield{author}{\bibinfo{person}{Katherine Blondon}, \bibinfo{person}{Rolf Wipfli}, {and} \bibinfo{person}{Christian Lovis}.} \bibinfo{year}{2015}\natexlab{}.
\newblock \showarticletitle{Use of eye-tracking technology in clinical reasoning: a systematic review}.
\newblock \bibinfo{journal}{\emph{Digital Healthcare Empowering Europeans}} (\bibinfo{year}{2015}), \bibinfo{pages}{90--94}.
\newblock


\bibitem[Bollen et~al\mbox{.}(1986)]%
        {bollen1986horizontal}
\bibfield{author}{\bibinfo{person}{ELEM Bollen}, \bibinfo{person}{JPH Reulen}, \bibinfo{person}{JC Den~Heyer}, \bibinfo{person}{W Van~der Kamp}, \bibinfo{person}{RAC Roos}, {and} \bibinfo{person}{OJS Buruma}.} \bibinfo{year}{1986}\natexlab{}.
\newblock \showarticletitle{Horizontal and vertical saccadic eye movement abnormalities in Huntington's chorea}.
\newblock \bibinfo{journal}{\emph{Journal of the neurological sciences}} \bibinfo{volume}{74}, \bibinfo{number}{1} (\bibinfo{year}{1986}), \bibinfo{pages}{11--22}.
\newblock


\bibitem[Bowman et~al\mbox{.}(2004)]%
        {bowman2004gaze}
\bibfield{author}{\bibinfo{person}{Sarah Bowman}, \bibinfo{person}{Lisa Hinkley}, \bibinfo{person}{Jim Barnes}, {and} \bibinfo{person}{Roger Lindsay}.} \bibinfo{year}{2004}\natexlab{}.
\newblock \showarticletitle{Gaze aversion and the primacy of emotional dysfunction in autism}.
\newblock \bibinfo{journal}{\emph{Cognition and technology. Amsterdam: John Benjamins Publishing Company}} (\bibinfo{year}{2004}), \bibinfo{pages}{267--301}.
\newblock


\bibitem[Boyd et~al\mbox{.}(2022)]%
        {boyd2022manipulating}
\bibfield{author}{\bibinfo{person}{LouAnne Boyd}, \bibinfo{person}{Vincent Berardi}, \bibinfo{person}{Deanna Hughes}, \bibinfo{person}{Franceli Cibrian}, \bibinfo{person}{Jazette Johnson}, \bibinfo{person}{Viseth Sean}, \bibinfo{person}{Eliza DelPizzo-Cheng}, \bibinfo{person}{Brandon Mackin}, \bibinfo{person}{Ayra Tusneem}, \bibinfo{person}{Riya Mody}, {et~al\mbox{.}}} \bibinfo{year}{2022}\natexlab{}.
\newblock \showarticletitle{Manipulating image luminance to improve eye gaze and verbal behavior in autistic children}.
\newblock \bibinfo{journal}{\emph{Humanities and Social Sciences Communications}} \bibinfo{volume}{9}, \bibinfo{number}{1} (\bibinfo{year}{2022}), \bibinfo{pages}{1--9}.
\newblock


\bibitem[Boz et~al\mbox{.}(2023)]%
        {boz2023examination}
\bibfield{author}{\bibinfo{person}{Hatice~Eraslan Boz}, \bibinfo{person}{Koray Ko{\c{c}}o{\u{g}}lu}, \bibinfo{person}{M{\"u}ge Akkoyun}, \bibinfo{person}{I{\c{s}}{\i}l~Ya{\u{g}}mur T{\"u}fekci}, \bibinfo{person}{Merve Ekin}, \bibinfo{person}{P{\i}nar {\"O}z{\c{c}}elik}, {and} \bibinfo{person}{G{\"u}lden Akdal}.} \bibinfo{year}{2023}\natexlab{}.
\newblock \showarticletitle{Examination of eye movements during visual scanning of real-world images in Alzheimer's disease and amnestic mild cognitive impairment}.
\newblock \bibinfo{journal}{\emph{International Journal of Psychophysiology}}  \bibinfo{volume}{190} (\bibinfo{year}{2023}), \bibinfo{pages}{84--93}.
\newblock


\bibitem[Bradley et~al\mbox{.}(2016)]%
        {bradley2016obsessive}
\bibfield{author}{\bibinfo{person}{Maria~C Bradley}, \bibinfo{person}{Donncha Hanna}, \bibinfo{person}{Paul Wilson}, \bibinfo{person}{Gareth Scott}, \bibinfo{person}{Paul Quinn}, {and} \bibinfo{person}{Kevin~FW Dyer}.} \bibinfo{year}{2016}\natexlab{}.
\newblock \showarticletitle{Obsessive--compulsive symptoms and attentional bias: An eye-tracking methodology}.
\newblock \bibinfo{journal}{\emph{Journal of behavior therapy and experimental psychiatry}}  \bibinfo{volume}{50} (\bibinfo{year}{2016}), \bibinfo{pages}{303--308}.
\newblock


\bibitem[Briand et~al\mbox{.}(2001)]%
        {briand2001automatic}
\bibfield{author}{\bibinfo{person}{Kevin~A Briand}, \bibinfo{person}{Wayne Hening}, \bibinfo{person}{Howard Poizner}, {and} \bibinfo{person}{Anne~B Sereno}.} \bibinfo{year}{2001}\natexlab{}.
\newblock \showarticletitle{Automatic orienting of visuospatial attention in Parkinson's disease}.
\newblock \bibinfo{journal}{\emph{Neuropsychologia}} \bibinfo{volume}{39}, \bibinfo{number}{11} (\bibinfo{year}{2001}), \bibinfo{pages}{1240--1249}.
\newblock


\bibitem[Brien et~al\mbox{.}(2023)]%
        {brien2023classification}
\bibfield{author}{\bibinfo{person}{Donald~C Brien}, \bibinfo{person}{Heidi~C Riek}, \bibinfo{person}{Rachel Yep}, \bibinfo{person}{Jeff Huang}, \bibinfo{person}{Brian Coe}, \bibinfo{person}{Corson Areshenkoff}, \bibinfo{person}{David Grimes}, \bibinfo{person}{Mandar Jog}, \bibinfo{person}{Anthony Lang}, \bibinfo{person}{Connie Marras}, {et~al\mbox{.}}} \bibinfo{year}{2023}\natexlab{}.
\newblock \showarticletitle{Classification and staging of Parkinson's disease using video-based eye tracking}.
\newblock \bibinfo{journal}{\emph{Parkinsonism \& Related Disorders}}  \bibinfo{volume}{110} (\bibinfo{year}{2023}), \bibinfo{pages}{105316}.
\newblock


\bibitem[Broder-Fingert et~al\mbox{.}(2018)]%
        {broder2018improving}
\bibfield{author}{\bibinfo{person}{Sarabeth Broder-Fingert}, \bibinfo{person}{Emily Feinberg}, {and} \bibinfo{person}{Michael Silverstein}.} \bibinfo{year}{2018}\natexlab{}.
\newblock \showarticletitle{Improving screening for autism spectrum disorder: is it time for something new?}
\newblock \bibinfo{journal}{\emph{Pediatrics}} \bibinfo{volume}{141}, \bibinfo{number}{6} (\bibinfo{year}{2018}).
\newblock


\bibitem[Bucci et~al\mbox{.}(2008)]%
        {bucci2008poor}
\bibfield{author}{\bibinfo{person}{Maria~Pia Bucci}, \bibinfo{person}{Dominique Br{\'e}mond-Gignac}, {and} \bibinfo{person}{Zo{\"\i} Kapoula}.} \bibinfo{year}{2008}\natexlab{}.
\newblock \showarticletitle{Poor binocular coordination of saccades in dyslexic children}.
\newblock \bibinfo{journal}{\emph{Graefe's archive for clinical and experimental ophthalmology}}  \bibinfo{volume}{246} (\bibinfo{year}{2008}), \bibinfo{pages}{417--428}.
\newblock


\bibitem[Burra and Vrti{\v{c}}ka(2023)]%
        {burra2023association}
\bibfield{author}{\bibinfo{person}{Nicolas Burra} {and} \bibinfo{person}{Pascal Vrti{\v{c}}ka}.} \bibinfo{year}{2023}\natexlab{}.
\newblock \showarticletitle{Association between attachment anxiety and the gaze direction-related N170}.
\newblock \bibinfo{journal}{\emph{Attachment \& Human Development}} \bibinfo{volume}{25}, \bibinfo{number}{1} (\bibinfo{year}{2023}), \bibinfo{pages}{181--198}.
\newblock


\bibitem[Burrell et~al\mbox{.}(2013)]%
        {burrell2013early}
\bibfield{author}{\bibinfo{person}{James~R Burrell}, \bibinfo{person}{Roger~HS Carpenter}, \bibinfo{person}{John~R Hodges}, {and} \bibinfo{person}{Matthew~C Kiernan}.} \bibinfo{year}{2013}\natexlab{}.
\newblock \showarticletitle{Early saccades in amyotrophic lateral sclerosis}.
\newblock \bibinfo{journal}{\emph{Amyotrophic Lateral Sclerosis and Frontotemporal Degeneration}} \bibinfo{volume}{14}, \bibinfo{number}{4} (\bibinfo{year}{2013}), \bibinfo{pages}{294--301}.
\newblock


\bibitem[Calvo and Avero(2005)]%
        {calvo2005time}
\bibfield{author}{\bibinfo{person}{Manuel~G Calvo} {and} \bibinfo{person}{Pedro Avero}.} \bibinfo{year}{2005}\natexlab{}.
\newblock \showarticletitle{Time course of attentional bias to emotional scenes in anxiety: Gaze direction and duration}.
\newblock \bibinfo{journal}{\emph{Cognition \& Emotion}} \bibinfo{volume}{19}, \bibinfo{number}{3} (\bibinfo{year}{2005}), \bibinfo{pages}{433--451}.
\newblock


\bibitem[Camero et~al\mbox{.}(2021)]%
        {camero2021gaze}
\bibfield{author}{\bibinfo{person}{Raquel Camero}, \bibinfo{person}{Ver{\'o}nica Mart{\'\i}nez}, {and} \bibinfo{person}{Carlos Gallego}.} \bibinfo{year}{2021}\natexlab{}.
\newblock \showarticletitle{Gaze following and pupil dilation as early diagnostic markers of autism in toddlers}.
\newblock \bibinfo{journal}{\emph{Children}} \bibinfo{volume}{8}, \bibinfo{number}{2} (\bibinfo{year}{2021}), \bibinfo{pages}{113}.
\newblock


\bibitem[Capriola-Hall et~al\mbox{.}(2020)]%
        {capriola2020gaze}
\bibfield{author}{\bibinfo{person}{Nicole~N Capriola-Hall}, \bibinfo{person}{Thomas~H Ollendick}, {and} \bibinfo{person}{Susan~W White}.} \bibinfo{year}{2020}\natexlab{}.
\newblock \showarticletitle{Gaze as an indicator of selective attention in adolescents with social anxiety disorder}.
\newblock \bibinfo{journal}{\emph{Cognitive Therapy and Research}}  \bibinfo{volume}{44} (\bibinfo{year}{2020}), \bibinfo{pages}{145--155}.
\newblock


\bibitem[Carelli et~al\mbox{.}(2022)]%
        {carelli2022gaze}
\bibfield{author}{\bibinfo{person}{Laura Carelli}, \bibinfo{person}{Federica Solca}, \bibinfo{person}{Sofia Tagini}, \bibinfo{person}{Silvia Torre}, \bibinfo{person}{Federico Verde}, \bibinfo{person}{Nicola Ticozzi}, \bibinfo{person}{Roberta Ferrucci}, \bibinfo{person}{Gabriella Pravettoni}, \bibinfo{person}{Edoardo~Nicol{\`o} Aiello}, \bibinfo{person}{Vincenzo Silani}, {et~al\mbox{.}}} \bibinfo{year}{2022}\natexlab{}.
\newblock \showarticletitle{Gaze-contingent eye-tracking training in brain disorders: a systematic review}.
\newblock \bibinfo{journal}{\emph{Brain Sciences}} \bibinfo{volume}{12}, \bibinfo{number}{7} (\bibinfo{year}{2022}), \bibinfo{pages}{931}.
\newblock


\bibitem[Castellani et~al\mbox{.}(2010)]%
        {castellani2010alzheimer}
\bibfield{author}{\bibinfo{person}{Rudy~J Castellani}, \bibinfo{person}{Raj~K Rolston}, {and} \bibinfo{person}{Mark~A Smith}.} \bibinfo{year}{2010}\natexlab{}.
\newblock \showarticletitle{Alzheimer disease}.
\newblock \bibinfo{journal}{\emph{Disease-a-month: DM}} \bibinfo{volume}{56}, \bibinfo{number}{9} (\bibinfo{year}{2010}), \bibinfo{pages}{484}.
\newblock


\bibitem[Cazzato et~al\mbox{.}(2019)]%
        {cazzato2019non}
\bibfield{author}{\bibinfo{person}{Dario Cazzato}, \bibinfo{person}{Silvia~M Castro}, \bibinfo{person}{Osvaldo Agamennoni}, \bibinfo{person}{Gerardo Fern{\'a}ndez}, {and} \bibinfo{person}{Holger Voos}.} \bibinfo{year}{2019}\natexlab{}.
\newblock \showarticletitle{A non-invasive tool for attention-deficit disorder analysis based on gaze tracks}. In \bibinfo{booktitle}{\emph{Proceedings of the 2nd International Conference on Applications of Intelligent Systems}}. \bibinfo{pages}{1--6}.
\newblock


\bibitem[Celebisoy et~al\mbox{.}(2016)]%
        {celebisoy2016evaluation}
\bibfield{author}{\bibinfo{person}{Mehmet Celebisoy}, \bibinfo{person}{Ne{\c{s}}e {\c{C}}ELEB{\.I}SOY}, \bibinfo{person}{Ay{\c{s}}en~S{\"u}zen EK{\.I}NC{\.I}}, \bibinfo{person}{Esra AKY{\"U}Z}, {and} \bibinfo{person}{Ahmet Acarer}.} \bibinfo{year}{2016}\natexlab{}.
\newblock \showarticletitle{Evaluation of Saccadic and Smooth Pursuit Eye Movements at an Early Stage of Essential Tremor.}
\newblock \bibinfo{journal}{\emph{Journal of Neurological Sciences}} \bibinfo{volume}{33}, \bibinfo{number}{4} (\bibinfo{year}{2016}).
\newblock


\bibitem[Chan et~al\mbox{.}(2023)]%
        {chan2023examining}
\bibfield{author}{\bibinfo{person}{Marko Ka-leung Chan}, \bibinfo{person}{Cho~Lee Wong}, \bibinfo{person}{King~Pong Yu}, {and} \bibinfo{person}{Raymond Kai-yu Tong}.} \bibinfo{year}{2023}\natexlab{}.
\newblock \showarticletitle{Examining Eye Tracking Metrics and Cognitive Function in Post-Stroke Individuals: A Comparison of Visual Searching Tasks Between Those with and without Cognitive Impairment}.
\newblock \bibinfo{journal}{\emph{Cerebrovasc Dis}} (\bibinfo{year}{2023}).
\newblock


\bibitem[Chang et~al\mbox{.}(2020)]%
        {chang2020accurate}
\bibfield{author}{\bibinfo{person}{Zhuoqing Chang}, \bibinfo{person}{Ziyu Chen}, \bibinfo{person}{Christopher~D Stephen}, \bibinfo{person}{Jeremy~D Schmahmann}, \bibinfo{person}{Hau-Tieng Wu}, \bibinfo{person}{Guillermo Sapiro}, {and} \bibinfo{person}{Anoopum~S Gupta}.} \bibinfo{year}{2020}\natexlab{}.
\newblock \showarticletitle{Accurate detection of cerebellar smooth pursuit eye movement abnormalities via mobile phone video and machine learning}.
\newblock \bibinfo{journal}{\emph{Scientific reports}} \bibinfo{volume}{10}, \bibinfo{number}{1} (\bibinfo{year}{2020}), \bibinfo{pages}{18641}.
\newblock


\bibitem[Chapman et~al\mbox{.}(2023)]%
        {chapman2023early}
\bibfield{author}{\bibinfo{person}{Elizabeth~A Chapman}, \bibinfo{person}{Stephanie Martinez}, \bibinfo{person}{Andreas Keil}, {and} \bibinfo{person}{Carol~A Mathews}.} \bibinfo{year}{2023}\natexlab{}.
\newblock \showarticletitle{Early visual perceptual processing is altered in obsessive--compulsive disorder}.
\newblock \bibinfo{journal}{\emph{Clinical Neurophysiology}}  \bibinfo{volume}{151} (\bibinfo{year}{2023}), \bibinfo{pages}{134--142}.
\newblock


\bibitem[Chen et~al\mbox{.}(2024)]%
        {chen2024does}
\bibfield{author}{\bibinfo{person}{Jiemiao Chen}, \bibinfo{person}{Esther van~den Bos}, {and} \bibinfo{person}{P~Michiel Westenberg}.} \bibinfo{year}{2024}\natexlab{}.
\newblock \showarticletitle{Does gaze anxiety predict actual gaze avoidance and is it more informative than social anxiety?}
\newblock \bibinfo{journal}{\emph{Journal of Behavior Therapy and Experimental Psychiatry}}  \bibinfo{volume}{82} (\bibinfo{year}{2024}), \bibinfo{pages}{101896}.
\newblock


\bibitem[Chen and Clarke(2017)]%
        {chen2017gaze}
\bibfield{author}{\bibinfo{person}{Nigel~TM Chen} {and} \bibinfo{person}{Patrick~JF Clarke}.} \bibinfo{year}{2017}\natexlab{}.
\newblock \showarticletitle{Gaze-based assessments of vigilance and avoidance in social anxiety: A review}.
\newblock \bibinfo{journal}{\emph{Current psychiatry reports}}  \bibinfo{volume}{19} (\bibinfo{year}{2017}), \bibinfo{pages}{1--9}.
\newblock


\bibitem[Chien et~al\mbox{.}(2022)]%
        {chien2022game}
\bibfield{author}{\bibinfo{person}{Yi-Ling Chien}, \bibinfo{person}{Chia-Hsin Lee}, \bibinfo{person}{Yen-Nan Chiu}, \bibinfo{person}{Wen-Che Tsai}, \bibinfo{person}{Yuan-Che Min}, \bibinfo{person}{Yang-Min Lin}, \bibinfo{person}{Jui-Shen Wong}, {and} \bibinfo{person}{Yi-Li Tseng}.} \bibinfo{year}{2022}\natexlab{}.
\newblock \showarticletitle{Game-based social interaction platform for cognitive assessment of autism using eye tracking}.
\newblock \bibinfo{journal}{\emph{IEEE Transactions on Neural Systems and Rehabilitation Engineering}}  \bibinfo{volume}{31} (\bibinfo{year}{2022}), \bibinfo{pages}{749--758}.
\newblock


\bibitem[Clarke et~al\mbox{.}(2012)]%
        {clarke2012assessing}
\bibfield{author}{\bibinfo{person}{Michael~T Clarke}, \bibinfo{person}{Deborah Loganathan}, {and} \bibinfo{person}{John Swettenham}.} \bibinfo{year}{2012}\natexlab{}.
\newblock \showarticletitle{Assessing true and false belief in young children with cerebral palsy through anticipatory gaze behaviours: A pilot study}.
\newblock \bibinfo{journal}{\emph{Research in developmental disabilities}} \bibinfo{volume}{33}, \bibinfo{number}{6} (\bibinfo{year}{2012}), \bibinfo{pages}{2058--2066}.
\newblock


\bibitem[Clauss et~al\mbox{.}(2022)]%
        {clauss2022eye}
\bibfield{author}{\bibinfo{person}{Kate Clauss}, \bibinfo{person}{Julia~Y Gorday}, {and} \bibinfo{person}{Joseph~R Bardeen}.} \bibinfo{year}{2022}\natexlab{}.
\newblock \showarticletitle{Eye tracking evidence of threat-related attentional bias in anxiety-and fear-related disorders: A systematic review and meta-analysis}.
\newblock \bibinfo{journal}{\emph{Clinical psychology review}}  \bibinfo{volume}{93} (\bibinfo{year}{2022}), \bibinfo{pages}{102142}.
\newblock


\bibitem[Clendaniel(2010)]%
        {clendaniel2010effects}
\bibfield{author}{\bibinfo{person}{Richard~A Clendaniel}.} \bibinfo{year}{2010}\natexlab{}.
\newblock \showarticletitle{The effects of habituation and gaze stability exercises in the treatment of unilateral vestibular hypofunction: a preliminary results}.
\newblock \bibinfo{journal}{\emph{Journal of Neurologic Physical Therapy}} \bibinfo{volume}{34}, \bibinfo{number}{2} (\bibinfo{year}{2010}), \bibinfo{pages}{111--116}.
\newblock


\bibitem[Cludius et~al\mbox{.}(2019)]%
        {cludius2019attentional}
\bibfield{author}{\bibinfo{person}{Barbara Cludius}, \bibinfo{person}{Frederike Wenzlaff}, \bibinfo{person}{Peer Briken}, {and} \bibinfo{person}{Charlotte~E Wittekind}.} \bibinfo{year}{2019}\natexlab{}.
\newblock \showarticletitle{Attentional biases of vigilance and maintenance in obsessive-compulsive disorder: an eye-tracking study}.
\newblock \bibinfo{journal}{\emph{Journal of Obsessive-Compulsive and Related Disorders}}  \bibinfo{volume}{20} (\bibinfo{year}{2019}), \bibinfo{pages}{30--38}.
\newblock


\bibitem[Coles(1977)]%
        {coles1977signs}
\bibfield{author}{\bibinfo{person}{William~H Coles}.} \bibinfo{year}{1977}\natexlab{}.
\newblock \showarticletitle{Signs of essential blepharospasm: A motion-picture analysis}.
\newblock \bibinfo{journal}{\emph{Archives of Ophthalmology}} \bibinfo{volume}{95}, \bibinfo{number}{6} (\bibinfo{year}{1977}), \bibinfo{pages}{1006--1009}.
\newblock


\bibitem[Comon et~al\mbox{.}(2022)]%
        {comon2022impaired}
\bibfield{author}{\bibinfo{person}{Martin Comon}, \bibinfo{person}{Isabelle Rouch}, \bibinfo{person}{Arlette Edjolo}, \bibinfo{person}{Catherine Padovan}, \bibinfo{person}{Pierre Krolak-Salmon}, {and} \bibinfo{person}{Jean-Michel Dorey}.} \bibinfo{year}{2022}\natexlab{}.
\newblock \showarticletitle{Impaired facial emotion recognition and gaze direction detection in mild Alzheimer’s disease: results from the PACO study}.
\newblock \bibinfo{journal}{\emph{Journal of Alzheimer's Disease}} \bibinfo{volume}{89}, \bibinfo{number}{4} (\bibinfo{year}{2022}), \bibinfo{pages}{1427--1437}.
\newblock


\bibitem[Conelea et~al\mbox{.}(2024)]%
        {conelea2024automated}
\bibfield{author}{\bibinfo{person}{Christine Conelea}, \bibinfo{person}{Hengyue Liang}, \bibinfo{person}{Megan DuBois}, \bibinfo{person}{Brittany Raab}, \bibinfo{person}{Mia Kellman}, \bibinfo{person}{Brianna Wellen}, \bibinfo{person}{Suma Jacob}, \bibinfo{person}{Sonya Wang}, \bibinfo{person}{Ju Sun}, {and} \bibinfo{person}{Kelvin Lim}.} \bibinfo{year}{2024}\natexlab{}.
\newblock \showarticletitle{Automated Quantification of Eye Tics Using Computer Vision and Deep Learning Techniques}.
\newblock \bibinfo{journal}{\emph{Movement Disorders}} \bibinfo{volume}{39}, \bibinfo{number}{1} (\bibinfo{year}{2024}), \bibinfo{pages}{183--191}.
\newblock


\bibitem[Constantino et~al\mbox{.}(2021)]%
        {constantino2021clinical}
\bibfield{author}{\bibinfo{person}{John~N Constantino}, \bibinfo{person}{Tony Charman}, {and} \bibinfo{person}{Emily~JH Jones}.} \bibinfo{year}{2021}\natexlab{}.
\newblock \showarticletitle{Clinical and translational implications of an emerging developmental substructure for autism}.
\newblock \bibinfo{journal}{\emph{Annual review of clinical psychology}} \bibinfo{volume}{17}, \bibinfo{number}{1} (\bibinfo{year}{2021}), \bibinfo{pages}{365--389}.
\newblock


\bibitem[Craven and Groom(2015)]%
        {craven2015computer}
\bibfield{author}{\bibinfo{person}{Michael~P Craven} {and} \bibinfo{person}{Madeleine~J Groom}.} \bibinfo{year}{2015}\natexlab{}.
\newblock \showarticletitle{Computer games for user engagement in Attention Deficit Hyperactivity Disorder (ADHD) monitoring and therapy}. In \bibinfo{booktitle}{\emph{2015 International Conference on Interactive Technologies and Games}}. IEEE, \bibinfo{pages}{34--40}.
\newblock


\bibitem[Crowe et~al\mbox{.}(2018)]%
        {crowe2018new}
\bibfield{author}{\bibinfo{person}{Emily~M Crowe}, \bibinfo{person}{Iain~D Gilchrist}, {and} \bibinfo{person}{Christopher Kent}.} \bibinfo{year}{2018}\natexlab{}.
\newblock \showarticletitle{New approaches to the analysis of eye movement behaviour across expertise while viewing brain MRIs}.
\newblock \bibinfo{journal}{\emph{Cognitive research: principles and implications}}  \bibinfo{volume}{3} (\bibinfo{year}{2018}), \bibinfo{pages}{1--14}.
\newblock


\bibitem[Dalmaso et~al\mbox{.}(2022)]%
        {dalmaso2022direct}
\bibfield{author}{\bibinfo{person}{Mario Dalmaso}, \bibinfo{person}{Lara Petri}, \bibinfo{person}{Elisabetta Patron}, \bibinfo{person}{Andrea Spoto}, {and} \bibinfo{person}{Michele Vicovaro}.} \bibinfo{year}{2022}\natexlab{}.
\newblock \showarticletitle{Direct Gaze Holds Attention, but Not in Individuals with Obsessive-Compulsive Disorder}.
\newblock \bibinfo{journal}{\emph{Brain Sciences}} \bibinfo{volume}{12}, \bibinfo{number}{2} (\bibinfo{year}{2022}), \bibinfo{pages}{288}.
\newblock


\bibitem[Dalton et~al\mbox{.}(2005)]%
        {dalton2005gaze}
\bibfield{author}{\bibinfo{person}{Kim~M Dalton}, \bibinfo{person}{Brendon~M Nacewicz}, \bibinfo{person}{Tom Johnstone}, \bibinfo{person}{Hillary~S Schaefer}, \bibinfo{person}{Morton~Ann Gernsbacher}, \bibinfo{person}{Hill~H Goldsmith}, \bibinfo{person}{Andrew~L Alexander}, {and} \bibinfo{person}{Richard~J Davidson}.} \bibinfo{year}{2005}\natexlab{}.
\newblock \showarticletitle{Gaze fixation and the neural circuitry of face processing in autism}.
\newblock \bibinfo{journal}{\emph{Nature neuroscience}} \bibinfo{volume}{8}, \bibinfo{number}{4} (\bibinfo{year}{2005}), \bibinfo{pages}{519--526}.
\newblock


\bibitem[David-John et~al\mbox{.}(2022)]%
        {david2022your}
\bibfield{author}{\bibinfo{person}{Brendan David-John}, \bibinfo{person}{Kevin Butler}, {and} \bibinfo{person}{Eakta Jain}.} \bibinfo{year}{2022}\natexlab{}.
\newblock \showarticletitle{For your eyes only: Privacy-preserving eye-tracking datasets}. In \bibinfo{booktitle}{\emph{2022 Symposium on Eye Tracking Research and Applications}}. \bibinfo{pages}{1--6}.
\newblock


\bibitem[Dawson(2023)]%
        {dawson2023could}
\bibfield{author}{\bibinfo{person}{Geraldine Dawson}.} \bibinfo{year}{2023}\natexlab{}.
\newblock \showarticletitle{Could an Eye-Tracking Test Aid Clinicians in Making an Autism Diagnosis?: New Findings and a Look to the Future}.
\newblock \bibinfo{journal}{\emph{JAMA}} \bibinfo{volume}{330}, \bibinfo{number}{9} (\bibinfo{year}{2023}), \bibinfo{pages}{815--817}.
\newblock


\bibitem[de~Belen et~al\mbox{.}(2023)]%
        {de2023eye}
\bibfield{author}{\bibinfo{person}{Ryan~Anthony de Belen}, \bibinfo{person}{Hannah Pincham}, \bibinfo{person}{Antoinette Hodge}, \bibinfo{person}{Natalie Silove}, \bibinfo{person}{Arcot Sowmya}, \bibinfo{person}{Tomasz Bednarz}, {and} \bibinfo{person}{Valsamma Eapen}.} \bibinfo{year}{2023}\natexlab{}.
\newblock \showarticletitle{Eye-tracking correlates of response to joint attention in preschool children with autism spectrum disorder}.
\newblock \bibinfo{journal}{\emph{BMC psychiatry}} \bibinfo{volume}{23}, \bibinfo{number}{1} (\bibinfo{year}{2023}), \bibinfo{pages}{211}.
\newblock


\bibitem[de~Rodez~Benavent et~al\mbox{.}(2017)]%
        {de2017fatigue}
\bibfield{author}{\bibinfo{person}{Sigrid~A de Rodez~Benavent}, \bibinfo{person}{Gro~O Nygaard}, \bibinfo{person}{Hanne~F Harbo}, \bibinfo{person}{Siren T{\o}nnesen}, \bibinfo{person}{Piotr Sowa}, \bibinfo{person}{Nils~I Landr{\o}}, \bibinfo{person}{Marte Wendel-Haga}, \bibinfo{person}{Lars Etholm}, \bibinfo{person}{Kristian~B Nilsen}, \bibinfo{person}{Liv Drolsum}, {et~al\mbox{.}}} \bibinfo{year}{2017}\natexlab{}.
\newblock \showarticletitle{Fatigue and cognition: Pupillary responses to problem-solving in early multiple sclerosis patients}.
\newblock \bibinfo{journal}{\emph{Brain and Behavior}} \bibinfo{volume}{7}, \bibinfo{number}{7} (\bibinfo{year}{2017}), \bibinfo{pages}{e00717}.
\newblock


\bibitem[De~Santi et~al\mbox{.}(2011)]%
        {de2011pursuit}
\bibfield{author}{\bibinfo{person}{Lorenzo De~Santi}, \bibinfo{person}{Pietro Lanzafame}, \bibinfo{person}{Barbara Spano}, \bibinfo{person}{Giangaetano D’Aleo}, \bibinfo{person}{Alessia Bramanti}, \bibinfo{person}{Placido Bramanti}, {and} \bibinfo{person}{Silvia Marino}.} \bibinfo{year}{2011}\natexlab{}.
\newblock \showarticletitle{Pursuit ocular movements in multiple sclerosis: a video-based eye-tracking study}.
\newblock \bibinfo{journal}{\emph{Neurological Sciences}}  \bibinfo{volume}{32} (\bibinfo{year}{2011}), \bibinfo{pages}{67--71}.
\newblock


\bibitem[De~Seze et~al\mbox{.}(2001)]%
        {de2001pupillary}
\bibfield{author}{\bibinfo{person}{J De~Seze}, \bibinfo{person}{C Arndt}, \bibinfo{person}{T Stojkovic}, \bibinfo{person}{M Ayachi}, \bibinfo{person}{JY Gauvrit}, \bibinfo{person}{M Bughin}, \bibinfo{person}{T Saint~Michel}, \bibinfo{person}{JP Pruvo}, \bibinfo{person}{JC Hache}, {and} \bibinfo{person}{P Vermersch}.} \bibinfo{year}{2001}\natexlab{}.
\newblock \showarticletitle{Pupillary disturbances in multiple sclerosis: correlation with MRI findings}.
\newblock \bibinfo{journal}{\emph{Journal of the neurological sciences}} \bibinfo{volume}{188}, \bibinfo{number}{1-2} (\bibinfo{year}{2001}), \bibinfo{pages}{37--41}.
\newblock


\bibitem[de~Villers-Sidani et~al\mbox{.}(2023a)]%
        {de2023oculomotor}
\bibfield{author}{\bibinfo{person}{{\'E}tienne de Villers-Sidani}, \bibinfo{person}{Patrice Voss}, \bibinfo{person}{Natacha Bastien}, \bibinfo{person}{J~Miguel Cisneros-Franco}, \bibinfo{person}{Shamiza Hussein}, \bibinfo{person}{Nancy~E Mayo}, \bibinfo{person}{Nils~A Koch}, \bibinfo{person}{Alexandre Drouin-Picaro}, \bibinfo{person}{Fran{\c{c}}ois Blanchette}, \bibinfo{person}{Daniel Guitton}, {et~al\mbox{.}}} \bibinfo{year}{2023}\natexlab{a}.
\newblock \showarticletitle{Oculomotor analysis to assess brain health: preliminary findings from a longitudinal study of multiple sclerosis using novel tablet-based eye-tracking software}.
\newblock \bibinfo{journal}{\emph{Frontiers in Neurology}}  \bibinfo{volume}{14} (\bibinfo{year}{2023}), \bibinfo{pages}{1243594}.
\newblock


\bibitem[de~Villers-Sidani et~al\mbox{.}(2023b)]%
        {de2023novel}
\bibfield{author}{\bibinfo{person}{{\'E}tienne de Villers-Sidani}, \bibinfo{person}{Patrice Voss}, \bibinfo{person}{Daniel Guitton}, \bibinfo{person}{J~Miguel Cisneros-Franco}, \bibinfo{person}{Nils~A Koch}, {and} \bibinfo{person}{Simon Ducharme}.} \bibinfo{year}{2023}\natexlab{b}.
\newblock \showarticletitle{A novel tablet-based software for the acquisition and analysis of gaze and eye movement parameters: a preliminary validation study in Parkinson’s disease}.
\newblock \bibinfo{journal}{\emph{Frontiers in Neurology}}  \bibinfo{volume}{14} (\bibinfo{year}{2023}), \bibinfo{pages}{1204733}.
\newblock


\bibitem[DeJong and Jones(1971)]%
        {dejong1971akinesia}
\bibfield{author}{\bibinfo{person}{J~David DeJong} {and} \bibinfo{person}{G~Melvill Jones}.} \bibinfo{year}{1971}\natexlab{}.
\newblock \showarticletitle{Akinesia, hypokinesia, and bradykinesia in the oculomotor system of patients with Parkinson's disease}.
\newblock \bibinfo{journal}{\emph{Experimental Neurology}} \bibinfo{volume}{32}, \bibinfo{number}{1} (\bibinfo{year}{1971}), \bibinfo{pages}{58--68}.
\newblock


\bibitem[Demer et~al\mbox{.}(1990)]%
        {demer1990ocular}
\bibfield{author}{\bibinfo{person}{Joseph~L Demer}, \bibinfo{person}{John~B Holds}, {and} \bibinfo{person}{Laura~A Hovis}.} \bibinfo{year}{1990}\natexlab{}.
\newblock \showarticletitle{Ocular movements in essential blepharospasm}.
\newblock \bibinfo{journal}{\emph{American journal of ophthalmology}} \bibinfo{volume}{110}, \bibinfo{number}{6} (\bibinfo{year}{1990}), \bibinfo{pages}{674--682}.
\newblock


\bibitem[Dereu et~al\mbox{.}(2010)]%
        {dereu2010screening}
\bibfield{author}{\bibinfo{person}{Mieke Dereu}, \bibinfo{person}{Petra Warreyn}, \bibinfo{person}{Ruth Raymaekers}, \bibinfo{person}{Mieke Meirsschaut}, \bibinfo{person}{Griet Pattyn}, \bibinfo{person}{Inge Schietecatte}, {and} \bibinfo{person}{Herbert Roeyers}.} \bibinfo{year}{2010}\natexlab{}.
\newblock \showarticletitle{Screening for autism spectrum disorders in Flemish day-care centres with the checklist for early signs of developmental disorders}.
\newblock \bibinfo{journal}{\emph{Journal of autism and developmental disorders}}  \bibinfo{volume}{40} (\bibinfo{year}{2010}), \bibinfo{pages}{1247--1258}.
\newblock


\bibitem[Desrochers et~al\mbox{.}(2019)]%
        {desrochers2019sensorimotor}
\bibfield{author}{\bibinfo{person}{Phillip Desrochers}, \bibinfo{person}{Alexander Brunfeldt}, \bibinfo{person}{Christos Sidiropoulos}, {and} \bibinfo{person}{Florian Kagerer}.} \bibinfo{year}{2019}\natexlab{}.
\newblock \showarticletitle{Sensorimotor control in dystonia}.
\newblock \bibinfo{journal}{\emph{Brain Sciences}} \bibinfo{volume}{9}, \bibinfo{number}{4} (\bibinfo{year}{2019}), \bibinfo{pages}{79}.
\newblock


\bibitem[Deuschl and Goddemeier(1998)]%
        {deuschl1998spontaneous}
\bibfield{author}{\bibinfo{person}{G{\"u}nther Deuschl} {and} \bibinfo{person}{Christof Goddemeier}.} \bibinfo{year}{1998}\natexlab{}.
\newblock \showarticletitle{Spontaneous and reflex activity of facial muscles in dystonia, Parkinson’s disease, and in normal subjects}.
\newblock \bibinfo{journal}{\emph{Journal of Neurology, Neurosurgery \& Psychiatry}} \bibinfo{volume}{64}, \bibinfo{number}{3} (\bibinfo{year}{1998}), \bibinfo{pages}{320--324}.
\newblock


\bibitem[digital~science solution({[n.\,d.]})]%
        {dimensions}
\bibfield{author}{\bibinfo{person}{A digital~science solution}.} \bibinfo{year}{[n.\,d.]}\natexlab{}.
\newblock \bibinfo{booktitle}{\emph{dimensions.ai}}.
\newblock
\newblock
\shownote{Accessed: 2024-01-30}.


\bibitem[Drewes et~al\mbox{.}(2014)]%
        {drewes2014smaller}
\bibfield{author}{\bibinfo{person}{Jan Drewes}, \bibinfo{person}{Weina Zhu}, \bibinfo{person}{Yingzhou Hu}, {and} \bibinfo{person}{Xintian Hu}.} \bibinfo{year}{2014}\natexlab{}.
\newblock \showarticletitle{Smaller is better: Drift in gaze measurements due to pupil dynamics}.
\newblock \bibinfo{journal}{\emph{PloS one}} \bibinfo{volume}{9}, \bibinfo{number}{10} (\bibinfo{year}{2014}), \bibinfo{pages}{e111197}.
\newblock


\bibitem[Dryad(2024)]%
        {dryad2024}
\bibfield{author}{\bibinfo{person}{Dryad}.} \bibinfo{year}{2024}\natexlab{}.
\newblock \bibinfo{title}{Dryad Digital Repository}.
\newblock \bibinfo{howpublished}{\url{https://datadryad.org/}}.
\newblock
\newblock
\shownote{Accessed: 2024-01-30}.


\bibitem[Dubey et~al\mbox{.}(2023)]%
        {dubey2023psychosocial}
\bibfield{author}{\bibinfo{person}{Souvik Dubey}, \bibinfo{person}{Ritwik Ghosh}, \bibinfo{person}{Mahua~Jana Dubey}, \bibinfo{person}{Shambaditya Das}, \bibinfo{person}{Arka~Prava Chakraborty}, \bibinfo{person}{Arindam Santra}, \bibinfo{person}{Ajitava Dutta}, \bibinfo{person}{Dipayan Roy}, \bibinfo{person}{Alak Pandit}, \bibinfo{person}{Biman~Kanti Roy}, {et~al\mbox{.}}} \bibinfo{year}{2023}\natexlab{}.
\newblock \showarticletitle{Psychosocial basis of human sufferings and poverty in patients with neurological and psychiatric disorders}.
\newblock \bibinfo{journal}{\emph{Medical research archives}} \bibinfo{volume}{11}, \bibinfo{number}{5} (\bibinfo{year}{2023}).
\newblock


\bibitem[Duncan et~al\mbox{.}(1991)]%
        {duncan1991gaze}
\bibfield{author}{\bibinfo{person}{Max~B Duncan}, \bibinfo{person}{Bahman Jabbari}, {and} \bibinfo{person}{Michael~L Rosenberg}.} \bibinfo{year}{1991}\natexlab{}.
\newblock \showarticletitle{Gaze-evoked visual seizures in nonketotic hyperglycemia}.
\newblock \bibinfo{journal}{\emph{Epilepsia}} \bibinfo{volume}{32}, \bibinfo{number}{2} (\bibinfo{year}{1991}), \bibinfo{pages}{221--224}.
\newblock


\bibitem[Dunn et~al\mbox{.}(2024)]%
        {dunn2024minimal}
\bibfield{author}{\bibinfo{person}{Matt~J Dunn}, \bibinfo{person}{Robert~G Alexander}, \bibinfo{person}{Onyekachukwu~M Amiebenomo}, \bibinfo{person}{Gemma Arblaster}, \bibinfo{person}{Denize Atan}, \bibinfo{person}{Jonathan~T Erichsen}, \bibinfo{person}{Ulrich Ettinger}, \bibinfo{person}{Mario~E Giardini}, \bibinfo{person}{Iain~D Gilchrist}, \bibinfo{person}{Ruth Hamilton}, {et~al\mbox{.}}} \bibinfo{year}{2024}\natexlab{}.
\newblock \showarticletitle{Minimal reporting guideline for research involving eye tracking (2023 edition)}.
\newblock \bibinfo{journal}{\emph{Behavior research methods}} \bibinfo{volume}{56}, \bibinfo{number}{5} (\bibinfo{year}{2024}), \bibinfo{pages}{4351--4357}.
\newblock


\bibitem[Dursun et~al\mbox{.}(2000)]%
        {dursun2000antisaccade}
\bibfield{author}{\bibinfo{person}{Serdar~M Dursun}, \bibinfo{person}{John~G Burke}, {and} \bibinfo{person}{Michael~A Reveley}.} \bibinfo{year}{2000}\natexlab{}.
\newblock \showarticletitle{Antisaccade eye movement abnormalities in Tourette syndrome: evidence for cortico-striatal network dysfunction?}
\newblock \bibinfo{journal}{\emph{Journal of Psychopharmacology}} \bibinfo{volume}{14}, \bibinfo{number}{1} (\bibinfo{year}{2000}), \bibinfo{pages}{37--39}.
\newblock


\bibitem[Dwyer(2022)]%
        {dwyer2022neurodiversity}
\bibfield{author}{\bibinfo{person}{Patrick Dwyer}.} \bibinfo{year}{2022}\natexlab{}.
\newblock \showarticletitle{The neurodiversity approach (es): What are they and what do they mean for researchers?}
\newblock \bibinfo{journal}{\emph{Human development}} \bibinfo{volume}{66}, \bibinfo{number}{2} (\bibinfo{year}{2022}), \bibinfo{pages}{73--92}.
\newblock


\bibitem[D’Antonio et~al\mbox{.}(2021)]%
        {d2021blink}
\bibfield{author}{\bibinfo{person}{Fabrizia D’Antonio}, \bibinfo{person}{Maria I~De Bartolo}, \bibinfo{person}{Gina Ferrazzano}, \bibinfo{person}{Micaela~S Monti}, \bibinfo{person}{Letizia Imbriano}, \bibinfo{person}{Alessandro Trebbastoni}, \bibinfo{person}{Alfredo Berardelli}, {and} \bibinfo{person}{Antonella Conte}.} \bibinfo{year}{2021}\natexlab{}.
\newblock \showarticletitle{Blink Rate Study in Patients with Alzheimer's Disease, Mild Cognitive Impairment and Subjective Cognitive Decline}.
\newblock \bibinfo{journal}{\emph{Current Alzheimer Research}} \bibinfo{volume}{18}, \bibinfo{number}{14} (\bibinfo{year}{2021}), \bibinfo{pages}{1104--1110}.
\newblock


\bibitem[Ehinger(2025)]%
        {Ehinger}
\bibfield{author}{\bibinfo{person}{Benedikt Ehinger}.} \bibinfo{year}{2025}\natexlab{}.
\newblock \bibinfo{title}{consentform\_gh\_action}.
\newblock
\urldef\tempurl%
\url{https://github.com/s-ccs/consentform_gh_action}
\showURL{%
\tempurl}
\newblock
\shownote{Accessed: 2025-02-10}.


\bibitem[El~Hmimdi et~al\mbox{.}(2022)]%
        {el2022predicting}
\bibfield{author}{\bibinfo{person}{Alae~Eddine El~Hmimdi}, \bibinfo{person}{Lindsey~M Ward}, \bibinfo{person}{Themis Palpanas}, \bibinfo{person}{Vivien Sainte Fare~Garnot}, {and} \bibinfo{person}{Zo{\"\i} Kapoula}.} \bibinfo{year}{2022}\natexlab{}.
\newblock \showarticletitle{Predicting Dyslexia in Adolescents from Eye Movements during Free Painting Viewing}.
\newblock \bibinfo{journal}{\emph{Brain Sciences}} \bibinfo{volume}{12}, \bibinfo{number}{8} (\bibinfo{year}{2022}), \bibinfo{pages}{1031}.
\newblock


\bibitem[Elfares et~al\mbox{.}(2023)]%
        {elfares2023federated}
\bibfield{author}{\bibinfo{person}{Mayar Elfares}, \bibinfo{person}{Zhiming Hu}, \bibinfo{person}{Pascal Reisert}, \bibinfo{person}{Andreas Bulling}, {and} \bibinfo{person}{Ralf K{\"u}sters}.} \bibinfo{year}{2023}\natexlab{}.
\newblock \showarticletitle{Federated learning for appearance-based gaze estimation in the wild}. In \bibinfo{booktitle}{\emph{Gaze Meets Machine Learning Workshop}}. PMLR, \bibinfo{pages}{20--36}.
\newblock


\bibitem[Elfares et~al\mbox{.}(2024)]%
        {elfares2024privateyes}
\bibfield{author}{\bibinfo{person}{Mayar Elfares}, \bibinfo{person}{Pascal Reisert}, \bibinfo{person}{Zhiming Hu}, \bibinfo{person}{Wenwu Tang}, \bibinfo{person}{Ralf K{\"u}sters}, {and} \bibinfo{person}{Andreas Bulling}.} \bibinfo{year}{2024}\natexlab{}.
\newblock \showarticletitle{PrivatEyes: Appearance-based Gaze Estimation Using Federated Secure Multi-Party Computation}.
\newblock \bibinfo{journal}{\emph{Proceedings of the ACM on Human-Computer Interaction}} \bibinfo{volume}{8}, \bibinfo{number}{ETRA} (\bibinfo{year}{2024}), \bibinfo{pages}{1--23}.
\newblock


\bibitem[Ellmers et~al\mbox{.}(2020)]%
        {ellmers2020evidence}
\bibfield{author}{\bibinfo{person}{Toby~J Ellmers}, \bibinfo{person}{Adam~J Cocks}, {and} \bibinfo{person}{William~R Young}.} \bibinfo{year}{2020}\natexlab{}.
\newblock \showarticletitle{Evidence of a link between fall-related anxiety and high-risk patterns of visual search in older adults during adaptive locomotion}.
\newblock \bibinfo{journal}{\emph{The Journals of Gerontology: Series A}} \bibinfo{volume}{75}, \bibinfo{number}{5} (\bibinfo{year}{2020}), \bibinfo{pages}{961--967}.
\newblock


\bibitem[Emery(2000)]%
        {emery2000eyes}
\bibfield{author}{\bibinfo{person}{Nathan~J Emery}.} \bibinfo{year}{2000}\natexlab{}.
\newblock \showarticletitle{The eyes have it: the neuroethology, function and evolution of social gaze}.
\newblock \bibinfo{journal}{\emph{Neuroscience \& biobehavioral reviews}} \bibinfo{volume}{24}, \bibinfo{number}{6} (\bibinfo{year}{2000}), \bibinfo{pages}{581--604}.
\newblock


\bibitem[Eriksson and Papanikotopoulos(1997)]%
        {eriksson1997eye}
\bibfield{author}{\bibinfo{person}{Martin Eriksson} {and} \bibinfo{person}{Nikolaos~P Papanikotopoulos}.} \bibinfo{year}{1997}\natexlab{}.
\newblock \showarticletitle{Eye-tracking for detection of driver fatigue}. In \bibinfo{booktitle}{\emph{Proceedings of Conference on Intelligent Transportation Systems}}. IEEE, \bibinfo{pages}{314--319}.
\newblock


\bibitem[Espay et~al\mbox{.}(2018)]%
        {espay2018current}
\bibfield{author}{\bibinfo{person}{Alberto~J Espay}, \bibinfo{person}{Selma Aybek}, \bibinfo{person}{Alan Carson}, \bibinfo{person}{Mark~J Edwards}, \bibinfo{person}{Laura~H Goldstein}, \bibinfo{person}{Mark Hallett}, \bibinfo{person}{Kathrin LaFaver}, \bibinfo{person}{W~Curt LaFrance}, \bibinfo{person}{Anthony~E Lang}, \bibinfo{person}{Tim Nicholson}, {et~al\mbox{.}}} \bibinfo{year}{2018}\natexlab{}.
\newblock \showarticletitle{Current concepts in diagnosis and treatment of functional neurological disorders}.
\newblock \bibinfo{journal}{\emph{JAMA neurology}} \bibinfo{volume}{75}, \bibinfo{number}{9} (\bibinfo{year}{2018}), \bibinfo{pages}{1132--1141}.
\newblock


\bibitem[Falaki et~al\mbox{.}(2023)]%
        {falaki2023multi}
\bibfield{author}{\bibinfo{person}{Ali Falaki}, \bibinfo{person}{Cristian Cuadra}, \bibinfo{person}{Mechelle~M Lewis}, \bibinfo{person}{Janina~M Prado-Rico}, \bibinfo{person}{Xuemei Huang}, {and} \bibinfo{person}{Mark~L Latash}.} \bibinfo{year}{2023}\natexlab{}.
\newblock \showarticletitle{Multi-muscle synergies in preparation for gait initiation in Parkinson’s disease}.
\newblock \bibinfo{journal}{\emph{Clinical Neurophysiology}}  \bibinfo{volume}{154} (\bibinfo{year}{2023}), \bibinfo{pages}{12--24}.
\newblock


\bibitem[Feit et~al\mbox{.}(2017)]%
        {feit2017toward}
\bibfield{author}{\bibinfo{person}{Anna~Maria Feit}, \bibinfo{person}{Shane Williams}, \bibinfo{person}{Arturo Toledo}, \bibinfo{person}{Ann Paradiso}, \bibinfo{person}{Harish Kulkarni}, \bibinfo{person}{Shaun Kane}, {and} \bibinfo{person}{Meredith~Ringel Morris}.} \bibinfo{year}{2017}\natexlab{}.
\newblock \showarticletitle{Toward everyday gaze input: Accuracy and precision of eye tracking and implications for design}. In \bibinfo{booktitle}{\emph{Proceedings of the 2017 Chi conference on human factors in computing systems}}. \bibinfo{pages}{1118--1130}.
\newblock


\bibitem[Fenoglio et~al\mbox{.}(2023)]%
        {fenoglio2023federated}
\bibfield{author}{\bibinfo{person}{Dario Fenoglio}, \bibinfo{person}{Daniel Josifovski}, \bibinfo{person}{Alessandro Gobbetti}, \bibinfo{person}{Mattias Formo}, \bibinfo{person}{Hristijan Gjoreski}, \bibinfo{person}{Martin Gjoreski}, {and} \bibinfo{person}{Marc Langheinrich}.} \bibinfo{year}{2023}\natexlab{}.
\newblock \showarticletitle{Federated Learning for Privacy-aware Cognitive Workload Estimation}. In \bibinfo{booktitle}{\emph{Proceedings of the 22nd International Conference on Mobile and Ubiquitous Multimedia}}. \bibinfo{pages}{25--36}.
\newblock


\bibitem[Fernandes et~al\mbox{.}(2023)]%
        {fernandes2023digital}
\bibfield{author}{\bibinfo{person}{Felipe Fernandes}, \bibinfo{person}{Ingridy Barbalho}, \bibinfo{person}{Arnaldo Bispo~Junior}, \bibinfo{person}{Luca Alves}, \bibinfo{person}{Danilo Nagem}, \bibinfo{person}{Hertz Lins}, \bibinfo{person}{Ernano Arrais~Junior}, \bibinfo{person}{Karilany~D Coutinho}, \bibinfo{person}{Ant{\^o}nio~HF Morais}, \bibinfo{person}{Jo{\~a}o Paulo~Q Santos}, {et~al\mbox{.}}} \bibinfo{year}{2023}\natexlab{}.
\newblock \showarticletitle{Digital Alternative Communication for Individuals with amyotrophic lateral sclerosis: what we have}.
\newblock \bibinfo{journal}{\emph{Journal of Clinical Medicine}} \bibinfo{volume}{12}, \bibinfo{number}{16} (\bibinfo{year}{2023}), \bibinfo{pages}{5235}.
\newblock


\bibitem[Fern{\'a}ndez et~al\mbox{.}(2015)]%
        {fernandez2015diagnosis}
\bibfield{author}{\bibinfo{person}{Gerardo Fern{\'a}ndez}, \bibinfo{person}{Liliana~R Castro}, \bibinfo{person}{Marcela Schumacher}, {and} \bibinfo{person}{Osvaldo~E Agamennoni}.} \bibinfo{year}{2015}\natexlab{}.
\newblock \showarticletitle{Diagnosis of mild Alzheimer disease through the analysis of eye movements during reading}.
\newblock \bibinfo{journal}{\emph{Journal of integrative neuroscience}} \bibinfo{volume}{14}, \bibinfo{number}{01} (\bibinfo{year}{2015}), \bibinfo{pages}{121--133}.
\newblock


\bibitem[Fern{\'a}ndez et~al\mbox{.}(2014)]%
        {fernandez2014lack}
\bibfield{author}{\bibinfo{person}{Gerardo Fern{\'a}ndez}, \bibinfo{person}{Facundo Manes}, \bibinfo{person}{Nora~P Rotstein}, \bibinfo{person}{Oscar Colombo}, \bibinfo{person}{Pablo Mandolesi}, \bibinfo{person}{Luis~E Politi}, {and} \bibinfo{person}{Osvaldo Agamennoni}.} \bibinfo{year}{2014}\natexlab{}.
\newblock \showarticletitle{Lack of contextual-word predictability during reading in patients with mild Alzheimer disease}.
\newblock \bibinfo{journal}{\emph{Neuropsychologia}}  \bibinfo{volume}{62} (\bibinfo{year}{2014}), \bibinfo{pages}{143--151}.
\newblock


\bibitem[Ferrazzano et~al\mbox{.}(2015)]%
        {ferrazzano2015botulinum}
\bibfield{author}{\bibinfo{person}{Gina Ferrazzano}, \bibinfo{person}{Antonella Conte}, \bibinfo{person}{Giovanni Fabbrini}, \bibinfo{person}{Matteo Bologna}, \bibinfo{person}{Antonella Macerollo}, \bibinfo{person}{Giovanni Defazio}, \bibinfo{person}{Mark Hallett}, {and} \bibinfo{person}{Alfredo Berardelli}.} \bibinfo{year}{2015}\natexlab{}.
\newblock \showarticletitle{Botulinum toxin and blink rate in patients with blepharospasm and increased blinking}.
\newblock \bibinfo{journal}{\emph{Journal of Neurology, Neurosurgery \& Psychiatry}} \bibinfo{volume}{86}, \bibinfo{number}{3} (\bibinfo{year}{2015}), \bibinfo{pages}{336--340}.
\newblock


\bibitem[Feys et~al\mbox{.}(2008)]%
        {feys2008unsteady}
\bibfield{author}{\bibinfo{person}{P Feys}, \bibinfo{person}{W Helsen}, \bibinfo{person}{B Nuttin}, \bibinfo{person}{A Lavrysen}, \bibinfo{person}{P Ketelaer}, \bibinfo{person}{S Swinnen}, {and} \bibinfo{person}{X Liu}.} \bibinfo{year}{2008}\natexlab{}.
\newblock \showarticletitle{Unsteady gaze fixation enhances the severity of MS intention tremor}.
\newblock \bibinfo{journal}{\emph{Neurology}} \bibinfo{volume}{70}, \bibinfo{number}{2} (\bibinfo{year}{2008}), \bibinfo{pages}{106--113}.
\newblock


\bibitem[Fiedler et~al\mbox{.}(2020)]%
        {fiedler2020guideline}
\bibfield{author}{\bibinfo{person}{Susann Fiedler}, \bibinfo{person}{Michael Schulte-Mecklenbeck}, \bibinfo{person}{Frank Renkewitz}, {and} \bibinfo{person}{Jacob~L Orquin}.} \bibinfo{year}{2020}\natexlab{}.
\newblock \showarticletitle{Guideline for reporting standards of eye-tracking research in decision sciences}.
\newblock  (\bibinfo{year}{2020}).
\newblock


\bibitem[Fink-Lamotte et~al\mbox{.}(2022)]%
        {fink2022you}
\bibfield{author}{\bibinfo{person}{Jakob Fink-Lamotte}, \bibinfo{person}{Frederike Svensson}, \bibinfo{person}{Julian Schmitz}, {and} \bibinfo{person}{Cornelia Exner}.} \bibinfo{year}{2022}\natexlab{}.
\newblock \showarticletitle{Are you looking or looking away? Visual exploration and avoidance of disgust-and fear-stimuli: An eye-tracking study.}
\newblock \bibinfo{journal}{\emph{Emotion}} \bibinfo{volume}{22}, \bibinfo{number}{8} (\bibinfo{year}{2022}), \bibinfo{pages}{1909}.
\newblock


\bibitem[Fischer and Hartnegg(2000)]%
        {fischer2000stability}
\bibfield{author}{\bibinfo{person}{Burkhart Fischer} {and} \bibinfo{person}{Klaus Hartnegg}.} \bibinfo{year}{2000}\natexlab{}.
\newblock \showarticletitle{Stability of gaze control in dyslexia}.
\newblock \bibinfo{journal}{\emph{Strabismus}} \bibinfo{volume}{8}, \bibinfo{number}{2} (\bibinfo{year}{2000}), \bibinfo{pages}{119--122}.
\newblock


\bibitem[Fisher et~al\mbox{.}(2022)]%
        {fisher2022visually}
\bibfield{author}{\bibinfo{person}{Robert~S Fisher}, \bibinfo{person}{Jayant~N Acharya}, \bibinfo{person}{Fiona~Mitchell Baumer}, \bibinfo{person}{Jacqueline~A French}, \bibinfo{person}{Pasquale Parisi}, \bibinfo{person}{Jessica~H Solodar}, \bibinfo{person}{Jerzy~P Szaflarski}, \bibinfo{person}{Liu~Lin Thio}, \bibinfo{person}{Benjamin Tolchin}, \bibinfo{person}{Arnold~J Wilkins}, {et~al\mbox{.}}} \bibinfo{year}{2022}\natexlab{}.
\newblock \showarticletitle{Visually sensitive seizures: An updated review by the Epilepsy Foundation}.
\newblock \bibinfo{journal}{\emph{Epilepsia}} \bibinfo{volume}{63}, \bibinfo{number}{4} (\bibinfo{year}{2022}), \bibinfo{pages}{739--768}.
\newblock


\bibitem[Fisher and Schachter(2000)]%
        {fisher2000postictal}
\bibfield{author}{\bibinfo{person}{Robert~S Fisher} {and} \bibinfo{person}{Steven~C Schachter}.} \bibinfo{year}{2000}\natexlab{}.
\newblock \showarticletitle{The postictal state: a neglected entity in the management of epilepsy}.
\newblock \bibinfo{journal}{\emph{Epilepsy \& Behavior}} \bibinfo{volume}{1}, \bibinfo{number}{1} (\bibinfo{year}{2000}), \bibinfo{pages}{52--59}.
\newblock


\bibitem[Fletcher and Sharpe(1986)]%
        {fletcher1986saccadic}
\bibfield{author}{\bibinfo{person}{William~A Fletcher} {and} \bibinfo{person}{James~A Sharpe}.} \bibinfo{year}{1986}\natexlab{}.
\newblock \showarticletitle{Saccadic eye movement dysfunction in Alzheimer's disease}.
\newblock \bibinfo{journal}{\emph{Annals of Neurology: Official Journal of the American Neurological Association and the Child Neurology Society}} \bibinfo{volume}{20}, \bibinfo{number}{4} (\bibinfo{year}{1986}), \bibinfo{pages}{464--471}.
\newblock


\bibitem[Fletcher and Sharpe(1988)]%
        {fletcher1988smooth}
\bibfield{author}{\bibinfo{person}{William~A Fletcher} {and} \bibinfo{person}{James~A Sharpe}.} \bibinfo{year}{1988}\natexlab{}.
\newblock \showarticletitle{Smooth pursuit dysfunction in Alzheimer's disease}.
\newblock \bibinfo{journal}{\emph{Neurology}} \bibinfo{volume}{38}, \bibinfo{number}{2} (\bibinfo{year}{1988}), \bibinfo{pages}{272--272}.
\newblock


\bibitem[Food and Administration(2024)]%
        {fda2024}
\bibfield{author}{\bibinfo{person}{U.S. Food} {and} \bibinfo{person}{Drug Administration}.} \bibinfo{year}{2024}\natexlab{}.
\newblock \bibinfo{title}{Food and Drug Administration (FDA)}.
\newblock \bibinfo{howpublished}{\url{https://www.fda.gov}}.
\newblock
\newblock
\shownote{Accessed: 2024-01-30}.


\bibitem[for Nuclear Research~(CERN)(2024)]%
        {zenodo2024}
\bibfield{author}{\bibinfo{person}{European~Organization for Nuclear Research~(CERN)}.} \bibinfo{year}{2024}\natexlab{}.
\newblock \bibinfo{title}{Zenodo}.
\newblock \bibinfo{howpublished}{\url{https://zenodo.org}}.
\newblock
\newblock
\shownote{Accessed: 2024-01-30}.


\bibitem[Foster and Deardorff(2017)]%
        {osf_project2024}
\bibfield{author}{\bibinfo{person}{Erin~D Foster} {and} \bibinfo{person}{Ariel Deardorff}.} \bibinfo{year}{2017}\natexlab{}.
\newblock \showarticletitle{Open science framework (OSF)}.
\newblock \bibinfo{journal}{\emph{Journal of the Medical Library Association: JMLA}} \bibinfo{volume}{105}, \bibinfo{number}{2} (\bibinfo{year}{2017}), \bibinfo{pages}{203}.
\newblock


\bibitem[Foster et~al\mbox{.}({[n.\,d.]})]%
        {fosterpre}
\bibfield{author}{\bibinfo{person}{Joshua~J Foster}, \bibinfo{person}{Amelia~H Harrison}, {and} \bibinfo{person}{Sam Ling}.} \bibinfo{year}{[n.\,d.]}\natexlab{}.
\newblock \showarticletitle{Pre-Registration: Does divided attention impair detection of simple visual features?}
\newblock  (\bibinfo{year}{[n.\,d.]}).
\newblock


\bibitem[Fotiou et~al\mbox{.}(2007)]%
        {fotiou2007pupil}
\bibfield{author}{\bibinfo{person}{Dimitris~F Fotiou}, \bibinfo{person}{Catherine~G Brozou}, \bibinfo{person}{Anna-Bettina Haidich}, \bibinfo{person}{Dimitris Tsiptsios}, \bibinfo{person}{Maria Nakou}, \bibinfo{person}{Anastasia Kabitsi}, \bibinfo{person}{Charalambos Giantselidis}, {and} \bibinfo{person}{Fotis Fotiou}.} \bibinfo{year}{2007}\natexlab{}.
\newblock \showarticletitle{Pupil reaction to light in Alzheimer’s disease: evaluation of pupil size changes and mobility}.
\newblock \bibinfo{journal}{\emph{Aging clinical and experimental research}}  \bibinfo{volume}{19} (\bibinfo{year}{2007}), \bibinfo{pages}{364--371}.
\newblock


\bibitem[Fox et~al\mbox{.}(2001)]%
        {fox2001threatening}
\bibfield{author}{\bibinfo{person}{Elaine Fox}, \bibinfo{person}{Riccardo Russo}, \bibinfo{person}{Robert Bowles}, {and} \bibinfo{person}{Kevin Dutton}.} \bibinfo{year}{2001}\natexlab{}.
\newblock \showarticletitle{Do threatening stimuli draw or hold visual attention in subclinical anxiety?}
\newblock \bibinfo{journal}{\emph{Journal of experimental psychology: General}} \bibinfo{volume}{130}, \bibinfo{number}{4} (\bibinfo{year}{2001}), \bibinfo{pages}{681}.
\newblock


\bibitem[Frankel and Cummings(1984)]%
        {frankel1984neuro}
\bibfield{author}{\bibinfo{person}{Michael Frankel} {and} \bibinfo{person}{Jeffrey~L Cummings}.} \bibinfo{year}{1984}\natexlab{}.
\newblock \showarticletitle{Neuro-opthalmic abnormalities in Tourette's syndrome: Functional and automic implications}.
\newblock \bibinfo{journal}{\emph{Neurology}} \bibinfo{volume}{34}, \bibinfo{number}{3} (\bibinfo{year}{1984}), \bibinfo{pages}{359--359}.
\newblock


\bibitem[Frei(2021)]%
        {frei2021abnormalities}
\bibfield{author}{\bibinfo{person}{Karen Frei}.} \bibinfo{year}{2021}\natexlab{}.
\newblock \showarticletitle{Abnormalities of smooth pursuit in Parkinson’s disease: A systematic review}.
\newblock \bibinfo{journal}{\emph{Clinical parkinsonism \& related disorders}}  \bibinfo{volume}{4} (\bibinfo{year}{2021}), \bibinfo{pages}{100085}.
\newblock


\bibitem[Fried et~al\mbox{.}(2014)]%
        {fried2014adhd}
\bibfield{author}{\bibinfo{person}{Moshe Fried}, \bibinfo{person}{Eteri Tsitsiashvili}, \bibinfo{person}{Yoram~S Bonneh}, \bibinfo{person}{Anna Sterkin}, \bibinfo{person}{Tamara Wygnanski-Jaffe}, \bibinfo{person}{Tamir Epstein}, {and} \bibinfo{person}{Uri Polat}.} \bibinfo{year}{2014}\natexlab{}.
\newblock \showarticletitle{ADHD subjects fail to suppress eye blinks and microsaccades while anticipating visual stimuli but recover with medication}.
\newblock \bibinfo{journal}{\emph{Vision research}}  \bibinfo{volume}{101} (\bibinfo{year}{2014}), \bibinfo{pages}{62--72}.
\newblock


\bibitem[Froment~Tilikete(2022)]%
        {froment2022assess}
\bibfield{author}{\bibinfo{person}{Caroline Froment~Tilikete}.} \bibinfo{year}{2022}\natexlab{}.
\newblock \showarticletitle{How to assess eye movements clinically}.
\newblock \bibinfo{journal}{\emph{Neurological Sciences}} \bibinfo{volume}{43}, \bibinfo{number}{5} (\bibinfo{year}{2022}), \bibinfo{pages}{2969--2981}.
\newblock


\bibitem[Fu et~al\mbox{.}(2024)]%
        {fu2024recognition}
\bibfield{author}{\bibinfo{person}{Yanlu Fu}, \bibinfo{person}{Jingxin Zhang}, \bibinfo{person}{Yina Cao}, \bibinfo{person}{Linmei Ye}, \bibinfo{person}{Runze Zheng}, \bibinfo{person}{Qiwei Li}, \bibinfo{person}{Beibei Shen}, \bibinfo{person}{Yi Shi}, \bibinfo{person}{Jiuwen Cao}, {and} \bibinfo{person}{Jiajia Fang}.} \bibinfo{year}{2024}\natexlab{}.
\newblock \showarticletitle{Recognition memory deficits detected through eye-tracking in well-controlled children with self-limited epilepsy with centrotemporal spikes}.
\newblock \bibinfo{journal}{\emph{Epilepsia}} \bibinfo{volume}{65}, \bibinfo{number}{4} (\bibinfo{year}{2024}), \bibinfo{pages}{1128--1140}.
\newblock


\bibitem[Fujita et~al\mbox{.}(2023)]%
        {fujita2023istradefylline}
\bibfield{author}{\bibinfo{person}{Youshi Fujita}, \bibinfo{person}{Emi Kawaguchi}, \bibinfo{person}{Takashi Toyomoto}, {and} \bibinfo{person}{Hirotaka Shirasaki}.} \bibinfo{year}{2023}\natexlab{}.
\newblock \showarticletitle{Istradefylline Improves Impaired Smooth Pursuit Eye Movements in Parkinson’s Disease}.
\newblock \bibinfo{journal}{\emph{Neurology and Therapy}} \bibinfo{volume}{12}, \bibinfo{number}{5} (\bibinfo{year}{2023}), \bibinfo{pages}{1791--1798}.
\newblock


\bibitem[Galante and Menezes(2012)]%
        {galante2012gaze}
\bibfield{author}{\bibinfo{person}{Adriano Galante} {and} \bibinfo{person}{Paulo Menezes}.} \bibinfo{year}{2012}\natexlab{}.
\newblock \showarticletitle{A gaze-based interaction system for people with cerebral palsy}.
\newblock \bibinfo{journal}{\emph{Procedia Technology}}  \bibinfo{volume}{5} (\bibinfo{year}{2012}), \bibinfo{pages}{895--902}.
\newblock


\bibitem[Galazka et~al\mbox{.}(2024)]%
        {galazka2024pupil}
\bibfield{author}{\bibinfo{person}{Martyna~A Galazka}, \bibinfo{person}{Max Thorsson}, \bibinfo{person}{Johan Lundin~Kleberg}, \bibinfo{person}{Nouchine Hadjikhani}, {and} \bibinfo{person}{Jakob {\AA}sberg~Johnels}.} \bibinfo{year}{2024}\natexlab{}.
\newblock \showarticletitle{Pupil contagion variation with gaze, arousal, and autistic traits}.
\newblock \bibinfo{journal}{\emph{Scientific Reports}} \bibinfo{volume}{14}, \bibinfo{number}{1} (\bibinfo{year}{2024}), \bibinfo{pages}{18282}.
\newblock


\bibitem[Gandolfi et~al\mbox{.}(2024)]%
        {gandolfi2024window}
\bibfield{author}{\bibinfo{person}{Marialuisa Gandolfi}, \bibinfo{person}{Angela Sandri}, \bibinfo{person}{Sara Mariotto}, \bibinfo{person}{Stefano Tamburin}, \bibinfo{person}{Anna Paolicelli}, \bibinfo{person}{Mirta Fiorio}, \bibinfo{person}{Giulia Pedrotti}, \bibinfo{person}{Paolo Barone}, \bibinfo{person}{Maria~Teresa Pellecchia}, \bibinfo{person}{Roberto Erro}, {et~al\mbox{.}}} \bibinfo{year}{2024}\natexlab{}.
\newblock \showarticletitle{A window into the mind-brain-body interplay: Development of diagnostic, prognostic biomarkers, and rehabilitation strategies in functional motor disorders}.
\newblock \bibinfo{journal}{\emph{Plos one}} \bibinfo{volume}{19}, \bibinfo{number}{9} (\bibinfo{year}{2024}), \bibinfo{pages}{e0309408}.
\newblock


\bibitem[Gao and Sabel(2017)]%
        {gao2017microsaccade}
\bibfield{author}{\bibinfo{person}{Ying Gao} {and} \bibinfo{person}{Bernhard~A Sabel}.} \bibinfo{year}{2017}\natexlab{}.
\newblock \showarticletitle{Microsaccade dysfunction and adaptation in hemianopia after stroke}.
\newblock \bibinfo{journal}{\emph{Restorative neurology and neuroscience}} \bibinfo{volume}{35}, \bibinfo{number}{4} (\bibinfo{year}{2017}), \bibinfo{pages}{365--376}.
\newblock


\bibitem[Garc{\'\i}a~Cena et~al\mbox{.}(2022)]%
        {garcia2022toward}
\bibfield{author}{\bibinfo{person}{Cecilia~E Garc{\'\i}a~Cena}, \bibinfo{person}{David G{\'o}mez-Andr{\'e}s}, \bibinfo{person}{Irene Pulido-Valdeolivas}, \bibinfo{person}{Victoria~Gal{\'a}n S{\'a}nchez-Seco}, \bibinfo{person}{Angela Domingo-Santos}, \bibinfo{person}{Sara Moreno-Garc{\'\i}a}, {and} \bibinfo{person}{Juli{\'a}n Benito-Le{\'o}n}.} \bibinfo{year}{2022}\natexlab{}.
\newblock \showarticletitle{Toward an automatic assessment of cognitive dysfunction in relapsing--remitting multiple sclerosis patients using eye movement analysis}.
\newblock \bibinfo{journal}{\emph{Sensors}} \bibinfo{volume}{22}, \bibinfo{number}{21} (\bibinfo{year}{2022}), \bibinfo{pages}{8220}.
\newblock


\bibitem[Garg et~al\mbox{.}(2018)]%
        {garg2018gaze}
\bibfield{author}{\bibinfo{person}{Hina Garg}, \bibinfo{person}{Leland~E Dibble}, \bibinfo{person}{Michael~C Schubert}, \bibinfo{person}{Jim Sibthorp}, \bibinfo{person}{K~Bo Foreman}, {and} \bibinfo{person}{Eduard Gappmaier}.} \bibinfo{year}{2018}\natexlab{}.
\newblock \showarticletitle{Gaze Stability, Dynamic Balance and Participation Deficits in People with Multiple Sclerosis at Fall-Risk}.
\newblock \bibinfo{journal}{\emph{The Anatomical Record}} \bibinfo{volume}{301}, \bibinfo{number}{11} (\bibinfo{year}{2018}), \bibinfo{pages}{1852--1860}.
\newblock


\bibitem[Garner et~al\mbox{.}(2006)]%
        {garner2006orienting}
\bibfield{author}{\bibinfo{person}{Matthew Garner}, \bibinfo{person}{Karin Mogg}, {and} \bibinfo{person}{Brendan~P Bradley}.} \bibinfo{year}{2006}\natexlab{}.
\newblock \showarticletitle{Orienting and maintenance of gaze to facial expressions in social anxiety.}
\newblock \bibinfo{journal}{\emph{Journal of abnormal psychology}} \bibinfo{volume}{115}, \bibinfo{number}{4} (\bibinfo{year}{2006}), \bibinfo{pages}{760}.
\newblock


\bibitem[Gauthier and Hofferer(1983)]%
        {gauthier1983visual}
\bibfield{author}{\bibinfo{person}{Gabriel~M Gauthier} {and} \bibinfo{person}{Jean~Marie Hofferer}.} \bibinfo{year}{1983}\natexlab{}.
\newblock \showarticletitle{Visual motor rehabilitation in children with cerebral palsy}.
\newblock \bibinfo{journal}{\emph{International rehabilitation medicine}} \bibinfo{volume}{5}, \bibinfo{number}{3} (\bibinfo{year}{1983}), \bibinfo{pages}{118--127}.
\newblock


\bibitem[Gaze et~al\mbox{.}(2006)]%
        {gaze2006co}
\bibfield{author}{\bibinfo{person}{Catherine Gaze}, \bibinfo{person}{Hayden~O Kepley}, {and} \bibinfo{person}{John~T Walkup}.} \bibinfo{year}{2006}\natexlab{}.
\newblock \showarticletitle{Co-occurring psychiatric disorders in children and adolescents with Tourette syndrome}.
\newblock \bibinfo{journal}{\emph{Journal of child neurology}} \bibinfo{volume}{21}, \bibinfo{number}{8} (\bibinfo{year}{2006}), \bibinfo{pages}{657--664}.
\newblock


\bibitem[Georgiou et~al\mbox{.}(1998)]%
        {georgiou1998directed}
\bibfield{author}{\bibinfo{person}{Nellie Georgiou}, \bibinfo{person}{John~L Bradshaw}, {and} \bibinfo{person}{Jim~G Phillips}.} \bibinfo{year}{1998}\natexlab{}.
\newblock \showarticletitle{Directed attention in Gilles de la Tourette syndrome}.
\newblock \bibinfo{journal}{\emph{Behavioural Neurology}} \bibinfo{volume}{11}, \bibinfo{number}{2} (\bibinfo{year}{1998}), \bibinfo{pages}{85--91}.
\newblock


\bibitem[Gerardo et~al\mbox{.}(2024)]%
        {gerardo2024abnormal}
\bibfield{author}{\bibinfo{person}{Fern{\'a}ndez Gerardo}, \bibinfo{person}{Eizaguirre B{\'a}rbara}, \bibinfo{person}{Gonzalez Cecilia}, \bibinfo{person}{Marinangeli Aldana}, \bibinfo{person}{Ciufia Natalia}, \bibinfo{person}{Bacigalupe Lucia}, \bibinfo{person}{Berenice Silva}, \bibinfo{person}{Cohen Leila}, \bibinfo{person}{Pita Cecilia}, \bibinfo{person}{Garcea Orlando}, {et~al\mbox{.}}} \bibinfo{year}{2024}\natexlab{}.
\newblock \showarticletitle{Abnormal eye movements increase as motor disabilities and cognitive impairments become more evident in Multiple Sclerosis: A novel eye-tracking study}.
\newblock \bibinfo{journal}{\emph{Multiple Sclerosis Journal--Experimental, Translational and Clinical}} \bibinfo{volume}{10}, \bibinfo{number}{2} (\bibinfo{year}{2024}), \bibinfo{pages}{20552173241255008}.
\newblock


\bibitem[Ghose et~al\mbox{.}(2020)]%
        {ghose2020pytrack}
\bibfield{author}{\bibinfo{person}{Upamanyu Ghose}, \bibinfo{person}{Arvind~A Srinivasan}, \bibinfo{person}{W~Paul Boyce}, \bibinfo{person}{Hong Xu}, {and} \bibinfo{person}{Eng~Siong Chng}.} \bibinfo{year}{2020}\natexlab{}.
\newblock \showarticletitle{PyTrack: An end-to-end analysis toolkit for eye tracking}.
\newblock \bibinfo{journal}{\emph{Behavior research methods}}  \bibinfo{volume}{52} (\bibinfo{year}{2020}), \bibinfo{pages}{2588--2603}.
\newblock


\bibitem[Giarelli et~al\mbox{.}(2014)]%
        {giarelli2014sensory}
\bibfield{author}{\bibinfo{person}{Ellen Giarelli}, \bibinfo{person}{Romy Nocera}, \bibinfo{person}{Renee Turchi}, \bibinfo{person}{Thomas~L Hardie}, \bibinfo{person}{Rachel Pagano}, {and} \bibinfo{person}{Ce Yuan}.} \bibinfo{year}{2014}\natexlab{}.
\newblock \showarticletitle{Sensory stimuli as obstacles to emergency care for children with autism spectrum disorder}.
\newblock \bibinfo{journal}{\emph{Advanced Emergency Nursing Journal}} \bibinfo{volume}{36}, \bibinfo{number}{2} (\bibinfo{year}{2014}), \bibinfo{pages}{145--163}.
\newblock


\bibitem[Gillespie-Lynch et~al\mbox{.}(2013)]%
        {gillespie2013atypical}
\bibfield{author}{\bibinfo{person}{Kristen Gillespie-Lynch}, \bibinfo{person}{Rebecca Elias}, \bibinfo{person}{Paola Escudero}, \bibinfo{person}{Ted Hutman}, {and} \bibinfo{person}{Scott~P Johnson}.} \bibinfo{year}{2013}\natexlab{}.
\newblock \showarticletitle{Atypical gaze following in autism: A comparison of three potential mechanisms}.
\newblock \bibinfo{journal}{\emph{Journal of autism and developmental disorders}}  \bibinfo{volume}{43} (\bibinfo{year}{2013}), \bibinfo{pages}{2779--2792}.
\newblock


\bibitem[Gitchel et~al\mbox{.}(2013)]%
        {gitchel2013slowed}
\bibfield{author}{\bibinfo{person}{George~T Gitchel}, \bibinfo{person}{Paul~A Wetzel}, {and} \bibinfo{person}{Mark~S Baron}.} \bibinfo{year}{2013}\natexlab{}.
\newblock \showarticletitle{Slowed saccades and increased square wave jerks in essential tremor}.
\newblock \bibinfo{journal}{\emph{Tremor and other hyperkinetic movements}}  \bibinfo{volume}{3} (\bibinfo{year}{2013}).
\newblock


\bibitem[GitHub(2024)]%
        {github2024}
\bibfield{author}{\bibinfo{person}{Inc. GitHub}.} \bibinfo{year}{2024}\natexlab{}.
\newblock \bibinfo{title}{GitHub}.
\newblock \bibinfo{howpublished}{\url{https://github.com}}.
\newblock
\newblock
\shownote{Accessed: 2024-01-30}.


\bibitem[Gomez-Iba{\~n}ez et~al\mbox{.}(2014)]%
        {gomez2014recognition}
\bibfield{author}{\bibinfo{person}{Asier Gomez-Iba{\~n}ez}, \bibinfo{person}{Elena Urrestarazu}, {and} \bibinfo{person}{Cesar Viteri}.} \bibinfo{year}{2014}\natexlab{}.
\newblock \showarticletitle{Recognition of facial emotions and identity in patients with mesial temporal lobe and idiopathic generalized epilepsy: an eye-tracking study}.
\newblock \bibinfo{journal}{\emph{Seizure}} \bibinfo{volume}{23}, \bibinfo{number}{10} (\bibinfo{year}{2014}), \bibinfo{pages}{892--898}.
\newblock


\bibitem[Goodwin et~al\mbox{.}(2021)]%
        {goodwin2021interstaars}
\bibfield{author}{\bibinfo{person}{Amy Goodwin}, \bibinfo{person}{Emily~JH Jones}, \bibinfo{person}{Simona Salomone}, \bibinfo{person}{Luke Mason}, \bibinfo{person}{Rebecca Holman}, \bibinfo{person}{Jannath Begum-Ali}, \bibinfo{person}{Anna Hunt}, \bibinfo{person}{Martin Ruddock}, \bibinfo{person}{George Vamvakas}, \bibinfo{person}{Emily Robinson}, {et~al\mbox{.}}} \bibinfo{year}{2021}\natexlab{}.
\newblock \showarticletitle{INTERSTAARS: Attention training for infants with elevated likelihood of developing ADHD: A proof-of-concept randomised controlled trial}.
\newblock \bibinfo{journal}{\emph{Translational Psychiatry}} \bibinfo{volume}{11}, \bibinfo{number}{1} (\bibinfo{year}{2021}), \bibinfo{pages}{644}.
\newblock


\bibitem[Gotardi et~al\mbox{.}(2022)]%
        {gotardi2022parkinson}
\bibfield{author}{\bibinfo{person}{Gisele~C Gotardi}, \bibinfo{person}{Fabio~A Barbieri}, \bibinfo{person}{Rafael~O Sim{\~a}o}, \bibinfo{person}{Vinicius~A Pereira}, \bibinfo{person}{Andr{\'e}~M Baptista}, \bibinfo{person}{Luiz~F Imaizumi}, \bibinfo{person}{Gabriel Moretto}, \bibinfo{person}{Martina Navarro}, \bibinfo{person}{Paula~F Polastri}, {and} \bibinfo{person}{S{\'e}rgio~T Rodrigues}.} \bibinfo{year}{2022}\natexlab{}.
\newblock \showarticletitle{Parkinson’s disease affects gaze behaviour and performance of drivers}.
\newblock \bibinfo{journal}{\emph{Ergonomics}} \bibinfo{volume}{65}, \bibinfo{number}{9} (\bibinfo{year}{2022}), \bibinfo{pages}{1302--1311}.
\newblock


\bibitem[Grabska et~al\mbox{.}(2024)]%
        {grabska2024reflexive}
\bibfield{author}{\bibinfo{person}{Natalia Grabska}, \bibinfo{person}{Magdalena Wojcik-Pkedziwiatr}, \bibinfo{person}{Jaroslaw Slawek}, \bibinfo{person}{Witold Soltan}, \bibinfo{person}{Justyna Gawryluk}, \bibinfo{person}{Marcin Rudzinski}, \bibinfo{person}{Andrzej Szczudlik}, {and} \bibinfo{person}{Monika Rudzinska-Bar}.} \bibinfo{year}{2024}\natexlab{}.
\newblock \showarticletitle{Reflexive and voluntary saccadic eye movements as biomarker of Huntington’s Disease}.
\newblock \bibinfo{journal}{\emph{Neurologia i Neurochirurgia Polska}} (\bibinfo{year}{2024}).
\newblock


\bibitem[Granholm et~al\mbox{.}(2017)]%
        {granholm2017pupillary}
\bibfield{author}{\bibinfo{person}{Eric~L Granholm}, \bibinfo{person}{Matthew~S Panizzon}, \bibinfo{person}{Jeremy~A Elman}, \bibinfo{person}{Amy~J Jak}, \bibinfo{person}{Richard~L Hauger}, \bibinfo{person}{Mark~W Bondi}, \bibinfo{person}{Michael~J Lyons}, \bibinfo{person}{Carol~E Franz}, {and} \bibinfo{person}{William~S Kremen}.} \bibinfo{year}{2017}\natexlab{}.
\newblock \showarticletitle{Pupillary responses as a biomarker of early risk for Alzheimer’s disease}.
\newblock \bibinfo{journal}{\emph{Journal of Alzheimer's disease}} \bibinfo{volume}{56}, \bibinfo{number}{4} (\bibinfo{year}{2017}), \bibinfo{pages}{1419--1428}.
\newblock


\bibitem[Gregory et~al\mbox{.}(2019)]%
        {gregory2019attention}
\bibfield{author}{\bibinfo{person}{Nicola~J Gregory}, \bibinfo{person}{Helen Bolderston}, {and} \bibinfo{person}{Jastine~V Antolin}.} \bibinfo{year}{2019}\natexlab{}.
\newblock \showarticletitle{Attention to faces and gaze-following in social anxiety: Preliminary evidence from a naturalistic eye-tracking investigation}.
\newblock \bibinfo{journal}{\emph{Cognition and emotion}} \bibinfo{volume}{33}, \bibinfo{number}{5} (\bibinfo{year}{2019}), \bibinfo{pages}{931--942}.
\newblock


\bibitem[Groom et~al\mbox{.}(2017)]%
        {groom2017atypical}
\bibfield{author}{\bibinfo{person}{Madeleine~J Groom}, \bibinfo{person}{Puja Kochhar}, \bibinfo{person}{Antonia Hamilton}, \bibinfo{person}{Elizabeth~B Liddle}, \bibinfo{person}{Marina Simeou}, {and} \bibinfo{person}{Chris Hollis}.} \bibinfo{year}{2017}\natexlab{}.
\newblock \showarticletitle{Atypical processing of gaze cues and faces explains comorbidity between autism spectrum disorder (ASD) and attention deficit/hyperactivity disorder (ADHD)}.
\newblock \bibinfo{journal}{\emph{Journal of autism and developmental disorders}}  \bibinfo{volume}{47} (\bibinfo{year}{2017}), \bibinfo{pages}{1496--1509}.
\newblock


\bibitem[Grossman et~al\mbox{.}(2019b)]%
        {grossman2019facetime}
\bibfield{author}{\bibinfo{person}{RB Grossman}, \bibinfo{person}{E Zane}, \bibinfo{person}{J Mertens}, {and} \bibinfo{person}{T Mitchell}.} \bibinfo{year}{2019}\natexlab{b}.
\newblock \showarticletitle{Facetime vs. screentime: Gaze patterns to live and video social stimuli in adolescents with ASD}.
\newblock \bibinfo{journal}{\emph{Scientific reports}} \bibinfo{volume}{9}, \bibinfo{number}{1} (\bibinfo{year}{2019}), \bibinfo{pages}{12643}.
\newblock


\bibitem[Grossman et~al\mbox{.}(2019a)]%
        {grossman2019perceptions}
\bibfield{author}{\bibinfo{person}{Ruth~B Grossman}, \bibinfo{person}{Julia Mertens}, {and} \bibinfo{person}{Emily Zane}.} \bibinfo{year}{2019}\natexlab{a}.
\newblock \showarticletitle{Perceptions of self and other: Social judgments and gaze patterns to videos of adolescents with and without autism spectrum disorder}.
\newblock \bibinfo{journal}{\emph{Autism}} \bibinfo{volume}{23}, \bibinfo{number}{4} (\bibinfo{year}{2019}), \bibinfo{pages}{846--857}.
\newblock


\bibitem[Groznik et~al\mbox{.}(2021)]%
        {groznik2021gaze}
\bibfield{author}{\bibinfo{person}{Vida Groznik}, \bibinfo{person}{Martin Mo{\v{z}}ina}, \bibinfo{person}{Timotej Lazar}, \bibinfo{person}{Dejan Georgiev}, {and} \bibinfo{person}{Aleksander Sadikov}.} \bibinfo{year}{2021}\natexlab{}.
\newblock \showarticletitle{Gaze behaviour during reading as a predictor of mild cognitive impairment}. In \bibinfo{booktitle}{\emph{2021 IEEE EMBS International Conference on Biomedical and Health Informatics (BHI)}}. IEEE, \bibinfo{pages}{1--4}.
\newblock


\bibitem[Guantay(2024)]%
        {guantay2024usefulness}
\bibfield{author}{\bibinfo{person}{Carla~Daniela Guantay}.} \bibinfo{year}{2024}\natexlab{}.
\newblock \showarticletitle{Usefulness of eye tracking systems in multiple sclerosis}.
\newblock \bibinfo{journal}{\emph{Acta Ophthalmologica}}  \bibinfo{volume}{102} (\bibinfo{year}{2024}).
\newblock


\bibitem[Guantay et~al\mbox{.}(2024)]%
        {guantay2024accounting}
\bibfield{author}{\bibinfo{person}{Carla~Daniela Guantay}, \bibinfo{person}{Laura Mena-Garc{\'\i}a}, \bibinfo{person}{Miguel~Angel Tola-Arribas}, \bibinfo{person}{Mar{\'\i}a Jos{\'e}~Garea Garc{\'\i}a-Malvar}, \bibinfo{person}{Marta Para-Prieto}, \bibinfo{person}{Mar{\'\i}a~Gloria Gonz{\'a}lez-Fern{\'a}ndez}, \bibinfo{person}{Agust{\'\i}n Mayo-Iscar}, {and} \bibinfo{person}{Jos{\'e}~Carlos Pastor-Jimeno}.} \bibinfo{year}{2024}\natexlab{}.
\newblock \showarticletitle{Accounting for visual field abnormalities when using eye-tracking to diagnose reading problems in neurological degeneration}.
\newblock \bibinfo{journal}{\emph{Journal of Eye Movement Research}} \bibinfo{volume}{17}, \bibinfo{number}{2} (\bibinfo{year}{2024}).
\newblock


\bibitem[Guerrero-Molina et~al\mbox{.}(2021)]%
        {guerrero2021antisaccades}
\bibfield{author}{\bibinfo{person}{Mar{\'\i}a~Paz Guerrero-Molina}, \bibinfo{person}{Claudia Rodriguez-L{\'o}pez}, \bibinfo{person}{Luisa Panad{\'e}s-de Oliveira}, \bibinfo{person}{David Uriarte-P{\'e}rez de Urabayen}, \bibinfo{person}{Nicol{\'a}s Garzo-Caldas}, \bibinfo{person}{Cecilia~E Garc{\'\i}a-Cena}, \bibinfo{person}{Rosa~A Saiz-D{\'\i}az}, \bibinfo{person}{Juli{\'a}n Benito-Le{\'o}n}, {and} \bibinfo{person}{Jes{\'u}s~Gonzalez de~la Aleja}.} \bibinfo{year}{2021}\natexlab{}.
\newblock \showarticletitle{Antisaccades and memory-guided saccades in genetic generalized epilepsy and temporal lobe epilepsy}.
\newblock \bibinfo{journal}{\emph{Epilepsy \& Behavior}}  \bibinfo{volume}{123} (\bibinfo{year}{2021}), \bibinfo{pages}{108236}.
\newblock


\bibitem[Guillon et~al\mbox{.}(2014)]%
        {guillon2014visual}
\bibfield{author}{\bibinfo{person}{Quentin Guillon}, \bibinfo{person}{Nouchine Hadjikhani}, \bibinfo{person}{Sophie Baduel}, {and} \bibinfo{person}{Bernadette Rog{\'e}}.} \bibinfo{year}{2014}\natexlab{}.
\newblock \showarticletitle{Visual social attention in autism spectrum disorder: Insights from eye tracking studies}.
\newblock \bibinfo{journal}{\emph{Neuroscience \& Biobehavioral Reviews}}  \bibinfo{volume}{42} (\bibinfo{year}{2014}), \bibinfo{pages}{279--297}.
\newblock


\bibitem[Guo et~al\mbox{.}(2022)]%
        {guo2022eye}
\bibfield{author}{\bibinfo{person}{Xintong Guo}, \bibinfo{person}{Xiaoxuan Liu}, \bibinfo{person}{Shan Ye}, \bibinfo{person}{Xiangyi Liu}, \bibinfo{person}{Xu Yang}, {and} \bibinfo{person}{Dongsheng Fan}.} \bibinfo{year}{2022}\natexlab{}.
\newblock \showarticletitle{Eye movement abnormalities in amyotrophic lateral sclerosis}.
\newblock \bibinfo{journal}{\emph{Brain Sciences}} \bibinfo{volume}{12}, \bibinfo{number}{4} (\bibinfo{year}{2022}), \bibinfo{pages}{489}.
\newblock


\bibitem[Haider and von Oertzen(2013)]%
        {haider2013neurological}
\bibfield{author}{\bibinfo{person}{Bernhard Haider} {and} \bibinfo{person}{Joachim von Oertzen}.} \bibinfo{year}{2013}\natexlab{}.
\newblock \showarticletitle{Neurological disorders}.
\newblock \bibinfo{journal}{\emph{Best practice \& research Clinical obstetrics \& gynaecology}} \bibinfo{volume}{27}, \bibinfo{number}{6} (\bibinfo{year}{2013}), \bibinfo{pages}{867--875}.
\newblock


\bibitem[Halchenko et~al\mbox{.}(2022)]%
        {openbrainconsent}
\bibfield{author}{\bibinfo{person}{Yaroslav Halchenko}, \bibinfo{person}{Cyril Pernet}, \bibinfo{person}{vborghesani}, \bibinfo{person}{Adina Wagner}, \bibinfo{person}{Alexandre Sayal}, \bibinfo{person}{Dimitri~Papadopoulos Orfanos}, \bibinfo{person}{John Pellman}, \bibinfo{person}{Julia Guiomar~Niso Galán}, \bibinfo{person}{Marcin Koculak}, \bibinfo{person}{Chris Gorgolewski}, \bibinfo{person}{Marko Havu}, \bibinfo{person}{Cassandra~Gould van Praag}, \bibinfo{person}{Stefan Appelhoff}, \bibinfo{person}{AsykaKolbacyka}, \bibinfo{person}{Hu Chuan-Peng}, \bibinfo{person}{Marie-Luise Kieseler}, \bibinfo{person}{pjtoussaint}, \bibinfo{person}{Peer Herholz}, \bibinfo{person}{Robert Oostenveld}, \bibinfo{person}{Satrajit Ghosh}, \bibinfo{person}{Stephan Heunis}, {and} \bibinfo{person}{yarikoptic private}.} \bibinfo{year}{2022}\natexlab{}.
\newblock \bibinfo{booktitle}{\emph{con/open-brain-consent: 1.2.0}}.
\newblock
\href{https://doi.org/10.5281/zenodo.7106031}{doi:\nolinkurl{10.5281/zenodo.7106031}}


\bibitem[Hallowell(2012)]%
        {hallowell2012exploiting}
\bibfield{author}{\bibinfo{person}{Brooke Hallowell}.} \bibinfo{year}{2012}\natexlab{}.
\newblock \showarticletitle{Exploiting Eye-Mind}.
\newblock \bibinfo{journal}{\emph{Translational Speech-Language Pathology and Audiology: Essays in Honor of Dr. Sadanand Singh}} (\bibinfo{year}{2012}), \bibinfo{pages}{335}.
\newblock


\bibitem[Hapakova et~al\mbox{.}(2024)]%
        {hapakova2024antisaccadic}
\bibfield{author}{\bibinfo{person}{Lenka Hapakova}, \bibinfo{person}{Jan Necpal}, {and} \bibinfo{person}{Zuzana Kosutzka}.} \bibinfo{year}{2024}\natexlab{}.
\newblock \showarticletitle{The antisaccadic paradigm: a complementary neuropsychological tool in basal ganglia disorders}.
\newblock \bibinfo{journal}{\emph{Cortex}}  \bibinfo{volume}{178} (\bibinfo{year}{2024}), \bibinfo{pages}{116--140}.
\newblock


\bibitem[HARDEMAN et~al\mbox{.}(2006)]%
        {hardeman2006saccadic}
\bibfield{author}{\bibinfo{person}{HANS~H HARDEMAN}, \bibinfo{person}{NICK~F RAMSEY}, \bibinfo{person}{MATHIJS RAEMAEKERS}, \bibinfo{person}{HAROLD~J VAN~MEGEN}, \bibinfo{person}{DAMIAAN~A DENYS}, \bibinfo{person}{HERMAN~G WESTENBERG}, \bibinfo{person}{REN{\'E}~S KAHN}, {et~al\mbox{.}}} \bibinfo{year}{2006}\natexlab{}.
\newblock \showarticletitle{Saccadic abnormalities in psychotropic-naive obsessive--compulsive disorder without co-morbidity}.
\newblock \bibinfo{journal}{\emph{Psychological medicine}} \bibinfo{volume}{36}, \bibinfo{number}{9} (\bibinfo{year}{2006}), \bibinfo{pages}{1321--1326}.
\newblock


\bibitem[Harding et~al\mbox{.}(2008)]%
        {harding2008wind}
\bibfield{author}{\bibinfo{person}{Graham Harding}, \bibinfo{person}{Pamela Harding}, {and} \bibinfo{person}{Arnold Wilkins}.} \bibinfo{year}{2008}\natexlab{}.
\newblock \showarticletitle{Wind turbines, flicker, and photosensitive epilepsy: Characterizing the flashing that may precipitate seizures and optimizing guidelines to prevent them}.
\newblock \bibinfo{journal}{\emph{Epilepsia}} \bibinfo{volume}{49}, \bibinfo{number}{6} (\bibinfo{year}{2008}), \bibinfo{pages}{1095--1098}.
\newblock


\bibitem[Hardwicke and Wagenmakers(2023)]%
        {hardwicke2023reducing}
\bibfield{author}{\bibinfo{person}{Tom~E Hardwicke} {and} \bibinfo{person}{Eric-Jan Wagenmakers}.} \bibinfo{year}{2023}\natexlab{}.
\newblock \showarticletitle{Reducing bias, increasing transparency and calibrating confidence with preregistration}.
\newblock \bibinfo{journal}{\emph{Nature Human Behaviour}} \bibinfo{volume}{7}, \bibinfo{number}{1} (\bibinfo{year}{2023}), \bibinfo{pages}{15--26}.
\newblock


\bibitem[Hassan et~al\mbox{.}(2022)]%
        {hassan2022approach}
\bibfield{author}{\bibinfo{person}{Mohamed~Abul Hassan}, \bibinfo{person}{Chad~M Aldridge}, \bibinfo{person}{Yan Zhuang}, \bibinfo{person}{Xuwang Yin}, \bibinfo{person}{Timothy McMurry}, \bibinfo{person}{Gustavo~K Rohde}, {and} \bibinfo{person}{Andrew~M Southerland}.} \bibinfo{year}{2022}\natexlab{}.
\newblock \showarticletitle{Approach to quantify eye movements to augment stroke diagnosis with a non-calibrated eye-tracker}.
\newblock \bibinfo{journal}{\emph{IEEE Transactions on Biomedical Engineering}} \bibinfo{volume}{70}, \bibinfo{number}{6} (\bibinfo{year}{2022}), \bibinfo{pages}{1750--1757}.
\newblock


\bibitem[Hasse and Bruder(2015)]%
        {hasse2015eye}
\bibfield{author}{\bibinfo{person}{Catrin Hasse} {and} \bibinfo{person}{Carmen Bruder}.} \bibinfo{year}{2015}\natexlab{}.
\newblock \showarticletitle{Eye-tracking measurements and their link to a normative model of monitoring behaviour}.
\newblock \bibinfo{journal}{\emph{Ergonomics}} \bibinfo{volume}{58}, \bibinfo{number}{3} (\bibinfo{year}{2015}), \bibinfo{pages}{355--367}.
\newblock


\bibitem[Heidarzadeh and Ratt{\'e}(2023)]%
        {heidarzadeh2023eye}
\bibfield{author}{\bibinfo{person}{Neda Heidarzadeh} {and} \bibinfo{person}{Sylvie Ratt{\'e}}.} \bibinfo{year}{2023}\natexlab{}.
\newblock \showarticletitle{‘Eye-Tracking’with Words for Alzheimer’s Disease Detection: Time Alignment of Words Enunciation with Image Regions During Image Description Tasks}.
\newblock \bibinfo{journal}{\emph{Journal of Alzheimer's Disease}} \bibinfo{number}{Preprint} (\bibinfo{year}{2023}), \bibinfo{pages}{1--14}.
\newblock


\bibitem[Helmchen et~al\mbox{.}(2003)]%
        {helmchen2003eye}
\bibfield{author}{\bibinfo{person}{C Helmchen}, \bibinfo{person}{A Hagenow}, \bibinfo{person}{J Miesner}, \bibinfo{person}{A Sprenger}, \bibinfo{person}{H Rambold}, \bibinfo{person}{R Wenzelburger}, \bibinfo{person}{W Heide}, {and} \bibinfo{person}{G Deuschl}.} \bibinfo{year}{2003}\natexlab{}.
\newblock \showarticletitle{Eye movement abnormalities in essential tremor may indicate cerebellar dysfunction}.
\newblock \bibinfo{journal}{\emph{Brain}} \bibinfo{volume}{126}, \bibinfo{number}{6} (\bibinfo{year}{2003}), \bibinfo{pages}{1319--1332}.
\newblock


\bibitem[Helmchen et~al\mbox{.}(2012)]%
        {helmchen2012role}
\bibfield{author}{\bibinfo{person}{Christoph Helmchen}, \bibinfo{person}{Jonas Pohlmann}, \bibinfo{person}{Peter Trillenberg}, \bibinfo{person}{Rebekka Lencer}, \bibinfo{person}{Julia Graf}, {and} \bibinfo{person}{Andreas Sprenger}.} \bibinfo{year}{2012}\natexlab{}.
\newblock \showarticletitle{Role of anticipation and prediction in smooth pursuit eye movement control in Parkinson's disease}.
\newblock \bibinfo{journal}{\emph{Movement Disorders}} \bibinfo{volume}{27}, \bibinfo{number}{8} (\bibinfo{year}{2012}), \bibinfo{pages}{1012--1018}.
\newblock


\bibitem[Hirsig et~al\mbox{.}(2020)]%
        {hirsig2020oculomotor}
\bibfield{author}{\bibinfo{person}{Anna Hirsig}, \bibinfo{person}{Carolin Barbey}, \bibinfo{person}{Michael~WM Sch{\"u}pbach}, {and} \bibinfo{person}{Panagiotis Bargiotas}.} \bibinfo{year}{2020}\natexlab{}.
\newblock \showarticletitle{Oculomotor functions in focal dystonias: A systematic review}.
\newblock \bibinfo{journal}{\emph{Acta Neurologica Scandinavica}} \bibinfo{volume}{141}, \bibinfo{number}{5} (\bibinfo{year}{2020}), \bibinfo{pages}{359--367}.
\newblock


\bibitem[Hirt et~al\mbox{.}(2020)]%
        {hirt2020stress}
\bibfield{author}{\bibinfo{person}{Christian Hirt}, \bibinfo{person}{Marcel Eckard}, {and} \bibinfo{person}{Andreas Kunz}.} \bibinfo{year}{2020}\natexlab{}.
\newblock \showarticletitle{Stress generation and non-intrusive measurement in virtual environments using eye tracking}.
\newblock \bibinfo{journal}{\emph{Journal of Ambient Intelligence and Humanized Computing}}  \bibinfo{volume}{11} (\bibinfo{year}{2020}), \bibinfo{pages}{5977--5989}.
\newblock


\bibitem[Hodgson et~al\mbox{.}(2002)]%
        {hodgson2002abnormal}
\bibfield{author}{\bibinfo{person}{TL Hodgson}, \bibinfo{person}{B Tiesman}, \bibinfo{person}{AM Owen}, {and} \bibinfo{person}{C Kennard}.} \bibinfo{year}{2002}\natexlab{}.
\newblock \showarticletitle{Abnormal gaze strategies during problem solving in Parkinson's disease}.
\newblock \bibinfo{journal}{\emph{Neuropsychologia}} \bibinfo{volume}{40}, \bibinfo{number}{4} (\bibinfo{year}{2002}), \bibinfo{pages}{411--422}.
\newblock


\bibitem[Hodgson et~al\mbox{.}(2019)]%
        {hodgson2019eye}
\bibfield{author}{\bibinfo{person}{Timothy~L Hodgson}, \bibinfo{person}{Gemma Ezard}, {and} \bibinfo{person}{Frouke Hermens}.} \bibinfo{year}{2019}\natexlab{}.
\newblock \showarticletitle{Eye movements in neuropsychological tasks}.
\newblock \bibinfo{journal}{\emph{Processes of Visuospatial Attention and Working Memory}} (\bibinfo{year}{2019}), \bibinfo{pages}{393--418}.
\newblock


\bibitem[Hodgson et~al\mbox{.}(2024)]%
        {hodgson2024gaze}
\bibfield{author}{\bibinfo{person}{Timothy~L Hodgson}, \bibinfo{person}{Frouke Hermens}, {and} \bibinfo{person}{Gemma Ezard}.} \bibinfo{year}{2024}\natexlab{}.
\newblock \showarticletitle{Gaze-speech coordination during social interaction in Parkinson's disease}.
\newblock \bibinfo{journal}{\emph{International Journal of Language \& Communication Disorders}} \bibinfo{volume}{59}, \bibinfo{number}{2} (\bibinfo{year}{2024}), \bibinfo{pages}{715--727}.
\newblock


\bibitem[Hokken et~al\mbox{.}(2024)]%
        {hokken2024eyes}
\bibfield{author}{\bibinfo{person}{Marinke~J Hokken}, \bibinfo{person}{Niklas Stein}, \bibinfo{person}{Rob~Rodrigues Pereira}, \bibinfo{person}{Ingrid~GIJG Rours}, \bibinfo{person}{Maarten~A Frens}, \bibinfo{person}{Johannes van~der Steen}, \bibinfo{person}{Johan~JM Pel}, {and} \bibinfo{person}{Marlou~JG Kooiker}.} \bibinfo{year}{2024}\natexlab{}.
\newblock \showarticletitle{Eyes on CVI: Eye movements unveil distinct visual search patterns in Cerebral Visual Impairment compared to ADHD, dyslexia, and neurotypical children}.
\newblock \bibinfo{journal}{\emph{Research in Developmental Disabilities}}  \bibinfo{volume}{151} (\bibinfo{year}{2024}), \bibinfo{pages}{104767}.
\newblock


\bibitem[Holmes et~al\mbox{.}(2006)]%
        {holmes2006anxiety}
\bibfield{author}{\bibinfo{person}{Amanda Holmes}, \bibinfo{person}{Anne Richards}, {and} \bibinfo{person}{Simon Green}.} \bibinfo{year}{2006}\natexlab{}.
\newblock \showarticletitle{Anxiety and sensitivity to eye gaze in emotional faces}.
\newblock \bibinfo{journal}{\emph{Brain and cognition}} \bibinfo{volume}{60}, \bibinfo{number}{3} (\bibinfo{year}{2006}), \bibinfo{pages}{282--294}.
\newblock


\bibitem[Holmqvist et~al\mbox{.}(2011)]%
        {holmqvist2011eye}
\bibfield{author}{\bibinfo{person}{Kenneth Holmqvist}, \bibinfo{person}{Marcus Nystr{\"o}m}, \bibinfo{person}{Richard Andersson}, \bibinfo{person}{Richard Dewhurst}, \bibinfo{person}{Halszka Jarodzka}, {and} \bibinfo{person}{Joost Van~de Weijer}.} \bibinfo{year}{2011}\natexlab{}.
\newblock \bibinfo{booktitle}{\emph{Eye tracking: A comprehensive guide to methods and measures}}.
\newblock \bibinfo{publisher}{oup Oxford}.
\newblock


\bibitem[Holmqvist et~al\mbox{.}(2012)]%
        {holmqvist2012eye}
\bibfield{author}{\bibinfo{person}{Kenneth Holmqvist}, \bibinfo{person}{Marcus Nystr{\"o}m}, {and} \bibinfo{person}{Fiona Mulvey}.} \bibinfo{year}{2012}\natexlab{}.
\newblock \showarticletitle{Eye tracker data quality: What it is and how to measure it}. In \bibinfo{booktitle}{\emph{Proceedings of the symposium on eye tracking research and applications}}. \bibinfo{pages}{45--52}.
\newblock


\bibitem[Holmqvist et~al\mbox{.}(2023)]%
        {holmqvist2023retracted}
\bibfield{author}{\bibinfo{person}{Kenneth Holmqvist}, \bibinfo{person}{Saga~Lee {\"O}rbom}, \bibinfo{person}{Ignace~TC Hooge}, \bibinfo{person}{Diederick~C Niehorster}, \bibinfo{person}{Robert~G Alexander}, \bibinfo{person}{Richard Andersson}, \bibinfo{person}{Jeroen~S Benjamins}, \bibinfo{person}{Pieter Blignaut}, \bibinfo{person}{Anne-Marie Brouwer}, \bibinfo{person}{Lewis~L Chuang}, {et~al\mbox{.}}} \bibinfo{year}{2023}\natexlab{}.
\newblock \showarticletitle{RETRACTED ARTICLE: Eye tracking: empirical foundations for a minimal reporting guideline}.
\newblock \bibinfo{journal}{\emph{Behavior research methods}} \bibinfo{volume}{55}, \bibinfo{number}{1} (\bibinfo{year}{2023}), \bibinfo{pages}{364--416}.
\newblock


\bibitem[Hopstaken et~al\mbox{.}(2016)]%
        {hopstaken2016shifts}
\bibfield{author}{\bibinfo{person}{Jesper~F Hopstaken}, \bibinfo{person}{Dimitri Van Der~Linden}, \bibinfo{person}{Arnold~B Bakker}, \bibinfo{person}{Michiel~AJ Kompier}, {and} \bibinfo{person}{Yik~Kiu Leung}.} \bibinfo{year}{2016}\natexlab{}.
\newblock \showarticletitle{Shifts in attention during mental fatigue: Evidence from subjective, behavioral, physiological, and eye-tracking data.}
\newblock \bibinfo{journal}{\emph{Journal of Experimental Psychology: Human Perception and Performance}} \bibinfo{volume}{42}, \bibinfo{number}{6} (\bibinfo{year}{2016}), \bibinfo{pages}{878}.
\newblock


\bibitem[Hu et~al\mbox{.}(2023)]%
        {hu2023brain}
\bibfield{author}{\bibinfo{person}{Mengjiao Hu}, \bibinfo{person}{Haihong Zhang}, {and} \bibinfo{person}{Kai~Keng Ang}.} \bibinfo{year}{2023}\natexlab{}.
\newblock \showarticletitle{Brain Criticality EEG analysis for tracking neurodevelopment from Childhood to Adolescence}. In \bibinfo{booktitle}{\emph{2023 45th Annual International Conference of the IEEE Engineering in Medicine \& Biology Society (EMBC)}}. IEEE, \bibinfo{pages}{1--4}.
\newblock


\bibitem[Hu et~al\mbox{.}(2020)]%
        {hu2020investigating}
\bibfield{author}{\bibinfo{person}{Yixin Hu}, \bibinfo{person}{Rui Liao}, \bibinfo{person}{Weiling Chen}, \bibinfo{person}{Xiangwei Kong}, \bibinfo{person}{Jingyi Liu}, \bibinfo{person}{Dongxu Liu}, \bibinfo{person}{Phil Maguire}, \bibinfo{person}{Shengqi Zhou}, {and} \bibinfo{person}{Dawei Wang}.} \bibinfo{year}{2020}\natexlab{}.
\newblock \showarticletitle{Investigating behavior inhibition in obsessive-compulsive disorder: Evidence from eye movements}.
\newblock \bibinfo{journal}{\emph{Scandinavian Journal of Psychology}} \bibinfo{volume}{61}, \bibinfo{number}{5} (\bibinfo{year}{2020}), \bibinfo{pages}{634--641}.
\newblock


\bibitem[Huang et~al\mbox{.}(2024)]%
        {huang2024reduced}
\bibfield{author}{\bibinfo{person}{Kailing Huang}, \bibinfo{person}{Ziwei Tian}, \bibinfo{person}{Qiong Zhang}, \bibinfo{person}{Haojun Yang}, \bibinfo{person}{Shirui Wen}, \bibinfo{person}{Jie Feng}, \bibinfo{person}{Weiting Tang}, \bibinfo{person}{Quan Wang}, {and} \bibinfo{person}{Li Feng}.} \bibinfo{year}{2024}\natexlab{}.
\newblock \showarticletitle{Reduced eye gaze fixation during emotion recognition among patients with temporal lobe epilepsy}.
\newblock \bibinfo{journal}{\emph{Journal of Neurology}} (\bibinfo{year}{2024}), \bibinfo{pages}{1--13}.
\newblock


\bibitem[Hubel(1995)]%
        {hubel1995eye}
\bibfield{author}{\bibinfo{person}{David~H Hubel}.} \bibinfo{year}{1995}\natexlab{}.
\newblock \bibinfo{booktitle}{\emph{Eye, brain, and vision.}}
\newblock \bibinfo{publisher}{Scientific American Library/Scientific American Books}.
\newblock


\bibitem[Hummel et~al\mbox{.}(2017)]%
        {hummel2017influence}
\bibfield{author}{\bibinfo{person}{Gerrit Hummel}, \bibinfo{person}{Iris Zerweck}, \bibinfo{person}{Janine Ehret}, \bibinfo{person}{Sara~Salazar Winter}, {and} \bibinfo{person}{Nanette Stroebele-Benschop}.} \bibinfo{year}{2017}\natexlab{}.
\newblock \showarticletitle{The influence of the arrangement of different food images on participants’ attention: An experimental eye-tracking study}.
\newblock \bibinfo{journal}{\emph{Food Quality and Preference}}  \bibinfo{volume}{62} (\bibinfo{year}{2017}), \bibinfo{pages}{111--119}.
\newblock


\bibitem[Hunter and Chin(2021)]%
        {hunter2021impaired}
\bibfield{author}{\bibinfo{person}{Matthew~B Hunter} {and} \bibinfo{person}{Richard~FM Chin}.} \bibinfo{year}{2021}\natexlab{}.
\newblock \showarticletitle{Impaired social attention detected through eye movements in children with early-onset epilepsy}.
\newblock \bibinfo{journal}{\emph{Epilepsia}} \bibinfo{volume}{62}, \bibinfo{number}{8} (\bibinfo{year}{2021}), \bibinfo{pages}{1921--1930}.
\newblock


\bibitem[Hutchinson et~al\mbox{.}(2014)]%
        {hutchinson2014cervical}
\bibfield{author}{\bibinfo{person}{Michael Hutchinson}, \bibinfo{person}{Tadashi Isa}, \bibinfo{person}{Anna Molloy}, \bibinfo{person}{Okka Kimmich}, \bibinfo{person}{Laura Williams}, \bibinfo{person}{Fiona Molloy}, \bibinfo{person}{Helena Moore}, \bibinfo{person}{Daniel~G Healy}, \bibinfo{person}{Tim Lynch}, \bibinfo{person}{Cathal Walsh}, {et~al\mbox{.}}} \bibinfo{year}{2014}\natexlab{}.
\newblock \showarticletitle{Cervical dystonia: a disorder of the midbrain network for covert attentional orienting}.
\newblock \bibinfo{journal}{\emph{Frontiers in neurology}}  \bibinfo{volume}{5} (\bibinfo{year}{2014}), \bibinfo{pages}{54}.
\newblock


\bibitem[Huys et~al\mbox{.}(2021)]%
        {huys2021misdirected}
\bibfield{author}{\bibinfo{person}{Anne-Catherine~ML Huys}, \bibinfo{person}{Patrick Haggard}, \bibinfo{person}{Kailash~P Bhatia}, {and} \bibinfo{person}{Mark~J Edwards}.} \bibinfo{year}{2021}\natexlab{}.
\newblock \showarticletitle{Misdirected attentional focus in functional tremor}.
\newblock \bibinfo{journal}{\emph{Brain}} \bibinfo{volume}{144}, \bibinfo{number}{11} (\bibinfo{year}{2021}), \bibinfo{pages}{3436--3450}.
\newblock


\bibitem[Ikuse et~al\mbox{.}(2022)]%
        {ikuse2022effects}
\bibfield{author}{\bibinfo{person}{Daisuke Ikuse}, \bibinfo{person}{Wakaho Hayashi}, \bibinfo{person}{Youichi Hanawa}, \bibinfo{person}{Dan Nakamura}, \bibinfo{person}{Gousuke Arai}, \bibinfo{person}{Nobuyuki Saga}, {and} \bibinfo{person}{Akira Iwanami}.} \bibinfo{year}{2022}\natexlab{}.
\newblock \showarticletitle{Effects of gaze cues on distributive behavior of the ultimatum game in adults with attention deficit hyperactivity disorder}.
\newblock \bibinfo{journal}{\emph{The Journal of Nervous and Mental Disease}} \bibinfo{volume}{210}, \bibinfo{number}{7} (\bibinfo{year}{2022}), \bibinfo{pages}{525--531}.
\newblock


\bibitem[Insch et~al\mbox{.}(2017)]%
        {insch2017gaze}
\bibfield{author}{\bibinfo{person}{Pauline~M Insch}, \bibinfo{person}{Gillian Slessor}, \bibinfo{person}{Jill Warrington}, {and} \bibinfo{person}{Louise~H Phillips}.} \bibinfo{year}{2017}\natexlab{}.
\newblock \showarticletitle{Gaze detection and gaze cuing in Alzheimer’s disease}.
\newblock \bibinfo{journal}{\emph{Brain and cognition}}  \bibinfo{volume}{116} (\bibinfo{year}{2017}), \bibinfo{pages}{47--53}.
\newblock


\bibitem[Ioannou et~al\mbox{.}(2020)]%
        {ioannou2020comorbidity}
\bibfield{author}{\bibinfo{person}{Chara Ioannou}, \bibinfo{person}{Divya Seernani}, \bibinfo{person}{Maria~Elena Stefanou}, \bibinfo{person}{Andreas Riedel}, \bibinfo{person}{Ludger Tebartz~van Elst}, \bibinfo{person}{Nikolaos Smyrnis}, \bibinfo{person}{Christian Fleischhaker}, \bibinfo{person}{Monica Biscaldi-Schaefer}, \bibinfo{person}{Giuseppe Boccignone}, {and} \bibinfo{person}{Christoph Klein}.} \bibinfo{year}{2020}\natexlab{}.
\newblock \showarticletitle{Comorbidity matters: social visual attention in a comparative study of autism spectrum disorder, attention-deficit/hyperactivity disorder and their comorbidity}.
\newblock \bibinfo{journal}{\emph{Frontiers in psychiatry}}  \bibinfo{volume}{11} (\bibinfo{year}{2020}), \bibinfo{pages}{545567}.
\newblock


\bibitem[Ionescu et~al\mbox{.}(2023)]%
        {ionescu2023correlating}
\bibfield{author}{\bibinfo{person}{Alec Ionescu}, \bibinfo{person}{Emanuel Ștefuanescu}, \bibinfo{person}{Stefan Strilciuc}, \bibinfo{person}{Alexandru Rafila}, {and} \bibinfo{person}{Dafin Mureșanu}.} \bibinfo{year}{2023}\natexlab{}.
\newblock \showarticletitle{Correlating Eye-Tracking Fixation Metrics and Neuropsychological Assessment after Ischemic Stroke}.
\newblock \bibinfo{journal}{\emph{Medicina}} \bibinfo{volume}{59}, \bibinfo{number}{8} (\bibinfo{year}{2023}), \bibinfo{pages}{1361}.
\newblock


\bibitem[Istance and Hyrskykari(2017)]%
        {istance2017supporting}
\bibfield{author}{\bibinfo{person}{Howell Istance} {and} \bibinfo{person}{Aulikki~I Hyrskykari}.} \bibinfo{year}{2017}\natexlab{}.
\newblock \showarticletitle{Supporting making fixations and the effect on gaze gesture performance}. In \bibinfo{booktitle}{\emph{Proceedings of the 2017 CHI Conference on Human Factors in Computing Systems}}. \bibinfo{pages}{3022--3033}.
\newblock


\bibitem[Izadyar et~al\mbox{.}(2018)]%
        {izadyar2018comparison}
\bibfield{author}{\bibinfo{person}{Shahram Izadyar}, \bibinfo{person}{Vishal Shah}, {and} \bibinfo{person}{Brandon James}.} \bibinfo{year}{2018}\natexlab{}.
\newblock \showarticletitle{Comparison of postictal semiology and behavior in psychogenic nonepileptic and epileptic seizures}.
\newblock \bibinfo{journal}{\emph{Epilepsy \& Behavior}}  \bibinfo{volume}{88} (\bibinfo{year}{2018}), \bibinfo{pages}{123--129}.
\newblock


\bibitem[Jacobs et~al\mbox{.}(1981)]%
        {jacobs1981eye}
\bibfield{author}{\bibinfo{person}{Lawrence Jacobs}, \bibinfo{person}{Diana Bozian}, \bibinfo{person}{Reid~R Heffner~Jr}, {and} \bibinfo{person}{Stephen~A Barron}.} \bibinfo{year}{1981}\natexlab{}.
\newblock \showarticletitle{An eye movement disorder in amyotrophic lateral sclerosis}.
\newblock \bibinfo{journal}{\emph{Neurology}} \bibinfo{volume}{31}, \bibinfo{number}{10} (\bibinfo{year}{1981}), \bibinfo{pages}{1282--1282}.
\newblock


\bibitem[Jakobsen(1990)]%
        {jakobsen1990pupillary}
\bibfield{author}{\bibinfo{person}{J Jakobsen}.} \bibinfo{year}{1990}\natexlab{}.
\newblock \showarticletitle{Pupillary function in multiple sclerosis}.
\newblock \bibinfo{journal}{\emph{Acta neurologica scandinavica}} \bibinfo{volume}{82}, \bibinfo{number}{6} (\bibinfo{year}{1990}), \bibinfo{pages}{392--395}.
\newblock


\bibitem[Jankovic et~al\mbox{.}(1982)]%
        {jankovic1982blinking}
\bibfield{author}{\bibinfo{person}{Joseph Jankovic}, \bibinfo{person}{Weldon~E Havins}, {and} \bibinfo{person}{Robert~B Wilkins}.} \bibinfo{year}{1982}\natexlab{}.
\newblock \showarticletitle{Blinking and blepharospasm: mechanism, diagnosis, and management}.
\newblock \bibinfo{journal}{\emph{Jama}} \bibinfo{volume}{248}, \bibinfo{number}{23} (\bibinfo{year}{1982}), \bibinfo{pages}{3160--3164}.
\newblock


\bibitem[Jansen et~al\mbox{.}(2020)]%
        {jansen2020social}
\bibfield{author}{\bibinfo{person}{Myrthe Jansen}, \bibinfo{person}{Sandy Overgaauw}, {and} \bibinfo{person}{Ellen~RA De~Bruijn}.} \bibinfo{year}{2020}\natexlab{}.
\newblock \showarticletitle{Social cognition and obsessive-compulsive disorder: a review of subdomains of social functioning}.
\newblock \bibinfo{journal}{\emph{Frontiers in psychiatry}}  \bibinfo{volume}{11} (\bibinfo{year}{2020}), \bibinfo{pages}{118}.
\newblock


\bibitem[Janvier et~al\mbox{.}(2016)]%
        {janvier2016screening}
\bibfield{author}{\bibinfo{person}{Yvette~M Janvier}, \bibinfo{person}{Jill~F Harris}, \bibinfo{person}{Caroline~N Coffield}, \bibinfo{person}{Barbara Louis}, \bibinfo{person}{Ming Xie}, \bibinfo{person}{Zuleyha Cidav}, {and} \bibinfo{person}{David~S Mandell}.} \bibinfo{year}{2016}\natexlab{}.
\newblock \showarticletitle{Screening for autism spectrum disorder in underserved communities: Early childcare providers as reporters}.
\newblock \bibinfo{journal}{\emph{Autism}} \bibinfo{volume}{20}, \bibinfo{number}{3} (\bibinfo{year}{2016}), \bibinfo{pages}{364--373}.
\newblock


\bibitem[Jhunjhunwala et~al\mbox{.}(2017)]%
        {jhunjhunwala2017abnormalities}
\bibfield{author}{\bibinfo{person}{Ketan Jhunjhunwala}, \bibinfo{person}{Raviteja Kotikalapudi}, \bibinfo{person}{Abhishek Lenka}, \bibinfo{person}{Kandavel Thennarassu}, \bibinfo{person}{Ravi Yadav}, \bibinfo{person}{Jitender Saini}, {and} \bibinfo{person}{Pramod~Kumar Pal}.} \bibinfo{year}{2017}\natexlab{}.
\newblock \showarticletitle{Abnormalities of Eye--Hand Coordination in Patients with Writer’s Cramp: Possible Role of the Cerebellum}.
\newblock \bibinfo{journal}{\emph{Tremor and Other Hyperkinetic Movements}}  \bibinfo{volume}{7} (\bibinfo{year}{2017}).
\newblock


\bibitem[Jiang et~al\mbox{.}(2024)]%
        {jiang2024auxiliary}
\bibfield{author}{\bibinfo{person}{Maosong Jiang}, \bibinfo{person}{Yanzhi Liu}, \bibinfo{person}{Yanlu Cao}, \bibinfo{person}{Yuzhu Liu}, \bibinfo{person}{Jiatian Wang}, \bibinfo{person}{Peixue Li}, \bibinfo{person}{Shufeng Xia}, \bibinfo{person}{Yongzhong Lin}, {and} \bibinfo{person}{Wenlong Liu}.} \bibinfo{year}{2024}\natexlab{}.
\newblock \showarticletitle{Auxiliary diagnostic method of Parkinson’s disease based on eye movement analysis in a virtual reality environment}.
\newblock \bibinfo{journal}{\emph{Neuroscience Letters}}  \bibinfo{volume}{842} (\bibinfo{year}{2024}), \bibinfo{pages}{137956}.
\newblock


\bibitem[Jlassi(2022)]%
        {jlassi2022analysis}
\bibfield{author}{\bibinfo{person}{Sara Jlassi}.} \bibinfo{year}{2022}\natexlab{}.
\newblock \emph{\bibinfo{title}{Analysis of the pupillary response in Amyotrophic Lateral Sclerosis patients}}.
\newblock \bibinfo{thesistype}{Ph.\,D. Dissertation}. \bibinfo{school}{Politecnico di Torino}.
\newblock


\bibitem[Johansson et~al\mbox{.}(2021)]%
        {johansson2021methodological}
\bibfield{author}{\bibinfo{person}{Jan Johansson}, \bibinfo{person}{Kristina Franzon}, \bibinfo{person}{Alison k Godbolt}, {and} \bibinfo{person}{Marika~C M{\"o}ller}.} \bibinfo{year}{2021}\natexlab{}.
\newblock \showarticletitle{Methodological aspects of using a wearable eye-tracker to support diagnostic clinical evaluation of prolonged disorders of consciousness}.
\newblock \bibinfo{journal}{\emph{Journal of Rehabilitation Medicine}} \bibinfo{volume}{53}, \bibinfo{number}{7} (\bibinfo{year}{2021}), \bibinfo{pages}{jrm00213--jrm00213}.
\newblock


\bibitem[Johkura et~al\mbox{.}(2015)]%
        {johkura2015gaze}
\bibfield{author}{\bibinfo{person}{Ken Johkura}, \bibinfo{person}{Yosuke Kudo}, \bibinfo{person}{Yu Amano}, \bibinfo{person}{Hideyuki Kikyo}, \bibinfo{person}{Ryoko Imazeki}, \bibinfo{person}{Kazumitsu Amari}, {and} \bibinfo{person}{Masahiro Yamamoto}.} \bibinfo{year}{2015}\natexlab{}.
\newblock \showarticletitle{Gaze palsy and exotropia in internuclear ophthalmoplegia}.
\newblock \bibinfo{journal}{\emph{Journal of the neurological sciences}} \bibinfo{volume}{353}, \bibinfo{number}{1-2} (\bibinfo{year}{2015}), \bibinfo{pages}{158--160}.
\newblock


\bibitem[Johnson et~al\mbox{.}(2021)]%
        {johnson2021cognitive}
\bibfield{author}{\bibinfo{person}{Camille~N Johnson}, \bibinfo{person}{Bruce Ramphal}, \bibinfo{person}{Emily Koe}, \bibinfo{person}{Amarelis Raudales}, \bibinfo{person}{Jeff Goldsmith}, {and} \bibinfo{person}{Amy~E Margolis}.} \bibinfo{year}{2021}\natexlab{}.
\newblock \showarticletitle{Cognitive correlates of autism spectrum disorder symptoms}.
\newblock \bibinfo{journal}{\emph{Autism Research}} \bibinfo{volume}{14}, \bibinfo{number}{11} (\bibinfo{year}{2021}), \bibinfo{pages}{2405--2411}.
\newblock


\bibitem[Jones et~al\mbox{.}(2008)]%
        {jones2008absence}
\bibfield{author}{\bibinfo{person}{Warren Jones}, \bibinfo{person}{Katelin Carr}, {and} \bibinfo{person}{Ami Klin}.} \bibinfo{year}{2008}\natexlab{}.
\newblock \showarticletitle{Absence of preferential looking to the eyes of approaching adults predicts level of social disability in 2-year-old toddlers with autism spectrum disorder}.
\newblock \bibinfo{journal}{\emph{Archives of general psychiatry}} \bibinfo{volume}{65}, \bibinfo{number}{8} (\bibinfo{year}{2008}), \bibinfo{pages}{946--954}.
\newblock


\bibitem[J{\'o}nsd{\'o}ttir et~al\mbox{.}(2023)]%
        {jonsdottir2023autistic}
\bibfield{author}{\bibinfo{person}{Lilja~Krist{\'\i}n J{\'o}nsd{\'o}ttir}, \bibinfo{person}{Janina Neufeld}, \bibinfo{person}{Terje Falck-Ytter}, {and} \bibinfo{person}{Johan~Lundin Kleberg}.} \bibinfo{year}{2023}\natexlab{}.
\newblock \showarticletitle{Autistic children quickly orient away from both eyes and mouths during face observation}.
\newblock \bibinfo{journal}{\emph{Journal of autism and developmental disorders}} \bibinfo{volume}{53}, \bibinfo{number}{1} (\bibinfo{year}{2023}), \bibinfo{pages}{495--502}.
\newblock


\bibitem[Jording et~al\mbox{.}(2024)]%
        {jording2024impaired}
\bibfield{author}{\bibinfo{person}{Mathis Jording}, \bibinfo{person}{Arne Hartz}, \bibinfo{person}{David~HV Vogel}, \bibinfo{person}{Martin Schulte-R{\"u}ther}, {and} \bibinfo{person}{Kai Vogeley}.} \bibinfo{year}{2024}\natexlab{}.
\newblock \showarticletitle{Impaired recognition of interactive intentions in adults with autism spectrum disorder not attributable to differences in visual attention or coordination via eye contact and joint attention}.
\newblock \bibinfo{journal}{\emph{Scientific reports}} \bibinfo{volume}{14}, \bibinfo{number}{1} (\bibinfo{year}{2024}), \bibinfo{pages}{8297}.
\newblock


\bibitem[Jozefowicz-Korczynska and Pajor(2011)]%
        {jozefowicz2011evaluation}
\bibfield{author}{\bibinfo{person}{Magdalena Jozefowicz-Korczynska} {and} \bibinfo{person}{Anna~Maria Pajor}.} \bibinfo{year}{2011}\natexlab{}.
\newblock \showarticletitle{Evaluation of the smooth pursuit tests in multiple sclerosis patients}.
\newblock \bibinfo{journal}{\emph{Journal of neurology}}  \bibinfo{volume}{258} (\bibinfo{year}{2011}), \bibinfo{pages}{1795--1800}.
\newblock


\bibitem[Jung and Kim(2019)]%
        {jung2019abnormal}
\bibfield{author}{\bibinfo{person}{Ileok Jung} {and} \bibinfo{person}{Ji-Soo Kim}.} \bibinfo{year}{2019}\natexlab{}.
\newblock \showarticletitle{Abnormal eye movements in parkinsonism and movement disorders}.
\newblock \bibinfo{journal}{\emph{Journal of movement disorders}} \bibinfo{volume}{12}, \bibinfo{number}{1} (\bibinfo{year}{2019}), \bibinfo{pages}{1}.
\newblock


\bibitem[Jyotsna and Amudha(2018)]%
        {jyotsna2018eye}
\bibfield{author}{\bibinfo{person}{Chandran Jyotsna} {and} \bibinfo{person}{Joseph Amudha}.} \bibinfo{year}{2018}\natexlab{}.
\newblock \showarticletitle{Eye gaze as an indicator for stress level analysis in students}. In \bibinfo{booktitle}{\emph{2018 International conference on advances in computing, communications and informatics (ICACCI)}}. IEEE, \bibinfo{pages}{1588--1593}.
\newblock


\bibitem[Kane and Morris(2017)]%
        {kane2017let}
\bibfield{author}{\bibinfo{person}{Shaun~K Kane} {and} \bibinfo{person}{Meredith~Ringel Morris}.} \bibinfo{year}{2017}\natexlab{}.
\newblock \showarticletitle{Let's Talk about X: Combining image recognition and eye gaze to support conversation for people with ALS}. In \bibinfo{booktitle}{\emph{Proceedings of the 2017 Conference on Designing Interactive Systems}}. \bibinfo{pages}{129--134}.
\newblock


\bibitem[Kapoula et~al\mbox{.}(2014)]%
        {kapoula2014distinctive}
\bibfield{author}{\bibinfo{person}{Zoi Kapoula}, \bibinfo{person}{Qing Yang}, \bibinfo{person}{Jorge Otero-Millan}, \bibinfo{person}{Shifu Xiao}, \bibinfo{person}{Stephen~L Macknik}, \bibinfo{person}{Alexandre Lang}, \bibinfo{person}{Marc Verny}, {and} \bibinfo{person}{Susana Martinez-Conde}.} \bibinfo{year}{2014}\natexlab{}.
\newblock \showarticletitle{Distinctive features of microsaccades in Alzheimer’s disease and in mild cognitive impairment}.
\newblock \bibinfo{journal}{\emph{Age}}  \bibinfo{volume}{36} (\bibinfo{year}{2014}), \bibinfo{pages}{535--543}.
\newblock


\bibitem[Kashihara(2020)]%
        {kashihara2020microsaccadic}
\bibfield{author}{\bibinfo{person}{Koji Kashihara}.} \bibinfo{year}{2020}\natexlab{}.
\newblock \showarticletitle{Microsaccadic modulation evoked by emotional events}.
\newblock \bibinfo{journal}{\emph{Journal of Physiological Anthropology}} \bibinfo{volume}{39}, \bibinfo{number}{1} (\bibinfo{year}{2020}), \bibinfo{pages}{26}.
\newblock


\bibitem[Kasprowski and Harezlaky(2016)]%
        {kasprowski2016gaze}
\bibfield{author}{\bibinfo{person}{Pawel Kasprowski} {and} \bibinfo{person}{Katarzyna Harezlaky}.} \bibinfo{year}{2016}\natexlab{}.
\newblock \showarticletitle{Gaze self-similarity plots as a useful tool for eye movement characteristics analysis}. In \bibinfo{booktitle}{\emph{2016 IEEE Second Workshop on Eye Tracking and Visualization (ETVIS)}}. IEEE, \bibinfo{pages}{6--10}.
\newblock


\bibitem[Kassavetis et~al\mbox{.}(2022)]%
        {kassavetis2022eye}
\bibfield{author}{\bibinfo{person}{Panagiotis Kassavetis}, \bibinfo{person}{Diego Kaski}, \bibinfo{person}{Tim Anderson}, {and} \bibinfo{person}{Mark Hallett}.} \bibinfo{year}{2022}\natexlab{}.
\newblock \showarticletitle{Eye movement disorders in movement disorders}.
\newblock \bibinfo{journal}{\emph{Movement Disorders Clinical Practice}} \bibinfo{volume}{9}, \bibinfo{number}{3} (\bibinfo{year}{2022}), \bibinfo{pages}{284--295}.
\newblock


\bibitem[Katayama and Tamas(1987)]%
        {katayama1987saccadic}
\bibfield{author}{\bibinfo{person}{Mitsuko Katayama} {and} \bibinfo{person}{Laszlo~B Tamas}.} \bibinfo{year}{1987}\natexlab{}.
\newblock \showarticletitle{Saccadic eye-movements of children with cerebral palsy}.
\newblock \bibinfo{journal}{\emph{Developmental Medicine \& Child Neurology}} \bibinfo{volume}{29}, \bibinfo{number}{1} (\bibinfo{year}{1987}), \bibinfo{pages}{36--39}.
\newblock


\bibitem[Kattoulas et~al\mbox{.}(2011)]%
        {kattoulas2011predictive}
\bibfield{author}{\bibinfo{person}{Emmanouil Kattoulas}, \bibinfo{person}{Ioannis Evdokimidis}, \bibinfo{person}{Nicholas~C Stefanis}, \bibinfo{person}{Dimitrios Avramopoulos}, \bibinfo{person}{Costas~N Stefanis}, {and} \bibinfo{person}{Nikolaos Smyrnis}.} \bibinfo{year}{2011}\natexlab{}.
\newblock \showarticletitle{Predictive smooth eye pursuit in a population of young men: II. Effects of schizotypy, anxiety and depression}.
\newblock \bibinfo{journal}{\emph{Experimental brain research}}  \bibinfo{volume}{215} (\bibinfo{year}{2011}), \bibinfo{pages}{219--226}.
\newblock


\bibitem[Keles et~al\mbox{.}(2022)]%
        {keles2022atypical}
\bibfield{author}{\bibinfo{person}{Umit Keles}, \bibinfo{person}{Dorit Kliemann}, \bibinfo{person}{Lisa Byrge}, \bibinfo{person}{Heini Saarim{\"a}ki}, \bibinfo{person}{Lynn~K Paul}, \bibinfo{person}{Daniel~P Kennedy}, {and} \bibinfo{person}{Ralph Adolphs}.} \bibinfo{year}{2022}\natexlab{}.
\newblock \showarticletitle{Atypical gaze patterns in autistic adults are heterogeneous across but reliable within individuals}.
\newblock \bibinfo{journal}{\emph{Molecular autism}} \bibinfo{volume}{13}, \bibinfo{number}{1} (\bibinfo{year}{2022}), \bibinfo{pages}{39}.
\newblock


\bibitem[Keller et~al\mbox{.}(2015)]%
        {keller2015eye}
\bibfield{author}{\bibinfo{person}{J{\"u}rgen Keller}, \bibinfo{person}{Martin Gorges}, \bibinfo{person}{Hannah~T Horn}, \bibinfo{person}{Helena~EA Aho-{\"O}zhan}, \bibinfo{person}{Elmar~H Pinkhardt}, \bibinfo{person}{Ingo Uttner}, \bibinfo{person}{Jan Kassubek}, \bibinfo{person}{Albert~C Ludolph}, {and} \bibinfo{person}{Doroth{\'e}e Lul{\'e}}.} \bibinfo{year}{2015}\natexlab{}.
\newblock \showarticletitle{Eye-tracking controlled cognitive function tests in patients with amyotrophic lateral sclerosis: a controlled proof-of-principle study}.
\newblock \bibinfo{journal}{\emph{Journal of neurology}}  \bibinfo{volume}{262} (\bibinfo{year}{2015}), \bibinfo{pages}{1918--1926}.
\newblock


\bibitem[Kennard(2011)]%
        {kennard2011disorders}
\bibfield{author}{\bibinfo{person}{Christopher Kennard}.} \bibinfo{year}{2011}\natexlab{}.
\newblock \showarticletitle{Disorders of higher gaze control}.
\newblock \bibinfo{journal}{\emph{Handbook of clinical neurology}}  \bibinfo{volume}{102} (\bibinfo{year}{2011}), \bibinfo{pages}{379--402}.
\newblock


\bibitem[Kim et~al\mbox{.}(2014)]%
        {kim2014investigating}
\bibfield{author}{\bibinfo{person}{Sunjung Kim}, \bibinfo{person}{Linda~J Lombardino}, \bibinfo{person}{Wind Cowles}, {and} \bibinfo{person}{Lori~J Altmann}.} \bibinfo{year}{2014}\natexlab{}.
\newblock \showarticletitle{Investigating graph comprehension in students with dyslexia: An eye tracking study}.
\newblock \bibinfo{journal}{\emph{Research in developmental disabilities}} \bibinfo{volume}{35}, \bibinfo{number}{7} (\bibinfo{year}{2014}), \bibinfo{pages}{1609--1622}.
\newblock


\bibitem[Kim et~al\mbox{.}(2020)]%
        {kim2020neurological}
\bibfield{author}{\bibinfo{person}{Tae~Jung Kim}, \bibinfo{person}{Soo-Hyun Park}, \bibinfo{person}{Hae-Bong Jeong}, \bibinfo{person}{Eun~Jin Ha}, \bibinfo{person}{Won~Sang Cho}, \bibinfo{person}{Hyun-Seung Kang}, \bibinfo{person}{Jung~Eun Kim}, {and} \bibinfo{person}{Sang-Bae Ko}.} \bibinfo{year}{2020}\natexlab{}.
\newblock \showarticletitle{Neurological pupil index as an indicator of neurological worsening in large hemispheric strokes}.
\newblock \bibinfo{journal}{\emph{Neurocritical care}}  \bibinfo{volume}{33} (\bibinfo{year}{2020}), \bibinfo{pages}{575--581}.
\newblock


\bibitem[Kimmig et~al\mbox{.}(2002)]%
        {kimmig2002pathological}
\bibfield{author}{\bibinfo{person}{Hubert Kimmig}, \bibinfo{person}{Katja Hau{\ss}mann}, \bibinfo{person}{Thomas Mergner}, {and} \bibinfo{person}{Carl~H L{\"u}cking}.} \bibinfo{year}{2002}\natexlab{}.
\newblock \showarticletitle{What is pathological with gaze shift fragmentation in Parkinson's disease?}
\newblock \bibinfo{journal}{\emph{Journal of Neurology}} \bibinfo{volume}{249}, \bibinfo{number}{6} (\bibinfo{year}{2002}), \bibinfo{pages}{683--692}.
\newblock


\bibitem[Kitt et~al\mbox{.}(2024)]%
        {kitt2024using}
\bibfield{author}{\bibinfo{person}{Elizabeth~R Kitt}, \bibinfo{person}{Rany Abend}, \bibinfo{person}{Paia Amelio}, \bibinfo{person}{Jordan Galbraith}, \bibinfo{person}{Anjali~D Poe}, \bibinfo{person}{Dylan~G Gee}, \bibinfo{person}{Daniel~S Pine}, {and} \bibinfo{person}{Anita Harrewijn}.} \bibinfo{year}{2024}\natexlab{}.
\newblock \showarticletitle{Using mobile eye-tracking to evaluate gaze behavior during a speech in pediatric anxiety disorders}.
\newblock \bibinfo{journal}{\emph{Journal of Affective Disorders}} (\bibinfo{year}{2024}).
\newblock


\bibitem[Kleberg et~al\mbox{.}(2021)]%
        {kleberg2021increased}
\bibfield{author}{\bibinfo{person}{Johan~Lundin Kleberg}, \bibinfo{person}{Matilda~A Frick}, {and} \bibinfo{person}{Karin~C Brocki}.} \bibinfo{year}{2021}\natexlab{}.
\newblock \showarticletitle{Increased pupil dilation to happy faces in children with hyperactive/impulsive symptoms of ADHD}.
\newblock \bibinfo{journal}{\emph{Development and psychopathology}} \bibinfo{volume}{33}, \bibinfo{number}{3} (\bibinfo{year}{2021}), \bibinfo{pages}{767--777}.
\newblock


\bibitem[Klin et~al\mbox{.}(2020)]%
        {klin2020affording}
\bibfield{author}{\bibinfo{person}{Ami Klin}, \bibinfo{person}{Megan Micheletti}, \bibinfo{person}{Cheryl Klaiman}, \bibinfo{person}{Sarah Shultz}, \bibinfo{person}{John~N Constantino}, {and} \bibinfo{person}{Warren Jones}.} \bibinfo{year}{2020}\natexlab{}.
\newblock \showarticletitle{Affording autism an early brain development re-definition}.
\newblock \bibinfo{journal}{\emph{Development and Psychopathology}} \bibinfo{volume}{32}, \bibinfo{number}{4} (\bibinfo{year}{2020}), \bibinfo{pages}{1175--1189}.
\newblock


\bibitem[Ko et~al\mbox{.}(2023)]%
        {ko2023protocol}
\bibfield{author}{\bibinfo{person}{Chanyoung Ko}, \bibinfo{person}{Soyeon Kang}, \bibinfo{person}{Soon-Beom Hong}, {and} \bibinfo{person}{Yu~Rang Park}.} \bibinfo{year}{2023}\natexlab{}.
\newblock \showarticletitle{Protocol for the development of joint attention-based subclassification of autism spectrum disorder and validation using multi-modal data}.
\newblock \bibinfo{journal}{\emph{BMC psychiatry}} \bibinfo{volume}{23}, \bibinfo{number}{1} (\bibinfo{year}{2023}), \bibinfo{pages}{589}.
\newblock


\bibitem[Ko et~al\mbox{.}(2016)]%
        {ko2016eye}
\bibfield{author}{\bibinfo{person}{Hee-kyoung Ko}, \bibinfo{person}{D~Max Snodderly}, {and} \bibinfo{person}{Martina Poletti}.} \bibinfo{year}{2016}\natexlab{}.
\newblock \showarticletitle{Eye movements between saccades: Measuring ocular drift and tremor}.
\newblock \bibinfo{journal}{\emph{Vision research}}  \bibinfo{volume}{122} (\bibinfo{year}{2016}), \bibinfo{pages}{93--104}.
\newblock


\bibitem[Koch et~al\mbox{.}(2024)]%
        {koch2024eye}
\bibfield{author}{\bibinfo{person}{Nils~A Koch}, \bibinfo{person}{Patrice Voss}, \bibinfo{person}{J~Miguel Cisneros-Franco}, \bibinfo{person}{Alexandre Drouin-Picaro}, \bibinfo{person}{Fama Tounkara}, \bibinfo{person}{Simon Ducharme}, \bibinfo{person}{Daniel Guitton}, {and} \bibinfo{person}{{\'E}tienne de Villers-Sidani}.} \bibinfo{year}{2024}\natexlab{}.
\newblock \showarticletitle{Eye movement function captured via an electronic tablet informs on cognition and disease severity in Parkinson’s disease}.
\newblock \bibinfo{journal}{\emph{Scientific Reports}} \bibinfo{volume}{14}, \bibinfo{number}{1} (\bibinfo{year}{2024}), \bibinfo{pages}{9082}.
\newblock


\bibitem[Kojima et~al\mbox{.}(2002)]%
        {kojima2002blink}
\bibfield{author}{\bibinfo{person}{Maki Kojima}, \bibinfo{person}{Toshiki Shioiri}, \bibinfo{person}{Toshihiro Hosoki}, \bibinfo{person}{Miwako Sakai}, \bibinfo{person}{Takehiko Bando}, {and} \bibinfo{person}{Toshiyuki Someya}.} \bibinfo{year}{2002}\natexlab{}.
\newblock \showarticletitle{Blink rate variability in patients with panic disorder: new trial using audiovisual stimulation}.
\newblock \bibinfo{journal}{\emph{Psychiatry and clinical neurosciences}} \bibinfo{volume}{56}, \bibinfo{number}{5} (\bibinfo{year}{2002}), \bibinfo{pages}{545--549}.
\newblock


\bibitem[Kordsachia et~al\mbox{.}(2017)]%
        {kordsachia2017abnormal}
\bibfield{author}{\bibinfo{person}{Catarina~C Kordsachia}, \bibinfo{person}{Izelle Labuschagne}, {and} \bibinfo{person}{Julie~C Stout}.} \bibinfo{year}{2017}\natexlab{}.
\newblock \showarticletitle{Abnormal visual scanning of emotionally evocative natural scenes in Huntington’s disease}.
\newblock \bibinfo{journal}{\emph{Frontiers in Psychology}}  \bibinfo{volume}{8} (\bibinfo{year}{2017}), \bibinfo{pages}{405}.
\newblock


\bibitem[Kordsachia et~al\mbox{.}(2018)]%
        {kordsachia2018visual}
\bibfield{author}{\bibinfo{person}{Catarina~C Kordsachia}, \bibinfo{person}{Izelle Labuschagne}, {and} \bibinfo{person}{Julie~C Stout}.} \bibinfo{year}{2018}\natexlab{}.
\newblock \showarticletitle{Visual scanning of the eye region of human faces predicts emotion recognition performance in Huntington’s disease.}
\newblock \bibinfo{journal}{\emph{Neuropsychology}} \bibinfo{volume}{32}, \bibinfo{number}{3} (\bibinfo{year}{2018}), \bibinfo{pages}{356}.
\newblock


\bibitem[Kovacich(2008)]%
        {kovacich2008tourette}
\bibfield{author}{\bibinfo{person}{Susan Kovacich}.} \bibinfo{year}{2008}\natexlab{}.
\newblock \showarticletitle{Tourette syndrome and the eye}.
\newblock \bibinfo{journal}{\emph{Optometry-Journal of the American Optometric Association}} \bibinfo{volume}{79}, \bibinfo{number}{8} (\bibinfo{year}{2008}), \bibinfo{pages}{432--435}.
\newblock


\bibitem[Krishna et~al\mbox{.}(2014)]%
        {krishna2014long}
\bibfield{author}{\bibinfo{person}{Nithin Krishna}, \bibinfo{person}{Hugh O’Neill}, \bibinfo{person}{Eva~Mar{\'\i}a S{\'a}nchez-Morla}, {and} \bibinfo{person}{Gunvant~K Thaker}.} \bibinfo{year}{2014}\natexlab{}.
\newblock \showarticletitle{Long range frontal/posterior phase synchronization during remembered pursuit task is impaired in schizophrenia}.
\newblock \bibinfo{journal}{\emph{Schizophrenia research}} \bibinfo{volume}{157}, \bibinfo{number}{1-3} (\bibinfo{year}{2014}), \bibinfo{pages}{198--203}.
\newblock


\bibitem[Kr{\'o}lak and Strumi{\l}{\l}o(2012)]%
        {krolak2012eye}
\bibfield{author}{\bibinfo{person}{Aleksandra Kr{\'o}lak} {and} \bibinfo{person}{Pawe{\l} Strumi{\l}{\l}o}.} \bibinfo{year}{2012}\natexlab{}.
\newblock \showarticletitle{Eye-blink detection system for human--computer interaction}.
\newblock \bibinfo{journal}{\emph{Universal Access in the Information Society}}  \bibinfo{volume}{11} (\bibinfo{year}{2012}), \bibinfo{pages}{409--419}.
\newblock


\bibitem[Kuechenmeister et~al\mbox{.}(1977)]%
        {kuechenmeister1977eye}
\bibfield{author}{\bibinfo{person}{Craig~A Kuechenmeister}, \bibinfo{person}{Patrick~H Linton}, \bibinfo{person}{Thelma~V Mueller}, {and} \bibinfo{person}{Hilton~B White}.} \bibinfo{year}{1977}\natexlab{}.
\newblock \showarticletitle{Eye tracking in relation to age, sex, and illness}.
\newblock \bibinfo{journal}{\emph{Archives of General Psychiatry}} \bibinfo{volume}{34}, \bibinfo{number}{5} (\bibinfo{year}{1977}), \bibinfo{pages}{578--579}.
\newblock


\bibitem[Kullmann et~al\mbox{.}(2021a)]%
        {kullmann2021normative}
\bibfield{author}{\bibinfo{person}{Aura Kullmann}, \bibinfo{person}{Robin~C Ashmore}, \bibinfo{person}{Alexandr Braverman}, \bibinfo{person}{Christian Mazur}, \bibinfo{person}{Hillary Snapp}, \bibinfo{person}{Erin Williams}, \bibinfo{person}{Mikhaylo Szczupak}, \bibinfo{person}{Sara Murphy}, \bibinfo{person}{Kathryn Marshall}, \bibinfo{person}{James Crawford}, {et~al\mbox{.}}} \bibinfo{year}{2021}\natexlab{a}.
\newblock \showarticletitle{Normative data for ages 18-45 for ocular motor and vestibular testing using eye tracking}.
\newblock \bibinfo{journal}{\emph{Laryngoscope investigative otolaryngology}} \bibinfo{volume}{6}, \bibinfo{number}{5} (\bibinfo{year}{2021}), \bibinfo{pages}{1116--1127}.
\newblock


\bibitem[Kullmann et~al\mbox{.}(2021b)]%
        {kullmann2021portable}
\bibfield{author}{\bibinfo{person}{Aura Kullmann}, \bibinfo{person}{Robin~C Ashmore}, \bibinfo{person}{Alexandr Braverman}, \bibinfo{person}{Christian Mazur}, \bibinfo{person}{Hillary Snapp}, \bibinfo{person}{Erin Williams}, \bibinfo{person}{Mikhaylo Szczupak}, \bibinfo{person}{Sara Murphy}, \bibinfo{person}{Kathryn Marshall}, \bibinfo{person}{James Crawford}, {et~al\mbox{.}}} \bibinfo{year}{2021}\natexlab{b}.
\newblock \showarticletitle{Portable eye-tracking as a reliable assessment of oculomotor, cognitive and reaction time function: Normative data for 18--45 year old}.
\newblock \bibinfo{journal}{\emph{PLoS One}} \bibinfo{volume}{16}, \bibinfo{number}{11} (\bibinfo{year}{2021}), \bibinfo{pages}{e0260351}.
\newblock


\bibitem[Kumar et~al\mbox{.}(2016)]%
        {kumar2016smarteye}
\bibfield{author}{\bibinfo{person}{Deepesh Kumar}, \bibinfo{person}{Anirban Dutta}, \bibinfo{person}{Abhijit Das}, {and} \bibinfo{person}{Uttama Lahiri}.} \bibinfo{year}{2016}\natexlab{}.
\newblock \showarticletitle{SmartEye: Developing a novel eye tracking system for quantitative assessment of oculomotor abnormalities}.
\newblock \bibinfo{journal}{\emph{IEEE Transactions on neural systems and rehabilitation engineering}} \bibinfo{volume}{24}, \bibinfo{number}{10} (\bibinfo{year}{2016}), \bibinfo{pages}{1051--1059}.
\newblock


\bibitem[Kynast et~al\mbox{.}(2020)]%
        {kynast2020mindreading}
\bibfield{author}{\bibinfo{person}{Jana Kynast}, \bibinfo{person}{Eva~Maria Quinque}, \bibinfo{person}{Maryna Polyakova}, \bibinfo{person}{Tobias Luck}, \bibinfo{person}{Steffi~G Riedel-Heller}, \bibinfo{person}{Simon Baron-Cohen}, \bibinfo{person}{Andreas Hinz}, \bibinfo{person}{A~Veronica Witte}, \bibinfo{person}{Julia Sacher}, \bibinfo{person}{Arno Villringer}, {et~al\mbox{.}}} \bibinfo{year}{2020}\natexlab{}.
\newblock \showarticletitle{Mindreading from the eyes declines with aging--evidence from 1,603 subjects}.
\newblock \bibinfo{journal}{\emph{Frontiers in Aging Neuroscience}}  \bibinfo{volume}{12} (\bibinfo{year}{2020}), \bibinfo{pages}{550416}.
\newblock


\bibitem[Lampe et~al\mbox{.}(2014)]%
        {lampe2014eye}
\bibfield{author}{\bibinfo{person}{Ren{\'e}e Lampe}, \bibinfo{person}{Varvara Turova}, \bibinfo{person}{Tobias Blumenstein}, {and} \bibinfo{person}{Ana Alves-Pinto}.} \bibinfo{year}{2014}\natexlab{}.
\newblock \showarticletitle{Eye movement during reading in young adults with cerebral palsy measured with eye tracking}.
\newblock \bibinfo{journal}{\emph{Postgraduate Medicine}} \bibinfo{volume}{126}, \bibinfo{number}{5} (\bibinfo{year}{2014}), \bibinfo{pages}{146--158}.
\newblock


\bibitem[Laretzaki et~al\mbox{.}(2011)]%
        {laretzaki2011threat}
\bibfield{author}{\bibinfo{person}{Georgia Laretzaki}, \bibinfo{person}{Sotiris Plainis}, \bibinfo{person}{Ioannis Vrettos}, \bibinfo{person}{Anna Chrisoulakis}, \bibinfo{person}{Ioannis Pallikaris}, {and} \bibinfo{person}{Panos Bitsios}.} \bibinfo{year}{2011}\natexlab{}.
\newblock \showarticletitle{Threat and trait anxiety affect stability of gaze fixation}.
\newblock \bibinfo{journal}{\emph{Biological psychology}} \bibinfo{volume}{86}, \bibinfo{number}{3} (\bibinfo{year}{2011}), \bibinfo{pages}{330--336}.
\newblock


\bibitem[Larsen et~al\mbox{.}(2018)]%
        {larsen2018identification}
\bibfield{author}{\bibinfo{person}{Kenneth Larsen}, \bibinfo{person}{Astrid Aasland}, {and} \bibinfo{person}{Trond~H Diseth}.} \bibinfo{year}{2018}\natexlab{}.
\newblock \showarticletitle{Identification of symptoms of autism spectrum disorders in the second year of life at day-care centres by day-care staff: Step one in the development of a short observation list}.
\newblock \bibinfo{journal}{\emph{Journal of Autism and Developmental Disorders}}  \bibinfo{volume}{48} (\bibinfo{year}{2018}), \bibinfo{pages}{2267--2277}.
\newblock


\bibitem[Larsson et~al\mbox{.}(2016)]%
        {larsson2016head}
\bibfield{author}{\bibinfo{person}{Linn{\'e}a Larsson}, \bibinfo{person}{Andrea Schwaller}, \bibinfo{person}{Marcus Nystr{\"o}m}, {and} \bibinfo{person}{Martin Stridh}.} \bibinfo{year}{2016}\natexlab{}.
\newblock \showarticletitle{Head movement compensation and multi-modal event detection in eye-tracking data for unconstrained head movements}.
\newblock \bibinfo{journal}{\emph{Journal of neuroscience methods}}  \bibinfo{volume}{274} (\bibinfo{year}{2016}), \bibinfo{pages}{13--26}.
\newblock


\bibitem[Lasker et~al\mbox{.}(1987)]%
        {lasker1987saccades}
\bibfield{author}{\bibinfo{person}{AG Lasker}, \bibinfo{person}{DS Zee}, \bibinfo{person}{TC Hain}, \bibinfo{person}{Susan~E Folstein}, {and} \bibinfo{person}{HS Singer}.} \bibinfo{year}{1987}\natexlab{}.
\newblock \showarticletitle{Saccades in Huntington's disease: initiation defects and distractibility}.
\newblock \bibinfo{journal}{\emph{Neurology}} \bibinfo{volume}{37}, \bibinfo{number}{3} (\bibinfo{year}{1987}), \bibinfo{pages}{364--364}.
\newblock


\bibitem[Laurens et~al\mbox{.}(2019)]%
        {laurens2019spatial}
\bibfield{author}{\bibinfo{person}{Brice Laurens}, \bibinfo{person}{Vincent Planche}, \bibinfo{person}{St{\'e}phanie Cubizolle}, \bibinfo{person}{L{\'e}a Declerck}, \bibinfo{person}{Sandrine Dupouy}, \bibinfo{person}{Ma{\"\i}t{\'e} Formaglio}, \bibinfo{person}{Lejla Koric}, \bibinfo{person}{Magali Seassau}, \bibinfo{person}{Caroline Tilikete}, \bibinfo{person}{Alain Vighetto}, {et~al\mbox{.}}} \bibinfo{year}{2019}\natexlab{}.
\newblock \showarticletitle{A spatial decision eye-tracking task in patients with prodromal and mild Alzheimer’s disease}.
\newblock \bibinfo{journal}{\emph{Journal of Alzheimer's Disease}} \bibinfo{volume}{71}, \bibinfo{number}{2} (\bibinfo{year}{2019}), \bibinfo{pages}{613--621}.
\newblock


\bibitem[Laurent et~al\mbox{.}(2014)]%
        {laurent2014visual}
\bibfield{author}{\bibinfo{person}{Agathe Laurent}, \bibinfo{person}{Alexis Arzimanoglou}, \bibinfo{person}{Eleni Panagiotakaki}, \bibinfo{person}{Ignacio Sfaello}, \bibinfo{person}{Philippe Kahane}, \bibinfo{person}{Philippe Ryvlin}, \bibinfo{person}{Edouard Hirsch}, {and} \bibinfo{person}{Scania de Schonen}.} \bibinfo{year}{2014}\natexlab{}.
\newblock \showarticletitle{Visual and auditory socio-cognitive perception in unilateral temporal lobe epilepsy in children and adolescents: a prospective controlled study}.
\newblock \bibinfo{journal}{\emph{Epileptic Disorders}} \bibinfo{volume}{16}, \bibinfo{number}{4} (\bibinfo{year}{2014}), \bibinfo{pages}{456--470}.
\newblock


\bibitem[Lavalle et~al\mbox{.}({[n.\,d.]})]%
        {lavalle4713241gamifying}
\bibfield{author}{\bibinfo{person}{Ana Lavalle}, \bibinfo{person}{Miguel~A Teruel}, \bibinfo{person}{Javier Sanchis}, \bibinfo{person}{Nicol{\'a}s Ruiz-Robledillo}, \bibinfo{person}{Borja Costa-L{\'o}pez}, {and} \bibinfo{person}{Juan Trujillo}.} \bibinfo{year}{[n.\,d.]}\natexlab{}.
\newblock \showarticletitle{Gamifying Attention Assessment for Attention Deficit Hyperactivity Disorder}.
\newblock \bibinfo{journal}{\emph{Available at SSRN 4713241}} (\bibinfo{year}{[n.\,d.]}).
\newblock


\bibitem[Laycock et~al\mbox{.}(2020)]%
        {laycock2020blink}
\bibfield{author}{\bibinfo{person}{Robin Laycock}, \bibinfo{person}{Sheila~G Crewther}, {and} \bibinfo{person}{Philippe~A Chouinard}.} \bibinfo{year}{2020}\natexlab{}.
\newblock \showarticletitle{Blink and you will miss it: A core role for fast and dynamic visual processing in social impairments in autism spectrum disorder}.
\newblock \bibinfo{journal}{\emph{Current Developmental Disorders Reports}}  \bibinfo{volume}{7} (\bibinfo{year}{2020}), \bibinfo{pages}{237--248}.
\newblock


\bibitem[Lazarov et~al\mbox{.}(2016)]%
        {lazarov2016social}
\bibfield{author}{\bibinfo{person}{Amit Lazarov}, \bibinfo{person}{Rany Abend}, {and} \bibinfo{person}{Yair Bar-Haim}.} \bibinfo{year}{2016}\natexlab{}.
\newblock \showarticletitle{Social anxiety is related to increased dwell time on socially threatening faces}.
\newblock \bibinfo{journal}{\emph{Journal of affective disorders}}  \bibinfo{volume}{193} (\bibinfo{year}{2016}), \bibinfo{pages}{282--288}.
\newblock


\bibitem[Leary and Hill(1996)]%
        {leary1996moving}
\bibfield{author}{\bibinfo{person}{Martha~R Leary} {and} \bibinfo{person}{David~A Hill}.} \bibinfo{year}{1996}\natexlab{}.
\newblock \showarticletitle{Moving on: autism and movement disturbance}.
\newblock \bibinfo{journal}{\emph{Mental Retardation-Washington}} \bibinfo{volume}{34}, \bibinfo{number}{1} (\bibinfo{year}{1996}), \bibinfo{pages}{39--53}.
\newblock


\bibitem[Lech et~al\mbox{.}(2019)]%
        {lech2019human}
\bibfield{author}{\bibinfo{person}{Micha{\l} Lech}, \bibinfo{person}{Micha{\l}~T Kucewicz}, {and} \bibinfo{person}{Andrzej Czy{\.z}ewski}.} \bibinfo{year}{2019}\natexlab{}.
\newblock \showarticletitle{Human computer interface for tracking eye movements improves assessment and diagnosis of patients with acquired brain injuries}.
\newblock \bibinfo{journal}{\emph{Frontiers in Neurology}}  \bibinfo{volume}{10} (\bibinfo{year}{2019}), \bibinfo{pages}{6}.
\newblock


\bibitem[Lee et~al\mbox{.}(2010)]%
        {lee2010pursuit}
\bibfield{author}{\bibinfo{person}{I-Ching Lee}, \bibinfo{person}{Chuan-Cheng Hung}, \bibinfo{person}{Wen-Ho Yang}, \bibinfo{person}{M-J Chen-Sea}, {and} \bibinfo{person}{Chin-Liang Tsai}.} \bibinfo{year}{2010}\natexlab{}.
\newblock \showarticletitle{The pursuit flexibility of children with attention-deficit/hyperactive disorder}.
\newblock \bibinfo{journal}{\emph{Medical and Health Science Journal}}  \bibinfo{volume}{4} (\bibinfo{year}{2010}), \bibinfo{pages}{8--18}.
\newblock


\bibitem[Lee et~al\mbox{.}(2015)]%
        {lee2015saccadic}
\bibfield{author}{\bibinfo{person}{Yun-Jeong Lee}, \bibinfo{person}{Sangil Lee}, \bibinfo{person}{Munseon Chang}, {and} \bibinfo{person}{Ho-Wan Kwak}.} \bibinfo{year}{2015}\natexlab{}.
\newblock \showarticletitle{Saccadic movement deficiencies in adults with ADHD tendencies}.
\newblock \bibinfo{journal}{\emph{ADHD Attention Deficit and Hyperactivity Disorders}}  \bibinfo{volume}{7} (\bibinfo{year}{2015}), \bibinfo{pages}{271--280}.
\newblock


\bibitem[Leigh et~al\mbox{.}(1983)]%
        {leigh1983abnormal}
\bibfield{author}{\bibinfo{person}{R~John Leigh}, \bibinfo{person}{Steven~A Newman}, \bibinfo{person}{Susan~E Folstein}, \bibinfo{person}{Adrian~G Lasker}, {and} \bibinfo{person}{Barbara~A Jensen}.} \bibinfo{year}{1983}\natexlab{}.
\newblock \showarticletitle{Abnormal ocular motor control in Huntington's disease}.
\newblock \bibinfo{journal}{\emph{Neurology}} \bibinfo{volume}{33}, \bibinfo{number}{10} (\bibinfo{year}{1983}), \bibinfo{pages}{1268--1268}.
\newblock


\bibitem[Lev et~al\mbox{.}(2022)]%
        {lev2022eye}
\bibfield{author}{\bibinfo{person}{Astar Lev}, \bibinfo{person}{Yoram Braw}, \bibinfo{person}{Tomer Elbaum}, \bibinfo{person}{Michael Wagner}, {and} \bibinfo{person}{Yuri Rassovsky}.} \bibinfo{year}{2022}\natexlab{}.
\newblock \showarticletitle{Eye tracking during a continuous performance test: Utility for assessing ADHD patients}.
\newblock \bibinfo{journal}{\emph{Journal of Attention Disorders}} \bibinfo{volume}{26}, \bibinfo{number}{2} (\bibinfo{year}{2022}), \bibinfo{pages}{245--255}.
\newblock


\bibitem[LeVasseur et~al\mbox{.}(2001)]%
        {levasseur2001control}
\bibfield{author}{\bibinfo{person}{Adrienne~L LeVasseur}, \bibinfo{person}{J~Randall Flanagan}, \bibinfo{person}{Richard~J Riopelle}, {and} \bibinfo{person}{Douglas~P Munoz}.} \bibinfo{year}{2001}\natexlab{}.
\newblock \showarticletitle{Control of volitional and reflexive saccades in Tourette's syndrome}.
\newblock \bibinfo{journal}{\emph{Brain}} \bibinfo{volume}{124}, \bibinfo{number}{10} (\bibinfo{year}{2001}), \bibinfo{pages}{2045--2058}.
\newblock


\bibitem[Leveille et~al\mbox{.}(1982)]%
        {leveille1982eye}
\bibfield{author}{\bibinfo{person}{Albert Leveille}, \bibinfo{person}{Joseph Kiernan}, \bibinfo{person}{James~A Goodwin}, {and} \bibinfo{person}{Jack Antel}.} \bibinfo{year}{1982}\natexlab{}.
\newblock \showarticletitle{Eye movements in amyotrophic lateral sclerosis}.
\newblock \bibinfo{journal}{\emph{Archives of Neurology}} \bibinfo{volume}{39}, \bibinfo{number}{11} (\bibinfo{year}{1982}), \bibinfo{pages}{684--686}.
\newblock


\bibitem[Li et~al\mbox{.}(2024)]%
        {li2024automating}
\bibfield{author}{\bibinfo{person}{Deming Li}, \bibinfo{person}{Ankur~A Butala}, \bibinfo{person}{Laureano Moro-Velazquez}, \bibinfo{person}{Trevor Meyer}, \bibinfo{person}{Esther~S Oh}, \bibinfo{person}{Chelsey Motley}, \bibinfo{person}{Jes{\'u}s Villalba}, {and} \bibinfo{person}{Najim Dehak}.} \bibinfo{year}{2024}\natexlab{}.
\newblock \showarticletitle{Automating the analysis of eye movement for different neurodegenerative disorders}.
\newblock \bibinfo{journal}{\emph{Computers in Biology and Medicine}}  \bibinfo{volume}{170} (\bibinfo{year}{2024}), \bibinfo{pages}{107951}.
\newblock


\bibitem[Li et~al\mbox{.}(2023)]%
        {li2023abnormal}
\bibfield{author}{\bibinfo{person}{Han Li}, \bibinfo{person}{Xue Zhang}, \bibinfo{person}{Yong Yang}, {and} \bibinfo{person}{Anmu Xie}.} \bibinfo{year}{2023}\natexlab{}.
\newblock \showarticletitle{Abnormal eye movements in Parkinson's disease: From experimental study to clinical application}.
\newblock \bibinfo{journal}{\emph{Parkinsonism \& Related Disorders}} (\bibinfo{year}{2023}), \bibinfo{pages}{105791}.
\newblock


\bibitem[Li et~al\mbox{.}(2021)]%
        {li2021kalvarepsilonido}
\bibfield{author}{\bibinfo{person}{Jingjie Li}, \bibinfo{person}{Amrita~Roy Chowdhury}, \bibinfo{person}{Kassem Fawaz}, {and} \bibinfo{person}{Younghyun Kim}.} \bibinfo{year}{2021}\natexlab{}.
\newblock \showarticletitle{$\{$Kal$\varepsilon$ido$\}$:$\{$Real-Time$\}$ privacy control for $\{$Eye-Tracking$\}$ systems}. In \bibinfo{booktitle}{\emph{30th USENIX security symposium (USENIX security 21)}}. \bibinfo{pages}{1793--1810}.
\newblock


\bibitem[Li and Liang(2024)]%
        {li2024eye}
\bibfield{author}{\bibinfo{person}{Ting-Xun Li} {and} \bibinfo{person}{Chi-Wen Liang}.} \bibinfo{year}{2024}\natexlab{}.
\newblock \showarticletitle{Eye-Tracking Based Visual Search Training in Social Anxiety: Effects on Attentional Bias, Attentional Control, Gaze Behavior, and Anxious Responses to a Speech Task}.
\newblock \bibinfo{journal}{\emph{Cognitive Therapy and Research}} (\bibinfo{year}{2024}), \bibinfo{pages}{1--13}.
\newblock


\bibitem[Li et~al\mbox{.}(2020)]%
        {li2020eye}
\bibfield{author}{\bibinfo{person}{Wenjin Li}, \bibinfo{person}{Wenju Zhou}, \bibinfo{person}{Minrui Fei}, \bibinfo{person}{Yulin Xu}, {and} \bibinfo{person}{Erfu Yang}.} \bibinfo{year}{2020}\natexlab{}.
\newblock \showarticletitle{Eye Tracking methodology for diagnosing neurological diseases: a survey}. In \bibinfo{booktitle}{\emph{2020 Chinese Automation Congress (CAC)}}. IEEE, \bibinfo{pages}{2158--2162}.
\newblock


\bibitem[Liang et~al\mbox{.}(2017)]%
        {liang2017sustained}
\bibfield{author}{\bibinfo{person}{Chi-Wen Liang}, \bibinfo{person}{Jie-Li Tsai}, {and} \bibinfo{person}{Wen-Yau Hsu}.} \bibinfo{year}{2017}\natexlab{}.
\newblock \showarticletitle{Sustained visual attention for competing emotional stimuli in social anxiety: An eye tracking study}.
\newblock \bibinfo{journal}{\emph{Journal of behavior therapy and experimental psychiatry}}  \bibinfo{volume}{54} (\bibinfo{year}{2017}), \bibinfo{pages}{178--185}.
\newblock


\bibitem[Likitgorn et~al\mbox{.}(2021)]%
        {likitgorn2021freezing}
\bibfield{author}{\bibinfo{person}{Techawit Likitgorn}, \bibinfo{person}{Yan Yan}, {and} \bibinfo{person}{Yaping~Joyce Liao}.} \bibinfo{year}{2021}\natexlab{}.
\newblock \showarticletitle{Freezing of saccades in dopa-responsive parkinsonian syndrome}.
\newblock \bibinfo{journal}{\emph{American Journal of Ophthalmology Case Reports}}  \bibinfo{volume}{23} (\bibinfo{year}{2021}), \bibinfo{pages}{101124}.
\newblock


\bibitem[Lim et~al\mbox{.}(2001)]%
        {lim2001focal}
\bibfield{author}{\bibinfo{person}{Vanessa~K Lim}, \bibinfo{person}{Eckart Altenm{\"u}ller}, {and} \bibinfo{person}{John~L Bradshaw}.} \bibinfo{year}{2001}\natexlab{}.
\newblock \showarticletitle{Focal dystonia: current theories}.
\newblock \bibinfo{journal}{\emph{Human movement science}} \bibinfo{volume}{20}, \bibinfo{number}{6} (\bibinfo{year}{2001}), \bibinfo{pages}{875--914}.
\newblock


\bibitem[Lima et~al\mbox{.}(2022)]%
        {lima2022comprehensive}
\bibfield{author}{\bibinfo{person}{Aklima~Akter Lima}, \bibinfo{person}{M~Firoz Mridha}, \bibinfo{person}{Sujoy~Chandra Das}, \bibinfo{person}{Muhammad~Mohsin Kabir}, \bibinfo{person}{Md~Rashedul Islam}, {and} \bibinfo{person}{Yutaka Watanobe}.} \bibinfo{year}{2022}\natexlab{}.
\newblock \showarticletitle{A comprehensive survey on the detection, classification, and challenges of neurological disorders}.
\newblock \bibinfo{journal}{\emph{Biology}} \bibinfo{volume}{11}, \bibinfo{number}{3} (\bibinfo{year}{2022}), \bibinfo{pages}{469}.
\newblock


\bibitem[Liu et~al\mbox{.}(2019)]%
        {liu2019differential}
\bibfield{author}{\bibinfo{person}{Ao Liu}, \bibinfo{person}{Lirong Xia}, \bibinfo{person}{Andrew Duchowski}, \bibinfo{person}{Reynold Bailey}, \bibinfo{person}{Kenneth Holmqvist}, {and} \bibinfo{person}{Eakta Jain}.} \bibinfo{year}{2019}\natexlab{}.
\newblock \showarticletitle{Differential privacy for eye-tracking data}. In \bibinfo{booktitle}{\emph{Proceedings of the 11th ACM Symposium on Eye Tracking Research \& Applications}}. \bibinfo{pages}{1--10}.
\newblock


\bibitem[Lopis et~al\mbox{.}(2019)]%
        {lopis2019eye}
\bibfield{author}{\bibinfo{person}{D{\'e}sir{\'e}e Lopis}, \bibinfo{person}{Matias Baltazar}, \bibinfo{person}{Nikoletta Geronikola}, \bibinfo{person}{Virginie Beaucousin}, {and} \bibinfo{person}{Laurence Conty}.} \bibinfo{year}{2019}\natexlab{}.
\newblock \showarticletitle{Eye contact effects on social preference and face recognition in normal ageing and in Alzheimer’s disease}.
\newblock \bibinfo{journal}{\emph{Psychological research}} \bibinfo{volume}{83}, \bibinfo{number}{6} (\bibinfo{year}{2019}), \bibinfo{pages}{1292--1303}.
\newblock


\bibitem[Lopis and Conty(2019)]%
        {lopis2019investigating}
\bibfield{author}{\bibinfo{person}{Desir{\'e}e Lopis} {and} \bibinfo{person}{Laurence Conty}.} \bibinfo{year}{2019}\natexlab{}.
\newblock \showarticletitle{Investigating eye contact effect on people’s name retrieval in normal aging and in Alzheimer’s disease}.
\newblock \bibinfo{journal}{\emph{Frontiers in Psychology}}  \bibinfo{volume}{10} (\bibinfo{year}{2019}), \bibinfo{pages}{1218}.
\newblock


\bibitem[Loyd et~al\mbox{.}(2022)]%
        {loyd2022rehabilitation}
\bibfield{author}{\bibinfo{person}{Brian~J Loyd}, \bibinfo{person}{Annie Fangman}, \bibinfo{person}{Daniel~S Peterson}, \bibinfo{person}{Eduard Gappmaier}, \bibinfo{person}{Anne Thackeray}, \bibinfo{person}{Michael~C Schubert}, {and} \bibinfo{person}{Leland~E Dibble}.} \bibinfo{year}{2022}\natexlab{}.
\newblock \showarticletitle{Rehabilitation to improve gaze and postural stability in people with multiple sclerosis: a randomized clinical trial}.
\newblock \bibinfo{journal}{\emph{Neurorehabilitation and neural repair}} \bibinfo{volume}{36}, \bibinfo{number}{10-11} (\bibinfo{year}{2022}), \bibinfo{pages}{678--688}.
\newblock


\bibitem[Lunn et~al\mbox{.}(2016)]%
        {lunn2016saccadic}
\bibfield{author}{\bibinfo{person}{Judith Lunn}, \bibinfo{person}{Tim Donovan}, \bibinfo{person}{Damien Litchfield}, \bibinfo{person}{Charlie Lewis}, \bibinfo{person}{Robert Davies}, {and} \bibinfo{person}{Trevor Crawford}.} \bibinfo{year}{2016}\natexlab{}.
\newblock \showarticletitle{Saccadic eye movement abnormalities in children with epilepsy}.
\newblock \bibinfo{journal}{\emph{PloS one}} \bibinfo{volume}{11}, \bibinfo{number}{8} (\bibinfo{year}{2016}), \bibinfo{pages}{e0160508}.
\newblock


\bibitem[MacAskill and Myall(2020)]%
        {macaskill2020pervasive}
\bibfield{author}{\bibinfo{person}{Michael~R MacAskill} {and} \bibinfo{person}{Daniel~J Myall}.} \bibinfo{year}{2020}\natexlab{}.
\newblock \showarticletitle{“Pervasive ocular tremor of Parkinson’s” is not pervasive, ocular, or uniquely parkinsonian}.
\newblock  (\bibinfo{year}{2020}).
\newblock


\bibitem[Macinska et~al\mbox{.}(2024)]%
        {macinska2024visual}
\bibfield{author}{\bibinfo{person}{Sylwia Macinska}, \bibinfo{person}{Shane Lindsay}, {and} \bibinfo{person}{Tjeerd Jellema}.} \bibinfo{year}{2024}\natexlab{}.
\newblock \showarticletitle{Visual attention to dynamic emotional faces in adults on the autism spectrum}.
\newblock \bibinfo{journal}{\emph{Journal of autism and developmental disorders}} \bibinfo{volume}{54}, \bibinfo{number}{6} (\bibinfo{year}{2024}), \bibinfo{pages}{2211--2223}.
\newblock


\bibitem[Maguire et~al\mbox{.}(2020)]%
        {maguire2020normal}
\bibfield{author}{\bibinfo{person}{Fiachra Maguire}, \bibinfo{person}{Richard~B Reilly}, {and} \bibinfo{person}{Kristina Simonyan}.} \bibinfo{year}{2020}\natexlab{}.
\newblock \showarticletitle{Normal temporal discrimination in musician's dystonia is linked to aberrant sensorimotor processing}.
\newblock \bibinfo{journal}{\emph{Movement Disorders}} \bibinfo{volume}{35}, \bibinfo{number}{5} (\bibinfo{year}{2020}), \bibinfo{pages}{800--807}.
\newblock


\bibitem[Mahajan et~al\mbox{.}(2021)]%
        {mahajan2021impaired}
\bibfield{author}{\bibinfo{person}{Abhimanyu Mahajan}, \bibinfo{person}{Palak Gupta}, \bibinfo{person}{Jonathan Jacobs}, \bibinfo{person}{Luca Marsili}, \bibinfo{person}{Andrea Sturchio}, \bibinfo{person}{HA Jinnah}, \bibinfo{person}{Alberto~J Espay}, {and} \bibinfo{person}{Aasef~G Shaikh}.} \bibinfo{year}{2021}\natexlab{}.
\newblock \showarticletitle{Impaired saccade adaptation in tremor-dominant cervical dystonia—evidence for maladaptive cerebellum}.
\newblock \bibinfo{journal}{\emph{The Cerebellum}}  \bibinfo{volume}{20} (\bibinfo{year}{2021}), \bibinfo{pages}{678--686}.
\newblock


\bibitem[Maioli et~al\mbox{.}(2019)]%
        {maioli2019visuospatial}
\bibfield{author}{\bibinfo{person}{Claudio Maioli}, \bibinfo{person}{Luca Falciati}, \bibinfo{person}{Jessica Galli}, \bibinfo{person}{Serena Micheletti}, \bibinfo{person}{Luisa Turetti}, \bibinfo{person}{Michela Balconi}, {and} \bibinfo{person}{Elisa~M Fazzi}.} \bibinfo{year}{2019}\natexlab{}.
\newblock \showarticletitle{Visuospatial attention and saccadic inhibitory control in children with cerebral palsy}.
\newblock \bibinfo{journal}{\emph{Frontiers in human neuroscience}}  \bibinfo{volume}{13} (\bibinfo{year}{2019}), \bibinfo{pages}{392}.
\newblock


\bibitem[Mallery et~al\mbox{.}(2018)]%
        {mallery2018visual}
\bibfield{author}{\bibinfo{person}{Robert~M Mallery}, \bibinfo{person}{Pieter Poolman}, \bibinfo{person}{Matthew~J Thurtell}, \bibinfo{person}{Jan~M Full}, \bibinfo{person}{Johannes Ledolter}, \bibinfo{person}{Dorlan Kimbrough}, \bibinfo{person}{Elliot~M Frohman}, \bibinfo{person}{Teresa~C Frohman}, {and} \bibinfo{person}{Randy~H Kardon}.} \bibinfo{year}{2018}\natexlab{}.
\newblock \showarticletitle{Visual fixation instability in multiple sclerosis measured using SLO-OCT}.
\newblock \bibinfo{journal}{\emph{Investigative Ophthalmology \& Visual Science}} \bibinfo{volume}{59}, \bibinfo{number}{1} (\bibinfo{year}{2018}), \bibinfo{pages}{196--201}.
\newblock


\bibitem[Ma{\~n}ago et~al\mbox{.}(2016)]%
        {manago2016gaze}
\bibfield{author}{\bibinfo{person}{Mark~M Ma{\~n}ago}, \bibinfo{person}{Margaret Schenkman}, \bibinfo{person}{Jean Berliner}, {and} \bibinfo{person}{Jeffrey~R Hebert}.} \bibinfo{year}{2016}\natexlab{}.
\newblock \showarticletitle{Gaze stabilization and dynamic visual acuity in people with multiple sclerosis}.
\newblock \bibinfo{journal}{\emph{Journal of Vestibular Research}} \bibinfo{volume}{26}, \bibinfo{number}{5-6} (\bibinfo{year}{2016}), \bibinfo{pages}{469--477}.
\newblock


\bibitem[Mancini and Esposito(2021)]%
        {mancini2021lived}
\bibfield{author}{\bibinfo{person}{Milena Mancini} {and} \bibinfo{person}{Cecilia~Maria Esposito}.} \bibinfo{year}{2021}\natexlab{}.
\newblock \showarticletitle{Lived body and the Other’s gaze: a phenomenological perspective on feeding and eating disorders}.
\newblock \bibinfo{journal}{\emph{Eating and Weight Disorders-Studies on Anorexia, Bulimia and Obesity}} \bibinfo{volume}{26}, \bibinfo{number}{8} (\bibinfo{year}{2021}), \bibinfo{pages}{2523--2529}.
\newblock


\bibitem[Mandel(2012)]%
        {mandel2012statistical}
\bibfield{author}{\bibinfo{person}{John Mandel}.} \bibinfo{year}{2012}\natexlab{}.
\newblock \bibinfo{booktitle}{\emph{The statistical analysis of experimental data}}.
\newblock \bibinfo{publisher}{Courier Corporation}.
\newblock


\bibitem[Mapstone et~al\mbox{.}(2001)]%
        {mapstone2001dynamic}
\bibfield{author}{\bibinfo{person}{Mark Mapstone}, \bibinfo{person}{Alexander R{\"o}sler}, \bibinfo{person}{Alissa Hays}, \bibinfo{person}{Darren~R Gitelman}, {and} \bibinfo{person}{Sandra Weintraub}.} \bibinfo{year}{2001}\natexlab{}.
\newblock \showarticletitle{Dynamic allocation of attention in aging and Alzheimer disease: uncoupling of the eye and mind}.
\newblock \bibinfo{journal}{\emph{Archives of Neurology}} \bibinfo{volume}{58}, \bibinfo{number}{9} (\bibinfo{year}{2001}), \bibinfo{pages}{1443--1447}.
\newblock


\bibitem[Marino et~al\mbox{.}(2010)]%
        {marino2010effect}
\bibfield{author}{\bibinfo{person}{Silvia Marino}, \bibinfo{person}{Pietro Lanzafame}, \bibinfo{person}{Edoardo Sessa}, \bibinfo{person}{Alessia Bramanti}, {and} \bibinfo{person}{Placido Bramanti}.} \bibinfo{year}{2010}\natexlab{}.
\newblock \showarticletitle{The effect of L-Dopa administration on pursuit ocular movements in suspected Parkinson’s disease}.
\newblock \bibinfo{journal}{\emph{Neurological Sciences}}  \bibinfo{volume}{31} (\bibinfo{year}{2010}), \bibinfo{pages}{381--385}.
\newblock


\bibitem[Martinez-Conde et~al\mbox{.}(2013)]%
        {martinez2013impact}
\bibfield{author}{\bibinfo{person}{Susana Martinez-Conde}, \bibinfo{person}{Jorge Otero-Millan}, {and} \bibinfo{person}{Stephen~L Macknik}.} \bibinfo{year}{2013}\natexlab{}.
\newblock \showarticletitle{The impact of microsaccades on vision: towards a unified theory of saccadic function}.
\newblock \bibinfo{journal}{\emph{Nature Reviews Neuroscience}} \bibinfo{volume}{14}, \bibinfo{number}{2} (\bibinfo{year}{2013}), \bibinfo{pages}{83--96}.
\newblock


\bibitem[Martino et~al\mbox{.}(2012)]%
        {martino2012prevalence}
\bibfield{author}{\bibinfo{person}{Davide Martino}, \bibinfo{person}{Andrea~E Cavanna}, \bibinfo{person}{Mary~M Robertson}, {and} \bibinfo{person}{Michael Orth}.} \bibinfo{year}{2012}\natexlab{}.
\newblock \showarticletitle{Prevalence and phenomenology of eye tics in Gilles de la Tourette syndrome}.
\newblock \bibinfo{journal}{\emph{Journal of neurology}}  \bibinfo{volume}{259} (\bibinfo{year}{2012}), \bibinfo{pages}{2137--2140}.
\newblock


\bibitem[Maruff et~al\mbox{.}(1999)]%
        {maruff1999abnormalities}
\bibfield{author}{\bibinfo{person}{P Maruff}, \bibinfo{person}{R Purcell}, \bibinfo{person}{P Tyler}, \bibinfo{person}{C Pantelis}, {and} \bibinfo{person}{J Currie}.} \bibinfo{year}{1999}\natexlab{}.
\newblock \showarticletitle{Abnormalities of internally generated saccades in obsessive--compulsive disorder}.
\newblock \bibinfo{journal}{\emph{Psychological medicine}} \bibinfo{volume}{29}, \bibinfo{number}{6} (\bibinfo{year}{1999}), \bibinfo{pages}{1377--1385}.
\newblock


\bibitem[Maruta et~al\mbox{.}(2017)]%
        {maruta2017visual}
\bibfield{author}{\bibinfo{person}{Jun Maruta}, \bibinfo{person}{Lisa~A Spielman}, \bibinfo{person}{Umesh Rajashekar}, {and} \bibinfo{person}{Jamshid Ghajar}.} \bibinfo{year}{2017}\natexlab{}.
\newblock \showarticletitle{Visual tracking in development and aging}.
\newblock \bibinfo{journal}{\emph{Frontiers in neurology}}  \bibinfo{volume}{8} (\bibinfo{year}{2017}), \bibinfo{pages}{640}.
\newblock


\bibitem[Mathews et~al\mbox{.}(2003)]%
        {mathews2003face}
\bibfield{author}{\bibinfo{person}{Andrew Mathews}, \bibinfo{person}{Elaine Fox}, \bibinfo{person}{Jenny Yiend}, {and} \bibinfo{person}{Andy Calder}.} \bibinfo{year}{2003}\natexlab{}.
\newblock \showarticletitle{The face of fear: Effects of eye gaze and emotion on visual attention}.
\newblock \bibinfo{journal}{\emph{Visual cognition}} \bibinfo{volume}{10}, \bibinfo{number}{7} (\bibinfo{year}{2003}), \bibinfo{pages}{823--835}.
\newblock


\bibitem[Mauriello et~al\mbox{.}(2022)]%
        {mauriello2022dysfunctional}
\bibfield{author}{\bibinfo{person}{Cheyenne Mauriello}, \bibinfo{person}{Eleonore Pham}, \bibinfo{person}{Samika Kumar}, \bibinfo{person}{Camille Piguet}, \bibinfo{person}{Marie-Pierre Deiber}, \bibinfo{person}{Jean-Michel Aubry}, \bibinfo{person}{Alexandre Dayer}, \bibinfo{person}{Christoph~M Michel}, \bibinfo{person}{Nader Perroud}, {and} \bibinfo{person}{Cristina Berchio}.} \bibinfo{year}{2022}\natexlab{}.
\newblock \showarticletitle{Dysfunctional temporal stages of eye-gaze perception in adults with ADHD: A high-density EEG study}.
\newblock \bibinfo{journal}{\emph{Biological Psychology}}  \bibinfo{volume}{171} (\bibinfo{year}{2022}), \bibinfo{pages}{108351}.
\newblock


\bibitem[Maxton et~al\mbox{.}(2013)]%
        {maxton2013don}
\bibfield{author}{\bibinfo{person}{C Maxton}, \bibinfo{person}{RA Dineen}, \bibinfo{person}{RC Padamsey}, {and} \bibinfo{person}{SK Munshi}.} \bibinfo{year}{2013}\natexlab{}.
\newblock \showarticletitle{Don’t neglect ‘neglect’--an update on post stroke neglect}.
\newblock \bibinfo{journal}{\emph{International journal of clinical practice}} \bibinfo{volume}{67}, \bibinfo{number}{4} (\bibinfo{year}{2013}), \bibinfo{pages}{369--378}.
\newblock


\bibitem[Maza et~al\mbox{.}(2020)]%
        {maza2020visual}
\bibfield{author}{\bibinfo{person}{Anny Maza}, \bibinfo{person}{Bel{\'e}n Moliner}, \bibinfo{person}{Joan Ferri}, {and} \bibinfo{person}{Roberto Llorens}.} \bibinfo{year}{2020}\natexlab{}.
\newblock \showarticletitle{Visual behavior, pupil dilation, and ability to identify emotions from facial expressions after stroke}.
\newblock \bibinfo{journal}{\emph{Frontiers in neurology}}  \bibinfo{volume}{10} (\bibinfo{year}{2020}), \bibinfo{pages}{1415}.
\newblock


\bibitem[Mazidi et~al\mbox{.}(2021)]%
        {mazidi2021time}
\bibfield{author}{\bibinfo{person}{Mahdi Mazidi}, \bibinfo{person}{Mohsen Dehghani}, \bibinfo{person}{Louise Sharpe}, \bibinfo{person}{Behrooz Dolatshahi}, \bibinfo{person}{Seyran Ranjbar}, {and} \bibinfo{person}{Ali Khatibi}.} \bibinfo{year}{2021}\natexlab{}.
\newblock \showarticletitle{Time course of attentional bias to painful facial expressions and the moderating role of attentional control: an eye-tracking study}.
\newblock \bibinfo{journal}{\emph{British Journal of Pain}} \bibinfo{volume}{15}, \bibinfo{number}{1} (\bibinfo{year}{2021}), \bibinfo{pages}{5--15}.
\newblock


\bibitem[McChesney and Bond(2018)]%
        {mcchesney2018gaze}
\bibfield{author}{\bibinfo{person}{Ian McChesney} {and} \bibinfo{person}{Raymond Bond}.} \bibinfo{year}{2018}\natexlab{}.
\newblock \showarticletitle{Gaze behaviour in computer programmers with dyslexia: considerations regarding code style, layout and crowding}. In \bibinfo{booktitle}{\emph{Proceedings of the Workshop on Eye Movements in Programming}}. \bibinfo{pages}{1--5}.
\newblock


\bibitem[McDougal and Gamlin(2015)]%
        {mcdougal2015autonomic}
\bibfield{author}{\bibinfo{person}{David~H McDougal} {and} \bibinfo{person}{Paul~D Gamlin}.} \bibinfo{year}{2015}\natexlab{}.
\newblock \showarticletitle{Autonomic control of the eye}.
\newblock \bibinfo{journal}{\emph{Comprehensive physiology}} \bibinfo{volume}{5}, \bibinfo{number}{1} (\bibinfo{year}{2015}), \bibinfo{pages}{439}.
\newblock


\bibitem[McInnis(2014)]%
        {mcinnis2014microsaccades}
\bibfield{author}{\bibinfo{person}{Hailey McInnis}.} \bibinfo{year}{2014}\natexlab{}.
\newblock \bibinfo{booktitle}{\emph{Microsaccades in Parkinson's disease}}.
\newblock \bibinfo{publisher}{Queen's University (Canada)}.
\newblock


\bibitem[Mellor and Psouni(2021)]%
        {mellor2021study}
\bibfield{author}{\bibinfo{person}{Rebecca~Louise Mellor} {and} \bibinfo{person}{Elia Psouni}.} \bibinfo{year}{2021}\natexlab{}.
\newblock \showarticletitle{The study of security priming on avoidant attentional biases: combining microsaccadic eye-movement measurement with a dot-probe task}.
\newblock \bibinfo{journal}{\emph{Frontiers in Psychology}}  \bibinfo{volume}{12} (\bibinfo{year}{2021}), \bibinfo{pages}{726817}.
\newblock


\bibitem[Metternich et~al\mbox{.}(2022)]%
        {metternich2022eye}
\bibfield{author}{\bibinfo{person}{Birgitta Metternich}, \bibinfo{person}{Nina~A Gehrer}, \bibinfo{person}{Kathrin Wagner}, \bibinfo{person}{Maximilian~J Geiger}, \bibinfo{person}{Elisa Sch{\"u}tz}, \bibinfo{person}{Andreas Schulze-Bonhage}, \bibinfo{person}{Marcel Heers}, {and} \bibinfo{person}{Michael Sch{\"o}nenberg}.} \bibinfo{year}{2022}\natexlab{}.
\newblock \showarticletitle{Eye-movement patterns during emotion recognition in focal epilepsy: An exploratory investigation}.
\newblock \bibinfo{journal}{\emph{Seizure: European Journal of Epilepsy}}  \bibinfo{volume}{100} (\bibinfo{year}{2022}), \bibinfo{pages}{95--102}.
\newblock


\bibitem[Michalek et~al\mbox{.}(2019)]%
        {michalek2019predicting}
\bibfield{author}{\bibinfo{person}{Anne~MP Michalek}, \bibinfo{person}{Gavindya Jayawardena}, {and} \bibinfo{person}{Sampath Jayarathna}.} \bibinfo{year}{2019}\natexlab{}.
\newblock \showarticletitle{Predicting ADHD using eye gaze metrics indexing working memory capacity}.
\newblock In \bibinfo{booktitle}{\emph{Computational Models for Biomedical Reasoning and Problem Solving}}. \bibinfo{publisher}{IGI Global}, \bibinfo{pages}{66--88}.
\newblock


\bibitem[Michalska et~al\mbox{.}(2017)]%
        {michalska2017anxiety}
\bibfield{author}{\bibinfo{person}{Kalina~J Michalska}, \bibinfo{person}{Laura Machlin}, \bibinfo{person}{Elizabeth Moroney}, \bibinfo{person}{Daniel~S Lowet}, \bibinfo{person}{John~M Hettema}, \bibinfo{person}{Roxann Roberson-Nay}, \bibinfo{person}{Bruno~B Averbeck}, \bibinfo{person}{Melissa~A Brotman}, \bibinfo{person}{Eric~E Nelson}, \bibinfo{person}{Ellen Leibenluft}, {et~al\mbox{.}}} \bibinfo{year}{2017}\natexlab{}.
\newblock \showarticletitle{Anxiety symptoms and children's eye gaze during fear learning}.
\newblock \bibinfo{journal}{\emph{Journal of Child Psychology and Psychiatry}} \bibinfo{volume}{58}, \bibinfo{number}{11} (\bibinfo{year}{2017}), \bibinfo{pages}{1276--1286}.
\newblock


\bibitem[Miranda et~al\mbox{.}(2014)]%
        {miranda2014anxiety}
\bibfield{author}{\bibinfo{person}{Dari{\'e}n Miranda}, \bibinfo{person}{Marco Calder{\'o}n}, {and} \bibinfo{person}{Jesus Favela}.} \bibinfo{year}{2014}\natexlab{}.
\newblock \showarticletitle{Anxiety detection using wearable monitoring}. In \bibinfo{booktitle}{\emph{Proceedings of the 5th Mexican Conference on Human-computer Interaction}}. \bibinfo{pages}{34--41}.
\newblock


\bibitem[Mo et~al\mbox{.}(2019)]%
        {mo2019shifting}
\bibfield{author}{\bibinfo{person}{Shuliang Mo}, \bibinfo{person}{Liang Liang}, \bibinfo{person}{Nicole Bardikoff}, {and} \bibinfo{person}{Mark~A Sabbagh}.} \bibinfo{year}{2019}\natexlab{}.
\newblock \showarticletitle{Shifting visual attention to social and non-social stimuli in Autism Spectrum Disorders}.
\newblock \bibinfo{journal}{\emph{Research in Autism Spectrum Disorders}}  \bibinfo{volume}{65} (\bibinfo{year}{2019}), \bibinfo{pages}{56--64}.
\newblock


\bibitem[Moghadami et~al\mbox{.}(2021)]%
        {moghadami2021investigation}
\bibfield{author}{\bibinfo{person}{Malihe Moghadami}, \bibinfo{person}{Sahar Moghimi}, \bibinfo{person}{Ali Moghimi}, \bibinfo{person}{Gholam~Reza Malekzadeh}, {and} \bibinfo{person}{Javad~Salehi Fadardi}.} \bibinfo{year}{2021}\natexlab{}.
\newblock \showarticletitle{The investigation of simultaneous EEG and eye tracking characteristics during fixation task in mild Alzheimer’s disease}.
\newblock \bibinfo{journal}{\emph{Clinical EEG and Neuroscience}} \bibinfo{volume}{52}, \bibinfo{number}{3} (\bibinfo{year}{2021}), \bibinfo{pages}{211--220}.
\newblock


\bibitem[Mohammadian et~al\mbox{.}(2015)]%
        {mohammadian2015blink}
\bibfield{author}{\bibinfo{person}{Fatemeh Mohammadian}, \bibinfo{person}{Maryam Noroozian}, \bibinfo{person}{Shahriar Nafissi}, {and} \bibinfo{person}{Farzad Fatehi}.} \bibinfo{year}{2015}\natexlab{}.
\newblock \showarticletitle{Blink reflex may help discriminate Alzheimer disease from vascular dementia}.
\newblock \bibinfo{journal}{\emph{Journal of clinical neurophysiology}} \bibinfo{volume}{32}, \bibinfo{number}{6} (\bibinfo{year}{2015}), \bibinfo{pages}{505--511}.
\newblock


\bibitem[Molina et~al\mbox{.}(2024)]%
        {molina2024eye}
\bibfield{author}{\bibinfo{person}{Ana~I Molina}, \bibinfo{person}{Yoel Arroyo}, \bibinfo{person}{Carmen Lacave}, \bibinfo{person}{Miguel~A Redondo}, \bibinfo{person}{Crescencio Bravo}, {and} \bibinfo{person}{Manuel Ortega}.} \bibinfo{year}{2024}\natexlab{}.
\newblock \showarticletitle{Eye tracking-based evaluation of accessible and usable interactive systems: tool set of guidelines and methodological issues}.
\newblock \bibinfo{journal}{\emph{Universal Access in the Information Society}} (\bibinfo{year}{2024}), \bibinfo{pages}{1--24}.
\newblock


\bibitem[Moro et~al\mbox{.}(2007)]%
        {moro2007recovery}
\bibfield{author}{\bibinfo{person}{SI Moro}, \bibinfo{person}{ML Rodriguez-Carmona}, \bibinfo{person}{EC Frost}, \bibinfo{person}{GT Plant}, {and} \bibinfo{person}{JL Barbur}.} \bibinfo{year}{2007}\natexlab{}.
\newblock \showarticletitle{Recovery of vision and pupil responses in optic neuritis and multiple sclerosis}.
\newblock \bibinfo{journal}{\emph{Ophthalmic and Physiological Optics}} \bibinfo{volume}{27}, \bibinfo{number}{5} (\bibinfo{year}{2007}), \bibinfo{pages}{451--460}.
\newblock


\bibitem[Morris et~al\mbox{.}(2002)]%
        {morris2002blink}
\bibfield{author}{\bibinfo{person}{Tim Morris}, \bibinfo{person}{Paul Blenkhorn}, {and} \bibinfo{person}{Farhan Zaidi}.} \bibinfo{year}{2002}\natexlab{}.
\newblock \showarticletitle{Blink detection for real-time eye tracking}.
\newblock \bibinfo{journal}{\emph{Journal of Network and Computer Applications}} \bibinfo{volume}{25}, \bibinfo{number}{2} (\bibinfo{year}{2002}), \bibinfo{pages}{129--143}.
\newblock


\bibitem[Mostofsky et~al\mbox{.}(2001)]%
        {mostofsky2001oculomotor}
\bibfield{author}{\bibinfo{person}{Stewart~H Mostofsky}, \bibinfo{person}{Adrian~G Lasker}, \bibinfo{person}{Harvey~S Singer}, \bibinfo{person}{Martha~B Denckla}, {and} \bibinfo{person}{David~S Zee}.} \bibinfo{year}{2001}\natexlab{}.
\newblock \showarticletitle{Oculomotor abnormalities in boys with Tourette syndrome with and without ADHD}.
\newblock \bibinfo{journal}{\emph{Journal of the American Academy of Child \& Adolescent Psychiatry}} \bibinfo{volume}{40}, \bibinfo{number}{12} (\bibinfo{year}{2001}), \bibinfo{pages}{1464--1472}.
\newblock


\bibitem[Motomura et~al\mbox{.}(2023)]%
        {motomura2023effects}
\bibfield{author}{\bibinfo{person}{Yuki Motomura}, \bibinfo{person}{Sayuri Hayashi}, \bibinfo{person}{Ryousei Kurose}, \bibinfo{person}{Hiroki Yoshida}, \bibinfo{person}{Takashi Okada}, {and} \bibinfo{person}{Shigekazu Higuchi}.} \bibinfo{year}{2023}\natexlab{}.
\newblock \showarticletitle{Effects of others’ gaze and facial expression on an observer’s microsaccades and their association with ADHD tendencies}.
\newblock \bibinfo{journal}{\emph{Journal of physiological anthropology}} \bibinfo{volume}{42}, \bibinfo{number}{1} (\bibinfo{year}{2023}), \bibinfo{pages}{19}.
\newblock


\bibitem[Mrabet et~al\mbox{.}(2024)]%
        {mrabet2024study}
\bibfield{author}{\bibinfo{person}{S Mrabet}, \bibinfo{person}{I Abdelkefi}, \bibinfo{person}{I Sghaier}, \bibinfo{person}{A Atrous}, \bibinfo{person}{Y Abida}, \bibinfo{person}{A Souissi}, \bibinfo{person}{A Gharbi}, \bibinfo{person}{A Nasri}, \bibinfo{person}{A Gargouri-Berrechid}, \bibinfo{person}{I Kacem}, {et~al\mbox{.}}} \bibinfo{year}{2024}\natexlab{}.
\newblock \showarticletitle{Study of Eye Movements Abnormalities in Epilepsy}.
\newblock \bibinfo{journal}{\emph{Neuro-Ophthalmology}} (\bibinfo{year}{2024}), \bibinfo{pages}{1--10}.
\newblock


\bibitem[M{\"u}ller({[n.\,d.]})]%
        {muller19nystagmus}
\bibfield{author}{\bibinfo{person}{Zarib M{\"u}ller}.} \bibinfo{year}{[n.\,d.]}\natexlab{}.
\newblock \showarticletitle{Nystagmus Patterns from Eye-Tracker Data}.
\newblock \bibinfo{journal}{\emph{binocular vision}} \bibinfo{volume}{19}, \bibinfo{number}{10} (\bibinfo{year}{[n.\,d.]}), \bibinfo{pages}{9}.
\newblock


\bibitem[M{\"u}ri and Meienberg(1985)]%
        {muri1985clinical}
\bibfield{author}{\bibinfo{person}{Ren{\'e}~M M{\"u}ri} {and} \bibinfo{person}{Otmar Meienberg}.} \bibinfo{year}{1985}\natexlab{}.
\newblock \showarticletitle{The clinical spectrum of internuclear ophthalmoplegia in multiple sclerosis}.
\newblock \bibinfo{journal}{\emph{Archives of neurology}} \bibinfo{volume}{42}, \bibinfo{number}{9} (\bibinfo{year}{1985}), \bibinfo{pages}{851--855}.
\newblock


\bibitem[Murray and Janelle(2003)]%
        {murray2003anxiety}
\bibfield{author}{\bibinfo{person}{Nicholas~P Murray} {and} \bibinfo{person}{Christopher~M Janelle}.} \bibinfo{year}{2003}\natexlab{}.
\newblock \showarticletitle{Anxiety and performance: A visual search examination of the processing efficiency theory}.
\newblock \bibinfo{journal}{\emph{Journal of Sport and Exercise psychology}} \bibinfo{volume}{25}, \bibinfo{number}{2} (\bibinfo{year}{2003}), \bibinfo{pages}{171--187}.
\newblock


\bibitem[Nakamagoe et~al\mbox{.}(2023)]%
        {nakamagoe2023saccadic}
\bibfield{author}{\bibinfo{person}{Kiyotaka Nakamagoe}, \bibinfo{person}{Shunya Matsumoto}, \bibinfo{person}{Nozomi Touno}, \bibinfo{person}{Ikumi Tateno}, {and} \bibinfo{person}{Tadachika Koganezawa}.} \bibinfo{year}{2023}\natexlab{}.
\newblock \showarticletitle{Saccadic oscillations as a biomarker of clinical symptoms in amyotrophic lateral sclerosis}.
\newblock \bibinfo{journal}{\emph{Neurological Sciences}} \bibinfo{volume}{44}, \bibinfo{number}{8} (\bibinfo{year}{2023}), \bibinfo{pages}{2787--2793}.
\newblock


\bibitem[Nanning et~al\mbox{.}(2023)]%
        {nanning2023altered}
\bibfield{author}{\bibinfo{person}{Felix Nanning}, \bibinfo{person}{Katharina Braune}, \bibinfo{person}{Ingo Uttner}, \bibinfo{person}{Albert~Christian Ludolph}, \bibinfo{person}{Martin Gorges}, {and} \bibinfo{person}{Doroth{\'e}e Lul{\'e}}.} \bibinfo{year}{2023}\natexlab{}.
\newblock \showarticletitle{Altered Gaze Control During Emotional Face Exploration in Patients With Amyotrophic Lateral Sclerosis}.
\newblock \bibinfo{journal}{\emph{Neurology}} \bibinfo{volume}{101}, \bibinfo{number}{6} (\bibinfo{year}{2023}), \bibinfo{pages}{264--269}.
\newblock


\bibitem[{National Library of Medicine (U.S.)}(2024)]%
        {clinicaltrials2024}
\bibfield{author}{\bibinfo{person}{{National Library of Medicine (U.S.)}}.} \bibinfo{year}{2024}\natexlab{}.
\newblock \bibinfo{title}{ClinicalTrials.gov}.
\newblock \bibinfo{howpublished}{\url{https://clinicaltrials.gov}}.
\newblock
\newblock
\shownote{Accessed: 2024-01-30}.


\bibitem[Nerrant and Tilikete(2017)]%
        {nerrant2017ocular}
\bibfield{author}{\bibinfo{person}{Elodie Nerrant} {and} \bibinfo{person}{Caroline Tilikete}.} \bibinfo{year}{2017}\natexlab{}.
\newblock \showarticletitle{Ocular motor manifestations of multiple sclerosis}.
\newblock \bibinfo{journal}{\emph{Journal of Neuro-Ophthalmology}} \bibinfo{volume}{37}, \bibinfo{number}{3} (\bibinfo{year}{2017}), \bibinfo{pages}{332--340}.
\newblock


\bibitem[Neru{\v{s}}il et~al\mbox{.}(2021)]%
        {neruvsil2021eye}
\bibfield{author}{\bibinfo{person}{Boris Neru{\v{s}}il}, \bibinfo{person}{Jaroslav Polec}, \bibinfo{person}{Juraj {\v{S}}kunda}, {and} \bibinfo{person}{Juraj Ka{\v{c}}ur}.} \bibinfo{year}{2021}\natexlab{}.
\newblock \showarticletitle{Eye tracking based dyslexia detection using a holistic approach}.
\newblock \bibinfo{journal}{\emph{Scientific Reports}} \bibinfo{volume}{11}, \bibinfo{number}{1} (\bibinfo{year}{2021}), \bibinfo{pages}{15687}.
\newblock


\bibitem[Ngo et~al\mbox{.}(2024)]%
        {ngo2024eeg}
\bibfield{author}{\bibinfo{person}{Thi~Duyen Ngo}, \bibinfo{person}{Hai~Dang Kieu}, \bibinfo{person}{Minh~Hoa Nguyen}, \bibinfo{person}{The Hoang-Anh Nguyen}, \bibinfo{person}{Van~Mao Can}, \bibinfo{person}{Ba~Hung Nguyen}, {and} \bibinfo{person}{Thanh~Ha Le}.} \bibinfo{year}{2024}\natexlab{}.
\newblock \showarticletitle{An EEG \& eye-tracking dataset of ALS patients \& healthy people during eye-tracking-based spelling system usage}.
\newblock \bibinfo{journal}{\emph{Scientific Data}} \bibinfo{volume}{11}, \bibinfo{number}{1} (\bibinfo{year}{2024}), \bibinfo{pages}{664}.
\newblock


\bibitem[Noda(1991)]%
        {noda1991cerebellar}
\bibfield{author}{\bibinfo{person}{Hiroharu Noda}.} \bibinfo{year}{1991}\natexlab{}.
\newblock \showarticletitle{Cerebellar control of saccadic eye movements: its neural mechanisms and pathways}.
\newblock \bibinfo{journal}{\emph{The Japanese Journal of Physiology}} \bibinfo{volume}{41}, \bibinfo{number}{3} (\bibinfo{year}{1991}), \bibinfo{pages}{351--368}.
\newblock


\bibitem[Nonoka et~al\mbox{.}(2024)]%
        {nonoka2024study}
\bibfield{author}{\bibinfo{person}{Sawada Nonoka}, \bibinfo{person}{Akira Uehara}, \bibinfo{person}{Hiroaki Kawamoto}, {and} \bibinfo{person}{Yoshiyuki Sankai}.} \bibinfo{year}{2024}\natexlab{}.
\newblock \showarticletitle{Study on Wearable Gaze-Based Communication System for Patients with Amyotrophic Lateral Sclerosis (ALS)}. In \bibinfo{booktitle}{\emph{2024 IEEE/SICE International Symposium on System Integration (SII)}}. IEEE, \bibinfo{pages}{165--171}.
\newblock


\bibitem[Nosek et~al\mbox{.}(2018)]%
        {nosek2018preregistration}
\bibfield{author}{\bibinfo{person}{Brian~A Nosek}, \bibinfo{person}{Charles~R Ebersole}, \bibinfo{person}{Alexander~C DeHaven}, {and} \bibinfo{person}{David~T Mellor}.} \bibinfo{year}{2018}\natexlab{}.
\newblock \showarticletitle{The preregistration revolution}.
\newblock \bibinfo{journal}{\emph{Proceedings of the National Academy of Sciences}} \bibinfo{volume}{115}, \bibinfo{number}{11} (\bibinfo{year}{2018}), \bibinfo{pages}{2600--2606}.
\newblock


\bibitem[of~California(2018)]%
        {ccpa2018}
\bibfield{author}{\bibinfo{person}{State of California}.} \bibinfo{year}{2018}\natexlab{}.
\newblock \bibinfo{title}{California Consumer Privacy Act of 2018 (CCPA)}.
\newblock \bibinfo{howpublished}{\url{https://oag.ca.gov/privacy/ccpa}}.
\newblock
\newblock
\shownote{Accessed: 2024-01-30}.


\bibitem[of~Health and (HHS)(1996)]%
        {hipaa1996}
\bibfield{author}{\bibinfo{person}{U.S.~Department of Health} {and} \bibinfo{person}{Human~Services (HHS)}.} \bibinfo{year}{1996}\natexlab{}.
\newblock \bibinfo{title}{Health Insurance Portability and Accountability Act of 1996 (HIPAA)}.
\newblock \bibinfo{howpublished}{\url{https://www.hhs.gov/hipaa/index.html}}.
\newblock
\newblock
\shownote{Accessed: 2024-01-30}.


\bibitem[of~Health and (HHS)(2024a)]%
        {hhs_irbs_assurances}
\bibfield{author}{\bibinfo{person}{U.S.~Department of Health} {and} \bibinfo{person}{Human~Services (HHS)}.} \bibinfo{year}{2024}\natexlab{a}.
\newblock \bibinfo{title}{Institutional Review Boards (IRBs) and Assurances}.
\newblock \bibinfo{howpublished}{\url{https://www.hhs.gov/ohrp/irbs-and-assurances.html}}.
\newblock
\newblock
\shownote{Accessed: 2024-01-30}.


\bibitem[of~Health and (HHS)(2024b)]%
        {hhs_45cfr46}
\bibfield{author}{\bibinfo{person}{U.S.~Department of Health} {and} \bibinfo{person}{Human~Services (HHS)}.} \bibinfo{year}{2024}\natexlab{b}.
\newblock \bibinfo{title}{Protection of Human Subjects, 45 CFR 46}.
\newblock \bibinfo{howpublished}{\url{https://www.hhs.gov/ohrp/regulations-and-policy/regulations/45-cfr-46/index.html}}.
\newblock
\newblock
\shownote{Accessed: 2024-01-30}.


\bibitem[of~the ACM Symposium~on Eye Tracking~Research and Applications(2024)]%
        {etra2024}
\bibfield{author}{\bibinfo{person}{Proceedings of~the ACM Symposium~on Eye Tracking~Research} {and} \bibinfo{person}{Applications}.} \bibinfo{year}{2024}\natexlab{}.
\newblock \bibinfo{title}{(ETRA)}.
\newblock


\bibitem[Okruszek et~al\mbox{.}(2017)]%
        {okruszek2017gaze}
\bibfield{author}{\bibinfo{person}{{\L}ukasz Okruszek}, \bibinfo{person}{Aleksandra Bala}, \bibinfo{person}{Marcela Dziekan}, \bibinfo{person}{Marta Szantroch}, \bibinfo{person}{Andrzej Rysz}, \bibinfo{person}{Andrzej Marchel}, {and} \bibinfo{person}{Sylwia Hyniewska}.} \bibinfo{year}{2017}\natexlab{}.
\newblock \showarticletitle{Gaze matters! The effect of gaze direction on emotional enhancement of memory for faces in patients with mesial temporal lobe epilepsy}.
\newblock \bibinfo{journal}{\emph{Epilepsy \& Behavior}}  \bibinfo{volume}{72} (\bibinfo{year}{2017}), \bibinfo{pages}{35--38}.
\newblock


\bibitem[Olivetti~Belardinelli et~al\mbox{.}(2021)]%
        {olivetti2021abnormal}
\bibfield{author}{\bibinfo{person}{Marta Olivetti~Belardinelli}, \bibinfo{person}{Thomas H{\"u}nefeldt}, \bibinfo{person}{Roberta Meloni}, \bibinfo{person}{Ferdinando Squitieri}, \bibinfo{person}{Sabrina Maffi}, {and} \bibinfo{person}{Simone Migliore}.} \bibinfo{year}{2021}\natexlab{}.
\newblock \showarticletitle{Abnormal visual scanning and impaired mental state recognition in pre-manifest Huntington disease}.
\newblock \bibinfo{journal}{\emph{Experimental Brain Research}} \bibinfo{volume}{239}, \bibinfo{number}{1} (\bibinfo{year}{2021}), \bibinfo{pages}{141--150}.
\newblock


\bibitem[Oprisan and Popescu(2023)]%
        {oprisan2023dysautonomia}
\bibfield{author}{\bibinfo{person}{Alexandra~L Oprisan} {and} \bibinfo{person}{Bogdan~Ovidiu Popescu}.} \bibinfo{year}{2023}\natexlab{}.
\newblock \showarticletitle{Dysautonomia in amyotrophic lateral sclerosis}.
\newblock \bibinfo{journal}{\emph{International Journal of Molecular Sciences}} \bibinfo{volume}{24}, \bibinfo{number}{19} (\bibinfo{year}{2023}), \bibinfo{pages}{14927}.
\newblock


\bibitem[Organization(2006)]%
        {world2006neurological}
\bibfield{author}{\bibinfo{person}{World~Health Organization}.} \bibinfo{year}{2006}\natexlab{}.
\newblock \bibinfo{booktitle}{\emph{Neurological disorders: public health challenges}}.
\newblock \bibinfo{publisher}{World Health Organization}.
\newblock


\bibitem[Orlosky et~al\mbox{.}(2017)]%
        {orlosky2017emulation}
\bibfield{author}{\bibinfo{person}{Jason Orlosky}, \bibinfo{person}{Yuta Itoh}, \bibinfo{person}{Maud Ranchet}, \bibinfo{person}{Kiyoshi Kiyokawa}, \bibinfo{person}{John Morgan}, {and} \bibinfo{person}{Hannes Devos}.} \bibinfo{year}{2017}\natexlab{}.
\newblock \showarticletitle{Emulation of physician tasks in eye-tracked virtual reality for remote diagnosis of neurodegenerative disease}.
\newblock \bibinfo{journal}{\emph{IEEE transactions on visualization and computer graphics}} \bibinfo{volume}{23}, \bibinfo{number}{4} (\bibinfo{year}{2017}), \bibinfo{pages}{1302--1311}.
\newblock


\bibitem[Otero-Millan et~al\mbox{.}(2013)]%
        {otero2013saccades}
\bibfield{author}{\bibinfo{person}{Jorge Otero-Millan}, \bibinfo{person}{Rosalyn Schneider}, \bibinfo{person}{R~John Leigh}, \bibinfo{person}{Stephen~L Macknik}, {and} \bibinfo{person}{Susana Martinez-Conde}.} \bibinfo{year}{2013}\natexlab{}.
\newblock \showarticletitle{Saccades during attempted fixation in parkinsonian disorders and recessive ataxia: from microsaccades to square-wave jerks}.
\newblock \bibinfo{journal}{\emph{PLoS One}} \bibinfo{volume}{8}, \bibinfo{number}{3} (\bibinfo{year}{2013}), \bibinfo{pages}{e58535}.
\newblock


\bibitem[{\"O}zdel et~al\mbox{.}(2024)]%
        {ozdel2024privacy}
\bibfield{author}{\bibinfo{person}{S{\"u}leyman {\"O}zdel}, \bibinfo{person}{Efe Bozkir}, {and} \bibinfo{person}{Enkelejda Kasneci}.} \bibinfo{year}{2024}\natexlab{}.
\newblock \showarticletitle{Privacy-preserving Scanpath Comparison for Pervasive Eye Tracking}.
\newblock \bibinfo{journal}{\emph{Proceedings of the ACM on Human-Computer Interaction}} \bibinfo{volume}{8}, \bibinfo{number}{ETRA} (\bibinfo{year}{2024}), \bibinfo{pages}{1--28}.
\newblock


\bibitem[Palmer et~al\mbox{.}(2018)]%
        {palmer2018autistic}
\bibfield{author}{\bibinfo{person}{Colin~J Palmer}, \bibinfo{person}{Rebecca~P Lawson}, \bibinfo{person}{Shravanti Shankar}, \bibinfo{person}{Colin~WG Clifford}, {and} \bibinfo{person}{Geraint Rees}.} \bibinfo{year}{2018}\natexlab{}.
\newblock \showarticletitle{Autistic adults show preserved normalisation of sensory responses in gaze processing}.
\newblock \bibinfo{journal}{\emph{Cortex}}  \bibinfo{volume}{103} (\bibinfo{year}{2018}), \bibinfo{pages}{13--23}.
\newblock


\bibitem[Panagiotidi et~al\mbox{.}(2017)]%
        {panagiotidi2017increased}
\bibfield{author}{\bibinfo{person}{Maria Panagiotidi}, \bibinfo{person}{Overton Paul}, {and} \bibinfo{person}{Stafford Tom}.} \bibinfo{year}{2017}\natexlab{}.
\newblock \showarticletitle{Increased microsaccade rate in individuals with ADHD traits}.
\newblock \bibinfo{journal}{\emph{Journal of Eye Movement Research}} \bibinfo{volume}{10}, \bibinfo{number}{1} (\bibinfo{year}{2017}).
\newblock


\bibitem[Pantelis and Kennedy(2017)]%
        {pantelis2017deconstructing}
\bibfield{author}{\bibinfo{person}{Peter~C Pantelis} {and} \bibinfo{person}{Daniel~P Kennedy}.} \bibinfo{year}{2017}\natexlab{}.
\newblock \showarticletitle{Deconstructing atypical eye gaze perception in autism spectrum disorder}.
\newblock \bibinfo{journal}{\emph{Scientific reports}} \bibinfo{volume}{7}, \bibinfo{number}{1} (\bibinfo{year}{2017}), \bibinfo{pages}{1--10}.
\newblock


\bibitem[Parikh and Kalva(2018)]%
        {parikh2018predicting}
\bibfield{author}{\bibinfo{person}{Saurin Parikh} {and} \bibinfo{person}{Hari Kalva}.} \bibinfo{year}{2018}\natexlab{}.
\newblock \showarticletitle{Predicting learning difficulty based on gaze and pupil response}. In \bibinfo{booktitle}{\emph{Adjunct Publication of the 26th Conference on User Modeling, Adaptation and Personalization}}. \bibinfo{pages}{131--135}.
\newblock


\bibitem[Parliament and of~the European~Union(2016)]%
        {gdpr2016}
\bibfield{author}{\bibinfo{person}{European Parliament} {and} \bibinfo{person}{Council of~the European~Union}.} \bibinfo{year}{2016}\natexlab{}.
\newblock \bibinfo{title}{General Data Protection Regulation}.
\newblock
\newblock
\shownote{Accessed: 2024-01-30}.


\bibitem[Parliament and of~the European~Union(2017)]%
        {mdr2017}
\bibfield{author}{\bibinfo{person}{European Parliament} {and} \bibinfo{person}{Council of~the European~Union}.} \bibinfo{year}{2017}\natexlab{}.
\newblock \bibinfo{title}{Regulation (EU) 2017/745 of the European Parliament and of the Council on Medical Devices}.
\newblock \bibinfo{howpublished}{\url{https://eur-lex.europa.eu/eli/reg/2017/745/oj}}.
\newblock
\newblock
\shownote{Accessed: 2024-01-30}.


\bibitem[Patel et~al\mbox{.}(2022)]%
        {patel2022evolution}
\bibfield{author}{\bibinfo{person}{Khushboo Patel}, \bibinfo{person}{Nitish Kamble}, \bibinfo{person}{Vikram~V Holla}, \bibinfo{person}{Pramod~K Pal}, {and} \bibinfo{person}{Ravi Yadav}.} \bibinfo{year}{2022}\natexlab{}.
\newblock \showarticletitle{Evolution of eye movement abnormalities in Huntington’s disease}.
\newblock \bibinfo{journal}{\emph{Annals of Movement Disorders}} \bibinfo{volume}{5}, \bibinfo{number}{1} (\bibinfo{year}{2022}), \bibinfo{pages}{1--11}.
\newblock


\bibitem[Patel et~al\mbox{.}(2012)]%
        {patel2012reflexive}
\bibfield{author}{\bibinfo{person}{Saumil~S Patel}, \bibinfo{person}{Joseph Jankovic}, \bibinfo{person}{Ashley~J Hood}, \bibinfo{person}{Cameron~B Jeter}, {and} \bibinfo{person}{Anne~B Sereno}.} \bibinfo{year}{2012}\natexlab{}.
\newblock \showarticletitle{Reflexive and volitional saccades: biomarkers of Huntington disease severity and progression}.
\newblock \bibinfo{journal}{\emph{Journal of the neurological sciences}} \bibinfo{volume}{313}, \bibinfo{number}{1-2} (\bibinfo{year}{2012}), \bibinfo{pages}{35--41}.
\newblock


\bibitem[Pauletti et~al\mbox{.}(1993)]%
        {pauletti1993blink}
\bibfield{author}{\bibinfo{person}{G Pauletti}, \bibinfo{person}{A Berardelli}, \bibinfo{person}{Giorgio Cruccu}, \bibinfo{person}{R Agostino}, {and} \bibinfo{person}{Mario Manfredi}.} \bibinfo{year}{1993}\natexlab{}.
\newblock \showarticletitle{Blink reflex and the masseter inhibitory reflex in patients with dystonia}.
\newblock \bibinfo{journal}{\emph{Movement disorders: official journal of the Movement Disorder Society}} \bibinfo{volume}{8}, \bibinfo{number}{4} (\bibinfo{year}{1993}), \bibinfo{pages}{495--500}.
\newblock


\bibitem[Peltsch et~al\mbox{.}(2008)]%
        {peltsch2008saccadic}
\bibfield{author}{\bibinfo{person}{A Peltsch}, \bibinfo{person}{A Hoffman}, \bibinfo{person}{I Armstrong}, \bibinfo{person}{G Pari}, {and} \bibinfo{person}{DP Munoz}.} \bibinfo{year}{2008}\natexlab{}.
\newblock \showarticletitle{Saccadic impairments in Huntington’s disease}.
\newblock \bibinfo{journal}{\emph{Experimental brain research}}  \bibinfo{volume}{186} (\bibinfo{year}{2008}), \bibinfo{pages}{457--469}.
\newblock


\bibitem[Pereira et~al\mbox{.}(2020)]%
        {pereira2020visual}
\bibfield{author}{\bibinfo{person}{Marta Lu{\'\i}sa Gon{\c{c}}alves de~Freitas Pereira}, \bibinfo{person}{Marina von Zuben de~Arruda Camargo}, \bibinfo{person}{Ariella Fornachari~Ribeiro Bellan}, \bibinfo{person}{Ana~Carolina Tahira}, \bibinfo{person}{Bernardo Dos~Santos}, \bibinfo{person}{J{\'e}ssica Dos~Santos}, \bibinfo{person}{Ariane Machado-Lima}, \bibinfo{person}{F{\'a}tima~LS Nunes}, {and} \bibinfo{person}{Orestes~Vicente Forlenza}.} \bibinfo{year}{2020}\natexlab{}.
\newblock \showarticletitle{Visual search efficiency in mild cognitive impairment and Alzheimer’s disease: An eye movement study}.
\newblock \bibinfo{journal}{\emph{Journal of Alzheimer's Disease}} \bibinfo{volume}{75}, \bibinfo{number}{1} (\bibinfo{year}{2020}), \bibinfo{pages}{261--275}.
\newblock


\bibitem[Pesenti et~al\mbox{.}(2001)]%
        {pesenti2001transient}
\bibfield{author}{\bibinfo{person}{Alessandra Pesenti}, \bibinfo{person}{Alberto Priori}, \bibinfo{person}{Guglielmo Scarlato}, {and} \bibinfo{person}{Sergio Barbieri}.} \bibinfo{year}{2001}\natexlab{}.
\newblock \showarticletitle{Transient improvement induced by motor fatigue in focal occupational dystonia: the handgrip test}.
\newblock \bibinfo{journal}{\emph{Movement disorders: official journal of the Movement Disorder Society}} \bibinfo{volume}{16}, \bibinfo{number}{6} (\bibinfo{year}{2001}), \bibinfo{pages}{1143--1147}.
\newblock


\bibitem[Pimenta et~al\mbox{.}(2017)]%
        {pimenta2017effects}
\bibfield{author}{\bibinfo{person}{Carla Pimenta}, \bibinfo{person}{Anabela Correia}, \bibinfo{person}{Marta Alves}, {and} \bibinfo{person}{Daniel Virella}.} \bibinfo{year}{2017}\natexlab{}.
\newblock \showarticletitle{Effects of oculomotor and gaze stability exercises on balance after stroke: Clinical trial protocol}.
\newblock \bibinfo{journal}{\emph{Porto biomedical journal}} \bibinfo{volume}{2}, \bibinfo{number}{3} (\bibinfo{year}{2017}), \bibinfo{pages}{76--80}.
\newblock


\bibitem[Pinkhardt et~al\mbox{.}(2009)]%
        {pinkhardt2009comparison}
\bibfield{author}{\bibinfo{person}{Elmar~H Pinkhardt}, \bibinfo{person}{Jan Kassubek}, \bibinfo{person}{Sigurd S{\"u}ssmuth}, \bibinfo{person}{Albert~C Ludolph}, \bibinfo{person}{Wolfgang Becker}, {and} \bibinfo{person}{Reinhart J{\"u}rgens}.} \bibinfo{year}{2009}\natexlab{}.
\newblock \showarticletitle{Comparison of smooth pursuit eye movement deficits in multiple system atrophy and Parkinson’s disease}.
\newblock \bibinfo{journal}{\emph{Journal of neurology}}  \bibinfo{volume}{256} (\bibinfo{year}{2009}), \bibinfo{pages}{1438--1446}.
\newblock


\bibitem[Pinnock et~al\mbox{.}(2010)]%
        {pinnock2010exploration}
\bibfield{author}{\bibinfo{person}{Ralph~Allen Pinnock}, \bibinfo{person}{Richard~Canice McGivern}, \bibinfo{person}{Raeburn Forbes}, {and} \bibinfo{person}{James~Mark Gibson}.} \bibinfo{year}{2010}\natexlab{}.
\newblock \showarticletitle{An exploration of ocular fixation in Parkinson’s disease, multiple system atrophy and progressive supranuclear palsy}.
\newblock \bibinfo{journal}{\emph{Journal of neurology}}  \bibinfo{volume}{257} (\bibinfo{year}{2010}), \bibinfo{pages}{533--539}.
\newblock


\bibitem[Pitigoi et~al\mbox{.}(2024)]%
        {pitigoi2024attentional}
\bibfield{author}{\bibinfo{person}{Isabell~C Pitigoi}, \bibinfo{person}{Brian~C Coe}, \bibinfo{person}{Olivia~G Calancie}, \bibinfo{person}{Donald~C Brien}, \bibinfo{person}{Rachel Yep}, \bibinfo{person}{Heidi~C Riek}, \bibinfo{person}{Ryan~H Kirkpatrick}, \bibinfo{person}{Blake~K Noyes}, \bibinfo{person}{Brian~J White}, \bibinfo{person}{Gunnar Blohm}, {et~al\mbox{.}}} \bibinfo{year}{2024}\natexlab{}.
\newblock \showarticletitle{Attentional modulation of eye blinking is altered by sex, age, and task structure}.
\newblock \bibinfo{journal}{\emph{Eneuro}} \bibinfo{volume}{11}, \bibinfo{number}{3} (\bibinfo{year}{2024}).
\newblock


\bibitem[P{\"o}hlchen et~al\mbox{.}(2021)]%
        {pohlchen2021examining}
\bibfield{author}{\bibinfo{person}{Dorothee P{\"o}hlchen}, \bibinfo{person}{Marthe Priouret}, \bibinfo{person}{Miriam~S Kraft}, \bibinfo{person}{Florian~P Binder}, \bibinfo{person}{Deniz~A G{\"u}rsel}, \bibinfo{person}{G{\"o}tz Berberich}, \bibinfo{person}{BeCOME~Working Group}, \bibinfo{person}{Kathrin Koch}, {and} \bibinfo{person}{Victor~I Spoormaker}.} \bibinfo{year}{2021}\natexlab{}.
\newblock \showarticletitle{Examining differences in fear learning in patients with obsessive-compulsive disorder with pupillometry, startle electromyography and skin conductance responses}.
\newblock \bibinfo{journal}{\emph{Frontiers in Psychiatry}}  \bibinfo{volume}{12} (\bibinfo{year}{2021}), \bibinfo{pages}{730742}.
\newblock


\bibitem[Polet et~al\mbox{.}(2023)]%
        {polet2023facial}
\bibfield{author}{\bibinfo{person}{Kevin Polet}, \bibinfo{person}{Solange Hesse}, \bibinfo{person}{Helo{\"\i}se Joly}, \bibinfo{person}{Mikael Cohen}, \bibinfo{person}{Adeline Morisot}, \bibinfo{person}{Benoit Kullmann}, \bibinfo{person}{Lydiane Mondot}, \bibinfo{person}{Sandrine~Louchart de~la Chapelle}, \bibinfo{person}{Alain Pesce}, {and} \bibinfo{person}{Christine Lebrun-Frenay}.} \bibinfo{year}{2023}\natexlab{}.
\newblock \showarticletitle{Facial emotion impairment in multiple sclerosis is linked to modifying observation strategies of emotional faces}.
\newblock \bibinfo{journal}{\emph{Multiple Sclerosis and Related Disorders}}  \bibinfo{volume}{69} (\bibinfo{year}{2023}), \bibinfo{pages}{104439}.
\newblock


\bibitem[Polet et~al\mbox{.}(2022)]%
        {polet2022eye}
\bibfield{author}{\bibinfo{person}{K{\'e}vin Polet}, \bibinfo{person}{Solange Hesse}, \bibinfo{person}{Adeline Morisot}, \bibinfo{person}{Beno{\^\i}t Kullmann}, \bibinfo{person}{Sandrine~Louchart de~la Chapelle}, \bibinfo{person}{Alain Pesce}, {and} \bibinfo{person}{Galina Iakimova}.} \bibinfo{year}{2022}\natexlab{}.
\newblock \showarticletitle{Eye-gaze strategies during facial emotion recognition in neurodegenerative diseases and links with neuropsychiatric disorders}.
\newblock \bibinfo{journal}{\emph{Cognitive and Behavioral Neurology}} \bibinfo{volume}{35}, \bibinfo{number}{1} (\bibinfo{year}{2022}), \bibinfo{pages}{14--31}.
\newblock


\bibitem[Poletti et~al\mbox{.}(2017)]%
        {poletti2017eye}
\bibfield{author}{\bibinfo{person}{Barbara Poletti}, \bibinfo{person}{Laura Carelli}, \bibinfo{person}{Federica Solca}, \bibinfo{person}{Annalisa Lafronza}, \bibinfo{person}{Elisa Pedroli}, \bibinfo{person}{Andrea Faini}, \bibinfo{person}{Stefano Zago}, \bibinfo{person}{Nicola Ticozzi}, \bibinfo{person}{Andrea Ciammola}, \bibinfo{person}{Claudia Morelli}, {et~al\mbox{.}}} \bibinfo{year}{2017}\natexlab{}.
\newblock \showarticletitle{An eye-tracking controlled neuropsychological battery for cognitive assessment in neurological diseases}.
\newblock \bibinfo{journal}{\emph{Neurological Sciences}}  \bibinfo{volume}{38} (\bibinfo{year}{2017}), \bibinfo{pages}{595--603}.
\newblock


\bibitem[Porciuncula et~al\mbox{.}(2020)]%
        {porciuncula2020quantifying}
\bibfield{author}{\bibinfo{person}{Franchino Porciuncula}, \bibinfo{person}{Paula Wasserman}, \bibinfo{person}{Karen~S Marder}, {and} \bibinfo{person}{Ashwini~K Rao}.} \bibinfo{year}{2020}\natexlab{}.
\newblock \showarticletitle{Quantifying postural control in premanifest and manifest huntington disease using wearable sensors}.
\newblock \bibinfo{journal}{\emph{Neurorehabilitation and neural repair}} \bibinfo{volume}{34}, \bibinfo{number}{9} (\bibinfo{year}{2020}), \bibinfo{pages}{771--783}.
\newblock


\bibitem[Portnova et~al\mbox{.}(2024)]%
        {portnova2024temporal}
\bibfield{author}{\bibinfo{person}{Galina Portnova}, \bibinfo{person}{Guzal Khayrullina}, {and} \bibinfo{person}{Olga Martynova}.} \bibinfo{year}{2024}\natexlab{}.
\newblock \showarticletitle{Temporal dynamics of autonomic nervous system responses under cognitive-emotional workload in obsessive-compulsive disorder}.
\newblock \bibinfo{journal}{\emph{Psychophysiology}} \bibinfo{volume}{61}, \bibinfo{number}{6} (\bibinfo{year}{2024}), \bibinfo{pages}{e14549}.
\newblock


\bibitem[Posner(1980)]%
        {posner1980orienting}
\bibfield{author}{\bibinfo{person}{Michael~I Posner}.} \bibinfo{year}{1980}\natexlab{}.
\newblock \showarticletitle{Orienting of attention}.
\newblock \bibinfo{journal}{\emph{Quarterly journal of experimental psychology}} \bibinfo{volume}{32}, \bibinfo{number}{1} (\bibinfo{year}{1980}), \bibinfo{pages}{3--25}.
\newblock


\bibitem[Prabha and Bhargavi(2020)]%
        {prabha2020predictive}
\bibfield{author}{\bibinfo{person}{A~Jothi Prabha} {and} \bibinfo{person}{Renta Bhargavi}.} \bibinfo{year}{2020}\natexlab{}.
\newblock \showarticletitle{Predictive model for dyslexia from fixations and saccadic eye movement events}.
\newblock \bibinfo{journal}{\emph{Computer Methods and Programs in Biomedicine}}  \bibinfo{volume}{195} (\bibinfo{year}{2020}), \bibinfo{pages}{105538}.
\newblock


\bibitem[Prasad and Galetta(2010)]%
        {prasad2010eye}
\bibfield{author}{\bibinfo{person}{Sashank Prasad} {and} \bibinfo{person}{Steven~L Galetta}.} \bibinfo{year}{2010}\natexlab{}.
\newblock \showarticletitle{Eye movement abnormalities in multiple sclerosis}.
\newblock \bibinfo{journal}{\emph{Neurologic clinics}} \bibinfo{volume}{28}, \bibinfo{number}{3} (\bibinfo{year}{2010}), \bibinfo{pages}{641--655}.
\newblock


\bibitem[Prein et~al\mbox{.}(2024)]%
        {prein2024variation}
\bibfield{author}{\bibinfo{person}{Julia~Christin Prein}, \bibinfo{person}{Luke Maurits}, \bibinfo{person}{Annika Werwach}, \bibinfo{person}{Daniel~BM Haun}, {and} \bibinfo{person}{Manuel Bohn}.} \bibinfo{year}{2024}\natexlab{}.
\newblock \showarticletitle{Variation in gaze following across the life span: A process-level perspective}.
\newblock \bibinfo{journal}{\emph{Developmental Science}} (\bibinfo{year}{2024}), \bibinfo{pages}{e13546}.
\newblock


\bibitem[Pretegiani and Optican(2017)]%
        {pretegiani2017eye}
\bibfield{author}{\bibinfo{person}{Elena Pretegiani} {and} \bibinfo{person}{Lance~M Optican}.} \bibinfo{year}{2017}\natexlab{}.
\newblock \showarticletitle{Eye movements in Parkinson’s disease and inherited parkinsonian syndromes}.
\newblock \bibinfo{journal}{\emph{Frontiers in neurology}}  \bibinfo{volume}{8} (\bibinfo{year}{2017}), \bibinfo{pages}{592}.
\newblock


\bibitem[Proudfoot et~al\mbox{.}(2016)]%
        {proudfoot2016eye}
\bibfield{author}{\bibinfo{person}{Malcolm Proudfoot}, \bibinfo{person}{Ricarda~AL Menke}, \bibinfo{person}{Rakesh Sharma}, \bibinfo{person}{Claire~M Berna}, \bibinfo{person}{Stephen~L Hicks}, \bibinfo{person}{Christopher Kennard}, \bibinfo{person}{Kevin Talbot}, {and} \bibinfo{person}{Martin~R Turner}.} \bibinfo{year}{2016}\natexlab{}.
\newblock \showarticletitle{Eye-tracking in amyotrophic lateral sclerosis: a longitudinal study of saccadic and cognitive tasks}.
\newblock \bibinfo{journal}{\emph{Amyotrophic Lateral Sclerosis and Frontotemporal Degeneration}} \bibinfo{volume}{17}, \bibinfo{number}{1-2} (\bibinfo{year}{2016}), \bibinfo{pages}{101--111}.
\newblock


\bibitem[Pruneti et~al\mbox{.}(2023)]%
        {pruneti2023systematic}
\bibfield{author}{\bibinfo{person}{Carlo Pruneti}, \bibinfo{person}{Gabriella Coscioni}, {and} \bibinfo{person}{Sara Guidotti}.} \bibinfo{year}{2023}\natexlab{}.
\newblock \showarticletitle{A Systematic Review of Clinical Psychophysiology of Obsessive--Compulsive Disorders: Does the Obsession with Diet Also Alter the Autonomic Imbalance of Orthorexic Patients?}
\newblock \bibinfo{journal}{\emph{Nutrients}} \bibinfo{volume}{15}, \bibinfo{number}{3} (\bibinfo{year}{2023}), \bibinfo{pages}{755}.
\newblock


\bibitem[Prystauka et~al\mbox{.}(2024)]%
        {prystauka2024online}
\bibfield{author}{\bibinfo{person}{Yanina Prystauka}, \bibinfo{person}{Gerry~TM Altmann}, {and} \bibinfo{person}{Jason Rothman}.} \bibinfo{year}{2024}\natexlab{}.
\newblock \showarticletitle{Online eye tracking and real-time sentence processing: On opportunities and efficacy for capturing psycholinguistic effects of different magnitudes and diversity}.
\newblock \bibinfo{journal}{\emph{Behavior Research Methods}} \bibinfo{volume}{56}, \bibinfo{number}{4} (\bibinfo{year}{2024}), \bibinfo{pages}{3504--3522}.
\newblock


\bibitem[Przybyszewski et~al\mbox{.}(2023)]%
        {przybyszewski2023machine}
\bibfield{author}{\bibinfo{person}{Andrzej~W Przybyszewski}, \bibinfo{person}{Albert {\'S}ledzianowski}, \bibinfo{person}{Artur Chudzik}, \bibinfo{person}{Stanis{\l}aw Szlufik}, {and} \bibinfo{person}{Dariusz Koziorowski}.} \bibinfo{year}{2023}\natexlab{}.
\newblock \showarticletitle{Machine learning and eye movements give insights into neurodegenerative disease mechanisms}.
\newblock \bibinfo{journal}{\emph{Sensors}} \bibinfo{volume}{23}, \bibinfo{number}{4} (\bibinfo{year}{2023}), \bibinfo{pages}{2145}.
\newblock


\bibitem[Putnam et~al\mbox{.}(2023)]%
        {putnam2023effects}
\bibfield{author}{\bibinfo{person}{Orla~C Putnam}, \bibinfo{person}{Noah Sasson}, \bibinfo{person}{Julia Parish-Morris}, {and} \bibinfo{person}{Clare Harrop}.} \bibinfo{year}{2023}\natexlab{}.
\newblock \showarticletitle{Effects of social complexity and gender on social and non-social attention in male and female autistic children: A comparison of four eye-tracking paradigms}.
\newblock \bibinfo{journal}{\emph{Autism Research}} \bibinfo{volume}{16}, \bibinfo{number}{2} (\bibinfo{year}{2023}), \bibinfo{pages}{315--326}.
\newblock


\bibitem[Putra et~al\mbox{.}(2021)]%
        {putra2021identifying}
\bibfield{author}{\bibinfo{person}{Prasetia~Utama Putra}, \bibinfo{person}{Keisuke Shima}, \bibinfo{person}{Sergio~A Alvarez}, {and} \bibinfo{person}{Koji Shimatani}.} \bibinfo{year}{2021}\natexlab{}.
\newblock \showarticletitle{Identifying autism spectrum disorder symptoms using response and gaze behavior during the Go/NoGo game CatChicken}.
\newblock \bibinfo{journal}{\emph{Scientific reports}} \bibinfo{volume}{11}, \bibinfo{number}{1} (\bibinfo{year}{2021}), \bibinfo{pages}{22012}.
\newblock


\bibitem[Quatieri et~al\mbox{.}(2017)]%
        {quatieri2017multimodal}
\bibfield{author}{\bibinfo{person}{Thomas~F Quatieri}, \bibinfo{person}{James~R Williamson}, \bibinfo{person}{Christopher~J Smalt}, \bibinfo{person}{Joey Perricone}, \bibinfo{person}{Tejash Patel}, \bibinfo{person}{Laura Brattain}, \bibinfo{person}{Brian Helfer}, \bibinfo{person}{Daryush Mehta}, \bibinfo{person}{Jeffrey Palmer}, \bibinfo{person}{Kristin Heaton}, {et~al\mbox{.}}} \bibinfo{year}{2017}\natexlab{}.
\newblock \showarticletitle{Multimodal biomarkers to discriminate cognitive state}.
\newblock \bibinfo{journal}{\emph{The role of technology in clinical neuropsychology}}  \bibinfo{volume}{409} (\bibinfo{year}{2017}).
\newblock


\bibitem[Ramzaoui et~al\mbox{.}(2018)]%
        {ramzaoui2018alzheimer}
\bibfield{author}{\bibinfo{person}{Hanane Ramzaoui}, \bibinfo{person}{Sylvane Faure}, {and} \bibinfo{person}{Sara Spotorno}.} \bibinfo{year}{2018}\natexlab{}.
\newblock \showarticletitle{Alzheimer’s disease, visual search, and instrumental activities of daily living: A review and a new perspective on attention and eye movements}.
\newblock \bibinfo{journal}{\emph{Journal of Alzheimer's Disease}} \bibinfo{volume}{66}, \bibinfo{number}{3} (\bibinfo{year}{2018}), \bibinfo{pages}{901--925}.
\newblock


\bibitem[Ranchet et~al\mbox{.}(2017)]%
        {ranchet2017pupillary}
\bibfield{author}{\bibinfo{person}{Maud Ranchet}, \bibinfo{person}{Jason Orlosky}, \bibinfo{person}{John Morgan}, \bibinfo{person}{Siraj Qadir}, \bibinfo{person}{Abiodun~Emmanuel Akinwuntan}, {and} \bibinfo{person}{Hannes Devos}.} \bibinfo{year}{2017}\natexlab{}.
\newblock \showarticletitle{Pupillary response to cognitive workload during saccadic tasks in Parkinson’s disease}.
\newblock \bibinfo{journal}{\emph{Behavioural brain research}}  \bibinfo{volume}{327} (\bibinfo{year}{2017}), \bibinfo{pages}{162--166}.
\newblock


\bibitem[Raschke et~al\mbox{.}(2014)]%
        {raschke2014visual}
\bibfield{author}{\bibinfo{person}{Michael Raschke}, \bibinfo{person}{Tanja Blascheck}, {and} \bibinfo{person}{Michael Burch}.} \bibinfo{year}{2014}\natexlab{}.
\newblock \showarticletitle{Visual analysis of eye tracking data}.
\newblock \bibinfo{journal}{\emph{Handbook of human centric visualization}} (\bibinfo{year}{2014}), \bibinfo{pages}{391--409}.
\newblock


\bibitem[Readman et~al\mbox{.}(2021)]%
        {readman2021potential}
\bibfield{author}{\bibinfo{person}{Megan~Rose Readman}, \bibinfo{person}{Megan Polden}, \bibinfo{person}{Melissa~Chloe Gibbs}, \bibinfo{person}{Lettie Wareing}, {and} \bibinfo{person}{Trevor~J Crawford}.} \bibinfo{year}{2021}\natexlab{}.
\newblock \showarticletitle{The potential of naturalistic eye movement tasks in the diagnosis of Alzheimer’s disease: a review}.
\newblock \bibinfo{journal}{\emph{Brain sciences}} \bibinfo{volume}{11}, \bibinfo{number}{11} (\bibinfo{year}{2021}), \bibinfo{pages}{1503}.
\newblock


\bibitem[Reichle and Reingold(2013)]%
        {reichle2013neurophysiological}
\bibfield{author}{\bibinfo{person}{Erik~D Reichle} {and} \bibinfo{person}{Eyal~M Reingold}.} \bibinfo{year}{2013}\natexlab{}.
\newblock \showarticletitle{Neurophysiological constraints on the eye-mind link}.
\newblock \bibinfo{journal}{\emph{Frontiers in Human Neuroscience}}  \bibinfo{volume}{7} (\bibinfo{year}{2013}), \bibinfo{pages}{361}.
\newblock


\bibitem[Rekik et~al\mbox{.}(2023)]%
        {rekik2023eye}
\bibfield{author}{\bibinfo{person}{Arwa Rekik}, \bibinfo{person}{Saloua Mrabet}, \bibinfo{person}{Amina Nasri}, \bibinfo{person}{Youssef Abida}, \bibinfo{person}{Alya Gharbi}, \bibinfo{person}{Amina Gargouri}, \bibinfo{person}{Imen Kacem}, {and} \bibinfo{person}{Riadh Gouider}.} \bibinfo{year}{2023}\natexlab{}.
\newblock \showarticletitle{Eye movement study in essential tremor patients and its clinical correlates}.
\newblock \bibinfo{journal}{\emph{Journal of Neural Transmission}} \bibinfo{volume}{130}, \bibinfo{number}{4} (\bibinfo{year}{2023}), \bibinfo{pages}{537--548}.
\newblock


\bibitem[Revankar et~al\mbox{.}(2020)]%
        {revankar2020ocular}
\bibfield{author}{\bibinfo{person}{Gajanan~S Revankar}, \bibinfo{person}{Noriaki Hattori}, \bibinfo{person}{Yuta Kajiyama}, \bibinfo{person}{Tomohito Nakano}, \bibinfo{person}{Masahito Mihara}, \bibinfo{person}{Etsuro Mori}, {and} \bibinfo{person}{Hideki Mochizuki}.} \bibinfo{year}{2020}\natexlab{}.
\newblock \showarticletitle{Ocular fixations and presaccadic potentials to explain pareidolias in Parkinson’s disease}.
\newblock \bibinfo{journal}{\emph{Brain communications}} \bibinfo{volume}{2}, \bibinfo{number}{1} (\bibinfo{year}{2020}), \bibinfo{pages}{fcaa073}.
\newblock


\bibitem[Reyes-Lopez et~al\mbox{.}(2024)]%
        {reyes2024saccades}
\bibfield{author}{\bibinfo{person}{Mariana Reyes-Lopez}, \bibinfo{person}{Israel Vaca-Palomares}, \bibinfo{person}{David Jos{\'e} D{\'a}vila-Ortiz de Montellano}, \bibinfo{person}{Brian~J White}, \bibinfo{person}{Donald~C Brien}, \bibinfo{person}{Brian~C Coe}, \bibinfo{person}{Douglas~P Munoz}, {and} \bibinfo{person}{Juan Fernandez-Ruiz}.} \bibinfo{year}{2024}\natexlab{}.
\newblock \showarticletitle{Saccades, pupil response and blink abnormalities in Huntington’s disease patients during free viewing}.
\newblock \bibinfo{journal}{\emph{Clinical Neurophysiology}}  \bibinfo{volume}{165} (\bibinfo{year}{2024}), \bibinfo{pages}{117--124}.
\newblock


\bibitem[Rizzo et~al\mbox{.}(2019)]%
        {rizzo2019eye}
\bibfield{author}{\bibinfo{person}{John-Ross Rizzo}, \bibinfo{person}{Mahya Beheshti}, \bibinfo{person}{Azadeh Shafieesabet}, \bibinfo{person}{James Fung}, \bibinfo{person}{Maryam Hosseini}, \bibinfo{person}{Janet~C Rucker}, \bibinfo{person}{Lawrence~H Snyder}, {and} \bibinfo{person}{Todd~E Hudson}.} \bibinfo{year}{2019}\natexlab{}.
\newblock \showarticletitle{Eye-hand re-coordination: a pilot investigation of gaze and reach biofeedback in chronic stroke}.
\newblock \bibinfo{journal}{\emph{Progress in Brain Research}}  \bibinfo{volume}{249} (\bibinfo{year}{2019}), \bibinfo{pages}{361--374}.
\newblock


\bibitem[Rojas-L{\'\i}bano et~al\mbox{.}(2019)]%
        {rojas2019pupil}
\bibfield{author}{\bibinfo{person}{Daniel Rojas-L{\'\i}bano}, \bibinfo{person}{Gabriel Wainstein}, \bibinfo{person}{Ximena Carrasco}, \bibinfo{person}{Francisco Aboitiz}, \bibinfo{person}{Nicol{\'a}s Crossley}, {and} \bibinfo{person}{Tom{\'a}s Ossand{\'o}n}.} \bibinfo{year}{2019}\natexlab{}.
\newblock \showarticletitle{A pupil size, eye-tracking and neuropsychological dataset from ADHD children during a cognitive task}.
\newblock \bibinfo{journal}{\emph{Scientific data}} \bibinfo{volume}{6}, \bibinfo{number}{1} (\bibinfo{year}{2019}), \bibinfo{pages}{25}.
\newblock


\bibitem[Rosa et~al\mbox{.}(2014)]%
        {rosa2014effects}
\bibfield{author}{\bibinfo{person}{Pedro~J Rosa}, \bibinfo{person}{Francisco Esteves}, {and} \bibinfo{person}{Patr{\'\i}cia Arriaga}.} \bibinfo{year}{2014}\natexlab{}.
\newblock \showarticletitle{Effects of fear-relevant stimuli on attention: integrating gaze data with subliminal exposure}. In \bibinfo{booktitle}{\emph{2014 IEEE international symposium on medical measurements and applications (MeMeA)}}. IEEE, \bibinfo{pages}{1--6}.
\newblock


\bibitem[Rosqvist et~al\mbox{.}(2020)]%
        {rosqvist2020neurodiversity}
\bibfield{author}{\bibinfo{person}{Hanna~Bertilsdotter Rosqvist}, \bibinfo{person}{Nick Chown}, {and} \bibinfo{person}{Anna Stenning}.} \bibinfo{year}{2020}\natexlab{}.
\newblock \bibinfo{booktitle}{\emph{Neurodiversity studies}}.
\newblock \bibinfo{publisher}{Routledge}.
\newblock


\bibitem[Ross et~al\mbox{.}(2000)]%
        {ross2000smooth}
\bibfield{author}{\bibinfo{person}{Randal~G Ross}, \bibinfo{person}{Ann Olincy}, \bibinfo{person}{Josette~G Harris}, \bibinfo{person}{Bernadette Sullivan}, {and} \bibinfo{person}{Allen Radant}.} \bibinfo{year}{2000}\natexlab{}.
\newblock \showarticletitle{Smooth pursuit eye movements in schizophrenia and attentional dysfunction: adults with schizophrenia, ADHD, and a normal comparison group}.
\newblock \bibinfo{journal}{\emph{Biological psychiatry}} \bibinfo{volume}{48}, \bibinfo{number}{3} (\bibinfo{year}{2000}), \bibinfo{pages}{197--203}.
\newblock


\bibitem[Rossano(2012)]%
        {rossano2012gaze}
\bibfield{author}{\bibinfo{person}{Federico Rossano}.} \bibinfo{year}{2012}\natexlab{}.
\newblock \showarticletitle{Gaze in conversation}.
\newblock \bibinfo{journal}{\emph{The handbook of conversation analysis}} (\bibinfo{year}{2012}), \bibinfo{pages}{308--329}.
\newblock


\bibitem[Rowe et~al\mbox{.}(2013)]%
        {rowe2013profile}
\bibfield{author}{\bibinfo{person}{Fiona~J Rowe}, \bibinfo{person}{David Wright}, \bibinfo{person}{Darren Brand}, \bibinfo{person}{Carole Jackson}, \bibinfo{person}{Shirley Harrison}, \bibinfo{person}{Tallat Maan}, \bibinfo{person}{Claire Scott}, \bibinfo{person}{Linda Vogwell}, \bibinfo{person}{Sarah Peel}, \bibinfo{person}{Nicola Akerman}, {et~al\mbox{.}}} \bibinfo{year}{2013}\natexlab{}.
\newblock \showarticletitle{Profile of gaze dysfunction following cerebrovascular accident}.
\newblock \bibinfo{journal}{\emph{International Scholarly Research Notices}} \bibinfo{volume}{2013}, \bibinfo{number}{1} (\bibinfo{year}{2013}), \bibinfo{pages}{264604}.
\newblock


\bibitem[Rubin et~al\mbox{.}(1993)]%
        {rubin1993quantitative}
\bibfield{author}{\bibinfo{person}{Allen~J Rubin}, \bibinfo{person}{W~Michael King}, \bibinfo{person}{Kirk~Alan Reinbold}, {and} \bibinfo{person}{Ira Shoulson}.} \bibinfo{year}{1993}\natexlab{}.
\newblock \showarticletitle{Quantitative longitudinal assessment of saccades in Huntington's disease}.
\newblock \bibinfo{journal}{\emph{Journal of Neuro-Ophthalmology}} \bibinfo{volume}{13}, \bibinfo{number}{1} (\bibinfo{year}{1993}), \bibinfo{pages}{59--66}.
\newblock


\bibitem[Rucker et~al\mbox{.}(2021)]%
        {rucker2021dysfunctional}
\bibfield{author}{\bibinfo{person}{Janet~C Rucker}, \bibinfo{person}{John-Ross Rizzo}, \bibinfo{person}{Todd~E Hudson}, \bibinfo{person}{Anja~KE Horn}, \bibinfo{person}{Jean~A Buettner-Ennever}, \bibinfo{person}{R~John Leigh}, {and} \bibinfo{person}{Lance~M Optican}.} \bibinfo{year}{2021}\natexlab{}.
\newblock \showarticletitle{Dysfunctional mode switching between fixation and saccades: collaborative insights into two unusual clinical disorders}.
\newblock \bibinfo{journal}{\emph{Journal of Computational Neuroscience}}  \bibinfo{volume}{49} (\bibinfo{year}{2021}), \bibinfo{pages}{283--293}.
\newblock


\bibitem[Rudling et~al\mbox{.}(2024)]%
        {rudling2024infant}
\bibfield{author}{\bibinfo{person}{Maja Rudling}, \bibinfo{person}{P{\"a}r Nystr{\"o}m}, \bibinfo{person}{Giorgia Bussu}, \bibinfo{person}{Sven B{\"o}lte}, {and} \bibinfo{person}{Terje Falck-Ytter}.} \bibinfo{year}{2024}\natexlab{}.
\newblock \showarticletitle{Infant responses to direct gaze and associations to autism: A live eye-tracking study}.
\newblock \bibinfo{journal}{\emph{Autism}} \bibinfo{volume}{28}, \bibinfo{number}{7} (\bibinfo{year}{2024}), \bibinfo{pages}{1677--1689}.
\newblock


\bibitem[Rupp et~al\mbox{.}(2011)]%
        {rupp2011abnormal}
\bibfield{author}{\bibinfo{person}{Jason Rupp}, \bibinfo{person}{Mario Dzemidzic}, \bibinfo{person}{Tanya Blekher}, \bibinfo{person}{Veronique Bragulat}, \bibinfo{person}{John West}, \bibinfo{person}{Jacqueline Jackson}, \bibinfo{person}{Siu Hui}, \bibinfo{person}{Joanne Wojcieszek}, \bibinfo{person}{Andrew~J Saykin}, \bibinfo{person}{David Kareken}, {et~al\mbox{.}}} \bibinfo{year}{2011}\natexlab{}.
\newblock \showarticletitle{Abnormal error-related antisaccade activation in premanifest and early manifest Huntington disease.}
\newblock \bibinfo{journal}{\emph{Neuropsychology}} \bibinfo{volume}{25}, \bibinfo{number}{3} (\bibinfo{year}{2011}), \bibinfo{pages}{306}.
\newblock


\bibitem[Saavedra et~al\mbox{.}(2009)]%
        {saavedra2009eye}
\bibfield{author}{\bibinfo{person}{Sandra Saavedra}, \bibinfo{person}{Aditi Joshi}, \bibinfo{person}{Marjorie Woollacott}, {and} \bibinfo{person}{Paul van Donkelaar}.} \bibinfo{year}{2009}\natexlab{}.
\newblock \showarticletitle{Eye hand coordination in children with cerebral palsy}.
\newblock \bibinfo{journal}{\emph{Experimental brain research}}  \bibinfo{volume}{192} (\bibinfo{year}{2009}), \bibinfo{pages}{155--165}.
\newblock


\bibitem[Salari et~al\mbox{.}(2023)]%
        {salari2023global}
\bibfield{author}{\bibinfo{person}{Nader Salari}, \bibinfo{person}{Hooman Ghasemi}, \bibinfo{person}{Nasrin Abdoli}, \bibinfo{person}{Adibeh Rahmani}, \bibinfo{person}{Mohammad~Hossain Shiri}, \bibinfo{person}{Amir~Hossein Hashemian}, \bibinfo{person}{Hakimeh Akbari}, {and} \bibinfo{person}{Masoud Mohammadi}.} \bibinfo{year}{2023}\natexlab{}.
\newblock \showarticletitle{The global prevalence of ADHD in children and adolescents: a systematic review and meta-analysis}.
\newblock \bibinfo{journal}{\emph{Italian journal of pediatrics}} \bibinfo{volume}{49}, \bibinfo{number}{1} (\bibinfo{year}{2023}), \bibinfo{pages}{48}.
\newblock


\bibitem[Saluja et~al\mbox{.}(2019)]%
        {saluja2019analyzing}
\bibfield{author}{\bibinfo{person}{Kamalpreet~Singh Saluja}, \bibinfo{person}{JeevithaShree Dv}, \bibinfo{person}{Somnath Arjun}, \bibinfo{person}{Pradipta Biswas}, {and} \bibinfo{person}{Teena Paul}.} \bibinfo{year}{2019}\natexlab{}.
\newblock \showarticletitle{Analyzing eye gaze of users with learning disability}. In \bibinfo{booktitle}{\emph{Proceedings of the 3rd International Conference on Graphics and Signal Processing}}. \bibinfo{pages}{95--99}.
\newblock


\bibitem[Santiago et~al\mbox{.}(2021)]%
        {santiago2021visual}
\bibfield{author}{\bibinfo{person}{Amada Uiritzikiri~Jim{\'e}nez Santiago}, \bibinfo{person}{Daniel~Hern{\'a}ndez Gonz{\'a}lez}, \bibinfo{person}{Eduardo Emmanuel~Rodr{\'\i}guez L{\'o}pez}, {and} \bibinfo{person}{Francisco Javier~{\'A}lvarez Rodr{\'\i}guez}.} \bibinfo{year}{2021}\natexlab{}.
\newblock \showarticletitle{Visual Preferences in children diagnosed with ASD by generating heatmaps with Eye Tracking}. In \bibinfo{booktitle}{\emph{2021 4th International Conference on Inclusive Technology and Education (CONTIE)}}. IEEE, \bibinfo{pages}{1--6}.
\newblock


\bibitem[Sargent et~al\mbox{.}(2013)]%
        {sargent2013use}
\bibfield{author}{\bibinfo{person}{Jenefer Sargent}, \bibinfo{person}{Michael Clarke}, \bibinfo{person}{Katie Price}, \bibinfo{person}{Tom Griffiths}, {and} \bibinfo{person}{John Swettenham}.} \bibinfo{year}{2013}\natexlab{}.
\newblock \showarticletitle{Use of eye-pointing by children with cerebral palsy: what are we looking at?}
\newblock \bibinfo{journal}{\emph{International journal of language \& communication disorders}} \bibinfo{volume}{48}, \bibinfo{number}{5} (\bibinfo{year}{2013}), \bibinfo{pages}{477--485}.
\newblock


\bibitem[Saxon et~al\mbox{.}(2020)]%
        {saxon2020cognition}
\bibfield{author}{\bibinfo{person}{Jennifer~A Saxon}, \bibinfo{person}{Jennifer~C Thompson}, \bibinfo{person}{Jennifer~M Harris}, \bibinfo{person}{Anna~M Richardson}, \bibinfo{person}{Tobias Langheinrich}, \bibinfo{person}{Sara Rollinson}, \bibinfo{person}{Stuart Pickering-Brown}, \bibinfo{person}{Amina Chaouch}, \bibinfo{person}{John Ealing}, \bibinfo{person}{Hisham Hamdalla}, {et~al\mbox{.}}} \bibinfo{year}{2020}\natexlab{}.
\newblock \showarticletitle{Cognition and behaviour in frontotemporal dementia with and without amyotrophic lateral sclerosis}.
\newblock \bibinfo{journal}{\emph{Journal of Neurology, Neurosurgery \& Psychiatry}} \bibinfo{volume}{91}, \bibinfo{number}{12} (\bibinfo{year}{2020}), \bibinfo{pages}{1304--1311}.
\newblock


\bibitem[Schauder et~al\mbox{.}(2019)]%
        {schauder2019initial}
\bibfield{author}{\bibinfo{person}{Kimberly~B Schauder}, \bibinfo{person}{Woon~Ju Park}, \bibinfo{person}{Yuliy Tsank}, \bibinfo{person}{Miguel~P Eckstein}, \bibinfo{person}{Duje Tadin}, {and} \bibinfo{person}{Loisa Bennetto}.} \bibinfo{year}{2019}\natexlab{}.
\newblock \showarticletitle{Initial eye gaze to faces and its functional consequence on face identification abilities in autism spectrum disorder}.
\newblock \bibinfo{journal}{\emph{Journal of neurodevelopmental disorders}}  \bibinfo{volume}{11} (\bibinfo{year}{2019}), \bibinfo{pages}{1--20}.
\newblock


\bibitem[Selaskowski et~al\mbox{.}(2023)]%
        {selaskowski2023gaze}
\bibfield{author}{\bibinfo{person}{Benjamin Selaskowski}, \bibinfo{person}{Laura~Marie Asch{\'e}}, \bibinfo{person}{Annika Wiebe}, \bibinfo{person}{Kyra Kannen}, \bibinfo{person}{Behrem Aslan}, \bibinfo{person}{Thiago~Morano Gerding}, \bibinfo{person}{Dario Sanchez}, \bibinfo{person}{Ulrich Ettinger}, \bibinfo{person}{Markus K{\"o}lle}, \bibinfo{person}{Silke Lux}, {et~al\mbox{.}}} \bibinfo{year}{2023}\natexlab{}.
\newblock \showarticletitle{Gaze-based attention refocusing training in virtual reality for adult attention-deficit/hyperactivity disorder}.
\newblock \bibinfo{journal}{\emph{BMC psychiatry}} \bibinfo{volume}{23}, \bibinfo{number}{1} (\bibinfo{year}{2023}), \bibinfo{pages}{74}.
\newblock


\bibitem[Severens et~al\mbox{.}(2014)]%
        {severens2014comparing}
\bibfield{author}{\bibinfo{person}{M Severens}, \bibinfo{person}{M Van~der Waal}, \bibinfo{person}{J Farquhar}, {and} \bibinfo{person}{P Desain}.} \bibinfo{year}{2014}\natexlab{}.
\newblock \showarticletitle{Comparing tactile and visual gaze-independent brain--computer interfaces in patients with amyotrophic lateral sclerosis and healthy users}.
\newblock \bibinfo{journal}{\emph{Clinical neurophysiology}} \bibinfo{volume}{125}, \bibinfo{number}{11} (\bibinfo{year}{2014}), \bibinfo{pages}{2297--2304}.
\newblock


\bibitem[Shaikh et~al\mbox{.}(2017)]%
        {shaikh2017fixational}
\bibfield{author}{\bibinfo{person}{Aasef~G Shaikh}, \bibinfo{person}{Shlomit~Ritz Finkelstein}, \bibinfo{person}{Ronald Schuchard}, \bibinfo{person}{Glen Ross}, {and} \bibinfo{person}{Jorge~L Juncos}.} \bibinfo{year}{2017}\natexlab{}.
\newblock \showarticletitle{Fixational eye movements in Tourette syndrome}.
\newblock \bibinfo{journal}{\emph{Neurological Sciences}}  \bibinfo{volume}{38} (\bibinfo{year}{2017}), \bibinfo{pages}{1977--1984}.
\newblock


\bibitem[Shaikh and Ghasia(2019)]%
        {shaikh2019saccades}
\bibfield{author}{\bibinfo{person}{Aasef~G Shaikh} {and} \bibinfo{person}{Fatema~F Ghasia}.} \bibinfo{year}{2019}\natexlab{}.
\newblock \showarticletitle{Saccades in Parkinson's disease: Hypometric, slow, and maladaptive}.
\newblock \bibinfo{journal}{\emph{Progress in Brain Research}}  \bibinfo{volume}{249} (\bibinfo{year}{2019}), \bibinfo{pages}{81--94}.
\newblock


\bibitem[Shaikh et~al\mbox{.}(2016)]%
        {shaikh2016cervical}
\bibfield{author}{\bibinfo{person}{Aasef~G Shaikh}, \bibinfo{person}{David~S Zee}, \bibinfo{person}{J~Douglas Crawford}, {and} \bibinfo{person}{Hyder~A Jinnah}.} \bibinfo{year}{2016}\natexlab{}.
\newblock \showarticletitle{Cervical dystonia: a neural integrator disorder}.
\newblock \bibinfo{journal}{\emph{Brain}} \bibinfo{volume}{139}, \bibinfo{number}{10} (\bibinfo{year}{2016}), \bibinfo{pages}{2590--2599}.
\newblock


\bibitem[Sharma et~al\mbox{.}(2011)]%
        {sharma2011oculomotor}
\bibfield{author}{\bibinfo{person}{Rakesh Sharma}, \bibinfo{person}{Stephen Hicks}, \bibinfo{person}{Claire~M Berna}, \bibinfo{person}{Christopher Kennard}, \bibinfo{person}{Kevin Talbot}, {and} \bibinfo{person}{Martin~R Turner}.} \bibinfo{year}{2011}\natexlab{}.
\newblock \showarticletitle{Oculomotor dysfunction in amyotrophic lateral sclerosis: a comprehensive review}.
\newblock \bibinfo{journal}{\emph{Archives of neurology}} \bibinfo{volume}{68}, \bibinfo{number}{7} (\bibinfo{year}{2011}), \bibinfo{pages}{857--861}.
\newblock


\bibitem[Sheehy et~al\mbox{.}(2018a)]%
        {sheehy2018methods}
\bibfield{author}{\bibinfo{person}{Christy~K Sheehy}, \bibinfo{person}{Alexandra Beaudry-Richard}, \bibinfo{person}{Ethan Bensinger}, \bibinfo{person}{Jacqueline Theis}, {and} \bibinfo{person}{Ari~J Green}.} \bibinfo{year}{2018}\natexlab{a}.
\newblock \showarticletitle{Methods to assess ocular motor dysfunction in multiple sclerosis}.
\newblock \bibinfo{journal}{\emph{Journal of Neuro-ophthalmology}} \bibinfo{volume}{38}, \bibinfo{number}{4} (\bibinfo{year}{2018}), \bibinfo{pages}{488--493}.
\newblock


\bibitem[Sheehy et~al\mbox{.}(2020)]%
        {sheehy2020fixational}
\bibfield{author}{\bibinfo{person}{Christy~K Sheehy}, \bibinfo{person}{Ethan~S Bensinger}, \bibinfo{person}{Andrew Romeo}, \bibinfo{person}{Lakshmisahithi Rani}, \bibinfo{person}{Natalie Stepien-Bernabe}, \bibinfo{person}{Bingyan Shi}, \bibinfo{person}{Zachary Helft}, \bibinfo{person}{Nicole Putnam}, \bibinfo{person}{Christian Cordano}, \bibinfo{person}{Jeffrey~M Gelfand}, {et~al\mbox{.}}} \bibinfo{year}{2020}\natexlab{}.
\newblock \showarticletitle{Fixational microsaccades: A quantitative and objective measure of disability in multiple sclerosis}.
\newblock \bibinfo{journal}{\emph{Multiple Sclerosis Journal}} \bibinfo{volume}{26}, \bibinfo{number}{3} (\bibinfo{year}{2020}), \bibinfo{pages}{343--353}.
\newblock


\bibitem[Sheehy et~al\mbox{.}(2018b)]%
        {sheehy2018validation}
\bibfield{author}{\bibinfo{person}{Christy~K Sheehy}, \bibinfo{person}{Bingyan Shi}, \bibinfo{person}{Ethan Bensinger}, \bibinfo{person}{Andrew Romeo}, \bibinfo{person}{Lakshmisahithi Rani}, \bibinfo{person}{Jeffrey Gelfand}, \bibinfo{person}{Scott~B Stevenson}, \bibinfo{person}{Ari Green}, {et~al\mbox{.}}} \bibinfo{year}{2018}\natexlab{b}.
\newblock \showarticletitle{Validation of microsaccades as a biomarker for disability in multiple sclerosis}.
\newblock \bibinfo{journal}{\emph{Investigative Ophthalmology \& Visual Science}} \bibinfo{volume}{59}, \bibinfo{number}{9} (\bibinfo{year}{2018}), \bibinfo{pages}{625--625}.
\newblock


\bibitem[Shen et~al\mbox{.}(2023)]%
        {shen2023evaluating}
\bibfield{author}{\bibinfo{person}{Haifeng Shen}, \bibinfo{person}{Othman Asiry}, \bibinfo{person}{M~Ali Babar}, {and} \bibinfo{person}{Tomasz Bednarz}.} \bibinfo{year}{2023}\natexlab{}.
\newblock \showarticletitle{Evaluating the efficacy of using a novel gaze-based attentive user interface to extend ADHD children’s attention span}.
\newblock \bibinfo{journal}{\emph{International Journal of Human-Computer Studies}}  \bibinfo{volume}{169} (\bibinfo{year}{2023}), \bibinfo{pages}{102927}.
\newblock


\bibitem[Shibasaki et~al\mbox{.}(1979)]%
        {shibasaki1979oculomotor}
\bibfield{author}{\bibinfo{person}{Hiroshi Shibasaki}, \bibinfo{person}{Sadatoshi Tsuji}, {and} \bibinfo{person}{Yoshigoro Kuroiwa}.} \bibinfo{year}{1979}\natexlab{}.
\newblock \showarticletitle{Oculomotor abnormalities in Parkinson's disease}.
\newblock \bibinfo{journal}{\emph{Archives of Neurology}} \bibinfo{volume}{36}, \bibinfo{number}{6} (\bibinfo{year}{1979}), \bibinfo{pages}{360--364}.
\newblock


\bibitem[Sierra et~al\mbox{.}(2024)]%
        {sierra2024deciphering}
\bibfield{author}{\bibinfo{person}{Luis~A Sierra}, \bibinfo{person}{Amy Wynn}, \bibinfo{person}{Ella Lanzaro}, \bibinfo{person}{Katya Dzekon}, \bibinfo{person}{Aine Russell}, \bibinfo{person}{Mark Halko}, \bibinfo{person}{Daniel~O Claassen}, \bibinfo{person}{Samuel Frank}, \bibinfo{person}{Ciaran~M Considine}, {and} \bibinfo{person}{Simon Laganiere}.} \bibinfo{year}{2024}\natexlab{}.
\newblock \showarticletitle{Deciphering Cognitive Impairments in Huntington’s Disease: A Comparative Study of Stroop Test Variations}.
\newblock \bibinfo{journal}{\emph{Journal of Huntington's Disease}} \bibinfo{number}{Preprint} (\bibinfo{year}{2024}), \bibinfo{pages}{1--9}.
\newblock


\bibitem[Silva et~al\mbox{.}(2020)]%
        {silva2020prognostic}
\bibfield{author}{\bibinfo{person}{Ana~Lima Silva}, \bibinfo{person}{Ana~Sofia Pessoa}, \bibinfo{person}{Renato Nogueira}, \bibinfo{person}{Jose~Manuel Araujo}, \bibinfo{person}{Jose~Nuno Alves}, \bibinfo{person}{Joao Pinho}, {and} \bibinfo{person}{Carla Ferreira}.} \bibinfo{year}{2020}\natexlab{}.
\newblock \showarticletitle{Prognostic information of gaze deviation in acute ischemic stroke patients}.
\newblock \bibinfo{journal}{\emph{Neurological Sciences}}  \bibinfo{volume}{41} (\bibinfo{year}{2020}), \bibinfo{pages}{435--440}.
\newblock


\bibitem[Simieli et~al\mbox{.}(2017)]%
        {simieli2017gaze}
\bibfield{author}{\bibinfo{person}{Lucas Simieli}, \bibinfo{person}{Rodrigo Vit{\'o}rio}, \bibinfo{person}{S{\'e}rgio~Tosi Rodrigues}, \bibinfo{person}{Paula F{\'a}varo~Polastri Zago}, \bibinfo{person}{Vin{\'\i}cius Alota~Ignacio Pereira}, \bibinfo{person}{Andr{\'e}~Macari Baptista}, \bibinfo{person}{Pedro Henrique~Alves de Paula}, \bibinfo{person}{Tiago Penedo}, \bibinfo{person}{Quincy~J Almeida}, {and} \bibinfo{person}{Fabio~Augusto Barbieri}.} \bibinfo{year}{2017}\natexlab{}.
\newblock \showarticletitle{Gaze and motor behavior of people with PD during obstacle circumvention}.
\newblock \bibinfo{journal}{\emph{Gait \& Posture}}  \bibinfo{volume}{58} (\bibinfo{year}{2017}), \bibinfo{pages}{504--509}.
\newblock


\bibitem[Simpson and Molloy(1971)]%
        {simpson1971effects}
\bibfield{author}{\bibinfo{person}{HM Simpson} {and} \bibinfo{person}{FM Molloy}.} \bibinfo{year}{1971}\natexlab{}.
\newblock \showarticletitle{Effects of audience anxiety on pupil size}.
\newblock \bibinfo{journal}{\emph{Psychophysiology}} \bibinfo{volume}{8}, \bibinfo{number}{4} (\bibinfo{year}{1971}), \bibinfo{pages}{491--496}.
\newblock


\bibitem[Smyrnis et~al\mbox{.}(2007)]%
        {smyrnis2007smooth}
\bibfield{author}{\bibinfo{person}{Nikolaos Smyrnis}, \bibinfo{person}{Ioannis Evdokimidis}, \bibinfo{person}{Asimakis Mantas}, \bibinfo{person}{Emmanouil Kattoulas}, \bibinfo{person}{Nicholas~C Stefanis}, \bibinfo{person}{Theodoros~S Constantinidis}, \bibinfo{person}{Dimitrios Avramopoulos}, {and} \bibinfo{person}{Costas~N Stefanis}.} \bibinfo{year}{2007}\natexlab{}.
\newblock \showarticletitle{Smooth pursuit eye movements in 1,087 men: effects of schizotypy, anxiety, and depression}.
\newblock \bibinfo{journal}{\emph{Experimental Brain Research}}  \bibinfo{volume}{179} (\bibinfo{year}{2007}), \bibinfo{pages}{397--408}.
\newblock


\bibitem[Snowden et~al\mbox{.}(2003)]%
        {snowden2003social}
\bibfield{author}{\bibinfo{person}{JS Snowden}, \bibinfo{person}{ZC Gibbons}, \bibinfo{person}{A Blackshaw}, \bibinfo{person}{E Doubleday}, \bibinfo{person}{Jennifer Thompson}, \bibinfo{person}{David Craufurd}, \bibinfo{person}{Jonathan Foster}, \bibinfo{person}{Francesca Happ{\'e}}, {and} \bibinfo{person}{David Neary}.} \bibinfo{year}{2003}\natexlab{}.
\newblock \showarticletitle{Social cognition in frontotemporal dementia and Huntington’s disease}.
\newblock \bibinfo{journal}{\emph{Neuropsychologia}} \bibinfo{volume}{41}, \bibinfo{number}{6} (\bibinfo{year}{2003}), \bibinfo{pages}{688--701}.
\newblock


\bibitem[Sorkhabi et~al\mbox{.}(2024)]%
        {sorkhabi2024frequency}
\bibfield{author}{\bibinfo{person}{Rana Sorkhabi}, \bibinfo{person}{Siamak Khavandi}, \bibinfo{person}{Hormoz Ayromlou}, \bibinfo{person}{Mohammad~Hosein Ahoor}, \bibinfo{person}{Mehdi Mohammadkhani}, {and} \bibinfo{person}{Elsa Tabibzadeh}.} \bibinfo{year}{2024}\natexlab{}.
\newblock \showarticletitle{Frequency of dry eye syndrome in patients with multiple sclerosis: A cross-sectional case-control study}.
\newblock \bibinfo{journal}{\emph{Journal of Research in Clinical Medicine}} \bibinfo{volume}{12}, \bibinfo{number}{1} (\bibinfo{year}{2024}), \bibinfo{pages}{7--7}.
\newblock


\bibitem[{\v{S}}pakov and Miniotas(2007)]%
        {vspakov2007visualization}
\bibfield{author}{\bibinfo{person}{Oleg {\v{S}}pakov} {and} \bibinfo{person}{Darius Miniotas}.} \bibinfo{year}{2007}\natexlab{}.
\newblock \showarticletitle{Visualization of eye gaze data using heat maps}.
\newblock  (\bibinfo{year}{2007}).
\newblock


\bibitem[Speer et~al\mbox{.}(2007)]%
        {speer2007face}
\bibfield{author}{\bibinfo{person}{Leslie~L Speer}, \bibinfo{person}{Anne~E Cook}, \bibinfo{person}{William~M McMahon}, {and} \bibinfo{person}{Elaine Clark}.} \bibinfo{year}{2007}\natexlab{}.
\newblock \showarticletitle{Face processing in children with autism: Effects of stimulus contents and type}.
\newblock \bibinfo{journal}{\emph{Autism}} \bibinfo{volume}{11}, \bibinfo{number}{3} (\bibinfo{year}{2007}), \bibinfo{pages}{265--277}.
\newblock


\bibitem[Spengler et~al\mbox{.}(2006)]%
        {spengler2006evidence}
\bibfield{author}{\bibinfo{person}{Dietmar Spengler}, \bibinfo{person}{Peter Trillenberg}, \bibinfo{person}{Andreas Sprenger}, \bibinfo{person}{Matthias Nagel}, \bibinfo{person}{Andreas Kordon}, \bibinfo{person}{Klaus Junghanns}, \bibinfo{person}{Wolfgang Heide}, \bibinfo{person}{Volker Arolt}, \bibinfo{person}{Fritz Hohagen}, {and} \bibinfo{person}{Rebekka Lencer}.} \bibinfo{year}{2006}\natexlab{}.
\newblock \showarticletitle{Evidence from increased anticipation of predictive saccades for a dysfunction of fronto-striatal circuits in obsessive--compulsive disorder}.
\newblock \bibinfo{journal}{\emph{Psychiatry research}} \bibinfo{volume}{143}, \bibinfo{number}{1} (\bibinfo{year}{2006}), \bibinfo{pages}{77--88}.
\newblock


\bibitem[Springer(2010)]%
        {springer2010essential}
\bibfield{author}{\bibinfo{person}{Utaka~S Springer}.} \bibinfo{year}{2010}\natexlab{}.
\newblock \emph{\bibinfo{title}{Essential tremor: Paradoxical visuospatial cognition}}.
\newblock \bibinfo{thesistype}{Ph.\,D. Dissertation}. \bibinfo{school}{University of Florida}.
\newblock


\bibitem[Srivastava and Garg(2020)]%
        {srivastava2020blink}
\bibfield{author}{\bibinfo{person}{Achal~K Srivastava} {and} \bibinfo{person}{Divyani Garg}.} \bibinfo{year}{2020}\natexlab{}.
\newblock \showarticletitle{Blink and Dont Miss it: The Role of Blink Reflex in Neurodegenerative Disorders}.
\newblock \bibinfo{journal}{\emph{Neurology India}} \bibinfo{volume}{68}, \bibinfo{number}{1} (\bibinfo{year}{2020}), \bibinfo{pages}{76--77}.
\newblock


\bibitem[Stadskleiv et~al\mbox{.}(2017)]%
        {stadskleiv2017executive}
\bibfield{author}{\bibinfo{person}{Kristine Stadskleiv}, \bibinfo{person}{Reidun Jahnsen}, \bibinfo{person}{Guro~L Andersen}, {and} \bibinfo{person}{Stephen von Tetzchner}.} \bibinfo{year}{2017}\natexlab{}.
\newblock \showarticletitle{Executive functioning in children aged 6--18 years with cerebral palsy}.
\newblock \bibinfo{journal}{\emph{Journal of Developmental and Physical Disabilities}}  \bibinfo{volume}{29} (\bibinfo{year}{2017}), \bibinfo{pages}{663--681}.
\newblock


\bibitem[STARR(1967)]%
        {starr1967disorder}
\bibfield{author}{\bibinfo{person}{ARNOLD STARR}.} \bibinfo{year}{1967}\natexlab{}.
\newblock \showarticletitle{A disorder of rapid eye movements in Huntington's chorea.}
\newblock \bibinfo{journal}{\emph{Brain}} \bibinfo{volume}{90}, \bibinfo{number}{3} (\bibinfo{year}{1967}), \bibinfo{pages}{545--564}.
\newblock


\bibitem[Stefanelli et~al\mbox{.}(2024)]%
        {stefanelli2024pupillary}
\bibfield{author}{\bibinfo{person}{Giulia Stefanelli}, \bibinfo{person}{Miriam~Paola Pili}, \bibinfo{person}{Giulia Crifaci}, \bibinfo{person}{Elena Capelli}, \bibinfo{person}{Carolina Beretta}, \bibinfo{person}{Elena~Maria Riboldi}, \bibinfo{person}{Lucia Billeci}, \bibinfo{person}{Chiara Cantiani}, \bibinfo{person}{Massimo Molteni}, {and} \bibinfo{person}{Valentina Riva}.} \bibinfo{year}{2024}\natexlab{}.
\newblock \showarticletitle{Pupillary responses for social versus non-social stimuli in autism: a systematic review and meta-analysis}.
\newblock \bibinfo{journal}{\emph{Neuroscience \& Biobehavioral Reviews}} (\bibinfo{year}{2024}), \bibinfo{pages}{105872}.
\newblock


\bibitem[{\c{S}}tef{\c{S}}nescu et~al\mbox{.}(2024)]%
        {cstefcsnescu2024eye}
\bibfield{author}{\bibinfo{person}{Emanuel {\c{S}}tef{\c{S}}nescu}, \bibinfo{person}{Vlad-Florin Chelaru}, \bibinfo{person}{Diana Chira}, \bibinfo{person}{Dafin Mure{\c{s}}anu}, {et~al\mbox{.}}} \bibinfo{year}{2024}\natexlab{}.
\newblock \showarticletitle{Eye tracking assessment of Parkinson's disease: a clinical retrospective analysis}.
\newblock \bibinfo{journal}{\emph{Journal of Medicine and Life}} \bibinfo{volume}{17}, \bibinfo{number}{3} (\bibinfo{year}{2024}), \bibinfo{pages}{360}.
\newblock


\bibitem[Steil et~al\mbox{.}(2019)]%
        {steil2019privacy}
\bibfield{author}{\bibinfo{person}{Julian Steil}, \bibinfo{person}{Inken Hagestedt}, \bibinfo{person}{Michael~Xuelin Huang}, {and} \bibinfo{person}{Andreas Bulling}.} \bibinfo{year}{2019}\natexlab{}.
\newblock \showarticletitle{Privacy-aware eye tracking using differential privacy}. In \bibinfo{booktitle}{\emph{Proceedings of the 11th ACM Symposium on Eye Tracking Research \& Applications}}. \bibinfo{pages}{1--9}.
\newblock


\bibitem[Stein et~al\mbox{.}(2019)]%
        {stein2019obsessive}
\bibfield{author}{\bibinfo{person}{Dan~J Stein}, \bibinfo{person}{Daniel~LC Costa}, \bibinfo{person}{Christine Lochner}, \bibinfo{person}{Euripedes~C Miguel}, \bibinfo{person}{YC~Janardhan Reddy}, \bibinfo{person}{Roseli~G Shavitt}, \bibinfo{person}{Odile~A van~den Heuvel}, {and} \bibinfo{person}{H~Blair Simpson}.} \bibinfo{year}{2019}\natexlab{}.
\newblock \showarticletitle{Obsessive--compulsive disorder}.
\newblock \bibinfo{journal}{\emph{Nature reviews Disease primers}} \bibinfo{volume}{5}, \bibinfo{number}{1} (\bibinfo{year}{2019}), \bibinfo{pages}{52}.
\newblock


\bibitem[Stewart(2018)]%
        {stewart2018pathology}
\bibfield{author}{\bibinfo{person}{David Stewart}.} \bibinfo{year}{2018}\natexlab{}.
\newblock \showarticletitle{Pathology, aetiology and pathogenesis}.
\newblock In \bibinfo{booktitle}{\emph{Parkinson's disease in the older patient}}. \bibinfo{publisher}{CRC Press}, \bibinfo{pages}{11--29}.
\newblock


\bibitem[Stokes et~al\mbox{.}(2022)]%
        {stokes2022measuring}
\bibfield{author}{\bibinfo{person}{Jared~D Stokes}, \bibinfo{person}{Albert Rizzo}, \bibinfo{person}{Joy~J Geng}, {and} \bibinfo{person}{Julie~B Schweitzer}.} \bibinfo{year}{2022}\natexlab{}.
\newblock \showarticletitle{Measuring attentional distraction in children with ADHD using virtual reality technology with eye-tracking}.
\newblock \bibinfo{journal}{\emph{Frontiers in virtual reality}}  \bibinfo{volume}{3} (\bibinfo{year}{2022}), \bibinfo{pages}{855895}.
\newblock


\bibitem[Stolze et~al\mbox{.}(2004)]%
        {stolze2004falls}
\bibfield{author}{\bibinfo{person}{Henning Stolze}, \bibinfo{person}{Stephan Klebe}, \bibinfo{person}{Christiane Zechlin}, \bibinfo{person}{Christoph Baecker}, \bibinfo{person}{Lars Friege}, {and} \bibinfo{person}{G{\"u}nther Deuschl}.} \bibinfo{year}{2004}\natexlab{}.
\newblock \showarticletitle{Falls in frequent neurological diseases: prevalence, risk factors and aetiology}.
\newblock \bibinfo{journal}{\emph{Journal of neurology}}  \bibinfo{volume}{251} (\bibinfo{year}{2004}), \bibinfo{pages}{79--84}.
\newblock


\bibitem[Stuart et~al\mbox{.}(2023)]%
        {stuart2023eye}
\bibfield{author}{\bibinfo{person}{Nicole Stuart}, \bibinfo{person}{Andrew Whitehouse}, \bibinfo{person}{Romina Palermo}, \bibinfo{person}{Ellen Bothe}, {and} \bibinfo{person}{Nicholas Badcock}.} \bibinfo{year}{2023}\natexlab{}.
\newblock \showarticletitle{Eye gaze in autism spectrum disorder: a review of neural evidence for the eye avoidance hypothesis}.
\newblock \bibinfo{journal}{\emph{Journal of Autism and Developmental Disorders}} \bibinfo{volume}{53}, \bibinfo{number}{5} (\bibinfo{year}{2023}), \bibinfo{pages}{1884--1905}.
\newblock


\bibitem[Stuart et~al\mbox{.}(2016)]%
        {stuart2016accuracy}
\bibfield{author}{\bibinfo{person}{Susan Stuart}, \bibinfo{person}{Lisa Alcock}, \bibinfo{person}{Alan Godfrey}, \bibinfo{person}{Stephen Lord}, \bibinfo{person}{Lynn Rochester}, {and} \bibinfo{person}{Brook Galna}.} \bibinfo{year}{2016}\natexlab{}.
\newblock \showarticletitle{Accuracy and re-test reliability of mobile eye-tracking in Parkinson's disease and older adults}.
\newblock \bibinfo{journal}{\emph{Medical engineering \& physics}} \bibinfo{volume}{38}, \bibinfo{number}{3} (\bibinfo{year}{2016}), \bibinfo{pages}{308--315}.
\newblock


\bibitem[Stuart et~al\mbox{.}(2017)]%
        {stuart2017direct}
\bibfield{author}{\bibinfo{person}{Samuel Stuart}, \bibinfo{person}{Brook Galna}, \bibinfo{person}{Louise~S Delicato}, \bibinfo{person}{Sue Lord}, {and} \bibinfo{person}{Lynn Rochester}.} \bibinfo{year}{2017}\natexlab{}.
\newblock \showarticletitle{Direct and indirect effects of attention and visual function on gait impairment in Parkinson's disease: influence of task and turning}.
\newblock \bibinfo{journal}{\emph{European Journal of Neuroscience}} \bibinfo{volume}{46}, \bibinfo{number}{1} (\bibinfo{year}{2017}), \bibinfo{pages}{1703--1716}.
\newblock


\bibitem[Sturm et~al\mbox{.}(2011)]%
        {sturm2011mutual}
\bibfield{author}{\bibinfo{person}{Virginia~E Sturm}, \bibinfo{person}{Megan~E McCarthy}, \bibinfo{person}{Ira Yun}, \bibinfo{person}{Anita Madan}, \bibinfo{person}{Joyce~W Yuan}, \bibinfo{person}{Sarah~R Holley}, \bibinfo{person}{Elizabeth~A Ascher}, \bibinfo{person}{Adam~L Boxer}, \bibinfo{person}{Bruce~L Miller}, {and} \bibinfo{person}{Robert~W Levenson}.} \bibinfo{year}{2011}\natexlab{}.
\newblock \showarticletitle{Mutual gaze in Alzheimer's disease, frontotemporal and semantic dementia couples}.
\newblock \bibinfo{journal}{\emph{Social Cognitive and Affective Neuroscience}} \bibinfo{volume}{6}, \bibinfo{number}{3} (\bibinfo{year}{2011}), \bibinfo{pages}{359--367}.
\newblock


\bibitem[Sui et~al\mbox{.}(2023)]%
        {sui2023geco}
\bibfield{author}{\bibinfo{person}{Longjiao Sui}, \bibinfo{person}{Nicolas Dirix}, \bibinfo{person}{Evy Woumans}, {and} \bibinfo{person}{Wouter Duyck}.} \bibinfo{year}{2023}\natexlab{}.
\newblock \showarticletitle{GECO-CN: Ghent Eye-tracking COrpus of sentence reading for Chinese-English bilinguals}.
\newblock \bibinfo{journal}{\emph{Behavior Research Methods}} \bibinfo{volume}{55}, \bibinfo{number}{6} (\bibinfo{year}{2023}), \bibinfo{pages}{2743--2763}.
\newblock


\bibitem[Sun et~al\mbox{.}(2022)]%
        {sun2022novel}
\bibfield{author}{\bibinfo{person}{Jinglin Sun}, \bibinfo{person}{Yu Liu}, \bibinfo{person}{Hao Wu}, \bibinfo{person}{Peiguang Jing}, {and} \bibinfo{person}{Yong Ji}.} \bibinfo{year}{2022}\natexlab{}.
\newblock \showarticletitle{A novel deep learning approach for diagnosing Alzheimer's disease based on eye-tracking data}.
\newblock \bibinfo{journal}{\emph{Frontiers in Human Neuroscience}}  \bibinfo{volume}{16} (\bibinfo{year}{2022}), \bibinfo{pages}{972773}.
\newblock


\bibitem[Surakka et~al\mbox{.}(2008)]%
        {surakka2008pupillary}
\bibfield{author}{\bibinfo{person}{Jukka Surakka}, \bibinfo{person}{Juhani Ruutiainen}, \bibinfo{person}{Anders Romberg}, \bibinfo{person}{Pauli Puukka}, \bibinfo{person}{Erkki Kronholm}, {and} \bibinfo{person}{Hannu Karanko}.} \bibinfo{year}{2008}\natexlab{}.
\newblock \showarticletitle{Pupillary function in early multiple sclerosis}.
\newblock \bibinfo{journal}{\emph{Clinical Autonomic Research}}  \bibinfo{volume}{18} (\bibinfo{year}{2008}), \bibinfo{pages}{150--154}.
\newblock


\bibitem[Suslow et~al\mbox{.}(2020)]%
        {suslow2020attentional}
\bibfield{author}{\bibinfo{person}{Thomas Suslow}, \bibinfo{person}{Anja Husslack}, \bibinfo{person}{Anette Kersting}, {and} \bibinfo{person}{Charlott~Maria Bodenschatz}.} \bibinfo{year}{2020}\natexlab{}.
\newblock \showarticletitle{Attentional biases to emotional information in clinical depression: a systematic and meta-analytic review of eye tracking findings}.
\newblock \bibinfo{journal}{\emph{Journal of Affective Disorders}}  \bibinfo{volume}{274} (\bibinfo{year}{2020}), \bibinfo{pages}{632--642}.
\newblock


\bibitem[Svenaeus(2014)]%
        {svenaeus2014diagnosing}
\bibfield{author}{\bibinfo{person}{Fredrik Svenaeus}.} \bibinfo{year}{2014}\natexlab{}.
\newblock \showarticletitle{Diagnosing mental disorders and saving the normal: American Psychiatric Association, 2013. Diagnostic and statistical manual of mental disorders, American Psychiatric Publishing: Washington, DC. 991 pp., ISBN: 978-0890425558.}
\newblock \bibinfo{journal}{\emph{Medicine, Health Care and Philosophy}}  \bibinfo{volume}{17} (\bibinfo{year}{2014}), \bibinfo{pages}{241--244}.
\newblock


\bibitem[Sweere et~al\mbox{.}(2024)]%
        {sweere2024efficacy}
\bibfield{author}{\bibinfo{person}{Dirk~JJ Sweere}, \bibinfo{person}{Jos~GM Hendriksen}, \bibinfo{person}{R~Jeroen Vermeulen}, {and} \bibinfo{person}{Sylvia Klinkenberg}.} \bibinfo{year}{2024}\natexlab{}.
\newblock \showarticletitle{Efficacy of methylphenidate treatment in childhood myotonic dystrophy type 1 and comorbid attention deficit hyperactivity disorder: A case report using eye tracking assessment}.
\newblock \bibinfo{journal}{\emph{Brain and Development}} \bibinfo{volume}{46}, \bibinfo{number}{2} (\bibinfo{year}{2024}), \bibinfo{pages}{118--121}.
\newblock


\bibitem[Sweere et~al\mbox{.}(2022)]%
        {sweere2022clinical}
\bibfield{author}{\bibinfo{person}{Dirk~JJ Sweere}, \bibinfo{person}{Johan~JM Pel}, \bibinfo{person}{Marlou~JG Kooiker}, \bibinfo{person}{Johannes~P van Dijk}, \bibinfo{person}{Elizabeth~JJM van Gemert}, \bibinfo{person}{Petra~PM Hurks}, \bibinfo{person}{Sylvia Klinkenberg}, \bibinfo{person}{R~Jeroen Vermeulen}, {and} \bibinfo{person}{Jos~GM Hendriksen}.} \bibinfo{year}{2022}\natexlab{}.
\newblock \showarticletitle{Clinical utility of eye tracking in assessing distractibility in children with neurological disorders or ADHD: A cross-sectional study}.
\newblock \bibinfo{journal}{\emph{Brain Sciences}} \bibinfo{volume}{12}, \bibinfo{number}{10} (\bibinfo{year}{2022}), \bibinfo{pages}{1369}.
\newblock


\bibitem[Tafasca et~al\mbox{.}(2023)]%
        {tafasca2023ai4autism}
\bibfield{author}{\bibinfo{person}{Samy Tafasca}, \bibinfo{person}{Anshul Gupta}, \bibinfo{person}{Nada Kojovic}, \bibinfo{person}{Mirko Gelsomini}, \bibinfo{person}{Thomas Maillart}, \bibinfo{person}{Michela Papandrea}, \bibinfo{person}{Marie Schaer}, {and} \bibinfo{person}{Jean-Marc Odobez}.} \bibinfo{year}{2023}\natexlab{}.
\newblock \showarticletitle{The AI4Autism Project: A Multimodal and Interdisciplinary Approach to Autism Diagnosis and Stratification}. In \bibinfo{booktitle}{\emph{Companion Publication of the 25th International Conference on Multimodal Interaction}}. \bibinfo{pages}{414--425}.
\newblock


\bibitem[Tajik-Parvinchi and Sandor(2011)]%
        {tajik2011smooth}
\bibfield{author}{\bibinfo{person}{Diana~J Tajik-Parvinchi} {and} \bibinfo{person}{Paul Sandor}.} \bibinfo{year}{2011}\natexlab{}.
\newblock \showarticletitle{Smooth pursuit and fixation ability in children with Tourette syndrome}.
\newblock \bibinfo{journal}{\emph{Cognitive and behavioral neurology}} \bibinfo{volume}{24}, \bibinfo{number}{4} (\bibinfo{year}{2011}), \bibinfo{pages}{174--186}.
\newblock


\bibitem[Tajik-Parvinchi and Sandor(2013)]%
        {tajik2013enhanced}
\bibfield{author}{\bibinfo{person}{Diana~J Tajik-Parvinchi} {and} \bibinfo{person}{Paul Sandor}.} \bibinfo{year}{2013}\natexlab{}.
\newblock \showarticletitle{Enhanced antisaccade abilities in children with Tourette syndrome: the Gap-effect Reversal}.
\newblock \bibinfo{journal}{\emph{Frontiers in Human Neuroscience}}  \bibinfo{volume}{7} (\bibinfo{year}{2013}), \bibinfo{pages}{768}.
\newblock


\bibitem[Takamura et~al\mbox{.}(2016)]%
        {takamura2016intentional}
\bibfield{author}{\bibinfo{person}{Yusaku Takamura}, \bibinfo{person}{Maho Imanishi}, \bibinfo{person}{Madoka Osaka}, \bibinfo{person}{Satoko Ohmatsu}, \bibinfo{person}{Takanori Tominaga}, \bibinfo{person}{Kentaro Yamanaka}, \bibinfo{person}{Shu Morioka}, {and} \bibinfo{person}{Noritaka Kawashima}.} \bibinfo{year}{2016}\natexlab{}.
\newblock \showarticletitle{Intentional gaze shift to neglected space: a compensatory strategy during recovery after unilateral spatial neglect}.
\newblock \bibinfo{journal}{\emph{Brain}} \bibinfo{volume}{139}, \bibinfo{number}{11} (\bibinfo{year}{2016}), \bibinfo{pages}{2970--2982}.
\newblock


\bibitem[Tao et~al\mbox{.}(2020)]%
        {tao2020eye}
\bibfield{author}{\bibinfo{person}{Ling Tao}, \bibinfo{person}{Quan Wang}, \bibinfo{person}{Ding Liu}, \bibinfo{person}{Jing Wang}, \bibinfo{person}{Ziqing Zhu}, {and} \bibinfo{person}{Li Feng}.} \bibinfo{year}{2020}\natexlab{}.
\newblock \showarticletitle{Eye tracking metrics to screen and assess cognitive impairment in patients with neurological disorders}.
\newblock \bibinfo{journal}{\emph{Neurological Sciences}}  \bibinfo{volume}{41} (\bibinfo{year}{2020}), \bibinfo{pages}{1697--1704}.
\newblock


\bibitem[Tata et~al\mbox{.}(1996)]%
        {tata1996attentional}
\bibfield{author}{\bibinfo{person}{Philip~R Tata}, \bibinfo{person}{Judy~A Leibowitz}, \bibinfo{person}{Mark~J Prunty}, \bibinfo{person}{Mary Cameron}, {and} \bibinfo{person}{Alan~D Pickering}.} \bibinfo{year}{1996}\natexlab{}.
\newblock \showarticletitle{Attentional bias in obsessional compulsive disorder}.
\newblock \bibinfo{journal}{\emph{Behaviour Research and Therapy}} \bibinfo{volume}{34}, \bibinfo{number}{1} (\bibinfo{year}{1996}), \bibinfo{pages}{53--60}.
\newblock


\bibitem[Tatlipinar et~al\mbox{.}(2001)]%
        {tatlipinar2001ophthalmic}
\bibfield{author}{\bibinfo{person}{S Tatlipinar}, \bibinfo{person}{EC Iener}, \bibinfo{person}{B Ilhan}, {and} \bibinfo{person}{B Semerci}.} \bibinfo{year}{2001}\natexlab{}.
\newblock \showarticletitle{Ophthalmic manifestations of Gilles de la Tourette syndrome.}
\newblock \bibinfo{journal}{\emph{European journal of ophthalmology}} \bibinfo{volume}{11}, \bibinfo{number}{3} (\bibinfo{year}{2001}), \bibinfo{pages}{223--226}.
\newblock


\bibitem[Taylor et~al\mbox{.}(2014)]%
        {taylor2014brief}
\bibfield{author}{\bibinfo{person}{Cora~M Taylor}, \bibinfo{person}{Alison Vehorn}, \bibinfo{person}{Hylan Noble}, \bibinfo{person}{Amy~S Weitlauf}, {and} \bibinfo{person}{Zachary~E Warren}.} \bibinfo{year}{2014}\natexlab{}.
\newblock \showarticletitle{Brief report: can metrics of reporting bias enhance early autism screening measures?}
\newblock \bibinfo{journal}{\emph{Journal of autism and developmental disorders}}  \bibinfo{volume}{44} (\bibinfo{year}{2014}), \bibinfo{pages}{2375--2380}.
\newblock


\bibitem[Temelturk and Ozer(2022)]%
        {temelturk2022binocular}
\bibfield{author}{\bibinfo{person}{Rahime~Duygu Temelturk} {and} \bibinfo{person}{Esmehan Ozer}.} \bibinfo{year}{2022}\natexlab{}.
\newblock \showarticletitle{Binocular coordination of children with dyslexia and typically developing children in linguistic and non-linguistic tasks: Evidence from eye movements}.
\newblock \bibinfo{journal}{\emph{Annals of Dyslexia}} \bibinfo{volume}{72}, \bibinfo{number}{3} (\bibinfo{year}{2022}), \bibinfo{pages}{426--444}.
\newblock


\bibitem[Tenenbaum et~al\mbox{.}(2021)]%
        {tenenbaum2021distance}
\bibfield{author}{\bibinfo{person}{Elena~J Tenenbaum}, \bibinfo{person}{Samantha Major}, \bibinfo{person}{Kimberly~LH Carpenter}, \bibinfo{person}{Jill Howard}, \bibinfo{person}{Michael Murias}, {and} \bibinfo{person}{Geraldine Dawson}.} \bibinfo{year}{2021}\natexlab{}.
\newblock \showarticletitle{Distance from typical scan path when viewing complex stimuli in children with autism spectrum disorder and its association with behavior}.
\newblock \bibinfo{journal}{\emph{Journal of Autism and Developmental Disorders}} (\bibinfo{year}{2021}), \bibinfo{pages}{1--14}.
\newblock


\bibitem[Terao et~al\mbox{.}(2023)]%
        {terao2023patients}
\bibfield{author}{\bibinfo{person}{Yasuo Terao}, \bibinfo{person}{Shin-ichi Tokushige}, \bibinfo{person}{Satomi Inomata-Terada}, \bibinfo{person}{Tai Miyazaki}, \bibinfo{person}{Naoki Kotsuki}, \bibinfo{person}{Francesco Fisicaro}, {and} \bibinfo{person}{Yoshikazu Ugawa}.} \bibinfo{year}{2023}\natexlab{}.
\newblock \showarticletitle{How do patients with Parkinson’s disease and cerebellar ataxia read aloud?-Eye--voice coordination in text reading}.
\newblock \bibinfo{journal}{\emph{Frontiers in Neuroscience}}  \bibinfo{volume}{17} (\bibinfo{year}{2023}), \bibinfo{pages}{1202404}.
\newblock


\bibitem[Thierfelder et~al\mbox{.}(2023)]%
        {thierfelder23_pervasiveh}
\bibfield{author}{\bibinfo{person}{Annika Thierfelder}, \bibinfo{person}{Björn Severitt}, \bibinfo{person}{Carolin~Sarah Klein}, \bibinfo{person}{Annika~Kristin Alt}, \bibinfo{person}{Karsten Hollmann}, \bibinfo{person}{Andreas Bulling}, {and} \bibinfo{person}{Winfried Ilg}.} \bibinfo{year}{2023}\natexlab{}.
\newblock \showarticletitle{Gaze Behaviour in Adolescents with Obsessive-compulsive Disorder During Exposure Within Cognitive-behavioural Therapy}. In \bibinfo{booktitle}{\emph{Proc. 17th EAI International Conference on Pervasive Computing Technologies for Healthcare (Pervasive Health)}}.
\newblock
\href{https://doi.org/10.13140/RG.2.2.30047.02721}{doi:\nolinkurl{10.13140/RG.2.2.30047.02721}}


\bibitem[Thorup et~al\mbox{.}(2016)]%
        {thorup2016altered}
\bibfield{author}{\bibinfo{person}{Emilia Thorup}, \bibinfo{person}{P{\"a}r Nystr{\"o}m}, \bibinfo{person}{Gustaf Gredeb{\"a}ck}, \bibinfo{person}{Sven B{\"o}lte}, \bibinfo{person}{Terje Falck-Ytter}, {and} \bibinfo{person}{EASE Team}.} \bibinfo{year}{2016}\natexlab{}.
\newblock \showarticletitle{Altered gaze following during live interaction in infants at risk for autism: an eye tracking study}.
\newblock \bibinfo{journal}{\emph{Molecular autism}}  \bibinfo{volume}{7} (\bibinfo{year}{2016}), \bibinfo{pages}{1--10}.
\newblock


\bibitem[Thurston et~al\mbox{.}(1985)]%
        {thurston1985epileptic}
\bibfield{author}{\bibinfo{person}{Stephen~E Thurston}, \bibinfo{person}{R~John Leigh}, {and} \bibinfo{person}{Ivan Osorio}.} \bibinfo{year}{1985}\natexlab{}.
\newblock \showarticletitle{Epileptic gaze deviation and nystagmus}.
\newblock \bibinfo{journal}{\emph{Neurology}} \bibinfo{volume}{35}, \bibinfo{number}{10} (\bibinfo{year}{1985}), \bibinfo{pages}{1518--1518}.
\newblock


\bibitem[Thurtell et~al\mbox{.}(2009)]%
        {thurtell2009evidence}
\bibfield{author}{\bibinfo{person}{MJ Thurtell}, \bibinfo{person}{A Mohamed}, \bibinfo{person}{HO L{\"u}ders}, {and} \bibinfo{person}{RJ Leigh}.} \bibinfo{year}{2009}\natexlab{}.
\newblock \showarticletitle{Evidence for three-dimensional cortical control of gaze from epileptic patients}.
\newblock \bibinfo{journal}{\emph{Journal of Neurology, Neurosurgery \& Psychiatry}} \bibinfo{volume}{80}, \bibinfo{number}{6} (\bibinfo{year}{2009}), \bibinfo{pages}{683--685}.
\newblock


\bibitem[Tian et~al\mbox{.}(1991)]%
        {tian1991saccades}
\bibfield{author}{\bibinfo{person}{JR Tian}, \bibinfo{person}{DS Zee}, \bibinfo{person}{AG Lasker}, {and} \bibinfo{person}{SE Folstein}.} \bibinfo{year}{1991}\natexlab{}.
\newblock \showarticletitle{Saccades in Huntington's disease: predictive tracking and interaction between release of fixation and initiation of saccades}.
\newblock \bibinfo{journal}{\emph{Neurology}} \bibinfo{volume}{41}, \bibinfo{number}{6} (\bibinfo{year}{1991}), \bibinfo{pages}{875--875}.
\newblock


\bibitem[Tien et~al\mbox{.}(2012)]%
        {tien2012measuring}
\bibfield{author}{\bibinfo{person}{Geoffrey Tien}, \bibinfo{person}{M~Stella Atkins}, {and} \bibinfo{person}{Bin Zheng}.} \bibinfo{year}{2012}\natexlab{}.
\newblock \showarticletitle{Measuring gaze overlap on videos between multiple observers}. In \bibinfo{booktitle}{\emph{Proceedings of the symposium on eye tracking research and applications}}. \bibinfo{pages}{309--312}.
\newblock


\bibitem[Ting et~al\mbox{.}(2011)]%
        {ting2011visual}
\bibfield{author}{\bibinfo{person}{Darren~SJ Ting}, \bibinfo{person}{Alex Pollock}, \bibinfo{person}{Gordon~N Dutton}, \bibinfo{person}{Fergus~N Doubal}, \bibinfo{person}{Daniel~SW Ting}, \bibinfo{person}{Michelle Thompson}, {and} \bibinfo{person}{Baljean Dhillon}.} \bibinfo{year}{2011}\natexlab{}.
\newblock \showarticletitle{Visual neglect following stroke: current concepts and future focus}.
\newblock \bibinfo{journal}{\emph{Survey of ophthalmology}} \bibinfo{volume}{56}, \bibinfo{number}{2} (\bibinfo{year}{2011}), \bibinfo{pages}{114--134}.
\newblock


\bibitem[Tipples and Pecchinenda(2019)]%
        {tipples2019closer}
\bibfield{author}{\bibinfo{person}{Jason Tipples} {and} \bibinfo{person}{Anna Pecchinenda}.} \bibinfo{year}{2019}\natexlab{}.
\newblock \showarticletitle{A closer look at the size of the gaze-liking effect: A preregistered replication}.
\newblock \bibinfo{journal}{\emph{Cognition and Emotion}} \bibinfo{volume}{33}, \bibinfo{number}{3} (\bibinfo{year}{2019}), \bibinfo{pages}{623--629}.
\newblock


\bibitem[Toh et~al\mbox{.}(2017)]%
        {toh2017attentional}
\bibfield{author}{\bibinfo{person}{Wei~Lin Toh}, \bibinfo{person}{David~J Castle}, {and} \bibinfo{person}{Susan~L Rossell}.} \bibinfo{year}{2017}\natexlab{}.
\newblock \showarticletitle{Attentional biases in body dysmorphic disorder (BDD): Eye-tracking using the emotional Stroop task}.
\newblock \bibinfo{journal}{\emph{Comprehensive Psychiatry}}  \bibinfo{volume}{74} (\bibinfo{year}{2017}), \bibinfo{pages}{151--161}.
\newblock


\bibitem[Toh et~al\mbox{.}(2011)]%
        {toh2011current}
\bibfield{author}{\bibinfo{person}{Wei~Lin Toh}, \bibinfo{person}{Susan~L Rossell}, {and} \bibinfo{person}{David~J Castle}.} \bibinfo{year}{2011}\natexlab{}.
\newblock \showarticletitle{Current visual scanpath research: a review of investigations into the psychotic, anxiety, and mood disorders}.
\newblock \bibinfo{journal}{\emph{Comprehensive psychiatry}} \bibinfo{volume}{52}, \bibinfo{number}{6} (\bibinfo{year}{2011}), \bibinfo{pages}{567--579}.
\newblock


\bibitem[Tokushige et~al\mbox{.}(2023)]%
        {tokushige2023early}
\bibfield{author}{\bibinfo{person}{Shin-ichi Tokushige}, \bibinfo{person}{Hideyuki Matsumoto}, \bibinfo{person}{Shun-ichi Matsuda}, \bibinfo{person}{Satomi Inomata-Terada}, \bibinfo{person}{Naoki Kotsuki}, \bibinfo{person}{Masashi Hamada}, \bibinfo{person}{Shoji Tsuji}, \bibinfo{person}{Yoshikazu Ugawa}, {and} \bibinfo{person}{Yasuo Terao}.} \bibinfo{year}{2023}\natexlab{}.
\newblock \showarticletitle{Early detection of cognitive decline in Alzheimer’s disease using eye tracking}.
\newblock \bibinfo{journal}{\emph{Frontiers in Aging Neuroscience}}  \bibinfo{volume}{15} (\bibinfo{year}{2023}), \bibinfo{pages}{1123456}.
\newblock


\bibitem[Torres et~al\mbox{.}(2013)]%
        {torres2013autism}
\bibfield{author}{\bibinfo{person}{Elizabeth~B Torres}, \bibinfo{person}{Maria Brincker}, \bibinfo{person}{Robert~W Isenhower}, \bibinfo{person}{Polina Yanovich}, \bibinfo{person}{Kimberly~A Stigler}, \bibinfo{person}{John~I Nurnberger}, \bibinfo{person}{Dimitris~N Metaxas}, {and} \bibinfo{person}{Jorge~V Jos{\'e}}.} \bibinfo{year}{2013}\natexlab{}.
\newblock \showarticletitle{Autism: the micro-movement perspective}.
\newblock \bibinfo{journal}{\emph{Frontiers in integrative neuroscience}}  \bibinfo{volume}{7} (\bibinfo{year}{2013}), \bibinfo{pages}{32}.
\newblock


\bibitem[Tramonti~Fantozzi et~al\mbox{.}(2023)]%
        {tramonti2023pointing}
\bibfield{author}{\bibinfo{person}{Maria~Paola Tramonti~Fantozzi}, \bibinfo{person}{Roberta Benedetti}, \bibinfo{person}{Alessandra Crecchi}, \bibinfo{person}{Lucia Briscese}, \bibinfo{person}{Paolo Andre}, \bibinfo{person}{Pieranna Arrighi}, \bibinfo{person}{Luca Bonfiglio}, \bibinfo{person}{Maria~Chiara Carboncini}, \bibinfo{person}{Luca Bruschini}, \bibinfo{person}{Paolo Bongioanni}, {et~al\mbox{.}}} \bibinfo{year}{2023}\natexlab{}.
\newblock \showarticletitle{Pointing in cervical dystonia patients}.
\newblock \bibinfo{journal}{\emph{Frontiers in Systems Neuroscience}}  \bibinfo{volume}{17} (\bibinfo{year}{2023}), \bibinfo{pages}{1306387}.
\newblock


\bibitem[Travers et~al\mbox{.}(2011)]%
        {travers2011attention}
\bibfield{author}{\bibinfo{person}{Brittany~G Travers}, \bibinfo{person}{Mark~R Klinger}, {and} \bibinfo{person}{Laura~Grofer Klinger}.} \bibinfo{year}{2011}\natexlab{}.
\newblock \showarticletitle{Attention and working memory in ASD}.
\newblock \bibinfo{journal}{\emph{The neuropsychology of autism}} (\bibinfo{year}{2011}), \bibinfo{pages}{161--184}.
\newblock


\bibitem[Trevisan et~al\mbox{.}(2017)]%
        {trevisan2017adults}
\bibfield{author}{\bibinfo{person}{Dominic~A Trevisan}, \bibinfo{person}{Nicole Roberts}, \bibinfo{person}{Cathy Lin}, {and} \bibinfo{person}{Elina Birmingham}.} \bibinfo{year}{2017}\natexlab{}.
\newblock \showarticletitle{How do adults and teens with self-declared Autism Spectrum Disorder experience eye contact? A qualitative analysis of first-hand accounts}.
\newblock \bibinfo{journal}{\emph{PloS one}} \bibinfo{volume}{12}, \bibinfo{number}{11} (\bibinfo{year}{2017}), \bibinfo{pages}{e0188446}.
\newblock


\bibitem[Tsitsi et~al\mbox{.}(2021)]%
        {tsitsi2021fixation}
\bibfield{author}{\bibinfo{person}{Panagiota Tsitsi}, \bibinfo{person}{Mattias~Nilsson Benfatto}, \bibinfo{person}{Gustaf~{\"O}qvist Seimyr}, \bibinfo{person}{Olof Larsson}, \bibinfo{person}{Per Svenningsson}, {and} \bibinfo{person}{Ioanna Markaki}.} \bibinfo{year}{2021}\natexlab{}.
\newblock \showarticletitle{Fixation duration and pupil size as diagnostic tools in Parkinson’s disease}.
\newblock \bibinfo{journal}{\emph{Journal of Parkinson's Disease}} \bibinfo{volume}{11}, \bibinfo{number}{2} (\bibinfo{year}{2021}), \bibinfo{pages}{865--875}.
\newblock


\bibitem[Tulen et~al\mbox{.}(1999)]%
        {tulen1999quantitative}
\bibfield{author}{\bibinfo{person}{JHM Tulen}, \bibinfo{person}{M Azzolini}, \bibinfo{person}{JA De~Vries}, \bibinfo{person}{WH Groeneveld}, \bibinfo{person}{Jan Passchier}, {and} \bibinfo{person}{BJM Van De~Wetering}.} \bibinfo{year}{1999}\natexlab{}.
\newblock \showarticletitle{Quantitative study of spontaneous eye blinks and eye tics in Gilles de la Tourette's syndrome}.
\newblock \bibinfo{journal}{\emph{Journal of Neurology, Neurosurgery \& Psychiatry}} \bibinfo{volume}{67}, \bibinfo{number}{6} (\bibinfo{year}{1999}), \bibinfo{pages}{800--802}.
\newblock


\bibitem[Turner et~al\mbox{.}(2011)]%
        {turner2011behavioral}
\bibfield{author}{\bibinfo{person}{Travis~H Turner}, \bibinfo{person}{Jody Goldstein}, \bibinfo{person}{Joanne~M Hamilton}, \bibinfo{person}{Mark Jacobson}, \bibinfo{person}{Eva Pirogovsky}, \bibinfo{person}{Guerry Peavy}, {and} \bibinfo{person}{Jody Corey-Bloom}.} \bibinfo{year}{2011}\natexlab{}.
\newblock \showarticletitle{Behavioral measures of saccade latency and inhibition in manifest and premanifest Huntington's disease}.
\newblock \bibinfo{journal}{\emph{Journal of Motor Behavior}} \bibinfo{volume}{43}, \bibinfo{number}{4} (\bibinfo{year}{2011}), \bibinfo{pages}{295--302}.
\newblock


\bibitem[Tye et~al\mbox{.}(2013)]%
        {tye2013neurophysiological}
\bibfield{author}{\bibinfo{person}{Charlotte Tye}, \bibinfo{person}{Evelyne Mercure}, \bibinfo{person}{Karen~L Ashwood}, \bibinfo{person}{Bahare Azadi}, \bibinfo{person}{Philip Asherson}, \bibinfo{person}{Mark~H Johnson}, \bibinfo{person}{Patrick Bolton}, {and} \bibinfo{person}{Gr{\'a}inne McLoughlin}.} \bibinfo{year}{2013}\natexlab{}.
\newblock \showarticletitle{Neurophysiological responses to faces and gaze direction differentiate children with ASD, ADHD and ASD+ ADHD}.
\newblock \bibinfo{journal}{\emph{Developmental cognitive neuroscience}}  \bibinfo{volume}{5} (\bibinfo{year}{2013}), \bibinfo{pages}{71--85}.
\newblock


\bibitem[Unoki et~al\mbox{.}(1999)]%
        {unoki1999attentional}
\bibfield{author}{\bibinfo{person}{Keiko Unoki}, \bibinfo{person}{Takashi Kasuga}, \bibinfo{person}{Eisuke Matsushima}, {and} \bibinfo{person}{Katsuya Ohta}.} \bibinfo{year}{1999}\natexlab{}.
\newblock \showarticletitle{Attentional processing of emotional information in obsessive--compulsive disorder}.
\newblock \bibinfo{journal}{\emph{Psychiatry and clinical neurosciences}} \bibinfo{volume}{53}, \bibinfo{number}{6} (\bibinfo{year}{1999}), \bibinfo{pages}{635--642}.
\newblock


\bibitem[Uomori et~al\mbox{.}(1993)]%
        {uomori1993analysis}
\bibfield{author}{\bibinfo{person}{Kenya Uomori}, \bibinfo{person}{Shinji Murakami}, \bibinfo{person}{Mitsuho Yamada}, \bibinfo{person}{Mitsuru Fujii}, \bibinfo{person}{Hiroshi Yoshimatsu}, \bibinfo{person}{Norihito Nakano}, \bibinfo{person}{Hitoshi Hongo}, \bibinfo{person}{Jiro Miyazawa}, \bibinfo{person}{Keiichi Ueno}, \bibinfo{person}{Ryo Fukatsu}, {et~al\mbox{.}}} \bibinfo{year}{1993}\natexlab{}.
\newblock \showarticletitle{Analysis of gaze shift in depth in Alzheimer's disease patients}.
\newblock \bibinfo{journal}{\emph{IEICE TRANSACTIONS on Information and Systems}} \bibinfo{volume}{76}, \bibinfo{number}{8} (\bibinfo{year}{1993}), \bibinfo{pages}{963--973}.
\newblock


\bibitem[Uono et~al\mbox{.}(2023)]%
        {uono2023reduced}
\bibfield{author}{\bibinfo{person}{Shota Uono}, \bibinfo{person}{Yuka Egashira}, \bibinfo{person}{Sayuri Hayashi}, \bibinfo{person}{Miki Takada}, \bibinfo{person}{Masatoshi Ukezono}, {and} \bibinfo{person}{Takashi Okada}.} \bibinfo{year}{2023}\natexlab{}.
\newblock \showarticletitle{Reduced gaze-cueing effect with neutral and emotional faces in adults with attention deficit/hyperactivity disorder}.
\newblock \bibinfo{journal}{\emph{Journal of Psychiatric Research}}  \bibinfo{volume}{168} (\bibinfo{year}{2023}), \bibinfo{pages}{310--317}.
\newblock


\bibitem[Vagge et~al\mbox{.}(2015)]%
        {vagge2015evaluation}
\bibfield{author}{\bibinfo{person}{Aldo Vagge}, \bibinfo{person}{Margherita Cavanna}, \bibinfo{person}{Carlo~Enrico Traverso}, {and} \bibinfo{person}{Michele Iester}.} \bibinfo{year}{2015}\natexlab{}.
\newblock \showarticletitle{Evaluation of ocular movements in patients with dyslexia}.
\newblock \bibinfo{journal}{\emph{Annals of dyslexia}}  \bibinfo{volume}{65} (\bibinfo{year}{2015}), \bibinfo{pages}{24--32}.
\newblock


\bibitem[Vajs et~al\mbox{.}(2023)]%
        {vajs2023accessible}
\bibfield{author}{\bibinfo{person}{Ivan Vajs}, \bibinfo{person}{Tamara Papi{\'c}}, \bibinfo{person}{Vanja Kovi{\'c}}, \bibinfo{person}{Andrej~M Savi{\'c}}, {and} \bibinfo{person}{Milica~M Jankovi{\'c}}.} \bibinfo{year}{2023}\natexlab{}.
\newblock \showarticletitle{Accessible dyslexia detection with real-time reading feedback through robust interpretable eye-tracking features}.
\newblock \bibinfo{journal}{\emph{Brain Sciences}} \bibinfo{volume}{13}, \bibinfo{number}{3} (\bibinfo{year}{2023}), \bibinfo{pages}{405}.
\newblock


\bibitem[Varela~Casal et~al\mbox{.}(2019)]%
        {varela2019clinical}
\bibfield{author}{\bibinfo{person}{Paloma Varela~Casal}, \bibinfo{person}{Flavia Lorena~Esposito}, \bibinfo{person}{Imanol Morata~Mart{\'\i}nez}, \bibinfo{person}{Alba Capdevila}, \bibinfo{person}{Maria Sol{\'e}~Puig}, \bibinfo{person}{N{\'u}ria de~la Osa}, \bibinfo{person}{Lourdes Ezpeleta}, \bibinfo{person}{Alexandre Perera~i Lluna}, \bibinfo{person}{Stephen~V Faraone}, \bibinfo{person}{Josep~Antoni Ramos-Quiroga}, {et~al\mbox{.}}} \bibinfo{year}{2019}\natexlab{}.
\newblock \showarticletitle{Clinical validation of eye vergence as an objective marker for diagnosis of ADHD in children}.
\newblock \bibinfo{journal}{\emph{Journal of Attention Disorders}} \bibinfo{volume}{23}, \bibinfo{number}{6} (\bibinfo{year}{2019}), \bibinfo{pages}{599--614}.
\newblock


\bibitem[Venjakob et~al\mbox{.}(2012)]%
        {venjakob2012radiologists}
\bibfield{author}{\bibinfo{person}{Antje Venjakob}, \bibinfo{person}{Tim Marnitz}, \bibinfo{person}{Jan Mahler}, \bibinfo{person}{Simone Sechelmann}, {and} \bibinfo{person}{Matthias Roetting}.} \bibinfo{year}{2012}\natexlab{}.
\newblock \showarticletitle{Radiologists' eye gaze when reading cranial CT images}. In \bibinfo{booktitle}{\emph{Medical imaging 2012: Image perception, observer performance, and technology assessment}}, Vol.~\bibinfo{volume}{8318}. SPIE, \bibinfo{pages}{78--87}.
\newblock


\bibitem[Vidal et~al\mbox{.}(2012)]%
        {vidal12_comcom}
\bibfield{author}{\bibinfo{person}{M{\'{e}}lodie Vidal}, \bibinfo{person}{Jayson Turner}, \bibinfo{person}{Andreas Bulling}, {and} \bibinfo{person}{Hans Gellersen}.} \bibinfo{year}{2012}\natexlab{}.
\newblock \showarticletitle{Wearable Eye Tracking for Mental Health Monitoring}.
\newblock \bibinfo{journal}{\emph{Computer Communications}} \bibinfo{volume}{35}, \bibinfo{number}{11} (\bibinfo{year}{2012}), \bibinfo{pages}{1306--1311}.
\newblock
\href{https://doi.org/10.1016/j.comcom.2011.11.002}{doi:\nolinkurl{10.1016/j.comcom.2011.11.002}}


\bibitem[Viktorsson et~al\mbox{.}(2023)]%
        {viktorsson202318}
\bibfield{author}{\bibinfo{person}{Charlotte Viktorsson}, \bibinfo{person}{Sven B{\"o}lte}, {and} \bibinfo{person}{Terje Falck-Ytter}.} \bibinfo{year}{2023}\natexlab{}.
\newblock \showarticletitle{How 18-month-olds with later autism look at other children interacting: The timing of gaze allocation}.
\newblock \bibinfo{journal}{\emph{Journal of autism and developmental disorders}} (\bibinfo{year}{2023}), \bibinfo{pages}{1--11}.
\newblock


\bibitem[Wagenmakers et~al\mbox{.}(2012)]%
        {wagenmakers2012agenda}
\bibfield{author}{\bibinfo{person}{Eric-Jan Wagenmakers}, \bibinfo{person}{Ruud Wetzels}, \bibinfo{person}{Denny Borsboom}, \bibinfo{person}{Han~LJ van~der Maas}, {and} \bibinfo{person}{Rogier~A Kievit}.} \bibinfo{year}{2012}\natexlab{}.
\newblock \showarticletitle{An agenda for purely confirmatory research}.
\newblock \bibinfo{journal}{\emph{Perspectives on psychological science}} \bibinfo{volume}{7}, \bibinfo{number}{6} (\bibinfo{year}{2012}), \bibinfo{pages}{632--638}.
\newblock


\bibitem[Waldthaler et~al\mbox{.}(2021)]%
        {waldthaler2021antisaccades}
\bibfield{author}{\bibinfo{person}{Josefine Waldthaler}, \bibinfo{person}{Lena Stock}, \bibinfo{person}{Justus Student}, \bibinfo{person}{Johanna Sommerkorn}, \bibinfo{person}{Stefan Dowiasch}, {and} \bibinfo{person}{Lars Timmermann}.} \bibinfo{year}{2021}\natexlab{}.
\newblock \showarticletitle{Antisaccades in Parkinson’s disease: a meta-analysis}.
\newblock \bibinfo{journal}{\emph{Neuropsychology Review}}  \bibinfo{volume}{31} (\bibinfo{year}{2021}), \bibinfo{pages}{628--642}.
\newblock


\bibitem[Walker(2014)]%
        {walker2014neurodiversity}
\bibfield{author}{\bibinfo{person}{Nick Walker}.} \bibinfo{year}{2014}\natexlab{}.
\newblock \bibinfo{title}{Neurodiversity: Some basic terms \& definitions}.
\newblock


\bibitem[Walton and Tiede(2020)]%
        {walton2020brief}
\bibfield{author}{\bibinfo{person}{Katherine~M Walton} {and} \bibinfo{person}{Gabrielle Tiede}.} \bibinfo{year}{2020}\natexlab{}.
\newblock \showarticletitle{Brief report: Does “healthy” family functioning look different for families who have a child with autism?}
\newblock \bibinfo{journal}{\emph{Research in autism spectrum disorders}}  \bibinfo{volume}{72} (\bibinfo{year}{2020}), \bibinfo{pages}{101527}.
\newblock


\bibitem[Wang et~al\mbox{.}(2019a)]%
        {wang2019gaze}
\bibfield{author}{\bibinfo{person}{Ruijie Wang}, \bibinfo{person}{Liming Chen}, \bibinfo{person}{Aladdin Ayesh}, \bibinfo{person}{Jethro Shell}, {and} \bibinfo{person}{Ivar Solheim}.} \bibinfo{year}{2019}\natexlab{a}.
\newblock \showarticletitle{Gaze-based assessment of dyslexic students' motivation within an e-learning environment}. In \bibinfo{booktitle}{\emph{2019 IEEE SmartWorld, Ubiquitous Intelligence \& Computing, Advanced \& Trusted Computing, Scalable Computing \& Communications, Cloud \& Big Data Computing, Internet of People and Smart City Innovation (SmartWorld/SCALCOM/UIC/ATC/CBDCom/IOP/SCI)}}. IEEE, \bibinfo{pages}{610--617}.
\newblock


\bibitem[Wang et~al\mbox{.}(2002)]%
        {wang2002neural}
\bibfield{author}{\bibinfo{person}{Xiaoxing Wang}, \bibinfo{person}{Jesse Jin}, {and} \bibinfo{person}{Marwan Jabri}.} \bibinfo{year}{2002}\natexlab{}.
\newblock \showarticletitle{Neural network models for the gaze shift system in the superior colliculus and cerebellum}.
\newblock \bibinfo{journal}{\emph{Neural Networks}} \bibinfo{volume}{15}, \bibinfo{number}{7} (\bibinfo{year}{2002}), \bibinfo{pages}{811--832}.
\newblock


\bibitem[Wang et~al\mbox{.}(2019b)]%
        {wang2019assessment}
\bibfield{author}{\bibinfo{person}{Yan Wang}, \bibinfo{person}{Guangtao Zhai}, \bibinfo{person}{Sichao Chen}, \bibinfo{person}{Xiongkuo Min}, \bibinfo{person}{Zhongpai Gao}, {and} \bibinfo{person}{Xuefei Song}.} \bibinfo{year}{2019}\natexlab{b}.
\newblock \showarticletitle{Assessment of eye fatigue caused by head-mounted displays using eye-tracking}.
\newblock \bibinfo{journal}{\emph{Biomedical engineering online}}  \bibinfo{volume}{18} (\bibinfo{year}{2019}), \bibinfo{pages}{1--19}.
\newblock


\bibitem[Wass et~al\mbox{.}(2015)]%
        {wass2015shorter}
\bibfield{author}{\bibinfo{person}{Sam~V Wass}, \bibinfo{person}{Emily~JH Jones}, \bibinfo{person}{Teodora Gliga}, \bibinfo{person}{Tim~J Smith}, \bibinfo{person}{Tony Charman}, \bibinfo{person}{Mark~H Johnson}, {and} \bibinfo{person}{BASIS team Baron-Cohen Simon 4 Bedford Rachael 5 Bolton Patrick 5 Chandler Susie 6 Davies Kim 7 Fernandes Janice 7 Garwood Holly 7 Hudry Kristelle 8 Maris Helen 8 Pasco Greg 5 Pickles Andrew 5 Ribiero Helena 7 Tucker Leslie 7 Volein Agnes~8}.} \bibinfo{year}{2015}\natexlab{}.
\newblock \showarticletitle{Shorter spontaneous fixation durations in infants with later emerging autism}.
\newblock \bibinfo{journal}{\emph{Scientific reports}} \bibinfo{volume}{5}, \bibinfo{number}{1} (\bibinfo{year}{2015}), \bibinfo{pages}{8284}.
\newblock


\bibitem[Wasserstein and Lazar(2016)]%
        {wasserstein2016asa}
\bibfield{author}{\bibinfo{person}{Ronald~L Wasserstein} {and} \bibinfo{person}{Nicole~A Lazar}.} \bibinfo{year}{2016}\natexlab{}.
\newblock \bibinfo{title}{The ASA statement on p-values: context, process, and purpose}.
\newblock \bibinfo{numpages}{129--133}~pages.
\newblock


\bibitem[Weeks et~al\mbox{.}(2013)]%
        {weeks2013gaze}
\bibfield{author}{\bibinfo{person}{Justin~W Weeks}, \bibinfo{person}{Ashley~N Howell}, {and} \bibinfo{person}{Philippe~R Goldin}.} \bibinfo{year}{2013}\natexlab{}.
\newblock \showarticletitle{Gaze avoidance in social anxiety disorder}.
\newblock \bibinfo{journal}{\emph{Depression and anxiety}} \bibinfo{volume}{30}, \bibinfo{number}{8} (\bibinfo{year}{2013}), \bibinfo{pages}{749--756}.
\newblock


\bibitem[Weeks et~al\mbox{.}(2019)]%
        {weeks2019fear}
\bibfield{author}{\bibinfo{person}{Justin~W Weeks}, \bibinfo{person}{Ashley~N Howell}, \bibinfo{person}{Akanksha Srivastav}, {and} \bibinfo{person}{Philippe~R Goldin}.} \bibinfo{year}{2019}\natexlab{}.
\newblock \showarticletitle{“Fear guides the eyes of the beholder”: Assessing gaze avoidance in social anxiety disorder via covert eye tracking of dynamic social stimuli}.
\newblock \bibinfo{journal}{\emph{Journal of anxiety disorders}}  \bibinfo{volume}{65} (\bibinfo{year}{2019}), \bibinfo{pages}{56--63}.
\newblock


\bibitem[Wetzel et~al\mbox{.}(2018)]%
        {wetzel2018eye}
\bibfield{author}{\bibinfo{person}{Paul~A Wetzel}, \bibinfo{person}{Anne~S Lindblad}, \bibinfo{person}{Hardik Raizada}, \bibinfo{person}{Nathan James}, \bibinfo{person}{Caroline Mulatya}, \bibinfo{person}{Mary~A Kannan}, \bibinfo{person}{Zoe Villamar}, \bibinfo{person}{George~T Gitchel}, {and} \bibinfo{person}{Lindell~K Weaver}.} \bibinfo{year}{2018}\natexlab{}.
\newblock \showarticletitle{Eye tracking results in postconcussive syndrome versus normative participants}.
\newblock \bibinfo{journal}{\emph{Investigative Ophthalmology \& Visual Science}} \bibinfo{volume}{59}, \bibinfo{number}{10} (\bibinfo{year}{2018}), \bibinfo{pages}{4011--4019}.
\newblock


\bibitem[White et~al\mbox{.}(1983)]%
        {white1983ocular}
\bibfield{author}{\bibinfo{person}{Owen~B White}, \bibinfo{person}{Jean~A Saint-Cyr}, \bibinfo{person}{R~David Tomlinson}, {and} \bibinfo{person}{James~A Sharpe}.} \bibinfo{year}{1983}\natexlab{}.
\newblock \showarticletitle{Ocular motor deficits in Parkinson's disease: II. Control of the saccadic and smooth pursuit systems}.
\newblock \bibinfo{journal}{\emph{Brain}} \bibinfo{volume}{106}, \bibinfo{number}{3} (\bibinfo{year}{1983}), \bibinfo{pages}{571--587}.
\newblock


\bibitem[White and Depue(1999)]%
        {white1999differential}
\bibfield{author}{\bibinfo{person}{Tara~L White} {and} \bibinfo{person}{Richard~A Depue}.} \bibinfo{year}{1999}\natexlab{}.
\newblock \showarticletitle{Differential association of traits of fear and anxiety with norepinephrine-and dark-induced pupil reactivity.}
\newblock \bibinfo{journal}{\emph{Journal of personality and social psychology}} \bibinfo{volume}{77}, \bibinfo{number}{4} (\bibinfo{year}{1999}), \bibinfo{pages}{863}.
\newblock


\bibitem[Whitford et~al\mbox{.}(2023)]%
        {whitford2023eye}
\bibfield{author}{\bibinfo{person}{Veronica Whitford}, \bibinfo{person}{Narissa Byers}, \bibinfo{person}{Gillian~A O'Driscoll}, {and} \bibinfo{person}{Debra Titone}.} \bibinfo{year}{2023}\natexlab{}.
\newblock \showarticletitle{Eye movements and the perceptual span in disordered reading: A comparison of schizophrenia and dyslexia}.
\newblock \bibinfo{journal}{\emph{Schizophrenia Research: Cognition}}  \bibinfo{volume}{34} (\bibinfo{year}{2023}), \bibinfo{pages}{100289}.
\newblock


\bibitem[Wieckowski and White(2017)]%
        {wieckowski2017eye}
\bibfield{author}{\bibinfo{person}{Andrea~Trubanova Wieckowski} {and} \bibinfo{person}{Susan~W White}.} \bibinfo{year}{2017}\natexlab{}.
\newblock \showarticletitle{Eye-gaze analysis of facial emotion recognition and expression in adolescents with ASD}.
\newblock \bibinfo{journal}{\emph{Journal of Clinical Child \& Adolescent Psychology}} \bibinfo{volume}{46}, \bibinfo{number}{1} (\bibinfo{year}{2017}), \bibinfo{pages}{110--124}.
\newblock


\bibitem[Wilkins et~al\mbox{.}(1984)]%
        {wilkins1984neurological}
\bibfield{author}{\bibinfo{person}{Arnold Wilkins}, \bibinfo{person}{IAN Nimmo-Smith}, \bibinfo{person}{Anne Tait}, \bibinfo{person}{Christopher Mcmanus}, \bibinfo{person}{Sergio~Della Sala}, \bibinfo{person}{Andrew Tilley}, \bibinfo{person}{KIM Arnold}, \bibinfo{person}{Margaret Barrie}, {and} \bibinfo{person}{Sydney Scott}.} \bibinfo{year}{1984}\natexlab{}.
\newblock \showarticletitle{A neurological basis for visual discomfort}.
\newblock \bibinfo{journal}{\emph{Brain}} \bibinfo{volume}{107}, \bibinfo{number}{4} (\bibinfo{year}{1984}), \bibinfo{pages}{989--1017}.
\newblock


\bibitem[Willard and Lueck(2014)]%
        {willard2014ocular}
\bibfield{author}{\bibinfo{person}{Anna Willard} {and} \bibinfo{person}{Christian~J Lueck}.} \bibinfo{year}{2014}\natexlab{}.
\newblock \showarticletitle{Ocular motor disorders}.
\newblock \bibinfo{journal}{\emph{Current opinion in neurology}} \bibinfo{volume}{27}, \bibinfo{number}{1} (\bibinfo{year}{2014}), \bibinfo{pages}{75--82}.
\newblock


\bibitem[Williams et~al\mbox{.}(2016)]%
        {williams2016divided}
\bibfield{author}{\bibinfo{person}{Isla~M Williams}, \bibinfo{person}{Peter Schofield}, \bibinfo{person}{Neha Khade}, {and} \bibinfo{person}{Larry~A Abel}.} \bibinfo{year}{2016}\natexlab{}.
\newblock \showarticletitle{Divided visual attention: a comparison of patients with multiple sclerosis and controls, assessed with an optokinetic nystagmus suppression task}.
\newblock \bibinfo{journal}{\emph{Journal of Clinical Neuroscience}}  \bibinfo{volume}{34} (\bibinfo{year}{2016}), \bibinfo{pages}{187--192}.
\newblock


\bibitem[Wojcik-Pkedziwiatr et~al\mbox{.}(2016)]%
        {wojcik2016eye}
\bibfield{author}{\bibinfo{person}{Magdalena Wojcik-Pkedziwiatr}, \bibinfo{person}{Klaudia Plinta}, \bibinfo{person}{Agnieszka Krzak-Kubica}, \bibinfo{person}{Katarzyna Zajdel}, \bibinfo{person}{Marcel Falkiewicz}, \bibinfo{person}{Jacek Dylak}, \bibinfo{person}{Jan Ober}, \bibinfo{person}{Andrzej Szczudlik}, {and} \bibinfo{person}{Monika Rudzinska}.} \bibinfo{year}{2016}\natexlab{}.
\newblock \showarticletitle{Eye movement abnormalities in essential tremor}.
\newblock \bibinfo{journal}{\emph{Journal of Human Kinetics}} \bibinfo{volume}{52}, \bibinfo{number}{1} (\bibinfo{year}{2016}), \bibinfo{pages}{53--64}.
\newblock


\bibitem[Wolf and Ueda(2021)]%
        {wolf2021contribution}
\bibfield{author}{\bibinfo{person}{Alexandra Wolf} {and} \bibinfo{person}{Kazuo Ueda}.} \bibinfo{year}{2021}\natexlab{}.
\newblock \showarticletitle{Contribution of eye-tracking to study cognitive impairments among clinical populations}.
\newblock \bibinfo{journal}{\emph{Frontiers in psychology}}  \bibinfo{volume}{12} (\bibinfo{year}{2021}), \bibinfo{pages}{590986}.
\newblock


\bibitem[Wu et~al\mbox{.}(2018)]%
        {wu2018eye}
\bibfield{author}{\bibinfo{person}{Chia-Chien Wu}, \bibinfo{person}{Bo Cao}, \bibinfo{person}{Veena Dali}, \bibinfo{person}{Celia Gagliardi}, \bibinfo{person}{Olivier~J Barthelemy}, \bibinfo{person}{Robert~D Salazar}, \bibinfo{person}{Marc Pomplun}, \bibinfo{person}{Alice Cronin-Golomb}, {and} \bibinfo{person}{Arash Yazdanbakhsh}.} \bibinfo{year}{2018}\natexlab{}.
\newblock \showarticletitle{Eye movement control during visual pursuit in Parkinson’s disease}.
\newblock \bibinfo{journal}{\emph{PeerJ}}  \bibinfo{volume}{6} (\bibinfo{year}{2018}), \bibinfo{pages}{e5442}.
\newblock


\bibitem[Wyder et~al\mbox{.}(2015)]%
        {wyder2015gaze}
\bibfield{author}{\bibinfo{person}{Stephan Wyder}, \bibinfo{person}{Fabian Hennings}, \bibinfo{person}{Simon Pezold}, \bibinfo{person}{Jan Hrbacek}, {and} \bibinfo{person}{Philippe~C Cattin}.} \bibinfo{year}{2015}\natexlab{}.
\newblock \showarticletitle{With gaze tracking toward noninvasive eye cancer treatment}.
\newblock \bibinfo{journal}{\emph{IEEE Transactions on Biomedical Engineering}} \bibinfo{volume}{63}, \bibinfo{number}{9} (\bibinfo{year}{2015}), \bibinfo{pages}{1914--1924}.
\newblock


\bibitem[Wyllie et~al\mbox{.}(1986)]%
        {wyllie1986lateralizing}
\bibfield{author}{\bibinfo{person}{Elaine Wyllie}, \bibinfo{person}{Hans Luders}, \bibinfo{person}{Harold~H Morris}, \bibinfo{person}{Ronald~P Lesser}, {and} \bibinfo{person}{Dudley~S Dinner}.} \bibinfo{year}{1986}\natexlab{}.
\newblock \showarticletitle{The lateralizing significance of versive head and eye movements during epileptic seizures}.
\newblock \bibinfo{journal}{\emph{Neurology}} \bibinfo{volume}{36}, \bibinfo{number}{5} (\bibinfo{year}{1986}), \bibinfo{pages}{606--606}.
\newblock


\bibitem[Yamada et~al\mbox{.}(2024)]%
        {yamada2024distinct}
\bibfield{author}{\bibinfo{person}{Yasunori Yamada}, \bibinfo{person}{Kaoru Shinkawa}, \bibinfo{person}{Masatomo Kobayashi}, \bibinfo{person}{Miyuki Nemoto}, \bibinfo{person}{Miho Ota}, \bibinfo{person}{Kiyotaka Nemoto}, {and} \bibinfo{person}{Tetsuaki Arai}.} \bibinfo{year}{2024}\natexlab{}.
\newblock \showarticletitle{Distinct eye movement patterns to complex scenes in Alzheimer’s disease and Lewy body disease}.
\newblock \bibinfo{journal}{\emph{Frontiers in Neuroscience}}  \bibinfo{volume}{18} (\bibinfo{year}{2024}), \bibinfo{pages}{1333894}.
\newblock


\bibitem[Yerys et~al\mbox{.}(2013)]%
        {yerys2013separate}
\bibfield{author}{\bibinfo{person}{Benjamin~E Yerys}, \bibinfo{person}{Lauren Kenworthy}, \bibinfo{person}{Kathryn~F Jankowski}, \bibinfo{person}{John Strang}, {and} \bibinfo{person}{Gregory~L Wallace}.} \bibinfo{year}{2013}\natexlab{}.
\newblock \showarticletitle{Separate components of emotional go/no-go performance relate to autism versus attention symptoms in children with autism.}
\newblock \bibinfo{journal}{\emph{Neuropsychology}} \bibinfo{volume}{27}, \bibinfo{number}{5} (\bibinfo{year}{2013}), \bibinfo{pages}{537}.
\newblock


\bibitem[Yoo et~al\mbox{.}(2021)]%
        {yoo2021gaze}
\bibfield{author}{\bibinfo{person}{Sangbong Yoo}, \bibinfo{person}{Seongmin Jeong}, {and} \bibinfo{person}{Yun Jang}.} \bibinfo{year}{2021}\natexlab{}.
\newblock \showarticletitle{Gaze behavior effect on gaze data visualization at different abstraction levels}.
\newblock \bibinfo{journal}{\emph{Sensors}} \bibinfo{volume}{21}, \bibinfo{number}{14} (\bibinfo{year}{2021}), \bibinfo{pages}{4686}.
\newblock


\bibitem[Yousefi et~al\mbox{.}(2022)]%
        {yousefi2022stress}
\bibfield{author}{\bibinfo{person}{Mansoureh~Seyed Yousefi}, \bibinfo{person}{Farnoush Reisi}, \bibinfo{person}{Mohammad~Reza Daliri}, {and} \bibinfo{person}{Vahid Shalchyan}.} \bibinfo{year}{2022}\natexlab{}.
\newblock \showarticletitle{Stress Detection Using Eye Tracking Data: An Evaluation of Full Parameters}.
\newblock \bibinfo{journal}{\emph{IEEE Access}}  \bibinfo{volume}{10} (\bibinfo{year}{2022}), \bibinfo{pages}{118941--118952}.
\newblock


\bibitem[Zaccara et~al\mbox{.}(1992)]%
        {zaccara1992smooth}
\bibfield{author}{\bibinfo{person}{G Zaccara}, \bibinfo{person}{PF Gangemi}, \bibinfo{person}{GC Muscas}, \bibinfo{person}{M Paganini}, \bibinfo{person}{Stefano Pallanti}, \bibinfo{person}{A Parigi}, \bibinfo{person}{A Messori}, {and} \bibinfo{person}{Graziano Arnetoli}.} \bibinfo{year}{1992}\natexlab{}.
\newblock \showarticletitle{Smooth-pursuit eye movements: alterations in Alzheimer's disease}.
\newblock \bibinfo{journal}{\emph{Journal of the neurological sciences}} \bibinfo{volume}{112}, \bibinfo{number}{1-2} (\bibinfo{year}{1992}), \bibinfo{pages}{81--89}.
\newblock


\bibitem[Zaino et~al\mbox{.}(2023)]%
        {zaino2023different}
\bibfield{author}{\bibinfo{person}{Domenica Zaino}, \bibinfo{person}{Valeria Serchi}, \bibinfo{person}{Fabio Giannini}, \bibinfo{person}{Barbara Pucci}, \bibinfo{person}{Giacomo Veneri}, \bibinfo{person}{Elena Pretegiani}, \bibinfo{person}{Francesca Rosini}, \bibinfo{person}{Lucia Monti}, {and} \bibinfo{person}{Alessandra Rufa}.} \bibinfo{year}{2023}\natexlab{}.
\newblock \showarticletitle{Different saccadic profile in bulbar versus spinal-onset amyotrophic lateral sclerosis}.
\newblock \bibinfo{journal}{\emph{Brain}} \bibinfo{volume}{146}, \bibinfo{number}{1} (\bibinfo{year}{2023}), \bibinfo{pages}{266--277}.
\newblock


\bibitem[Zalla et~al\mbox{.}(2018)]%
        {zalla2018saccadic}
\bibfield{author}{\bibinfo{person}{Tiziana Zalla}, \bibinfo{person}{Magali Seassau}, \bibinfo{person}{Fabienne Cazalis}, \bibinfo{person}{Doriane Gras}, {and} \bibinfo{person}{Marion Leboyer}.} \bibinfo{year}{2018}\natexlab{}.
\newblock \showarticletitle{Saccadic eye movements in adults with high-functioning autism spectrum disorder}.
\newblock \bibinfo{journal}{\emph{Autism}} \bibinfo{volume}{22}, \bibinfo{number}{2} (\bibinfo{year}{2018}), \bibinfo{pages}{195--204}.
\newblock


\bibitem[Zhou et~al\mbox{.}(2022)]%
        {zhou2022oculomotor}
\bibfield{author}{\bibinfo{person}{Meng-Xi Zhou}, \bibinfo{person}{Qin Wang}, \bibinfo{person}{Yin Lin}, \bibinfo{person}{Qian Xu}, \bibinfo{person}{Li Wu}, \bibinfo{person}{Ya-Jing Chen}, \bibinfo{person}{Yu-Han Jiang}, \bibinfo{person}{Qing He}, \bibinfo{person}{Lei Zhao}, \bibinfo{person}{You-Rong Dong}, {et~al\mbox{.}}} \bibinfo{year}{2022}\natexlab{}.
\newblock \showarticletitle{Oculomotor impairments in de novo Parkinson’s disease}.
\newblock \bibinfo{journal}{\emph{Frontiers in Aging Neuroscience}}  \bibinfo{volume}{14} (\bibinfo{year}{2022}), \bibinfo{pages}{985679}.
\newblock


\bibitem[Zhou et~al\mbox{.}(2023)]%
        {zhou2023pathogenesis}
\bibfield{author}{\bibinfo{person}{Wei Zhou}, \bibinfo{person}{Yi Fan}, \bibinfo{person}{Yulin Chang}, \bibinfo{person}{Wenjuan Liu}, \bibinfo{person}{Jiuju Wang}, {and} \bibinfo{person}{Yufeng Wang}.} \bibinfo{year}{2023}\natexlab{}.
\newblock \showarticletitle{Pathogenesis of Comorbid ADHD and Chinese Developmental Dyslexia: Evidence From Eye-Movement Tracking and Rapid Automatized Naming}.
\newblock \bibinfo{journal}{\emph{Journal of Attention Disorders}} \bibinfo{volume}{27}, \bibinfo{number}{3} (\bibinfo{year}{2023}), \bibinfo{pages}{294--306}.
\newblock


\bibitem[Zurek et~al\mbox{.}(2024)]%
        {zurek2024can}
\bibfield{author}{\bibinfo{person}{Grzegorz Zurek}, \bibinfo{person}{Marek Binder}, \bibinfo{person}{Bartosz Kunka}, \bibinfo{person}{Robert Kosikowski}, \bibinfo{person}{Malgorzata Rodzen}, \bibinfo{person}{Danuta Karas}, \bibinfo{person}{Gabriela Mucha}, \bibinfo{person}{Roman Olejniczak}, \bibinfo{person}{Agata Gorkaczko}, \bibinfo{person}{Katarzyna Kujawa}, {et~al\mbox{.}}} \bibinfo{year}{2024}\natexlab{}.
\newblock \showarticletitle{Can Eye Tracking Help Assess the State of Consciousness in Non-Verbal Brain Injury Patients?}
\newblock \bibinfo{journal}{\emph{Journal of Clinical Medicine}} \bibinfo{volume}{13}, \bibinfo{number}{20} (\bibinfo{year}{2024}), \bibinfo{pages}{6227}.
\newblock


\end{thebibliography}

\end{document}